\documentclass[a4paper,fleqn]{cas-sc}

\usepackage[authoryear,longnamesfirst]{natbib}
\usepackage{float}
\usepackage{enumitem}
\usepackage{tabularx}
\usepackage{graphicx}
\usepackage{subcaption}
\usepackage{multirow}

\def\tsc#1{\csdef{#1}{\textsc{\lowercase{#1}}\xspace}}
\tsc{WGM}
\tsc{QE}
\tsc{EP}
\tsc{PMS}
\tsc{BEC}
\tsc{DE}

\begin{document}
\let\WriteBookmarks\relax
\def\textpagefraction{.001}

\shorttitle{Multi-sectoral Impacts of H$_2$ and Synthetic Fuels Adoption for Heavy-duty Transportation Decarbonization}

\shortauthors{Shaker et~al.}

\title [mode = title]{Multi-sectoral Impacts of H$_2$ and Synthetic Fuels Adoption for Heavy-duty Transportation Decarbonization}                      


%
\author[1]{Youssef Shaker}[]
\ead{shaker@mit.edu}

\credit{Research framing, modeling, data collection, data visualization, writing}

\affiliation[1]{organization={MIT Energy Initiative, Massachusetts Institute of Technology},
    city={Cambridge},
    postcode={02139}, 
    state={MA},
    country={}}

\author[1]{Jun Wen Law}[]
\ead{junlaw@mit.edu}
\credit{Modeling, data collection, manuscript writing, manuscript feedback}

\author[2]{Audun Botterud}[]
\ead{audunb@mit.edu}
\credit{Conceptualization of this study, revision of results, manuscript feedback}

\affiliation[2]{organization={Laboratory for Information \& Decision Systems, Massachusetts Institute of Technology},
    city={Cambridge},
    postcode={02139}, 
    state={MA},
    country={}}

\author[3]{Dharik Mallapragada}[]
\ead{dharik.mallapragada@nyu.edu}
\cormark[1]

\credit{Conceptualization of this study, Data curation, manuscript writing, manuscript feedback}
\cortext[cor1]{Corresponding author}

\affiliation[3]{organization={Chemical and Biomolecular Engineering Department, Tandon School of Engineering, New
York University},
    city={Brooklyn},
    postcode={11201}, 
    state={NY},
    country={ }}

\begin{abstract}
Policies focused on deep decarbonization of regional economies tend to emphasize electricity sector decarbonization alongside electrification of end-uses. There is also growing interest in utilizing hydrogen (H$_2$) produced via electricity to displace fossil fuels in difficult-to-electrify sectors. One such use case is heavy-duty vehicles (HDV), which represents a substantial and growing share of global transport emissions due to electrification of light duty vehicles. Here, we assess the bulk energy system impact of decarbonizing the HDV segment via the use of either H$_2$, or drop-in synthetic liquid fuels produced from H$_2$ and CO$_2$. Our analysis soft-links two modeling approaches: a) a bottom-up model of transportation energy demand that produces variety of final energy demand scenarios for the same service demand and b) a multi-sectoral capacity expansion model that co-optimizes power, H$_2$ and CO$_2$ supply chains subjected to technological and policy constraints to meet exogenous final energy demands. Through a case study of Western European countries under deep decarbonization constraints in 2040, we quantify the energy system implications of different levels of H$_2$ and synthetic fuels adoption in the HDV sector under scenarios with and without CO$_2$ sequestration. In the absence of CO$_2$ storage, substitution of liquid fossil fuels in HDVs is essential to meet the deep decarbonization constraint across the modeled power, H$_2$ and transport sectors. Additionally, utilizing H$_2$ HDVs reduces decarbonization costs and fossil liquids demand, but could increase natural gas consumption. While H$_2$ HDV adoption reduces the need for direct air capture (DAC), synthetic fuel adoption increases DAC investments and total system costs. The study highlights the trade-offs associated with different transportation decarbonization pathways, and underscores the importance of multi-sectoral consideration in decarbonization studies.
\end{abstract}


\maketitle



\section{Introduction}
Modeled pathways for energy system decarbonization are generally based on a two-pronged strategy of: a) decarbonizing the power sector by expanding variable renewable energy (VRE) supply, and b) increasing use of electricity to displace fossil fuel energy use. To date, only the first part of this two-pronged strategy has yielded meaningful progress in major-emitting regions like the European Union and U.S. – for example, power sector greenhouse gas (GHG) emissions in the European Union (EU) for 2021 were 40\% lower than 2005, as VRE generation increased from 16\% to 38\% over the same period \citep{european_environment_agency_eea_2023, energy_-_eurostat_shares_2023}. In contrast, GHG emissions from the EU transportation sector have remained largely unchanged over the same period; electricity consumption as a share of total transportation energy consumption was less than 1\% in 2021 \citep{european_environment_agency_eea_2023, energy_-_eurostat_shares_2023}.

The electrification of light duty vehicles (LDVs) is seemingly well underway, as indicated by the share of plug-in hybrid (PHEVs) and battery electric vehicles (EVs) as a percentage of new car sales (PHEV and EV car sales in the EU increased from 3\% to 23 \% of all new car sales between 2019 and 2023) \citep{acea_fuel_2023}. However, the electrification of heavy-duty vehicles (HDV), which accounted for around 27\% of the transportation sector’s CO$_2$ emissions in EU in 2023, is uncertain due to several factors including concerns with payload reduction impacts, refueling time associated with state-of-art batteries, charging technologies, and grid impacts \citep{davis_net-zero_2018, bethoux_hydrogen_2020, camacho_hydrogen_2022, european_environment_agency_eea_2023, mowry_grid_2021}. 

Besides electrification, other decarbonization strategies considered for HDVs include: a) direct use of hydrogen (H$_2$) produced from low-carbon sources, b) use of so-called synthetic liquid fuels (SFs) or e-fuels produced using electricity, H$_2$ and captured CO$_2$, and c) continued use of petroleum-based liquid fuels that are offset by atmospheric CO$_2$ capture using negative emissions technologies such as direct air capture (DAC). Each of these pathways could have a far-reaching impact on the electricity grid, as well as the H$_2$ and CO$_2$ supply chains. For instance, the production of SFs require substantial quantities of low-carbon H$_2$, which in turn relies on the coordinated development of H$_2$ infrastructure and the electricity grid \citep{eddy_germany_2023, drunert_power--liquid_2020, alsunousi_role_2024, williams_carbonneutral_2021}. This is particularly important as electrolyzers are projected to supply a major share of H$_2$ under deep decarbonization scenarios, driving the need for additional investment in VRE resources \citep{law_role_2025, bodal_decarbonization_2020, jensen_hydrogen_2007}. Similarly, the CO$_2$ feedstock for SF production will require deployment of CO$_2$ capture and transport infrastructure to facilitate CO$_2$ utilization  and could also facilitate CO$_2$ sequestration where available \citep{nguyen_syngas_2015, ishaq_co2based_2023, karjunen_method_2017}. Besides technological coupling,  decarbonization efforts across sectors are also coupled through policy instruments like emissions trading schemes (e.g. EU ETS) that allow for emissions reduction strategies across sectors to directly compete with each other. Here, we systematically explore the multi-sector impacts of the above-mentioned strategies for HDV decarbonization, which as noted in the literature review, remains one of the lesser studied topics in the area of transportation decarbonization. 


Previous literature on transportation decarbonization can be categorized into a few broad themes: 1) the techno-economics and efficiency of EVs, H$_2$ vehicles, and SFs \citep{brynolf_electrofuels_2018, hanggi_review_2019, mohideen_techno-economic_2023, cremades_techno-environmental_2024, gough_vehicle--grid_2017} 2) fleet evolution and the impact of policy interventions \citep{harrison_policy_2017, martins_assessing_2023, pasaoglu_system_2016, kester_policy_2018} 3) scenario-based characterization of the energy demand and emissions associated with transportation decarbonization under various technology scenarios \citep{krause_eu_2020, siskos_implications_2018, karkatsoulis_simulating_2017} 4) energy system impacts of transportation decarbonization \citep{li_modeling_2021, li_modelling_2021, michalski_role_2019, nakano_impacts_2022, mccollum_transport_2014, van_vliet_combining_2011}. This study combines the latter two categories, by leveraging transportation energy demand scenarios as inputs into a multi-vector energy system model.

Some studies have investigated the use of SFs and H$_2$ in transportation \citep{brynolf_electrofuels_2018, ueckerdt_potential_2021, millinger_electrofuels_2021,zang_life_2021}, focusing on quantifying  the process-level efficiency, levelized costs and life cycle emissions impacts, where the grid interactions are generally treated in a static manner (i.e. exogenously determined emissions intensity and cost of grid electricity per scenario). These studies generally do not consider what the adoption of SFs or H$_2$ would imply for the power, H$_2$, or CO$_2$ infrastructure in a given region.  

Studies investigating the impacts of transportation decarbonization on wider energy systems are often focused on LDV \citep{heuberger_ev-olution_2020, li_modelling_2021, harrison_policy_2017, powell_charging_2022}. Additionally, because electrification of LDVs appears to be imminent, many of these studies focus on the impacts on the power sector, and studies that consider the impacts on adjacent H$_2$ and CO$_2$ infrastructure are limited \citep{colbertaldo_modelling_2018, li_modelling_2021, powell_charging_2022}. Given the interactions between the infrastructure for these vectors noted earlier, particularly in the case of SFs and H$_2$, it is important to consider relevant supply chains in an integrated manner. Multiple studies have emphasized that transport decarbonization strategies have system-wide impacts on electricity, fuel, and CO$_2$ infrastructure, highlighting the importance for integrated, cross-sectoral assessments \citep{millinger_are_2022, speizer_integrated_2024, wan_assessment_2025}. 

Studies that focus on the wider energy system impacts of transportation decarbonization incorporate transportation demand using three methods. The first approach assumes a set amount of transportation energy demand, and subsequently investigates the necessary supply infrastructure to meet this demand~\citep{heuberger_ev-olution_2020}. The second approach relies on specifying transportation service demand (e.g. vehicle km or tonne-km), and then endogeneously optimizing for both energy and drivetrain choice to meet this demand \citep{li_modeling_2021}. The third uses a multi-model approach, which determines transportation demand exogenously using a transportation demand model, and then uses said model's results as an input in a macro-energy systems model \citep{powell_charging_2022}. The latter approach, which is also considered in this study, enables the inclusion of non-economic factors influencing drivetrain adoption to meet transportation demand. It also allows for evaluating how such choices affect energy infrastructure investments and operations. Compared to modeling final energy demands alone (the first approach), the multi-model framework is able to capture differences in end-use efficiency across alternative fuels used to satisfy transportation energy demands.

In this study, we use a multi-model approach, consisting of a bottom-up transportation energy demand model and a multi-vector energy system model, to study the role for H$_2$ and SFs for HDV decarbonization and its bulk energy system impact (Figure \ref{Fig_modelling}). For the demand-side analysis, we develop a model to evaluate alternative energy demand scenarios for HDVs that accounts for vehicle-specific factors like vehicle efficiency and market share of each vehicle sub-type. We then use these resulting scenarios as inputs to our supply-side multi-vector energy infrastructure planning model, DOLPHYN, to investigate the bulk energy infrastructure impact of wide-scale heavy-duty transportation decarbonization for the identified transportation energy demand scenarios \citep{he_dolphyn_2023}. As part of this study, we update DOLPHYN to include a representation of the CO$_2$ infrastructure, including storage, transportation and utilization (as SFs), as well as a representation of competition between conventional fossil fuels and SFs. In this way, we are able to capture the interactions between H$_2$, CO$_2$, and liquid fuels supply chains and their impact on the power sector which includes: a) changes in electricity consumption, b) inducing competition for constrained resources like VRE capacity for renewable electricity generation and CO$_2$ storage sites and c) affecting the available emissions budget for power and H$_2$ production as part of multi-sectoral decarbonization efforts. The goal of our analysis is to highlight the key technology and policy drivers for H$_2$ and SFs adoption as part of cost-optimized deeply decarbonized power, H$_2$ and transportation sectors.


Our analysis focuses on a case study of Western Europe in 2040 under deep decarbonization scenarios, where policy deliberations have recently focused on a multi-sectoral decarbonization effort (e.g. via the EU ETS, The European Green Deal, the Fit for 55 Package, and the EU 2024 Climate Target), reducing fossil fuels reliance and advancing deployment of H$_2$ and SFs for transportation decarbonization \citep{fetting_constanze_european_2020, seibert_eu_2024, haas_decarbonizing_2020, bayer_european_2020, vivanco-martin_analysis_2023, ovaere_cost-effective_2022}. On the other hand, there is also considerable uncertainty on the potential role for CO$_2$ storage in deep decarbonization scenarios for this region, with considerable public opposition to projects, as well as political and legislative barriers \citep{holz_2050_2021, koukouzas_current_2022}. 

In this context, our analysis leads to a few key policy-relevant observations. First, we find that H$_2$ use for HDVs reduces bulk system (power-H$_2$ and transportation) cost of deep decarbonization and decreases demand for fossil liquids, but could increase overall natural gas (NG) consumption compared to equivalent decarbonization scenarios without H$_2$ use for HDVs. Part of the cost saving stems from the substitution of more expensive conventional fossil liquid fuels versus NG (on a per GJ of energy basis) for H$_2$ in end-use that also reduces need for atmospheric CO$_2$ removal via modeled DAC technologies. Second, limitations on CO$_2$ storage availability increase the bulk system cost savings (in absolute terms) of adopting H$_2$ use for HDVs. Third, the deployment of SFs results in substantial expansion of power and H$_2$ production capacity, with a preference for non-fossil fuel generation sources (electrolyzers for H$_2$ production and VRE for power generation) to maximize carbon abatement benefits of SF use. Fourth, while SF adoption generally increases bulk system costs, the cost increases vs. no SF adoption case are the lowest in case CO$_2$ storage availability is constrained and fossil fuels (NG and fossil liquids) are expensive.  The use of H$_2$ for transport decarbonization reduces the upstream burden on the power and H$_2$ sectors, but comes with the additional downstream challenges of deploying extensive distribution, refueling, and vehicular infrastructure. Fifth, our analysis highlights that the optimal-level of sectoral decarbonization is dependent on the technology pathways adopted and reinforces the use of multi-sector emissions reduction strategies similar to the European emissions trading system, rather than sector-specific decarbonization approaches.

The rest of the paper is organized as follows. Section \ref{sec:data_methods} outlines the methods and data used to answer key questions around heavy-duty transportation decarbonization, including an overview of the models, the case-study used, technology assumptions, and a summary of the scenarios analyzed. Further details are provided in the supporting information (SI). Section \ref{sec:system_impacts_h2} describes the results on the impact of using H$_2$ for transportation decarbonization, while Section \ref{sec:system_impacts_sf} explores the impact of using SFs in transportation decarbonization along with H$_2$. Section \ref{sec:sen_results} discusses the results for key sensitivity scenarios. Finally, Section \ref{sec:disc} delves into some key  takeaways associated with power and transportation decarbonization and highlights areas for further research.

\section{Data and Methods}\label{sec:data_methods}
\subsection{Modeling Approach}

\begin{figure}[pos = H]
\centering

\begin{subfigure}{0.8\linewidth}
    \includegraphics[width=\linewidth]{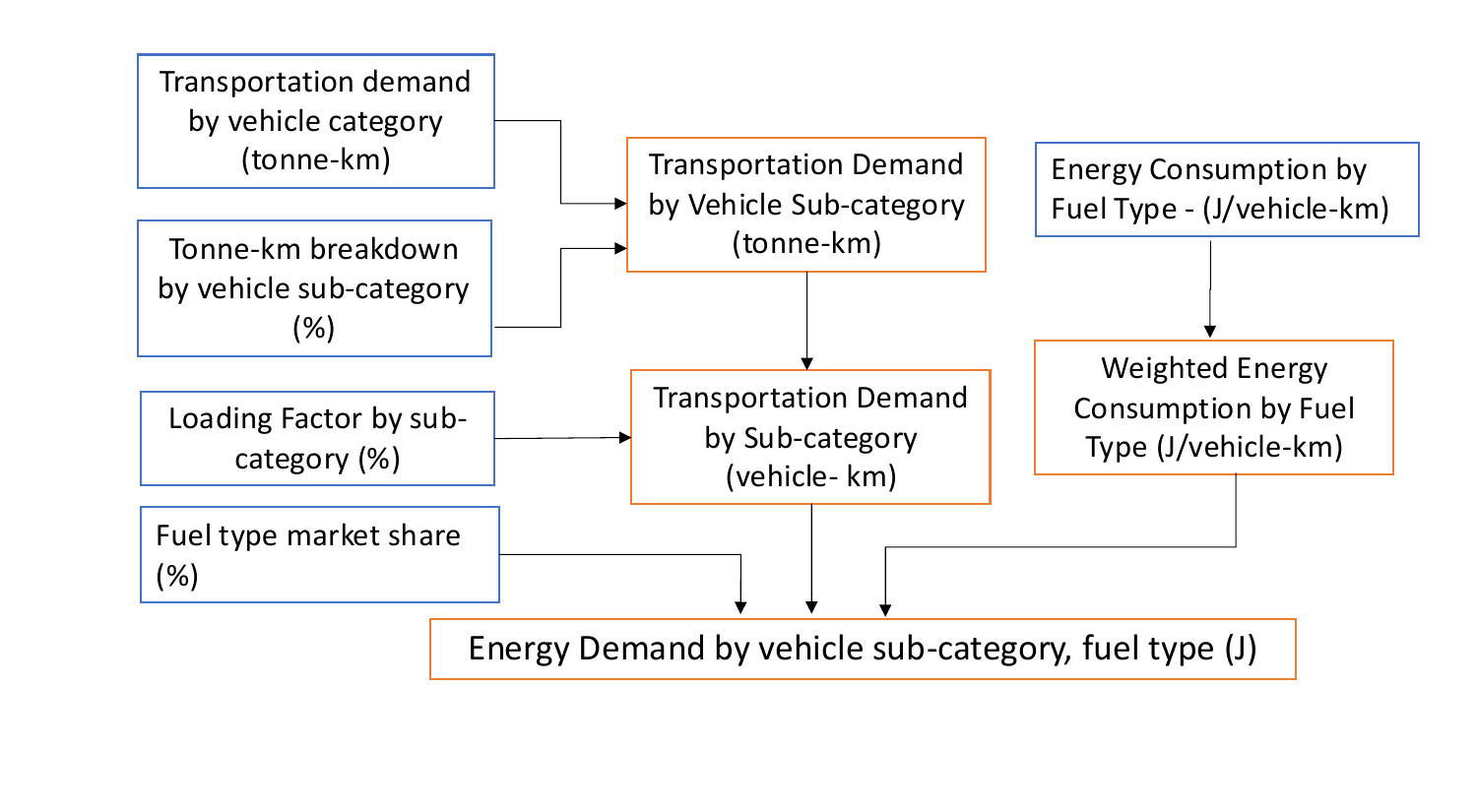}
    \caption{}
\end{subfigure}

\begin{subfigure}{0.8\linewidth}
    \includegraphics[width=\linewidth]{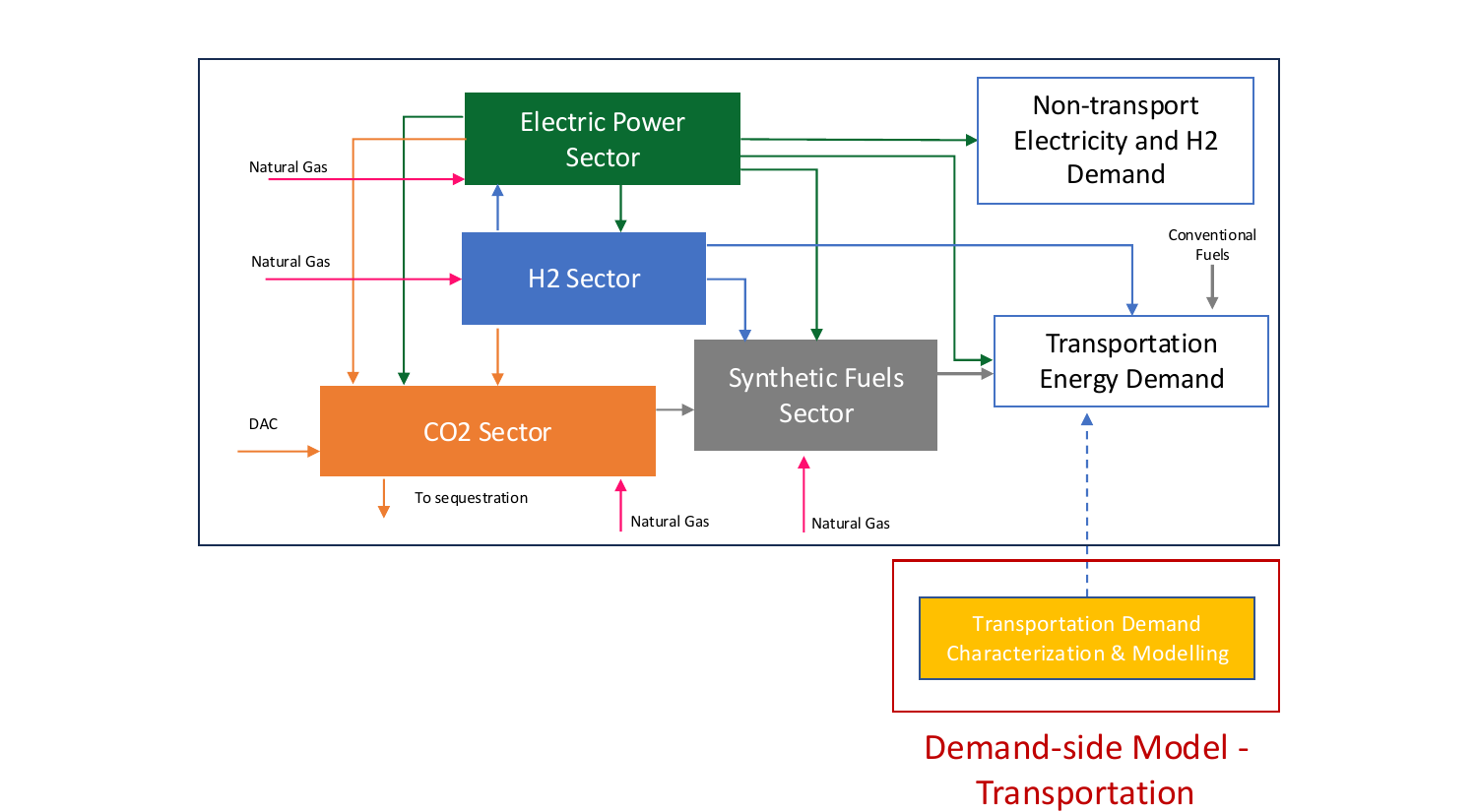}
    \caption{}
\end{subfigure}

\caption[Overview of Overall Modeling]{Overview of supply and demand-side modeling used for this study. a) Modeling approach to estimate transportation final energy demand by fuel type and vehicle sub-category, illustrated for the heavy-duty vehicle segment (HDV). Blue boxes are data inputs, while orange boxes are calculated values. Loading factor represents the fraction of vehicle loading capacity used on average. Input data is sourced from the 2020 Reference Scenario produced by the EU Commission \citep{european_commission_eu_2020}, Eurostat \citep{energy_-_eurostat_shares_2023}, a transportation survey of European countries \citep{emisia_traccs_2014}, and \citep{krause_eu_2020}. Modeling for LDV follows similar approach, but defines service demand in terms of passenger-km (pkm) instead of tonne-km (tkm) and occupancy rate instead of loading factor to convert pkm to vkm. b) Overview of DOLPHYN capacity expansion model used in this study and the link with the demand estimation model - available here: \citep{he_dolphyn_2023}. The color of arrows highlight various vectors – green: electricity, blue: H$_2$, orange: CO$_2$, grey: fuels; magenta: natural gas.}
\label{Fig_modelling}
\end{figure}

We developed a data-driven model to estimate transportation energy demand in 2040 across various end-use technology scenarios. The model, summarized in Figure \ref{Fig_modelling} (a) enables us to construct alternative fuel demand scenarios for the LDV and HDV transportation sector that are consistent in terms of delivering the same end-use service demand (e.g. vehicle-km (vkm) for LDVs or tonne-km (tkm) for HDVs). Through the demand-side model, we are able to modify the market share for different vehicle types, and by extension their energy consumption. Although the demand model allows for the creation of scenarios based on vehicle efficiency, modal shifts, and demand reduction, we focus on shifts in vehicle/fuel types for this study. To do so, we first disaggregate transportation energy demand into the following categories: 1) light-duty passenger vehicles, 2) buses and coaches, 3) 2-wheelers, and 4) heavy-duty and light commercial vehicles. This study focuses on decarbonization strategies for the HDV sub-category, defined as vehicles that have maximum gross weight greater than 7.5 tonnes,  which is part of the heavy-duty vehicle and light commercial vehicles category. This category accounted for approximately 25\% of transportation sector fuel consumption for the modeled region in 2015 \citep{european_commission_eu_2020}. Then, to construct the baseline energy demand for HDV transportation in 2040, we use the service demand (measured in tkm for HDVs) projections for 2040 as per the 2020 Reference Scenario produced by the EU Commission \citep{european_commission_eu_2020}. We decompose this projected service demand into service demand met by each vehicle sub-category using data from TRACCS, a transportation survey of European countries \citep{emisia_traccs_2014}. To calculate service demand in vkm, we combine country-specific loading factors (i.e. vehicle load as a \% of maximum-laden weight for each of the vehicle sub-categories as reported in Eurostat). Vehicle demand is then broken down by fuel type, the shares of which are determined using projected market shares. Baseline vehicles type market shares are based on projections by Krause et al. \citep{krause_eu_2020}. Vehicle-level energy consumption per functional unit (vkm or tkm) is based on the EU Reference Scenarios Technology Assumptions \citep{european_commission_eu_2020}. Modeling for LDVs follows a similar form, but utilizes passenger-km (pkm) instead of vkm to characterize service demand. Additionally, to convert service demand from pkm to vkm, we utilize occupancy rate data from \citep{emisia_traccs_2014}. Details on the demand model formulation and the demand model data can be found in sections B.1 and B.2 in the SI, respectively.

The supply-side modeling approach, summarized in Figure \ref{Fig_modelling} (b), is based on the DOLPHYN capacity expansion model (CEM), which evaluates the cost-optimal investments in electricity, H$_2$, carbon-capture utilization and sequestration (CCUS), and SF infrastructure, while adhering to a range of technology-specific and system-specific operational constraints, as well as imposed policy constraints \citep{he_dolphyn_2023,he_hydrogen_2021}. For this study, we expanded the DOLPHYN model in the following ways: 1) we added investment and operation of infrastructure for CO$_2$ sequestration, transmission and utilization (to produce SF), along with their energy and CO$_2$ interactions with the power and H$_2$ supply chains, and 2) allowed the possibility to meet exogenous liquid fuel demand through a combination of fossil-derived fuels and SF production. The resulting modeling framework allows us to, for example, identify the location and scale of fossil fuel power generation with CO$_2$ capture in the electricity sector, given inputs on the location and cost of CO$_2$ sequestration. Similarly, the liquid fuels supply chain allows for the production of SFs, which induces demand for H$_2$ and CO$_2$ that needs to be balanced in the model. Details on the formulation of the supply-side model can be found in Section C of the SI.

\subsection{Case study: Western European decarbonization scenarios for 2040}
\begin{figure}[pos = H]
\centering
\begin{subfigure}{0.55\linewidth}
    \includegraphics[width=\linewidth]{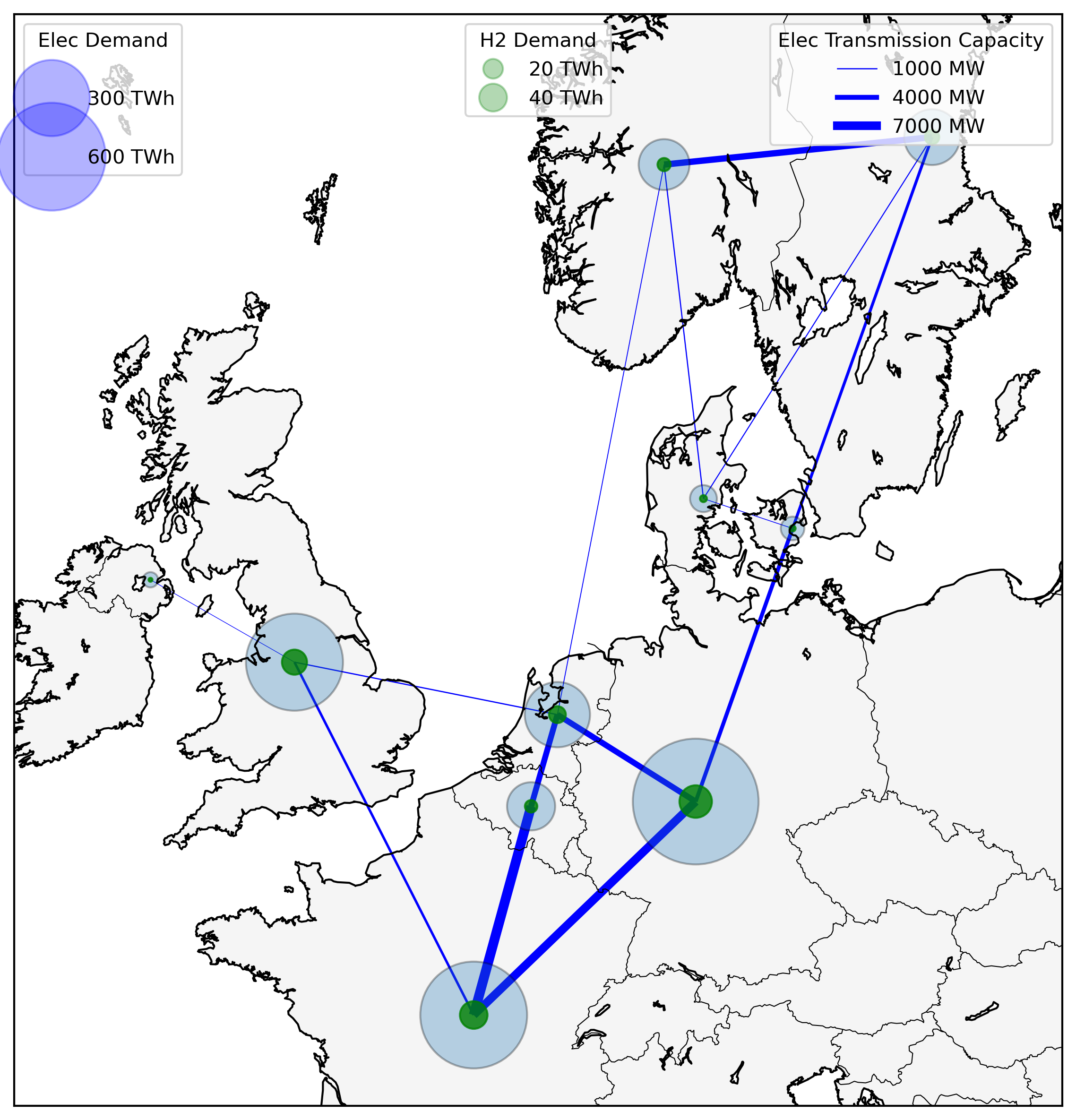}
    \caption{Energy network representation}
\end{subfigure}

\begin{subfigure}{0.55\linewidth}
    \includegraphics[width=\linewidth]{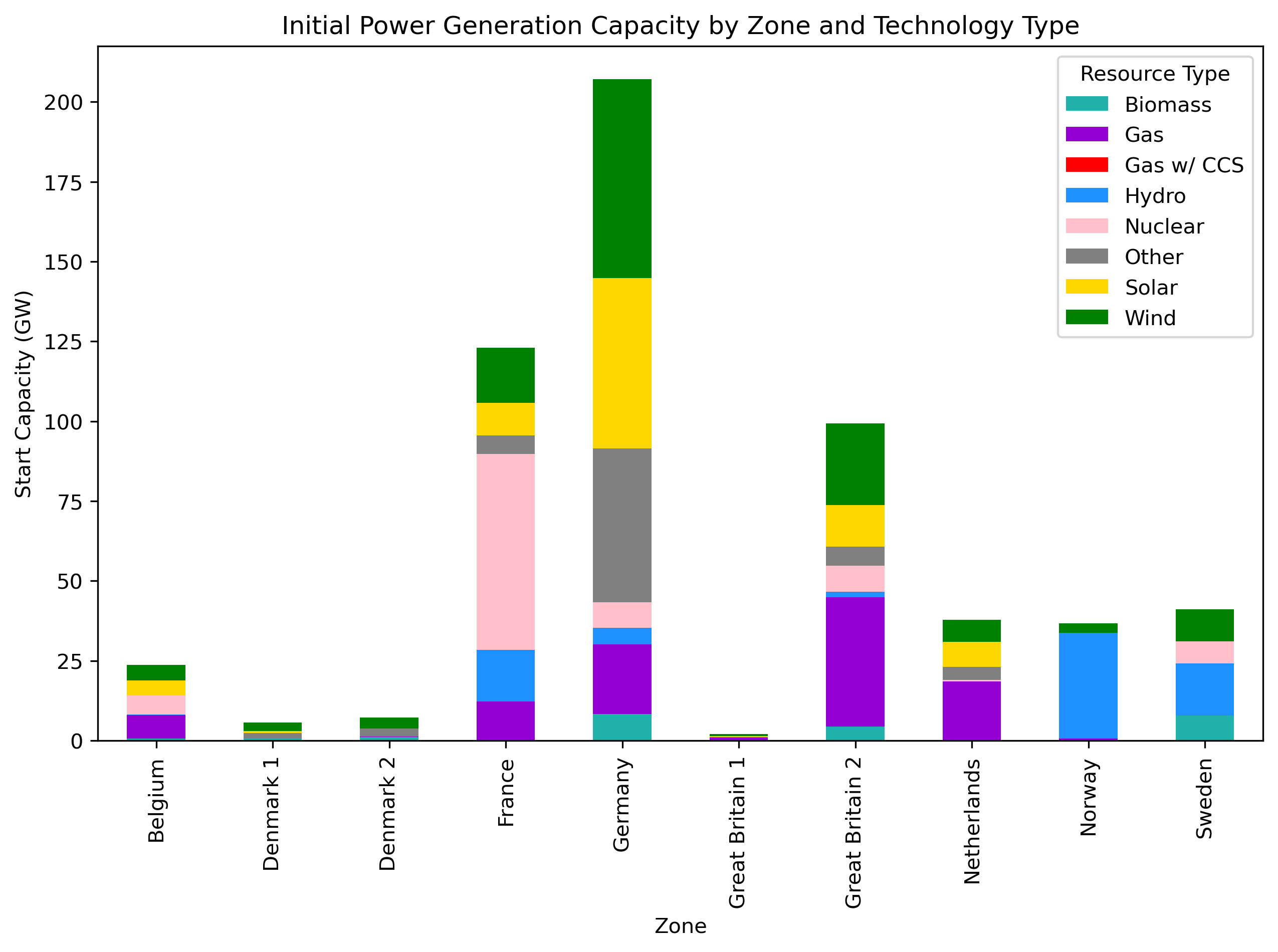}
    \caption{Initial Power Generation by Zone and Technology}
\end{subfigure}

\caption{a) 10-node model representation of the Western European region for the supply-side modeling, highlighting the initial power transfer capacities between the regions (as of 2020, which is assumed to be the built capacity in 2040) and regional distribution of non-transport electricity and H$_2$ demand for modeled 2040 demand scenarios. The size of each bubble represents the non-transportation demand for power (blue) and hydrogen (green), while the thickness of the edges between the nodes represents existing power transmission capacity. b) The second figure shows the initial power generation capacity by zone and technology type, corresponding to installed capacity in 2021. The existing available generation is based on data from the ENTSOE transparency platform for the year 2021 \citep{entsoe_entso-e_nodate}. The network is based on PYPSA-EUR CEM \citep{horsch_complete_2019}.  }
\label{Fig_network}
\end{figure}

The developed models are applied to the case study of Western Europe (Germany, France, the United Kingdom, Belgium, the Netherlands, Denmark, Sweden, and Norway) using a transportation energy demand model, and a 10-zone network representation for power, H$_2$, and CCUS supply chains (Figure \ref{Fig_network}a). This region is central to European decarbonization efforts due to the availability of high quality onshore and offshore wind resources, domestic NG supply, and CO$_2$ sequestration potential. Countries in the region have also shown a significant commitment towards decarbonization and a willingness to invest in a H$_2$ supply chain \citep{european_2019}. Starting with a brownfield power sector representation (i.e. existing power transfer capacity between zones and existing generation capacity), we explore the least-cost system outcomes for alternative technology, demand, and policy scenarios for 2040. Additionally, we use a time domain reduction method to reduce the number of days modeled from 365 days, to 100 representative days to reduce the computational intensity of the model. Since we are focused on transportation energy demand and its energy system impacts, we fix the annual non-transport electricity and H$_2$ demand for the region to be equal to the projections available from ENTSOE for their Distributed Energy Scenario at 2,081 TWh and 468 TWh, respectively \citep{entsoe_entso-e_nodate}. Table \ref{tbl:study_assumptions} summarizes the major case study assumptions. 

\begin{table}[h!]
\caption[Major case study assumptions]{Major case study assumptions. More details on demand-side input assumptions can be found in Section B.2. Details on supply-side assumptions can be found in Section C.2. Non-transportation demand is based on \citep{entsoe_tyndp_2022}. Transportation demand is an output of transportation demand estimation model highlighted in Figure \ref{Fig_modelling}. Fuel price details can be found in Section C.2.5. }
\label{tbl:study_assumptions}
\begin{tabularx}{0.9\linewidth}{@{}Xp{4cm}@{}}
\toprule
Parameter & Value \\
\midrule
2040 Non-transportation Electric Demand & 2,081 TWh \\
2040 Non-transportation H$_2$ Demand & 468 TWh \\
\midrule
Natural Gas Price &  8.56 EUR/GJ \\
Fossil Gasoline Price & 21.95 EUR/GJ \\
Fossil Diesel Price & 26.63 EUR/GJ \\
Jet Fuel Price & 15.18 EUR/GJ\\
\midrule
Discount Rate & 4.5\% \\
\bottomrule
\end{tabularx}
\end{table}

\subsection{Technology assumptions}
We model the region in DOLPHYN using a 10-node spatial representation as shown in Figure \ref{Fig_network}a, with existing infrastructure including power generation capacity (Figure \ref{Fig_network}b) and electricity transmission interconnections between the regions. 
The power system representation of this study is based on a brownfield representation of the European Grid, adapted from  the representation used in a prior European energy system analysis study by researchers at TU Berlin using the PYPSA-EUR CEM \citep{horsch_complete_2019}. The existing available generation is based on data from the ENTSOE transparency platform for the year 2021 \citep{entsoe_entso-e_nodate}. Costs and operational assumptions for power generation technologies are based on the NREL Annual Technology Baseline 2022 (using data for the year 2040, medium case, with a 1.11 EUR/USD conversion) and Sepulveda et al. \citep{nrel_nrel_2022, sepulveda_role_2018}. The maximum available generation capacity and temporally resolved capacity factors associated with the VRE generation technologies is based on PYPSA-EUR data set \citep{horsch_complete_2019}. Additionally, we assume that existing power transfer capacities between regions can be expanded by up to 4 times. New power lines are assumed to have a maximum capacity of up to 5,000 MW. A greenfield representation of H$_2$ and CO$_2$ infrastructure is utilized for this study. The candidate set of pipelines for CO$_2$ and H$_2$ is made up of all the possible combinations of zones (i.e. it is assumed that a pipeline could be built between any two modeled zones). Costs and assumptions for the main H$_2$ production technologies of electrolyzers, steam methane reforming (SMR), SMR with CCS, and autothermal reforming (ATR) with CCS are obtained from various technological reports in the literature \citep{iea_future_2019,lewis_comparison_2022}.

Since the focus of this study is road transportation, the supply-side model considers supply-demand balance for two liquid fuels; diesel and gasoline. We assume that liquid fuel demand can be met in one of two ways. The first is using conventional fossil hydrocarbons purchased at a specified price (see Table \ref{tbl:study_assumptions}), and the second is through SFs based on syngas production from CO$_2$ and H$_2$ followed by Fischer-Tropsch synthesis. Three SF plant configurations are modeled, summarized in Table \ref{synfuel_table}: A) baseline SF plant based on Zang et al. with 52\% energy efficiency and 47\% of the feed carbon recovered as liquid fuel, B) high CO$_2$ capture variant of the process described in Zang et al. where 90\% of the vented CO$_2$ is captured for sequestration purposes and C) high fuel production variant of the process where 90\% of the vented CO$_2$ is captured and recycled to the syngas generation unit to enhance liquid fuel production \citep{zang_life_2021}. Technology cost and performance assumptions for option A are sourced from Zang et al. \citep{zang_life_2021}. The cost and performance assumptions for option B are developed by combining those of the SF process in option A with the estimated cost and energy requirements of a CO$_2$ capture system to capture 90\% of the vented CO$_2$, adapted from literature studies on industrial point source carbon capture \citep{schmitt_cost_2022} The assumptions for option C are developed by assuming that the vented CO$_2$ is captured in a process similar to option B, but that the captured CO$_2$ is recycled to the syngas generation unit of the SF process directly instead of being stored. As such, this option also results in increased use of electricity and H$_2$ per tonne of external CO$_2$ feed input. The model is allowed to choose between these 3 options as part of the optimization. The SF production facility also produces jet fuel, for which we do not model an exogenous demand. Instead, we credit the market value of jet fuel in the objective function of the supply-side model, while including the emissions associated with its end-use in the emissions constraint.

\begin{table}[htbp]
\caption[Cost and performance assumptions for synthetic fuel production with different levels of CO$_2$ utilization]{Cost and performance assumptions for synthetic fuel production with different levels of CO$_2$ utilization. Process information for option B and C were obtained by accounting for the estimated additional energy requirements and costs of capturing 90\% of flue gas CO$_2$ produced by the baseline plant (option A), using data from industrial point source CO$_2$ capture \citep{zang_life_2021, schmitt_cost_2022}. Some minor modifications from original sources were made to ensure facility carbon balance is satisfied for the specified fuel carbon intensities. The carbon intensities of the diesel, gasoline, and jet fuel are 69.3, 67.7, and 68.4 kg of CO$_2$/GJ, respectively.
}
\centering
\begin{tabular}{@{}>{\raggedright\arraybackslash}p{9cm}>{\raggedright\arraybackslash}p{2cm}>{\raggedright\arraybackslash}p{2cm}>{\raggedright\arraybackslash}p{2cm}@{}}
\toprule
Synfuel Production Technology & Option A: Baseline Synfuel Plant & Option B: Synfuel Plant w/ Capture & Option C: Synfuel Plant w/ Capture and Recycling \\
\midrule
CAPEX (MEUR/MW of Fuel Out) & 1.90 & 2.48 & 2.48 \\
\midrule
CO$_2$ Emissions (Tonnes of CO$_2$ / Tonne of CO$_2$ Feed) & 0.52 & 0.05 & 0.10 \\
\midrule
H$_2$ Consumption (GJ of H$_2$ / Tonne of CO$_2$ Feed) & 11.2 & 11.2 & 21.3 \\
Electricity Consumption (GJ of Electricity / Tonne of CO$_2$ Feed) & 0.13 & 2.10 & 4.00 \\
\midrule
Diesel Out (GJ / Tonne of CO$_2$ Feed) & 1.889 & 1.889 & 3.596 \\
Gasoline Out (GJ / Tonne of CO$_2$ Feed) & 1.784 & 1.784 & 3.396 \\
Jet Fuel Out (GJ / Tonne of CO$_2$ Feed) & 3.225 & 3.225 & 6.138 \\
\bottomrule
\end{tabular}
\label{synfuel_table}
\end{table}

We model 3 DAC technologies based on solid-sorbent and solvent based schemes that use natural gas and electricity inputs as per the cost and performance assumptions from a recent study \citep{valentine_direct_2022, valentine_direct_2022-1} (See Section C.2.3). We model CO$_2$ geological sequestration sites and capacities as per the data available for CO$_2$ storage in saline aquifers from the EU GeoCapacity project \citep{vangkilde-pedersen_thomas_eu_2009}. The model is allowed to invest in CO$_2$ pipelines to connect CO$_2$ sources and sinks.

Table \ref{technologies} summarizes the list of technologies considered across various sectors in the supply-side modeling efforts. Further details of demand-side input assumptions can be found in Section B.2.

\begin{table}[ht]
\caption[Summary of technologies considered in the supply-side analysis]{Summary of technologies considered in the supply-side analysis. \textbf{Bolded resources} have spatially-resolved capacity deployment limits. \textit{Italicized resources} are not considered for expansion. CCGT w CCS, SMR w CCS, and ATR w CCS capture rates are 95.0\%, 96.2\%, and 94.5\%, respectively. Detailed technology cost and performance assumptions are provided in Section C.2. VRE = Variable Renewable Energy, CCGT = Combined Cycle Gas Turbine, CCS = Carbon Capture and Storage, OCGT = Open Cycle Gas Turbine.}
\label{technologies}
\begin{tabularx}{0.9\linewidth}{@{}LL@{}}
\toprule
Sector & Technologies considered  \\
\midrule
Power & \begin{tabular}[t]{@{}p{12cm}@{}}\textbf{Utility-scale VRE}, CCGT w \& w/o CCS, OCGT, Nuclear, Coal, Lignite, \textit{Hydro}, \textit{Pumped Hydro}, \textit{Biomass}, Li-ion storage, \textbf{Transmission}\end{tabular} \\
H$_2$ & \begin{tabular}[t]{@{}p{12cm}@{}}Steam Methane Reforming (SMR) w and w/o CCS, Autothermal Reforming (ATR) w CCS, Electrolyzer, Tank H$_2$ storage, CCGT-H$_2$, H$_2$ pipelines\end{tabular} \\
CO$_2$ & \begin{tabular}[t]{@{}p{12cm}@{}}Direct air capture (DAC), CO$_2$ transport pipelines, and \textbf{CO$_2$ geological storage}\end{tabular} \\
Liquid Fuels & \begin{tabular}[t]{@{}p{12cm}@{}}Conventional Fuels, Synthetic Fuels\end{tabular} \\
\bottomrule
\end{tabularx}
\end{table}

\subsection{Scenarios evaluated}\label{sec:scenarios}

\begin{figure}[pos = H]
    \centering
    \begin{subfigure}[t]{0.9\linewidth}
        \centering
        \includegraphics[width=\linewidth]{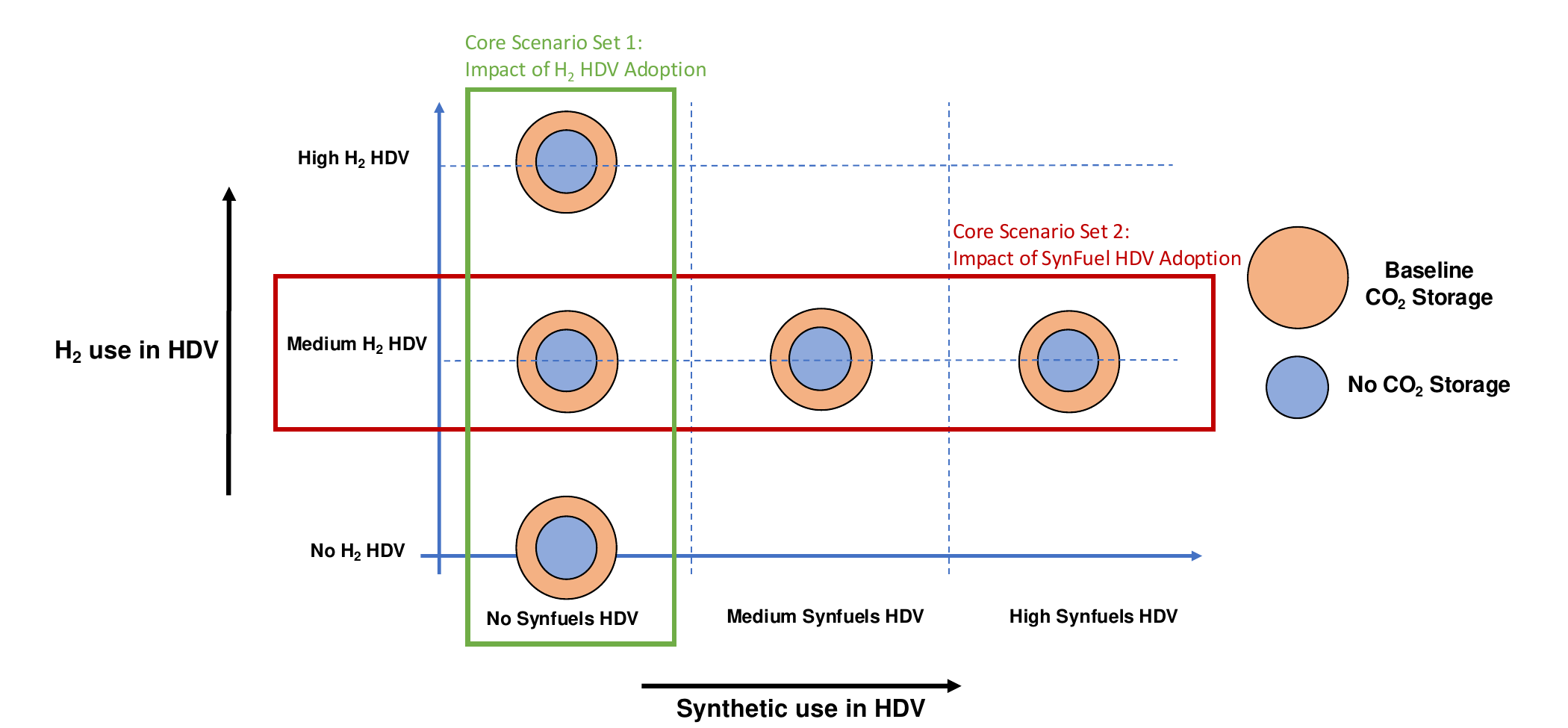}
        \caption{Core Scenario Sets}
        \label{fig:scenarios_plot}
    \end{subfigure}
    
    \vspace{0.5cm} 
    
    \begin{subfigure}[t]{0.9\linewidth}
        \centering
        \renewcommand{\arraystretch}{1.2}
        \begin{tabular}{|c|c|c|}
            \hline
            \textbf{Scenario} & \textbf{Input Modified} & \textbf{Value(s) Tested} \\
            \hline
            Sensitivity Set 1 & Emissions Constraint  & 258 MtCO$_2$/y  \\
            Sensitivity Set 2 & Natural Gas Price  & +/- 30\% of Base Scenario \\
            Sensitivity Set 3 & Uptake of Hydrogen Vehicles & 0\% Uptake of H$_2$ HDVs   \\
            Sensitivity Set 4 & Natural Gas Price & +/- 30\% of Base Scenario \\
            \hline
        \end{tabular}
        \caption{Sensitivity Scenario Set Summary}
        \label{fig:scenarios_table}
    \end{subfigure}
    
    \caption[Core Scenario Summary]{a) Summary of core scenarios evaluated. The y-axis represents varying levels of H$_2$ HDV adoption (between 0 and 142 TWh of H$_2$ consumption), while the x-axis represents varying levels of synthetic fuel adoption (between 0 and 128 TWh of synthetic Diesel consumption). All scenarios are equivalent from an emissions capping perspective with a cap of 103 MtCO$_2$/y. The HDV fleet represents all vehicle types with gross weight greater than 7.5 tonnes. Each bubble in (a) represents a scenario. Baseline CO$_2$ storage corresponds to maximum annual CO$_2$ storage injections equal to 650 MtCO$_2$/year. More details on the transportation demand scenarios can be found in Figures \ref{fig:trans_scenarios}, b) Shows sensitivity scenarios tested. Synfuel = SF.}
    \label{Fig_scenarios}
\end{figure}

We evaluated alternative scenarios spanning different assumptions for compressed H$_2$ use for HDV transportation, minimum levels of SF utilization, and CO$_2$ sequestration availability, as shown in Figure \ref{Fig_scenarios}. Unless otherwise stated, the annual CO$_2$ emissions constraint for the modeled system shown in Figure \ref{Fig_modelling}b was set to be 103 MtCO$_2$/y. This emissions constraint  corresponds to almost 90\% reduction in electricity and heat production, and road transportation sector emissions relative to 2019 levels. Emission limits are enforced jointly across sectors, which allows for emissions trading across sectors in the model.

To test the impact of CO$_2$ sequestration availability on model outcomes, we consider a baseline CO$_2$ sequestration potential scenario, which allows for up to 650 MtCO$_2$/year of sequestration distributed across the region (as shown in Table C.24), and an alternative scenario where no CO$_2$ sequestration is available. Widespread availability of CO$_2$ sequestration is likely to incentivize adoption of carbon capture technologies, while their limited availability could motivate greater reliance on renewable energy adoption and CO$_2$ utilization via SF \citep{mignone_drivers_2024, millinger_are_2022}. 

\begin{figure}[pos = H]
    \centering
    \includegraphics[width=\textwidth]{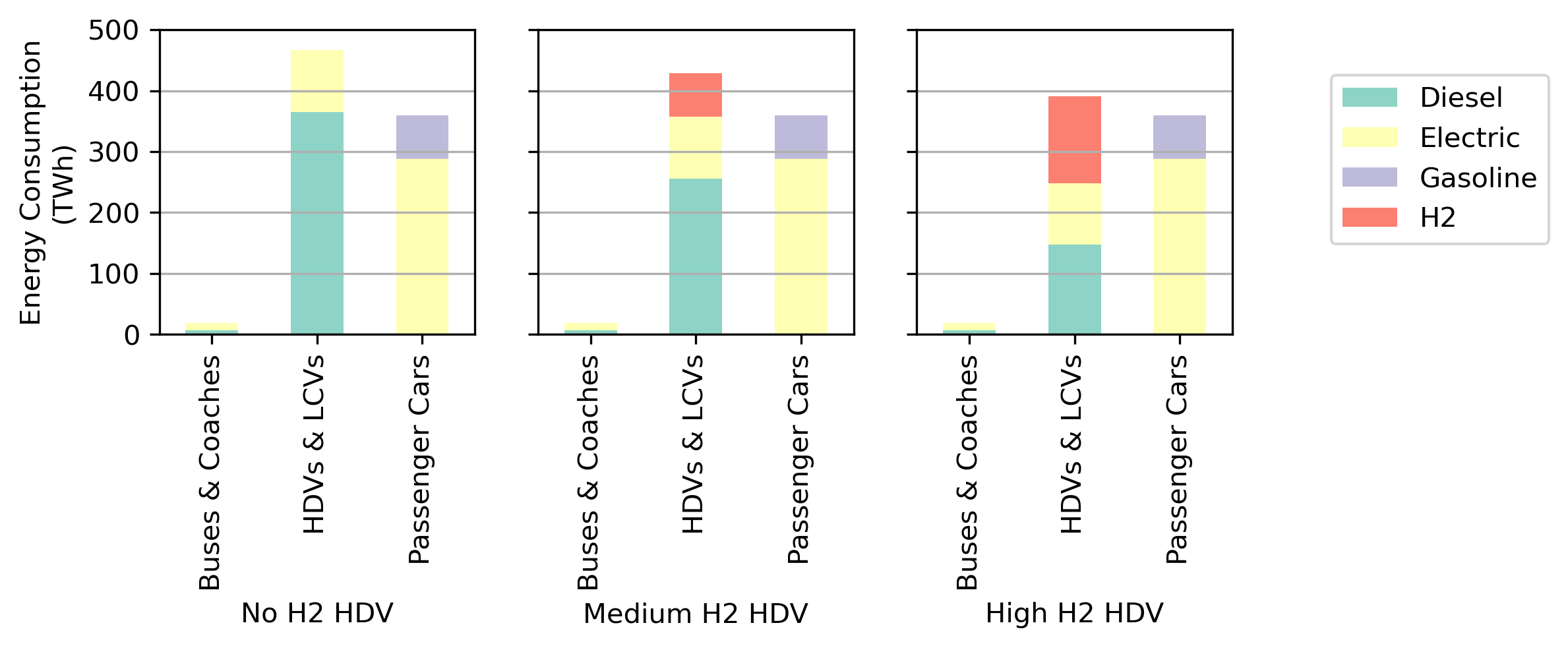}
    \caption{Transportation final energy consumption across different H$_2$ HDV adoption scenarios. HDV energy use, included in the category Heavy-duty Vehicles and Light Commercial Vehicles (HDVs \& LCVs), represents 71-76\% of the category’s energy consumption and 36-42\% of road transportation final energy consumption. Two-wheelers are excluded from the diagram as their demand is negligible compared to other vehicle categories. }
    \label{fig:trans_scenarios}
\end{figure}

To isolate the energy system impacts of H$_2$ and SF use for HDVs, we evaluated the model across different H$_2$ and SF adoption scenarios. We create a baseline case (no H$_2$ HDV, no SFs HDV, bottom left in Figure \ref{fig:scenarios_plot} where all the HDV demand is met through diesel either via internal combustion engine vehicles or PHEVs. For the purposes of this study, we assume that the percentage service demand of plug-in hybrid EV-diesel vehicles and other EVs is static. Scenario set 1 (green box in Figure \ref{fig:scenarios_plot}), evaluates impact of increasing H$_2$ use in HDVs, ranging from 0\% (baseline scenario), 50\% (medium), and 100\% (high) of the service demand satisfied by the diesel ICE vehicles in the baseline scenario. SF adoption is not considered in this scenario set. Figure \ref{fig:trans_scenarios} shows the results of the demand-side model for the varying levels of H$_2$ HDV adoption. The scenario set 2 (red box in Figure \ref{fig:scenarios_plot}) evaluates impact of increasing SF use in HDVs and includes scenarios with 0\%, 25\% (medium), and 50\% (high) of service demand met by diesel in the baseline scenario, while holding H$_2$ use in HDV at the medium level. This second scenario set is motivated by recent policy discussions in the EU that utilize SF production as part of a set of transportation sector decarbonization policies \citep{wacket_exclusive_2023}. Each scenario set is evaluated with and without CO$_2$ sequestration availability.

For these transportation energy demand scenarios, a few key points are to be noted: a) all scenarios assume a fixed amount of electricity use for LDV and some segments of HDV with a high degree of electrification (illustrated in Figure \ref{fig:trans_scenarios}). This also implies that gasoline consumption for road transportation also remains constant across the scenarios, b) scenarios with increasing H$_2$ adoption leads to reduced end-use diesel consumption, c) HDV energy consumption represents 36-42\% of modeled road transportation energy consumption. 

In addition to the core set of scenarios outlined above, we ran 4 additional sets of sensitivity scenarios as shown in Figure \ref{fig:scenarios_table}. The first is  based on Scenario Set 1, but with a less stringent emissions constraint (258 MtCO$_2$/y or corresponding to a 25\% reduction compared to 2019 emissions from transport, power and heat production). The second is based on scenario set 1, but with sensitivities around NG price (+/- 30\%) of the baseline NG price assumption. NG price is a key input as the relative price between NG and electricity determines the type of H$_2$ production, while the relative price between NG and liquid fuels determines the cost-effectiveness of H$_2$ and SF-based decarbonization solutions. 

The third sensitivity set focuses on the impact of SF adoption with zero H$_2$ adoption in the transport sector. Here, we use the same assumptions as scenario set 2, but with a lower level of H$_2$ adoption. Because the gross demand for diesel is higher in the absence of any H$_2$ adoption, the percentage of synthetic diesel out of total diesel demand was adjusted to maintain the same gross amount of synthetic diesel demand across Scenario Set 2 and sensitivity set 3. The final set of sensitivities is the same as scenario set 2, but with sensitivities around NG price, +/- 30\% of the baseline scenario NG price.

\section{System impacts of H$_2$ adoption in HDVs}\label{sec:system_impacts_h2}

\begin{figure}[pos = H]
    \centering
    \includegraphics[width=1\linewidth]{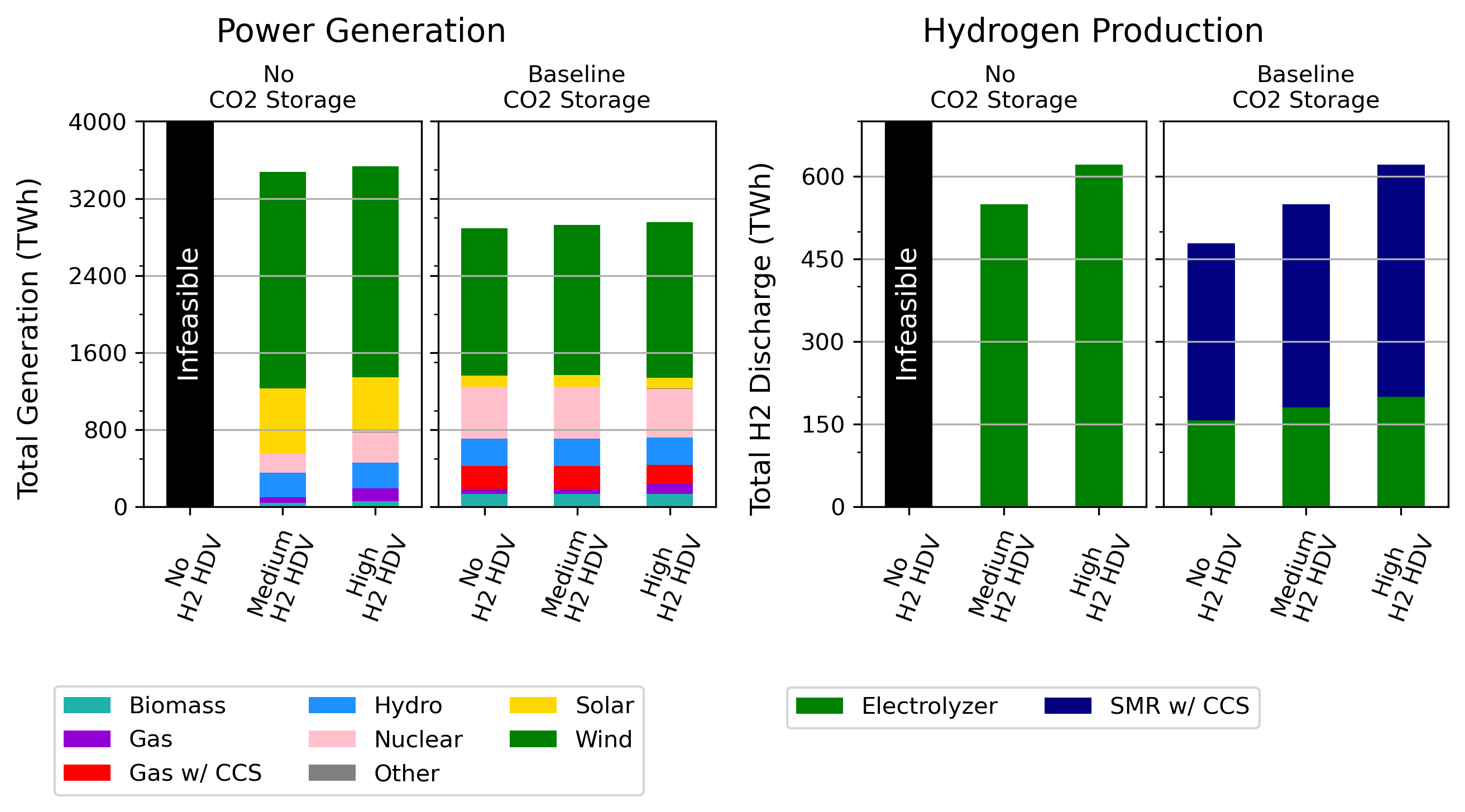}
    \caption[Power and H$_2$ Generation Scenario Set 1]{Power and H$_2$ generation for baseline and no CO$_2$ sequestration scenarios under no synthetic fuel adoption. The left set of charts shows power generation and the right set of charts shows H$_2$  generation. Within each panel, the amount of H$_2$ HDV adoption increases moving from left to right. Corresponding capacity charts are shown in Figure A.1.}
    \label{fig_h2_hdv_power_h2}
\end{figure}

The power and H$_2$ generation impacts resulting from increasing H$_2$ adoption for HDVs under the two CO$_2$ storage scenarios and without any SF utilization are highlighted in Figure \ref{fig_h2_hdv_power_h2} (see Figure A.1 for capacity outcomes). In the absence of CO$_2$ storage, Figure \ref{fig_h2_hdv_power_h2} shows that the model produces an infeasible supply-side solution in the case of no H$_2$ use in HDVs. This infeasibility is a result of the emissions limit being lower than the CO$_2$ emissions associated with liquid fuel consumption in the transportation sector, where there are no abatement options available in this scenario. Use of H$_2$ for HDVs, however, resolves the model infeasibility and leads to incremental H$_2$ supply via electrolytic hydrogen production that consumes 110.1 - 206.2 TWh of electricity or approximately 5.3 - 9.9\% of non-H$_2$ sector electricity demand. Increasing the share of H$_2$ in HDVs from medium to high level results in CO$_2$ emissions reduction in the transportation sector at the expense of increased power sector emissions through utilization of NG generation without CCS that lowers the share of VRE generation. For example, wind and solar generation share with high H$_2$ HDV is 75\% compared to 77\% in the medium H$_2$ HDV case. Interestingly, increasing H$_2$ use in HDVs from medium to high levels reduces the marginal emissions abatement cost (see Table A.1). This result stems from the reduced investment in battery storage and transmission capacity (see Figures A.3 and A.4) and leveraging electrolyzer operational flexibility to support power system operations. In all cases, the maximum level of capacity deployment for VREs (including on-shore wind) is not reached for all regions. Finally, we see that battery storage is higher when CO$_2$ storage is not available (394 - 1006 GWh), compared to when it is available (12 - 22 GWh), as shown in Figure A.3. H$_2$ storage is not deployed across scenarios. 

\begin{figure}[pos = H]
    \centering
    \includegraphics[width=1\linewidth]{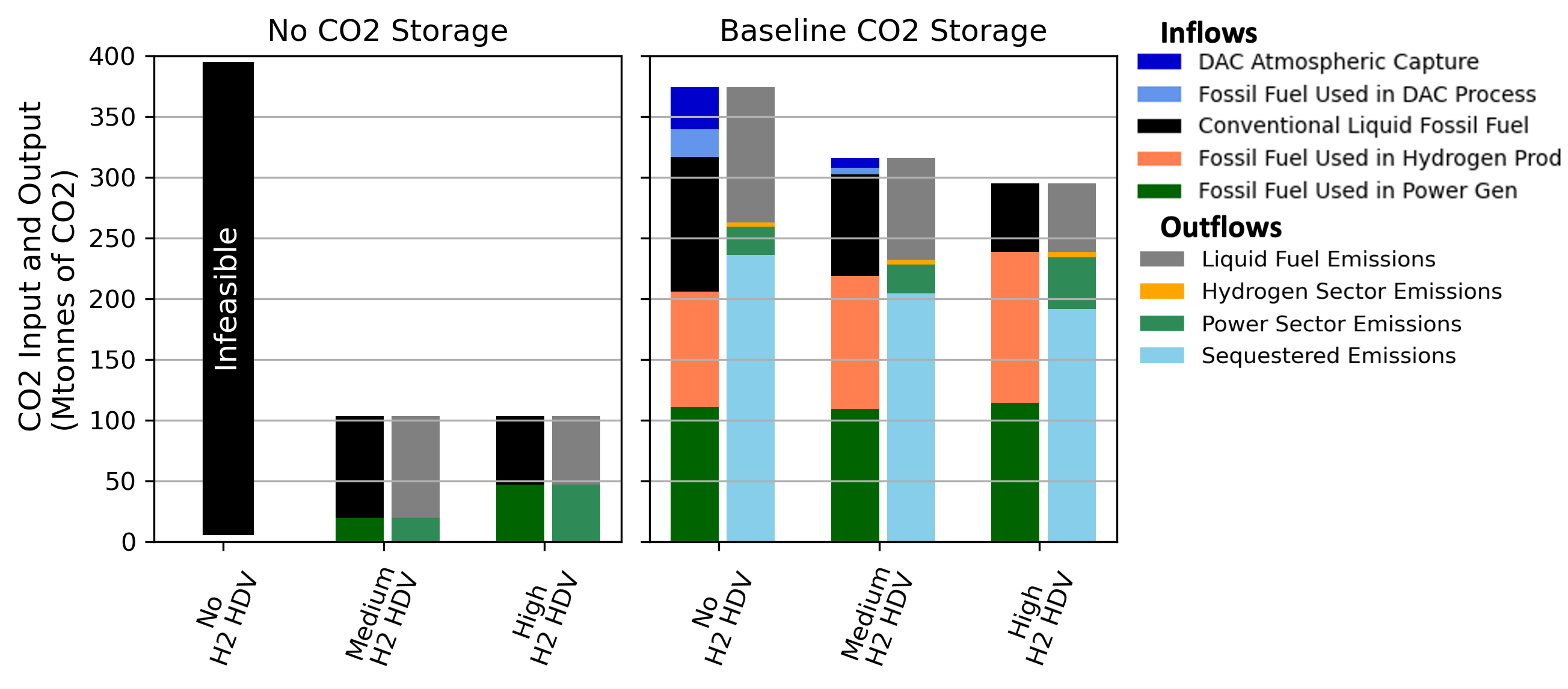}
    \caption[CO$_2$ Balance Scenario Set 1]{System CO$_2$ balance under varying levels of H$_2$ HDV adoption and no SF adoption (Scenario Set 1). The subfigure on the left shows the CO$_2$ balance under no CO$_2$ sequestration availability, while the one on the right shows the CO$_2$ balance under baseline CO$_2$ storage availability. Within each subplot the H$_2$ HDV adoption level increases left to right. The leftward column in each subfigure represents CO$_2$ input into the system, while the rightward column represents CO$_2$ outputted by the system. All scenarios adhere to the same emissions constraint of 103 MtCO$_2$/y. System net emissions can be calculated from the chart by subtracting sequestered emissions and DAC atmospheric capture from the emission outflows. It is worth noting that all CO$_2$ diagrams include all transportation emissions including non-HDV vehicle categories.}
    \label{fig_h2_hdv_co2}
\end{figure}

The availability of CO$_2$ storage results in deployment of CCS technologies for power and H$_2$ generation as shown in Figure \ref{fig_h2_hdv_power_h2} which reduces the power sector impacts of H$_2$ adoption in HDV and also leads to a feasible solution without H$_2$ use. This is achieved via the deployment of DAC and CCS technologies in conjunction with CO$_2$ storage, as indicated in the CO$_2$ inflows in Figure \ref{fig_h2_hdv_co2}. CO$_2$ storage availability also leads to deployment of CO$_2$ infrastructure (Figure A.2), that is utilized by both DAC and CCS technologies in the power and H$_2$ sectors. Similar to the case without CO$_2$ storage, increasing H$_2$ use for HDVs leads to a greater role for gas-based power generation without CCS that comes at the expense of reduced consumption of liquid fuels in transport, gas power generation with CCS (Figure A.3) and DAC deployment (Figure \ref{fig_h2_hdv_co2}). At the same time, we see an increase in CCS-based H$_2$ production (greater carbon inflows in Figure \ref{fig_h2_hdv_co2}), which highlights relative cost-effectiveness of deploying CCS in H$_2$ vs. power generation. 

Despite achieving the same emissions target, scenarios with CO$_2$ storage result in: a) greater overall CO$_2$ throughput as compared to scenarios without CO$_2$ storage, owing to greater total fossil fuel (i.e. NG) utilization (Figure \ref{fig_h2_hdv_co2} and \ref{fig:fuel_comp}) and b) lower marginal CO$_2$ abatement costs (Table A.1). Despite these substitution effects, it is important to note that NG consumption in the scenarios are 61 - 65\% lower than the 2019 levels. Additionally, the percentage of electricity produced by fossil fuel sources is 9.8 - 10.1\% in our scenarios compared to 26\% in 2015, despite a significant expansion of the power sector (467 GW - 593 GW vs. 217 GW peak demand in 2019). 


Comparing scenarios with and without CO$_2$ storage also reveals the complementary nature of VRE and electrolyzer deployment (Figure \ref{fig_h2_hdv_power_h2}). This result, previously noted by other studies, is a result of the capability to operate electrolyzer in a flexible manner in conjunction with H$_2$ storage and H$_2$ pipelines, so as to maximize H$_2$ production during times and locations of low electricity prices, synonymous with abundant VRE electricity supply \citep{law_role_2025, he_sector_2021, brown_synergies_2018}. 

\begin{figure}[pos = H]
    \centering
    \includegraphics[width=1\linewidth]{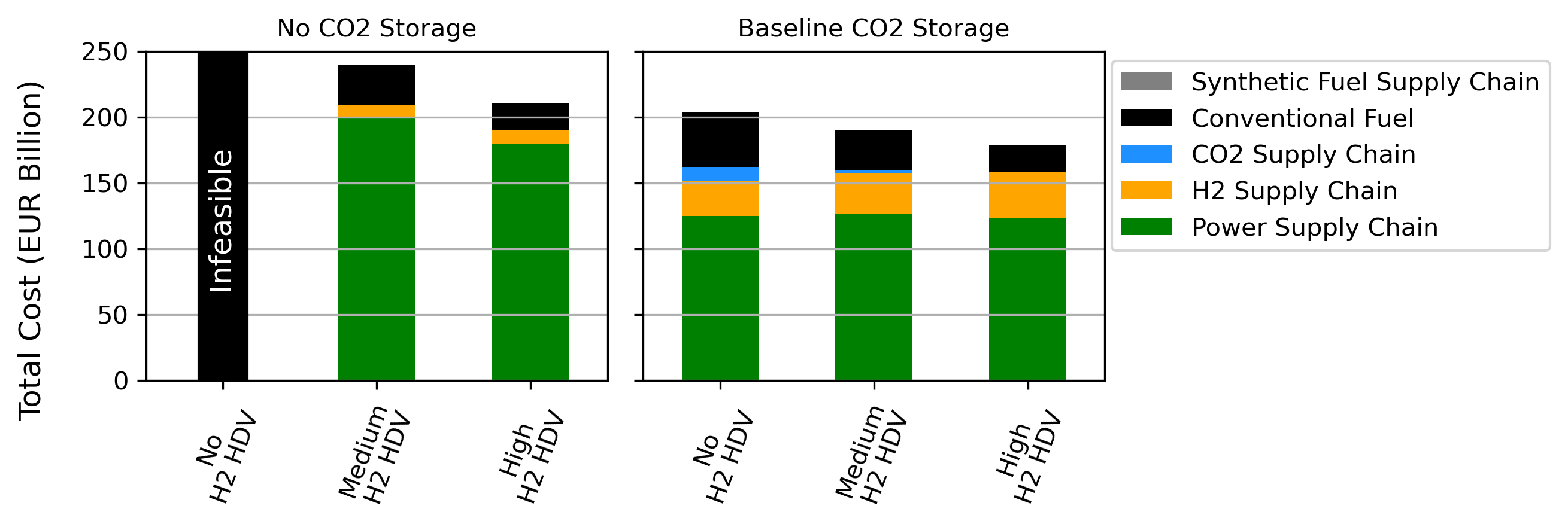}
    \caption[Costs Scenario Set 1]{Annualized bulk-system costs under varying levels of H$_2$ HDV adoption and no SF adoption. The subfigure on the left shows the cost breakdown under no CO$_2$ sequestration availability, while the one on the right shows the cost breakdown under baseline CO$_2$ sequestration availability. Within each subplot the H$_2$ HDV adoption level increases left to right. The costs do not include vehicle replacement or H$_2$ distribution costs.}
    \label{fig:h2_hdv_cost}
\end{figure}

Irrespective of CO$_2$ storage availability, increasing H$_2$ use in the transportation sector reduces bulk energy system costs, with reductions of up to 6\% observed when comparing the no H$_2$ HDV scenarios to the high H$_2$ HDV adoption scenarios (Annualized cost savings of 21  Billion EUR/year in the case with CO$_2$ storage). As seen in Figure \ref{fig:h2_hdv_cost}, the cost savings primarily stem from reduced liquid fuel consumption, but also lower power system costs due to reduced decarbonization of this sector and lower carbon supply chain costs due to the reduced reliance on DAC. These savings fully counteract the net increase in H$_2$ system costs associated with meeting the added H$_2$ demand. There are two important caveats to these findings. First, these results only represent bulk system costs and do not include the cost of distribution, refueling and vehicular infrastructure associated with H$_2$ use for HDV. 
H$_2$ use in HDV transportation would only be cost-effective, if the additional end-use infrastructure and equipment upgrades do not outweigh the bulk energy system cost savings estimated here. Second, because H$_2$ use displaces liquid fuels in lieu of increases in NG utilization in some scenarios, these results are sensitive to the relative price of NG and diesel, in the case with CO$_2$ storage available. In Figure A.31, we show how decreasing spread between NG and liquid fuel costs reduces the incentive for H$_2$ use in HDVs and vice versa (see section \ref{sec:sen_scenario_2_results}).

The results uncover trade-offs between the utilization of liquid fossil fuels and NG, as shown in Figure \ref{fig:fuel_comp}. In scenarios without CO$_2$ storage, the adoption of H$_2$ HDVs results in an increase of NG consumption. This occurs due to the reallocation of the emissions budget from the transportation to the power sector, allowing for expanded NG-based generation. The relationship is not as straight-forward in the scenarios with CO$_2$ storage: while NG consumption increases due to the expansion of NG-based H$_2$ production, it decreases due to the contraction of DAC, specifically solvent-based DAC technologies that use NG as energy input \citep{netl_comparison_2022}. The net change in NG consumption depends on the relative size of these two effects.


\begin{figure}[pos = H]
    \centering
    \includegraphics[width=0.7\linewidth]{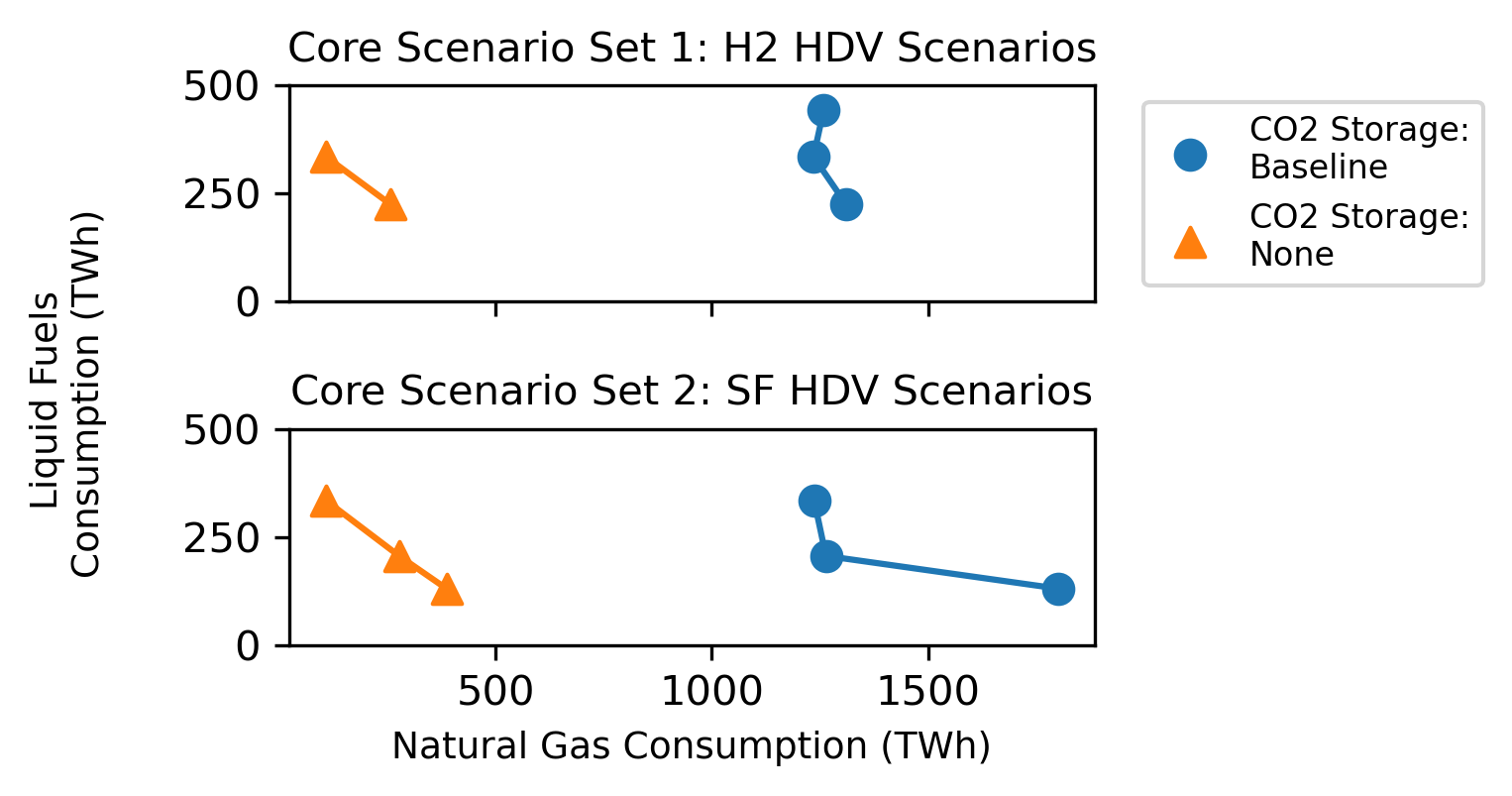}
    \caption[Natural Gas and Liquid Fuel Utilization Trade-off Core Scenarios]{Trade-off between natural gas (NG) and liquid fossil fuel utilization. The subfigure on the top shows the relationship for the H$_2$ HDV scenarios (i.e. scenario set 1), while the one on the bottom shows the relationship for SF adoption scenarios (i.e. scenario set 2). Within each subplot the amount of natural gas consumption can be examined on the x-axis, while the amount of liquid fossil fuel consumption can be examined on the y-axis. The amount of H$_2$ and SF HDV adoption increases from top to bottom. The liquid fossil fuel consumption represents the final energy demand for diesel and gasoline in the transport sector, and excludes jet fuel as well as excess fuel supply produced in some of the scenarios.}
    \label{fig:fuel_comp}
\end{figure}

\section{System Impacts of Synthetic Fuel Adoption}\label{sec:system_impacts_sf}


\begin{figure}[pos = H]
    \centering
    \includegraphics[width=1\linewidth]{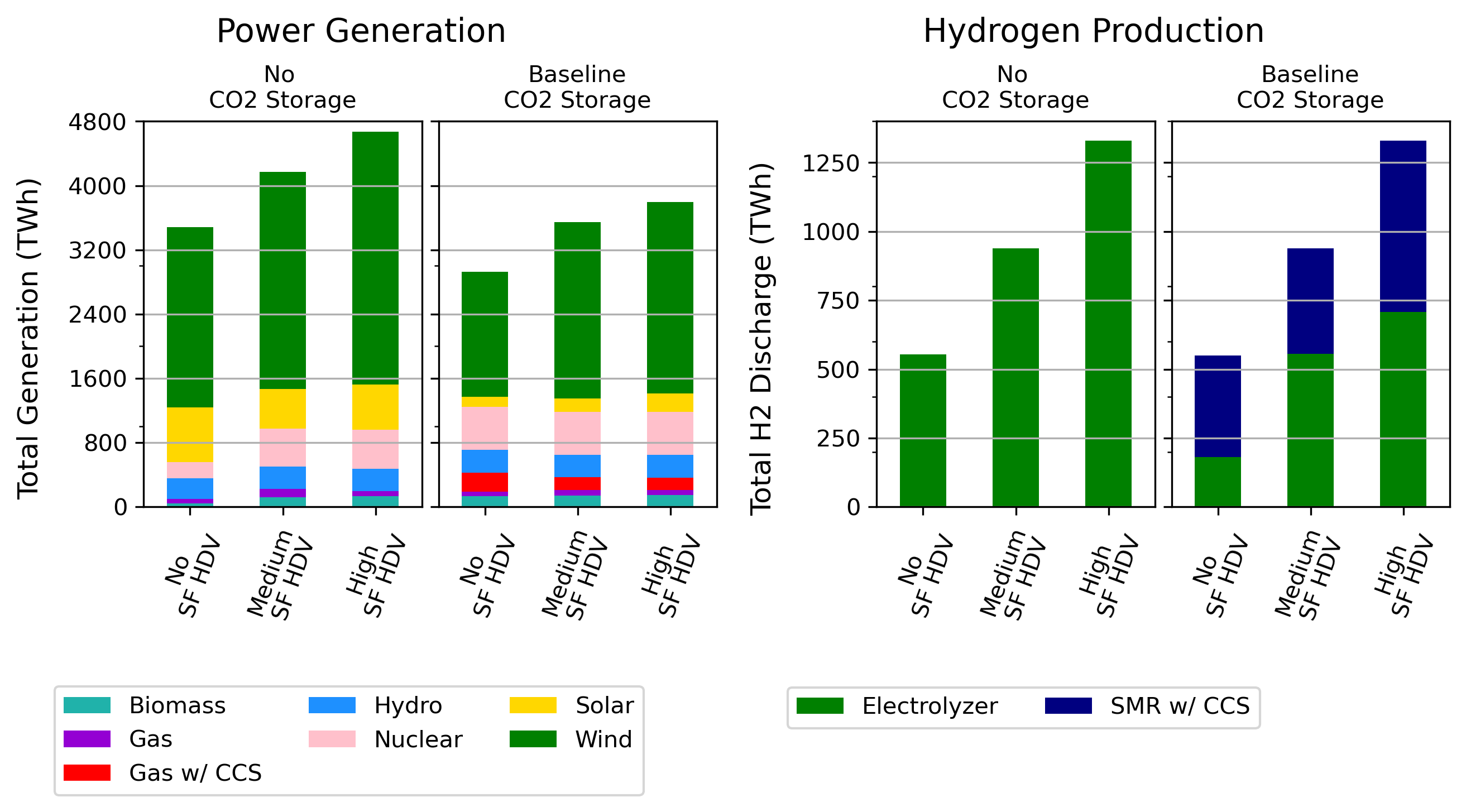}
    \caption[Power and H$_2$ Generation Scenario Set 2]{Power and hydrogen generation for baseline and no CO$_2$ sequestration scenarios under medium H$_2$ HDV adoption and varying scenarios of synthetic fuel adoption. The left set of charts shows power generation and the right set of charts shows H$_2$  generation.  Within each panel, the amount of synthetic fuel adoption increases moving from left to right. Total system emissions constrained to 103 MtCO$_2$/y.}
    \label{fig:sf_hdv_power_h2}
\end{figure}

The power and H$_2$ generation impacts resulting from SF production under various CO$_2$ storage scenarios and with medium H$_2$ HDV utilization are highlighted in Figure \ref{fig:sf_hdv_power_h2} (capacity results are shown in Figure A.7). The production of SFs requires three inputs: H$_2$, CO$_2$ and small quantities of electricity inputs (Table \ref{synfuel_table}). The maximum CO$_2$ abatement potential of SFs is realized when the H$_2$ and electricity supply are sourced from low-carbon sources while the CO$_2$ is sourced from the atmosphere. Consequently, we find that the hydrogen for medium levels of SFs production is sourced primarily via electrolysis (Figure \ref{fig:sf_hdv_power_h2}). However, increasing electrolyzer deployment raises electricity demand and consequently, the average electricity price seen by the electrolyzer \citep{mallapragada_electricity_2023}. For further increases in SF production, it is more cost-effective to produce H$_2$ via NG SMR with CCS rather than electrolysis when CO$_2$ storage is available, as shown in Figure \ref{fig:sf_hdv_power_h2}. Overall, SF production is accompanied by a 142\% increase in H$_2$ production vs. non-transport H$_2$ demand, as compared to 30\% increase in the case of H$_2$ use in HDVs (Figure \ref{fig_h2_hdv_power_h2}), reflecting the lower energy efficiency of SF based decarbonization strategies. By extension, and as a result of the expanded electrolyzer-based H$_2$ demand, a large incremental growth in the power sector is also required for SF production, with a growth in power generation of 44\% and 33\% vs. non-transport power demand in the no CO$_2$ and baseline CO$_2$ storage cases, respectively. As shown in Figure \ref{fig:sf_hdv_power_h2}, the growth in power sector generation resulting from SF adoption is dominated by VRE, primarily wind, reinforcing the synergies between electrolyzers and VREs noted earlier. 

\begin{figure}[pos = H]
    \centering
    \includegraphics[width=1\linewidth]{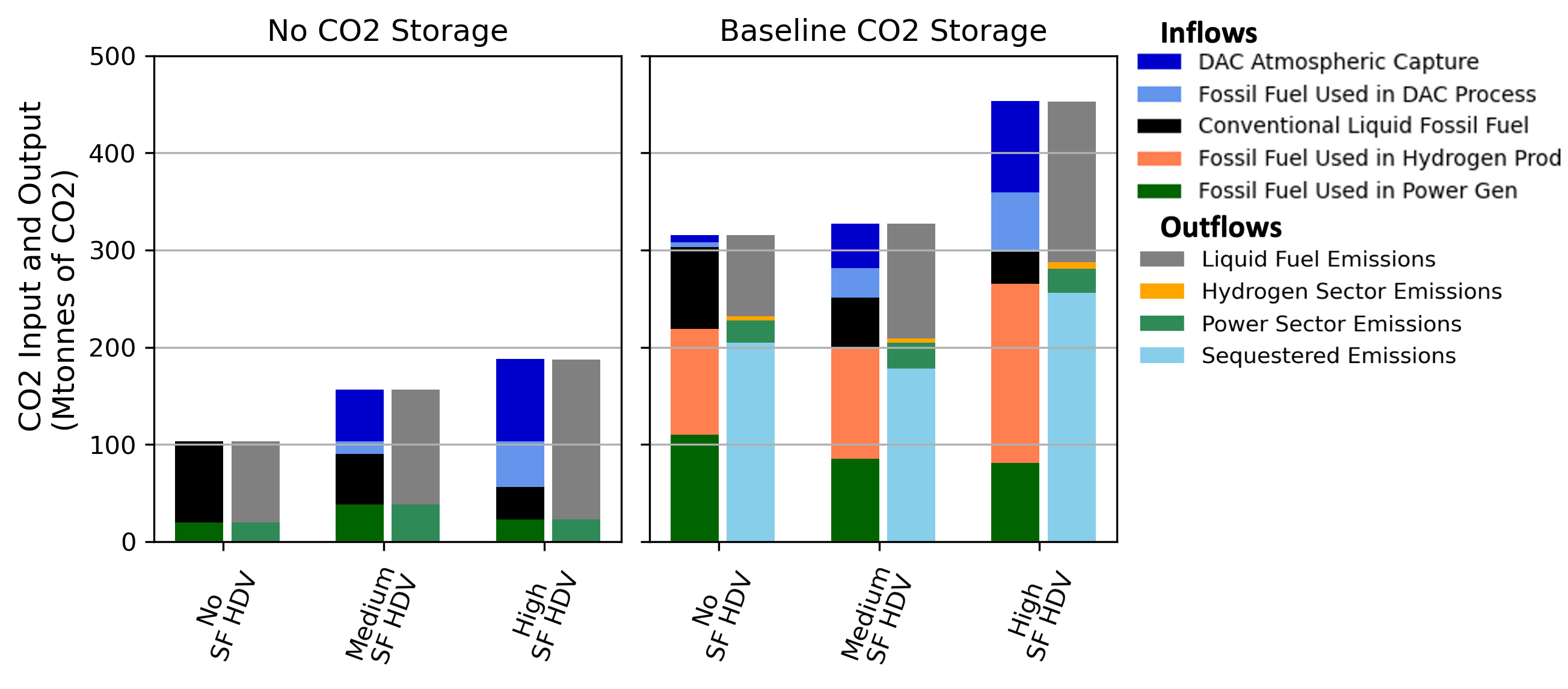}
    \caption[CO$_2$ Balance Scenario Set 2]{System CO$_2$ balance under varying levels of SF adoption and medium H$_2$ HDV adoption and varying scenarios of synthetic fuel adoption. The subfigure on the left shows the CO$_2$ balance under no CO$_2$ sequestration availability, while the one on the right shows the CO$_2$ balance under baseline CO$_2$ sequestration availability. Within each subplot the SF adoption level increases left to right. The leftward column represents CO$_2$ input into the system, while the rightward column represents CO$_2$ outputted by the system. System emissions can be calculated from the chart by subtracting sequestered emissions and DAC atmospheric capture from the emission outflows.}
    \label{fig:sf_hdv_co2}
\end{figure}

In both CO$_2$ storage scenarios, increasing SF production also leads to additional DAC deployment, as highlighted in Figures \ref{fig:sf_hdv_co2} and A.9. DAC deployment rises substantially with high SF adoption due to the need to meet increased diesel demand via synthetic fuels. However, the SF production facility has a fixed product distribution and this leads to excess gasoline production beyond system requirements. This excess gasoline represents lost carbon that reduces the facility’s overall carbon efficiency for producing valuable fuels. Consequently, high SF adoption leads to increased carbon inflows for the DAC process. In the baseline CO$_2$ storage case, CO$_2$ sequestration requirements are reduced in the medium SF adoption case, but are subsequently increased in the high SF adoption case to account for the expansion of NG based H$_2$ production with CCS. Among SF production processes considered, we see a consistent preference for the pathway with the lowest overall emissions as noted in Table \ref{synfuel_table}, option B. In other words, pathways that maximize feed carbon conversion into one of two co-products, either SF or captured CO$_2$, are preferred over pathways with lower carbon conversion. 


\begin{figure}[pos = H]
    \centering
    \includegraphics[width=1\linewidth]{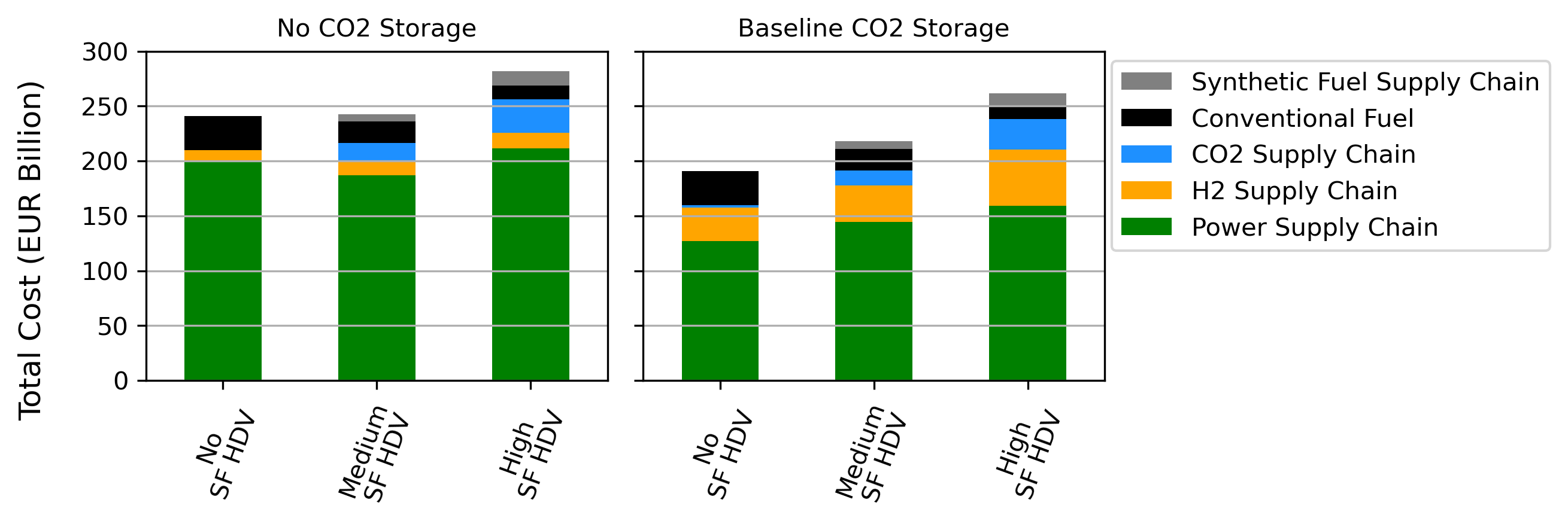}
    \caption[Costs Scenario Set 2]{Annualized bulk-system costs under varying levels of SF adoption and medium H$_2$ HDV adoption and varying scenarios of synthetic fuel adoption. The subfigure on the left shows the cost breakdown under no CO$_2$ sequestration availability, while the one on the right shows the cost breakdown under baseline CO$_2$ sequestration availability. Within each subplot the SF adoption level increases left to right. The costs do not include vehicle replacement or H$_2$ distribution costs.}
    \label{fig:sf_hdv_cost}
\end{figure}

The cost impacts of SF adoption are shown in Figure \ref{fig:sf_hdv_cost}, where we see that cost savings from reduced purchases of liquid fossil fuel are more than offset by cost increases resulting from expanding energy infrastructure of other vectors (H$_2$, CO$_2$, electricity) to produce SFs. Thus, per our modeled assumptions, SFs would not be cost-optimal to deploy across any of the scenarios evaluated here (note that we modeled SF adoption as a minimum requirement per Eqn. C.1.1). Interestingly, in both CO$_2$ storage cases, the adoption of SFs increases costs by differing magnitudes; in the no CO$_2$ storage scenario, system cost increase by 3\% and 14\% for the medium and high SF adoption cases, respectively. In the baseline case, the cost increase is more even; the costs increase by 14\% from the No SF case to the Medium SF case, and then 18\% from the Medium SF to the High SF case. As such, the system cost increase resulting from SF deployment is lower when no CO$_2$ storage is available due to the limited number of abatement options in the power generation and H$_2$ sectors.  Similar to scenario set 1, we also observe trade-offs between the utilization of liquid fossil fuels and NG in scenarios with SF adoption, (Figure \ref{fig:fuel_comp}). In all scenarios, we see an increase in the amount of NG consumed as a result of SF deployment. The increase results from the increased adoption of DAC (that uses NG for energy input), expansion of NG-based H$_2$ production needed for SF production processes. This is particularly true in the high SF case, where there is a large expansion of DAC, in addition to an expansion of NG-based H$_2$ production.

\section{Sensitivities of Emissions Constraints, Natural Gas Price, and Level of Fuel Substitution in HDV Sector}\label{sec:sen_results}
\subsection{Sensitivity Scenario Set 1: Sensitivity to Stringency of Emissions Constraint}
The first sensitivity scenario set is based on Scenario Set 1, but at a less stringent emissions constraint of 258 MtCO$_2$/y (detailed results can be found in Section A.3).

In this case, without CO$_2$ storage, the incremental H$_2$ demand is primarily met through grey H$_2$, while in baseline CO$_2$ storage scenario, the added H$_2$ demand from H$_2$ HDVs is met primarily via combination of H$_2$ production with and without CCS (Figure A.14). The substitution of fossil fuels in the transportation sector also reduces the need for fossil fuel substitution in the power sector (See Figure A.14), resulting in the expansion of fossil fuel power generation without CCS in the baseline scenario (i.e. zero H$_2$ use in HDV). In the baseline CO$_2$ sequestration scenario, we also see a substitution of less polluting fossil fuel power generation for more polluting fossil fuel generation (i.e. a substitution of NG with CCS to NG, and NG to coal, as seen in Figure A.14). 

 Relaxed CO$_2$ emissions constraints leads to no DAC deployment. In addition, the bulk system cost savings of H$_2$ use in HDV are lower compared to the stringent emissions cap case (See Figure A.20), which highlights the increased importance of transportation decarbonization under stringent emissions constraints.

\subsection{Sensitivity Scenario Set 2: Impact of Natural Gas Prices on H$_2$ HDV Adoption Scenarios}\label{sec:sen_scenario_2_results}
The second sensitivity scenario set investigates impact of NG prices (See Section A.4 for detailed results) on results of Scenario Set 1. This sensitivity analysis is motivated by the recent volatility in European NG prices \citep{iea_10-point_2022}. While our base case NG prices are based on the assumption that supply for NG in European context is based on liquefied NG (LNG), there is considerable uncertainty in the future evolution of the LNG market itself. For these reasons, our sensitivity focuses on testing how our model outcomes regarding H$_2$ and SF adoption change with changes in NG prices. 

Changes in NG prices have little effect under the no CO$_2$ sequestration scenario due to low levels of NG consumption in these cases (See Figure A.21). The impacts in baseline CO$_2$ storage case are more substantial. For instance, the amount of H$_2$ produced using electrolyzers increases from 101.75 TWh to 217.7 TWh in medium H$_2$ HDV case between the low NG price and the high NG price sensitivities (See Figure A.23). Additionally, a shift away from NG power generation towards VREs also occurs; in the medium H$_2$ HDV case for instance, the percentage of wind and solar generation increases from 52\% in the low NG price case to 61\% in the high NG price case. Additionally, the amount of gas-based power generation also decreases (See Figure A.21). This combination of shifts in H$_2$ generation and power production, as well as a reduction in DAC deployment, results in a reduction of CO$_2$ sequestration requirements from 273 MtCO$_2$/y of CO$_2$ to 163 MtCO$_2$/y of CO$_2$ in the medium H$_2$ HDV adoption case (See Figure A.30). The cost savings arising from the increased adoption of H$_2$ HDVs from none to high with baseline CO$_2$, decreases from 13\% in the low NG price case to 11\% in high NG price case (See Figure A.31). While the cost savings margin decreases with higher price of NG, the limited change suggest that the results are somewhat robust to the price of NG. Finally, the cost of NG has a large impact on the total amount of NG consumption as seen in Figure A.32.

\subsection{Sensitivity Scenario Set 3: Effect of zero H$_2$ use in HDVs on System Impacts of Synthetic Fuel Adoption}
The third sensitivity scenario set is based on second Scenario Set 2, but with no H$_2$ HDVs as opposed to a medium level of H$_2$ HDV adoption (See Section A.5 for detailed results). The motivation behind this scenario is to explore the robustness of the results to the base level of H$_2$ HDVs. As in Scenario Set 1, without the adoption of transportation fuel substitution measures and with the lack of CO$_2$ storage availability, the supply-system is infeasible. However, we see that the adoption of SFs allow for sufficient system decarbonization to meet the necessary emission constraints (See Figure A.33). This highlights the key role SFs can play in highly decarbonized systems. In the no CO$_2$ storage case, the expansion of the power sector to meet electrolyzer demand in H$_2$ sector is notable, mostly occurring with the expansion of VREs. Additionally, we see that the existence of a CO$_2$ utilization pathway allows for the deployment of DAC, even in the absence of CO$_2$ storage (See Figure A.41). In the baseline CO$_2$ storage case, the H$_2$ demand is met with a mixture of fossil H$_2$ with CCS and electrolytic H$_2$ (See Figure A.33).

\subsection{Sensitivity Scenario Set 4: Impact of Natural Gas Prices on SF HDV Adoption Scenarios}
The final sensitivity is based on Scenario Set 2, but with varying  NG prices (See Section A.6 for detailed results). The impact of the higher NG prices mirrors some of the findings from the Sensitivity Set 2 discussed earlier, including limited impact of NG prices on the power and H$_2$ supply mixes under the no CO$_2$ storage scenario. Further, as in the sensitivity set 2, higher NG prices result in a larger share of H$_2$ production from electrolyzers (See Figure A.44). Additionally, the cost increases from the increased adoption of SFs from none to high, increases from 35.9\% in the low NG price case to 36.4\% in high NG price case under baseline CO$_2$ storage, and increases from 16\% in the low NG price case to 18\% in high NG price case under the no CO$_2$ storage (See Figure A.54).

\section{Discussion and Policy Takeaways}\label{sec:disc}
%

\begin{figure}[pos = H]
 \centering
 \includegraphics[width=1\linewidth]{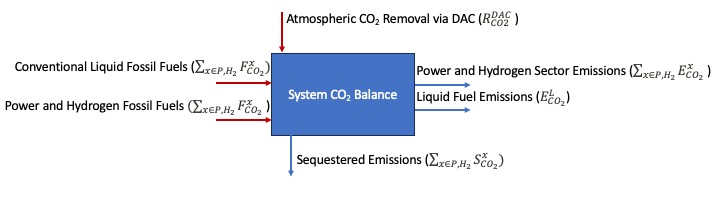}
 \caption[System Emissions Balance Diagram]{System emissions balance. All terms are assumed to be positive in value.}
 \label{Fig_carbon_balance}
 \end{figure}

Our analysis reveals the inter-dependence between different sectoral decarbonization strategies, resulting from the modeled system carbon balance shown in Figure \ref{Fig_carbon_balance}. These strategies include: 1) fossil fuel substitution in the transportation sector (e.g. use of H$_2$ or SFs) 2) fossil fuel emissions abatement via CO$_2$ sequestration, 3) fossil fuel substitution in power and H$_2$ production, and 4) atmospheric CO$_2$ removal. All these strategies change the inflows and outflows of CO$_2$ into the system as highlighted in Figures \ref{fig_h2_hdv_co2} and \ref{fig:sf_hdv_co2}. We see how the emphasis on each decarbonization strategy changes depending on CO$_2$ storage availability and emissions constraint (see sensitivity results in SI). For example, as H$_2$ use in HDVs increases, the reliance on fossil fuel substitution in the power and H$_2$ sector, carbon sequestration (when available), and atmospheric carbon removal decreases. In contrast, the adoption of SFs generally increases the reliance on atmospheric carbon removal and sequestration, when available, as well as increasing the role for fossil fuel substitution in the power and H$_2$ production mix. Below we discuss the policy implications of our findings, while considering the prevailing policy landscape in the European context that was the basis for our case study.

In the absence of CO$_2$ storage, deep decarbonization of power, H$_2$ and transportation sectors without liquid fossil fuel substitution (using H$_2$, SFs, or other methods not considered in this study) may not be viable, as illustrated by the infeasible outcomes from the modeled scenarios mimicking these conditions. This finding reinforces the importance for demand-side measures to enable transport decarbonization that reduce fossil fuel consumption, including HDV vehicle efficiency improvements, as well as regulations to phase out new ICE vehicle sales etc.. Such measures have been introduced by some of the countries in our case study, albeit to a limited extent so far \citep{shapps_uk_2021, berge_na_2023}.

Our study notes the bulk system cost savings from adopting H$_2$ directly in HDVs, as opposed to system cost increases resulting from its use to first produce SFs that are then used in HDVs. While the bulk system cost savings are one measure of relevance, several other factors need to be considered when comparing the two different fuel options. In the case of H$_2$, substantial investment in distribution, refueling and vehicular infrastructure will be needed that is not accounted in our analysis. For SFs, these costs are expected to be small, because SFs can use the existing vehicular, distribution, and refueling infrastructure developed for fossil-based liquid fuels, thereby minimizing impact on vehicle owners and operators. 

We find that SFs adoption tends to lead to lower system cost increases in the absence of CO$_2$ sequestration capacity and when deployed at a limited scale. The widespread adoption of SFs (high SF adoption scenario representing 50\% of HDV demand) results in much larger cost impacts irrespective of the CO$_2$ storage assumption. These cost increases stem from the significant investment in H$_2$ and by extension electricity supply chains, but also CO$_2$ supply chains; in particular deployment and substantial scale up of emerging DAC technologies, as well as CO$_2$ transportation infrastructure. Further, the system carbon balance (Figure \ref{fig:sf_hdv_co2}) with SFs has many more components than the system without SFs adoption. At an individual producer level, ascertaining the low-carbon nature of SFs will require establishing regulations and possibly new tracking mechanisms that account for the induced grid emissions associated with each new load. Recent discussions in the U.S. and European context on ascertaining the carbon intensity of low-carbon H$_2$ \citep{ricks_minimizing_2023, giovanniello_influence_2024} are a harbinger to the challenges facing carbon intensity assessment for SF producers, who also have to consider embodied emissions burdens of the CO$_2$ feedstock. 


This study also reinforces findings from other studies regarding the impact of CO$_2$ sequestration availability on energy system decarbonization pathways \citep{mignone_drivers_2024, millinger_are_2022}. While spatially-resolved estimates of CO$_2$ storage capacity have been developed, other factors like social acceptance might constrain practically available CO$_2$ sequestration capacity \citep{huijts_social_2007}. At the same time, limiting sequestration capacity results in increased reliance on electrolyzer based H$_2$ production and consequently VRE generation for power supply, highlighting societal trade-offs implicit in the choice of decarbonization pathway \citep{mignone_drivers_2024}. Another interesting trade-off revealed in our analysis is the increase in NG consumption in lieu of liquid fossil fuels displacement via H$_2$ or SFs adoption. This is particularly relevant for policy makers to consider given the precarious nature of European NG supply after the Russian invasion of Ukraine \citep{iea_how_2022}. In our standard cases, our NG prices are synonymous with long-term projections for liquefied NG prices. That assumption, combined with the deep decarbonization emissions constraint ensure that overall NG consumption levels for power and H$_2$ production are well below 2019 levels in our cases. At the same time, we find that substitution of fossil liquids with NG is part of many of the cost-optimized decarbonization scenarios studied here and potentially provides a roadmap for reducing fossil liquid fuel imports in the European context, at the cost of increasing NG imports and reliance.  

The integrated energy system modeling framework used here allows for co-optimizing supply chains for electricity, H$_2$, CO$_2$ and liquid fuels and thus evaluate potential cross-sectoral impacts of multi-sector deep decarbonization. Through the case of power-H$_2$-transportation interactions in the European context, we highlight  key enabling opportunities for cost-beneficial sector coupling across sectors. 

This study has several important limitations that present opportunities for future research. In particular, we did not explore the role of bioenergy pathways in meeting demands for liquid fuels, hydrogen, and electricity under emissions-constrained scenarios. Prior energy system modeling studies have shown that available biomass resources are typically fully utilized under deep decarbonization pathways, either for electricity generation or fuel production—depending on the relative costs of production technologies and competing alternatives. However, these studies generally do not account for the potential of hydrogen to displace liquid fuels in the transportation sector. A valuable extension of this work would be to assess the relative value of hydrogen for transportation decarbonization in the context of competing bioenergy uses \citep{mignone_drivers_2024, millinger_are_2022}. Another area of interest is exploring alternative synthetic fuel production pathways with varying fuel product slates, to assess how product flexibility influences the relative cost-effectiveness of these pathways. Finally, it would be interesting to undertake a distribution-level assessment to quantify the cost of switching from liquid fuels to H$_2$ to support HDV transportation. Such a study would complement the findings of this work that focused on the bulk energy system assessment and thus provide a holistic assessment of H$_2$-based transport decarbonization.


\section{Conclusions}
Here, we used a multi-model approach to study the energy system impacts of H$_2$ and synthetic fuel adoption for decarbonizing the HDV transportation sector in the context of a Western European decarbonization case study. Our analysis leads to a few key policy-relevant observations. First, we find that H$_2$  use for HDVs reduces bulk system (power-H$_2$ and transportation) cost of deep decarbonization and decreases demand for fossil liquids, but could increase overall NG consumption compared to equivalent decarbonization scenarios without H$_2$ use for HDVs. Part of the cost saving stems from the substitution of more expensive conventional fossil liquid fuels (vs. NG on a per GJ of energy basis) for H$_2$ in end-use that also reduces need for atmospheric CO$_2$ removal via modeled DAC technologies. Second, limitations on CO$_2$ storage availability increase the bulk system cost savings (in absolute terms) of adopting H$_2$ use for HDVs. Third, the deployment of SFs results in substantial expansion of power and H$_2$ production capacity, with a preference for low carbon fuel generation sources (electrolyzers and SMR w/ CCS for H$_2$ production and VRE for power generation) to maximize carbon abatement benefits of SF use. This suggests that the impacts of SF adoption on other sectors should be considered when creating policies that encourage SF adoption. Fourth, while SF adoption generally increases bulk system costs, the cost increases vs. no SF adoption case are the lowest in case CO$_2$ storage availability is constrained and fossil fuels (NG and fossil liquids) are expensive. The role for H$_2$ for transport decarbonization reduces the upstream burden on the power and H$_2$ sectors but comes with the additional downstream challenges of deploying extensive distribution, refueling and vehicular infrastructure. Finally, our analysis highlights that the optimal-level of sectoral decarbonization is dependent on the technology pathways adopted and reinforces the use of multi-sector emissions reduction strategies.

\section{Acknowledgements}
The authors acknowledge the MIT SuperCloud and Lincoln Laboratory Supercomputing Center for providing HPC resources that have contributed to the research results reported within this paper \citep{reuther_interactive_2018}. Funding for this project was provided by the MIT Energy Initiative Future Energy Systems Center. The authors acknowledge Anna Cybulsky's contribution in developing the European energy system data set and visualizations.


\printcredits

\bibliography{trans_decarb_refs}


\newpage
\appendix
\renewcommand\thefigure{\thesection.\arabic{figure}} 
\renewcommand\thetable{\thesection.\arabic{table}} 
\setcounter{figure}{0}
\setcounter{table}{0}
\section{SI 1: Detailed Scenario Results}
\subsection{Core Scenario Set 1 Detailed Results}\label{sec:core_sen_1}

\begin{figure}[pos = H]
    \centering
    \includegraphics[width=1\linewidth]{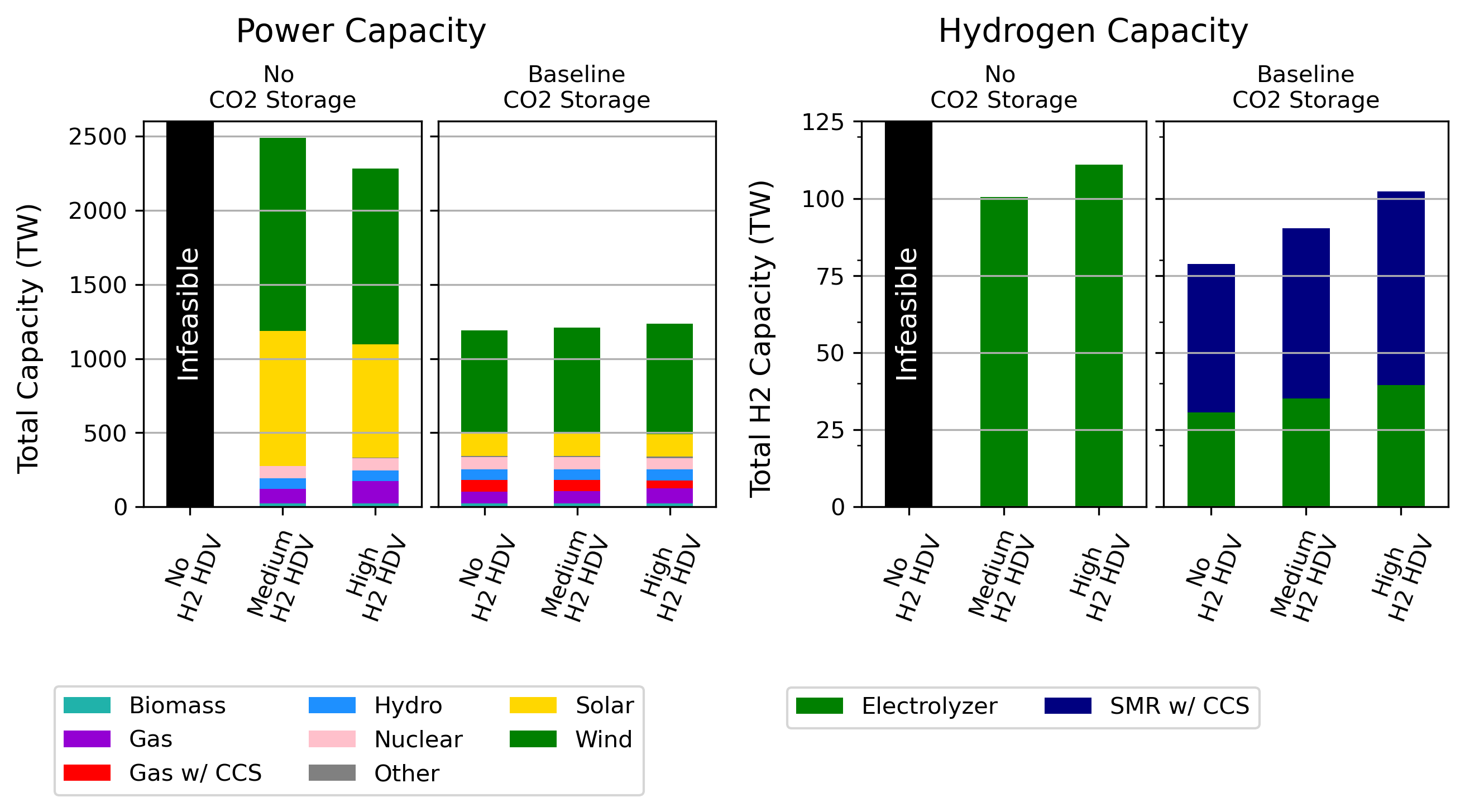}
    \caption[Power and H$_2$ Capacity Core Scenario Set 1]{Power and H$_2$ capacity for baseline and no CO$_2$ sequestration scenarios under no synthetic fuel adoption. The left set of charts shows power generation and the right set of charts shows H$_2$  generation. Within each panel, the amount of H$_2$ HDV adoption increases moving from left to right.}
    \label{fig_h2_hdv_power_h2_cap}
\end{figure}

\begin{figure}[pos = H]
    \centering
    \includegraphics[width=0.5\linewidth]{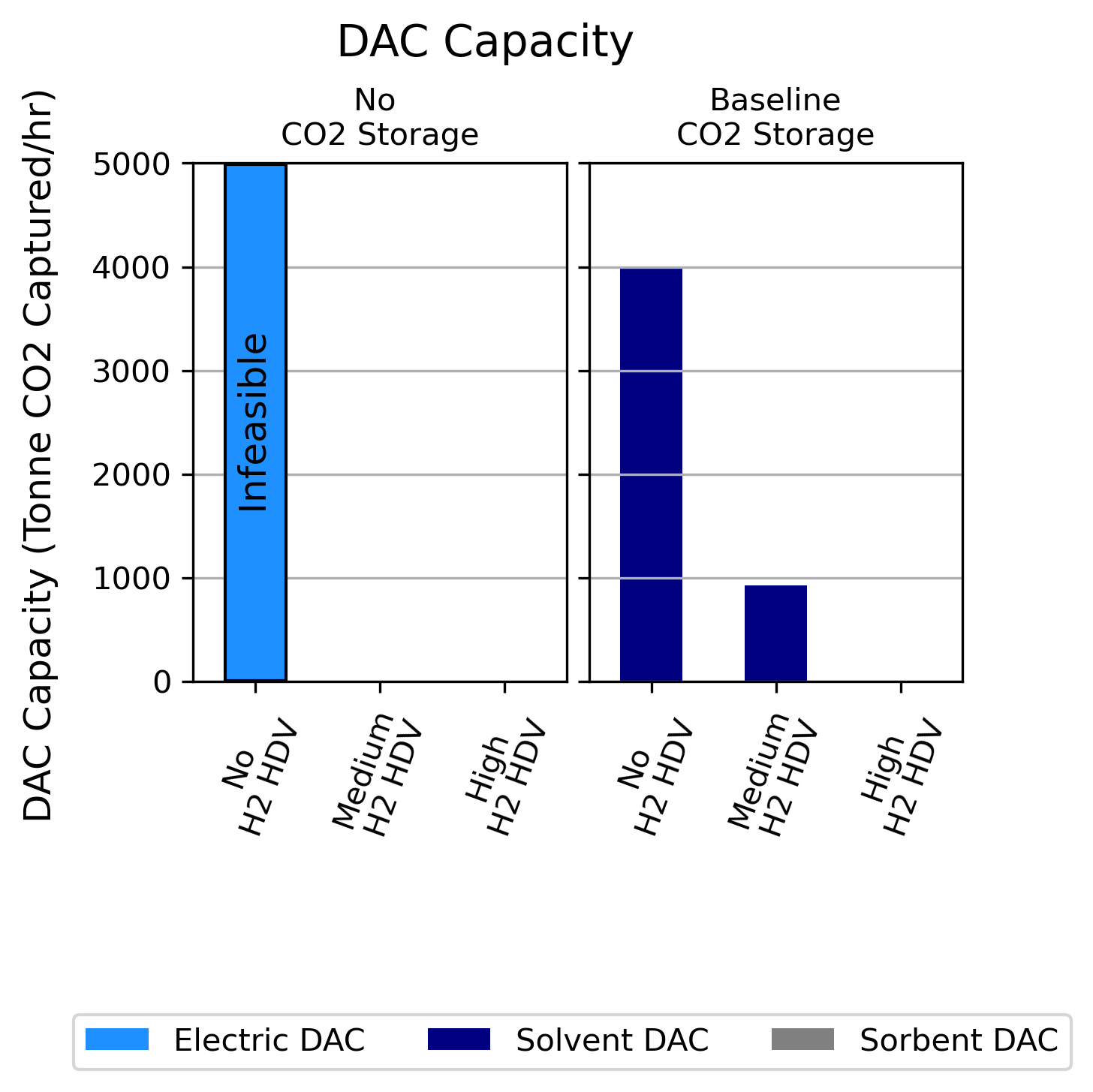}
    \caption[DAC Capacity Core Scenario Set 1]{Direct Air Capture capacity for baseline and no CO$_2$ sequestration scenarios under no synthetic fuel adoption. Within each panel, the amount of H$_2$ HDV adoption increases moving from left to right.}
    \label{fig_h2_hdv_dac_cap}
\end{figure}

\begin{figure}[pos = H]
    \centering
    \includegraphics[width=0.5\linewidth]{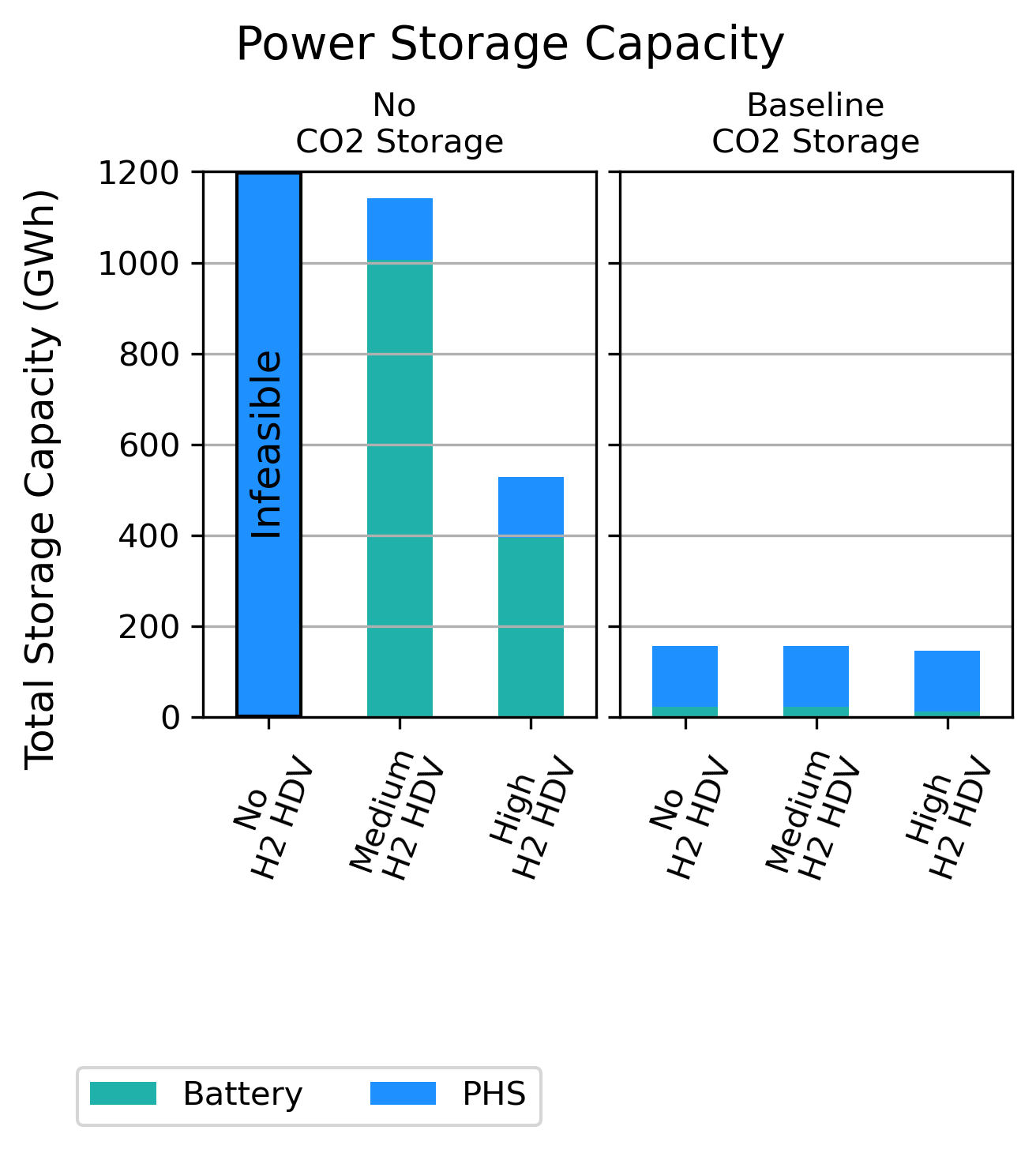}
    \caption[Power Storage Capacity Core Scenario Set 1]{Power storage capacity for baseline and no CO$_2$ sequestration scenarios under no synthetic fuel adoption. Within each panel, the amount of H$_2$ HDV adoption increases moving from left to right.}
    \label{fig_h2_hdv_power_cap_stor}
\end{figure}

\begin{figure}[pos = H]
    \centering
    \includegraphics[width=0.5\linewidth]{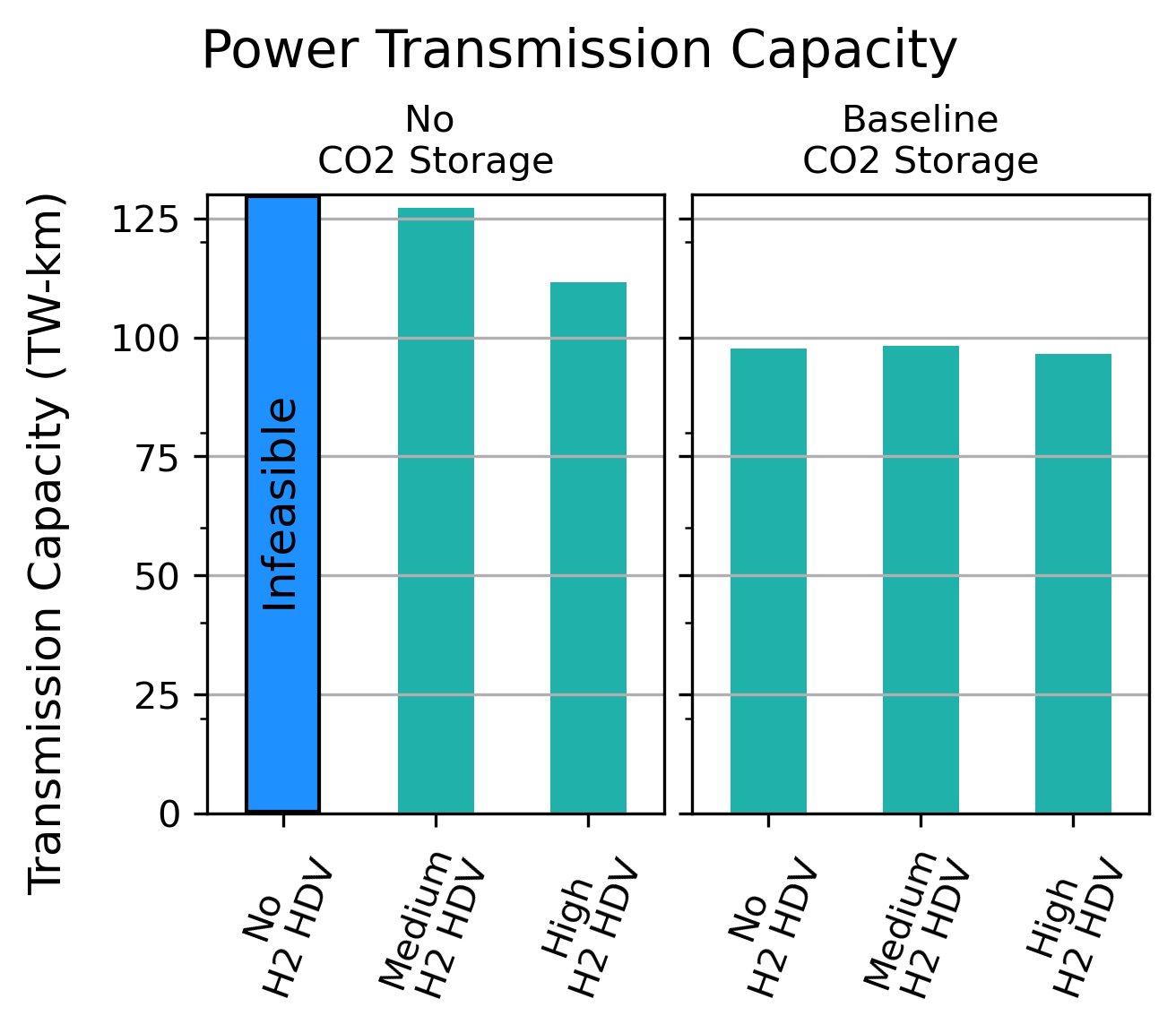}
    \caption[Power Transmission Capacity Core Scenario Set 1]{Power transmission capacity for baseline and no CO$_2$ sequestration scenarios under no synthetic fuel adoption. Within each panel, the amount of H$_2$ HDV adoption increases moving from left to right. }
    \label{fig_h2_hdv_power_trans}
\end{figure}

\begin{figure}[pos = H]
    \centering
    \includegraphics[width=0.5\linewidth]{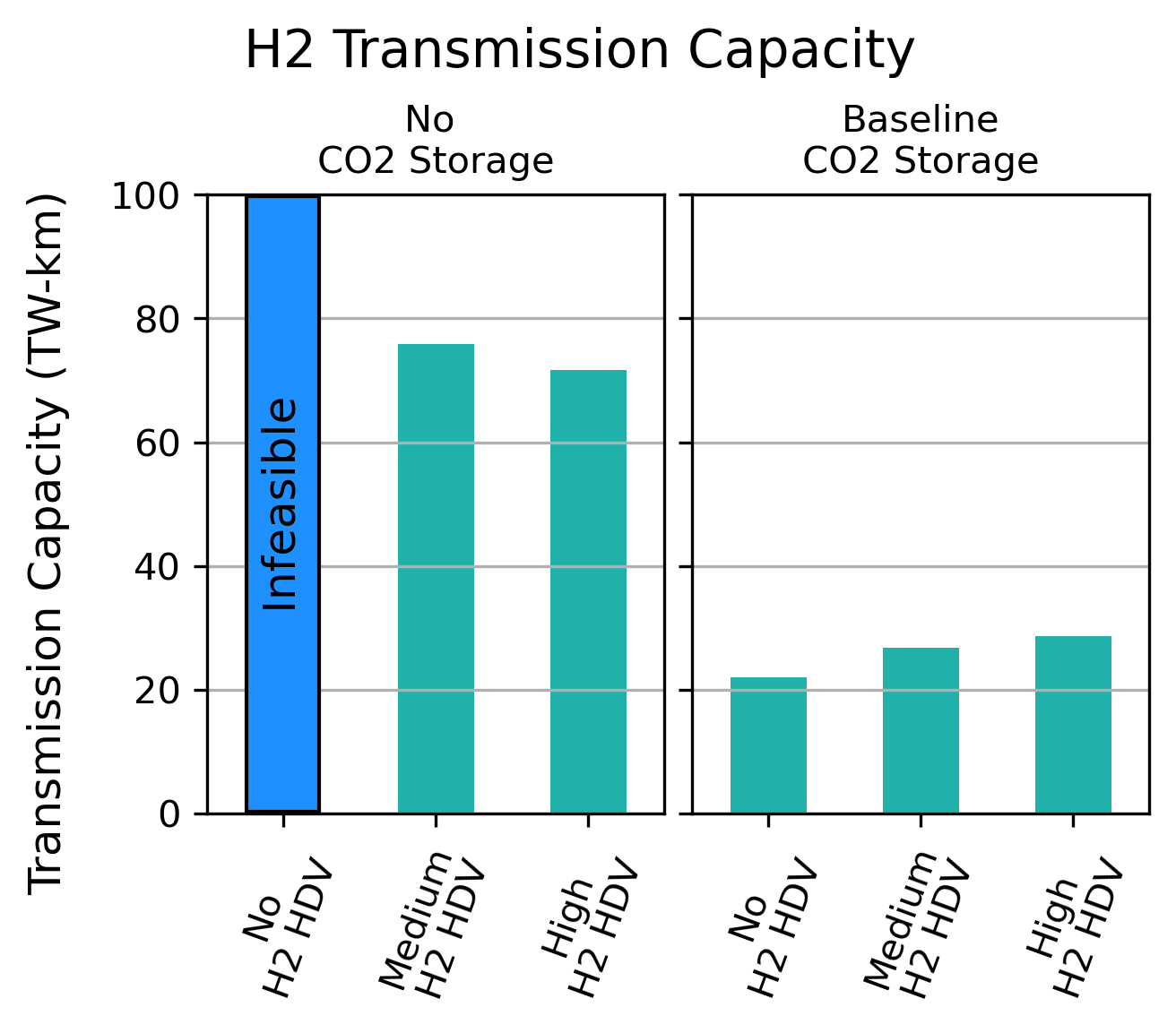}
    \caption[H$_2$ Transmission Capacity Core Scenario Set 1]{H$_2$ transmission capacity for baseline and no CO$_2$ sequestration scenarios under no synthetic fuel adoption. Within each panel, the amount of H$_2$ HDV adoption increases moving from left to right. }
    \label{fig_h2_hdv_h2_trans}
\end{figure}

\begin{figure}[pos = H]
    \centering
    \includegraphics[width=0.5\linewidth]{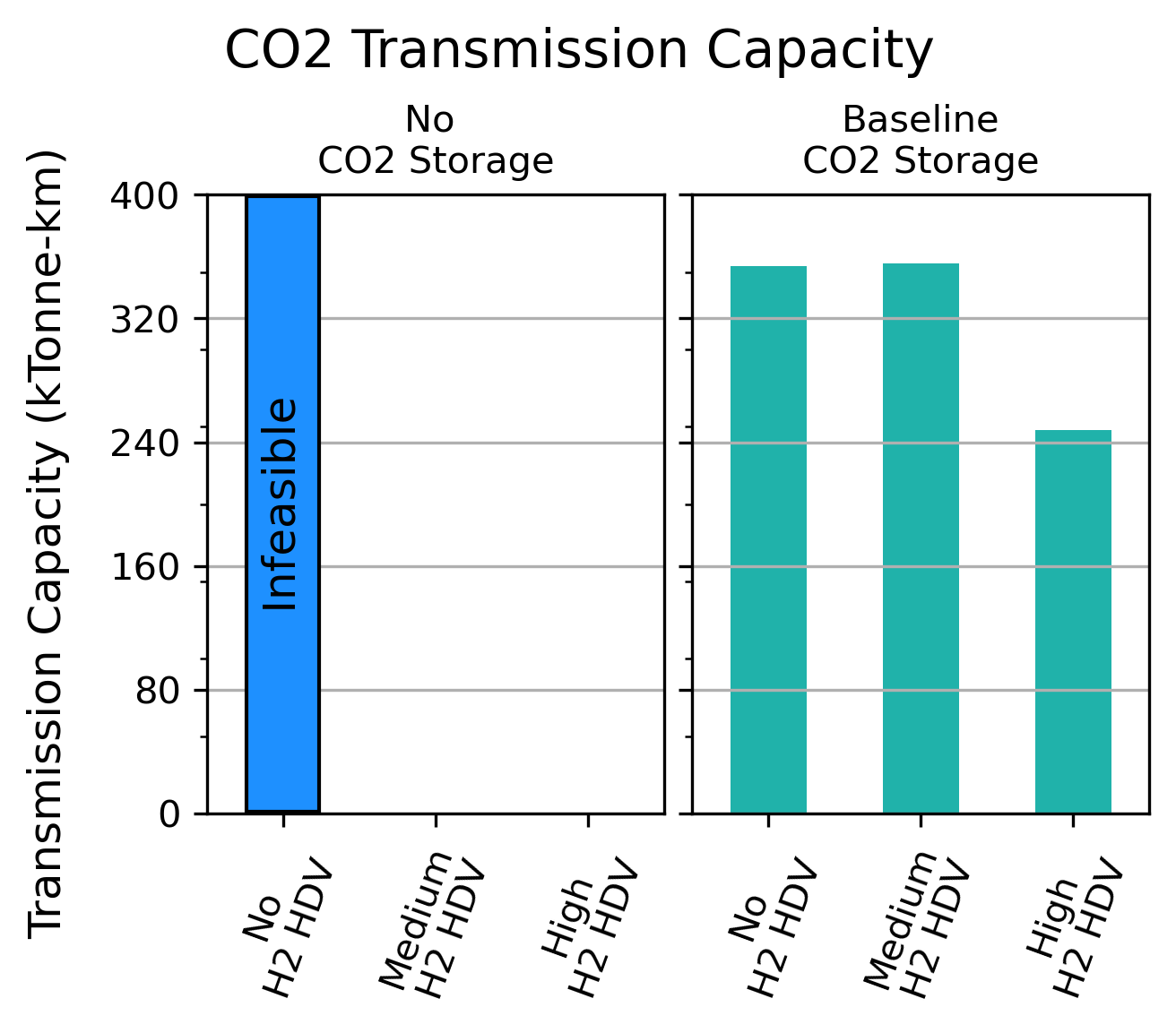}
    \caption[CO$_2$ Transmission Capacity Core Scenario Set 1]{CO$_2$ transmission capacity for baseline and no CO$_2$ sequestration scenarios under no synthetic fuel adoption. Within each panel, the amount of H$_2$ HDV adoption increases moving from left to right. }
    \label{fig_co2_hdv_h2_trans}
\end{figure}

\begin{table}[pos = H]
\caption{This tables shows the marginal price of abatement of CO$_2$ for Core Scenario Set 1}
\label{co2_price_core_set_1}
\begin{tabular}{lllr}
\toprule
CO$_2$ Storage & H$_2$ HDV Level & Synthetic Fuel HDV Level & CO$_2$ Marginal Cost of Abatement \\
\midrule
Baseline & None & None & 293.92 \\
Baseline & Medium & None & 293.92 \\
Baseline & High & None & 149.79 \\
None & Medium & None & 1523.03 \\
None & High & None & 746.25 \\
\bottomrule
\end{tabular}
\end{table}

\newpage

\subsection{Core Scenario Set 2 Detailed Results}\label{sec:core_sen_2}

\begin{figure}[pos = H]
    \centering
    \includegraphics[width=1\linewidth]{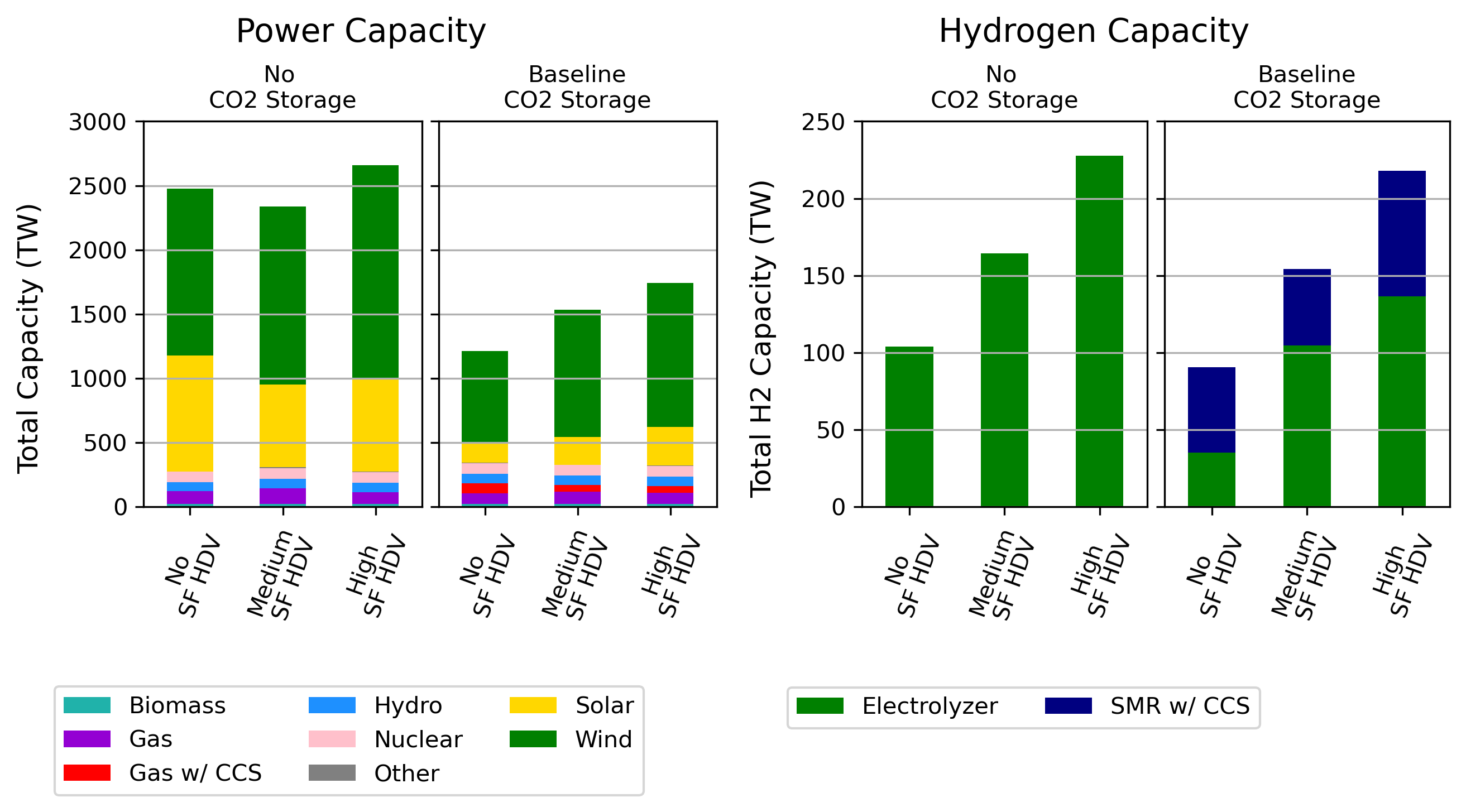}
    \caption[Power and H$_2$ Capacity Core Scenario Set 2]{Power and H$_2$ capacity for baseline and no CO$_2$ sequestration scenarios under medium H$_2$ HDV adoption and varying scenarios of synthetic fuel adoption. The left set of charts shows power generation and the right set of charts shows H$_2$ generation. Within each panel, the amount of synthetic fuel adoption increases moving from left to right.}
    \label{fig_sf_hdv_power_h2_cap}
\end{figure}

\begin{figure}[pos = H]
    \centering
    \includegraphics[width=0.5\linewidth]{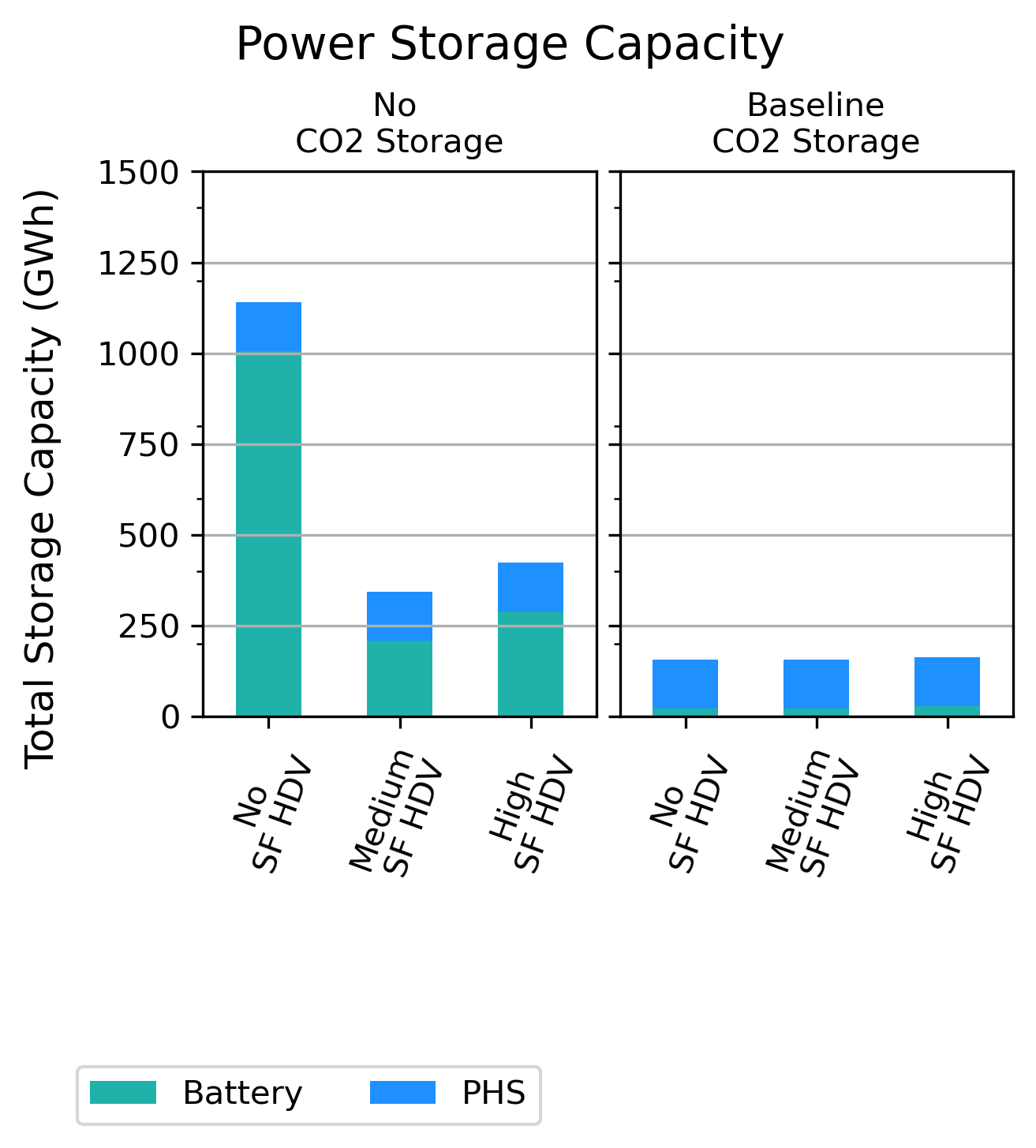}
    \caption[Power Storage Capacity Core Scenario Set 2]{Power storage capacity for baseline and no CO$_2$ sequestration scenarios under medium H$_2$ HDV adoption and varying scenarios of synthetic fuel adoption. Within each panel, the amount of synthetic fuel adoption increases moving from left to right.}
    \label{fig_sf_hdv_power_cap_stor}
\end{figure}

\begin{figure}[pos = H]
    \centering
    \includegraphics[width=0.5\linewidth]{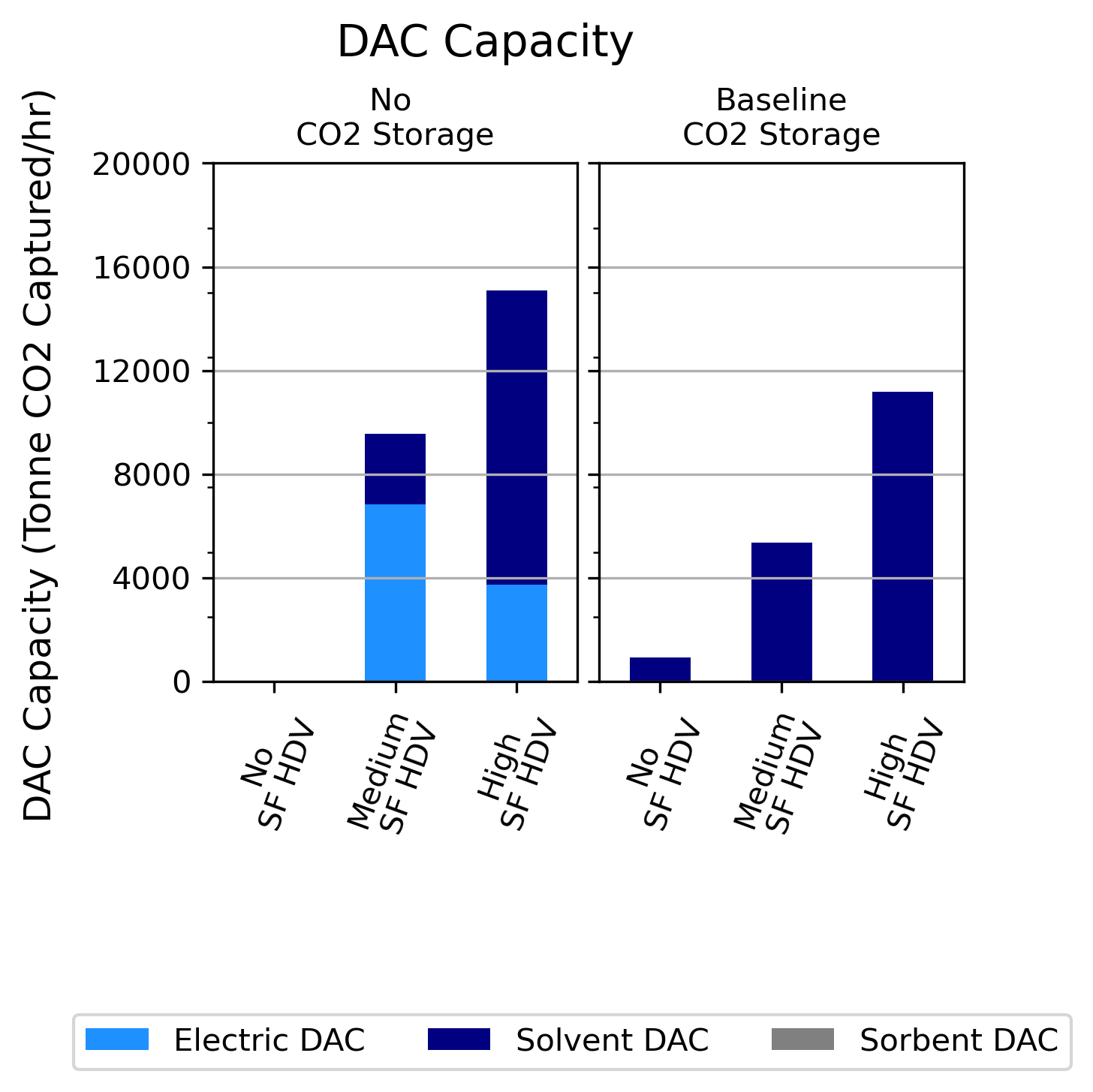}
    \caption[DAC Capacity Core Scenario Set 1]{Direct Air Capture capacity for baseline and no CO$_2$ sequestration scenarios under medium H$_2$ HDV adoption and varying scenarios of synthetic fuel adoption. Within each panel, the amount of synthetic fuel adoption increases moving from left to right.}
    \label{fig_sf_hdv_dac_cap}
\end{figure}

\begin{figure}[pos = H]
    \centering
    \includegraphics[width=0.5\linewidth]{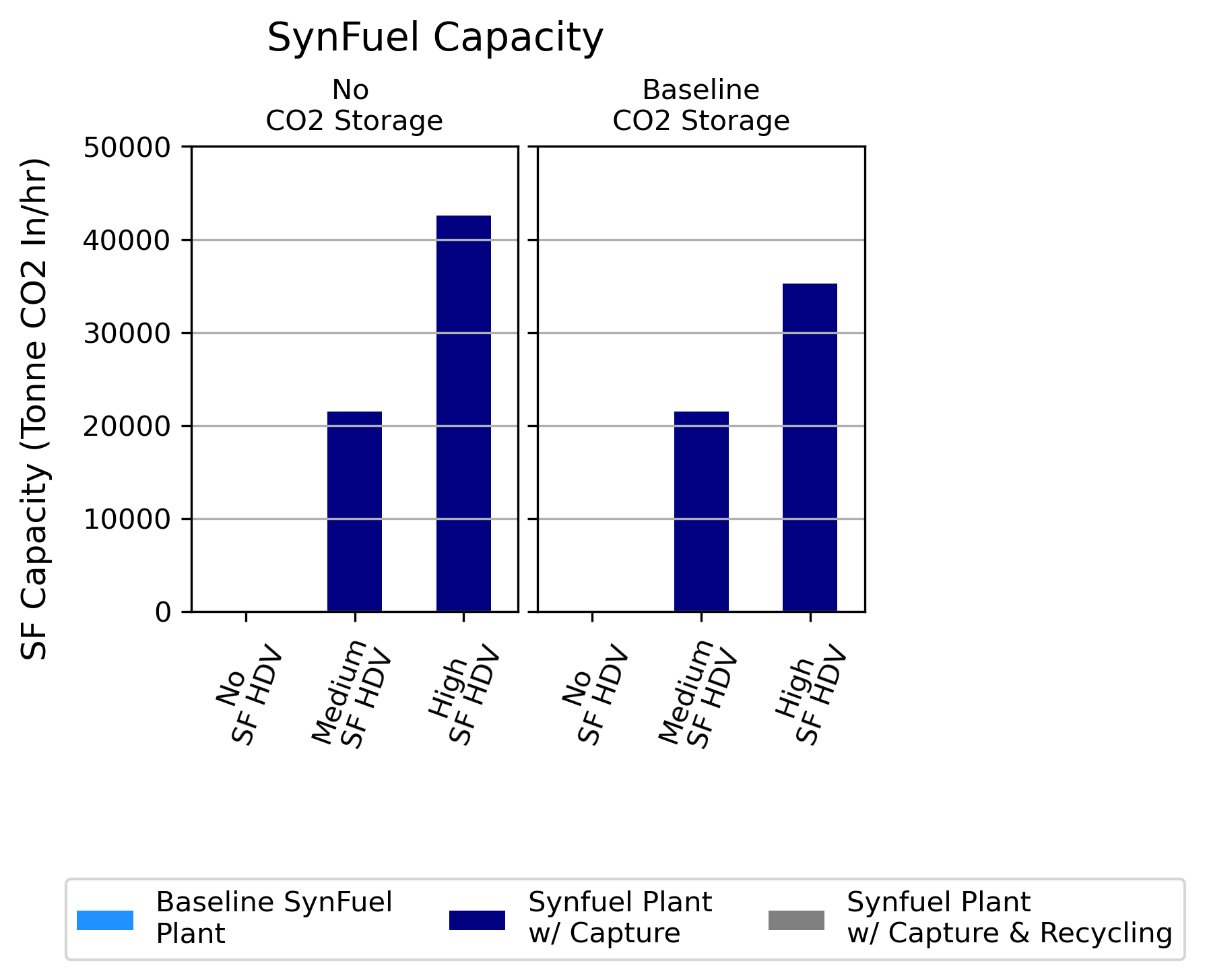}
    \caption[synthetic fuel Capacity Core Scenario Set 1]{synthetic fuel capacity for baseline and no CO$_2$ sequestration scenarios under medium H$_2$ HDV adoption and varying scenarios of synthetic fuel adoption. Within each panel, the amount of synthetic fuel adoption increases moving from left to right.}
    \label{fig_sf_hdv_sf_cap}
\end{figure}

\begin{figure}[pos = H]
    \centering
    \includegraphics[width=0.5\linewidth]{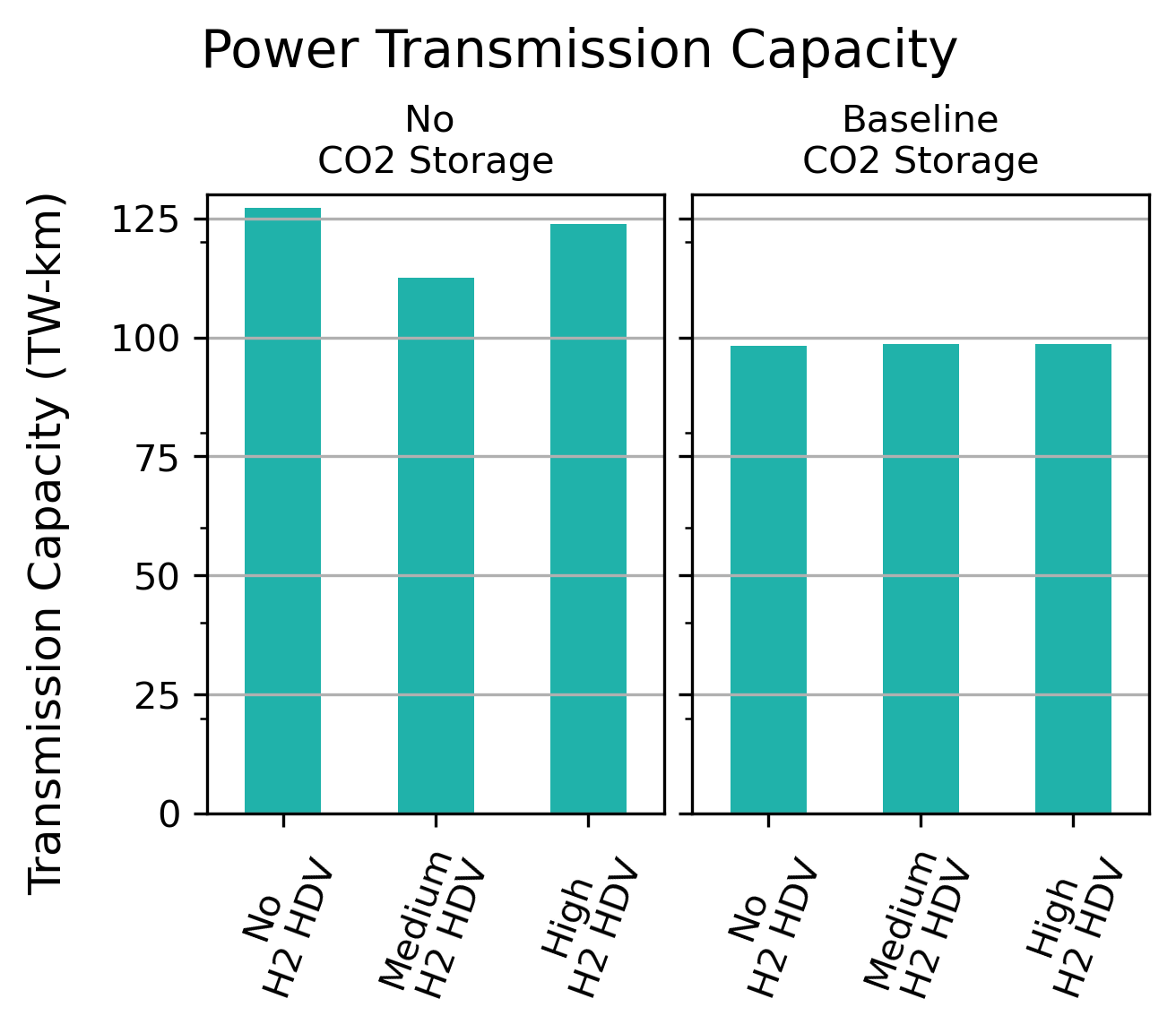}
    \caption[Power Transmission Capacity Core Scenario Set 2]{Power transmission capacity for baseline and no CO$_2$ sequestration scenarios under medium H$_2$ HDV adoption and varying scenarios of synthetic fuel adoption. Within each panel, the amount of synthetic fuel adoption increases moving from left to right. }
    \label{fig_sf_hdv_power_trans}
\end{figure}

\begin{figure}[pos = H]
    \centering
    \includegraphics[width=0.5\linewidth]{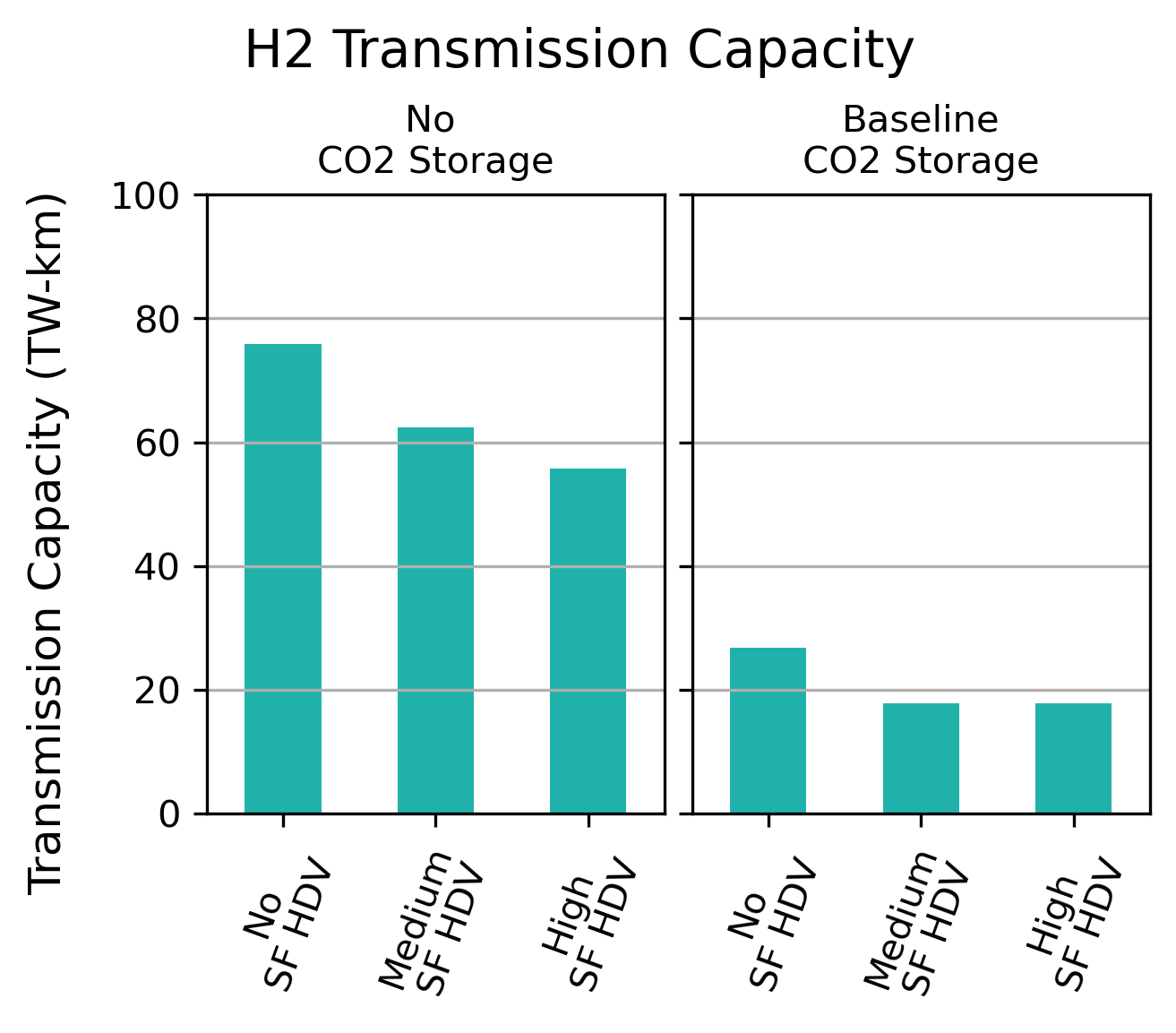}
    \caption[H$_2$ Transmission Capacity Core Scenario Set 2]{H$_2$ transmission capacity for baseline and no CO$_2$ sequestration scenarios under medium H$_2$ HDV adoption and varying scenarios of synthetic fuel adoption. Within each panel, the amount of synthetic fuel adoption increases moving from left to right. }
    \label{fig_sf_hdv_h2_trans}
\end{figure}

\begin{figure}[pos = H]
    \centering
    \includegraphics[width=0.5\linewidth]{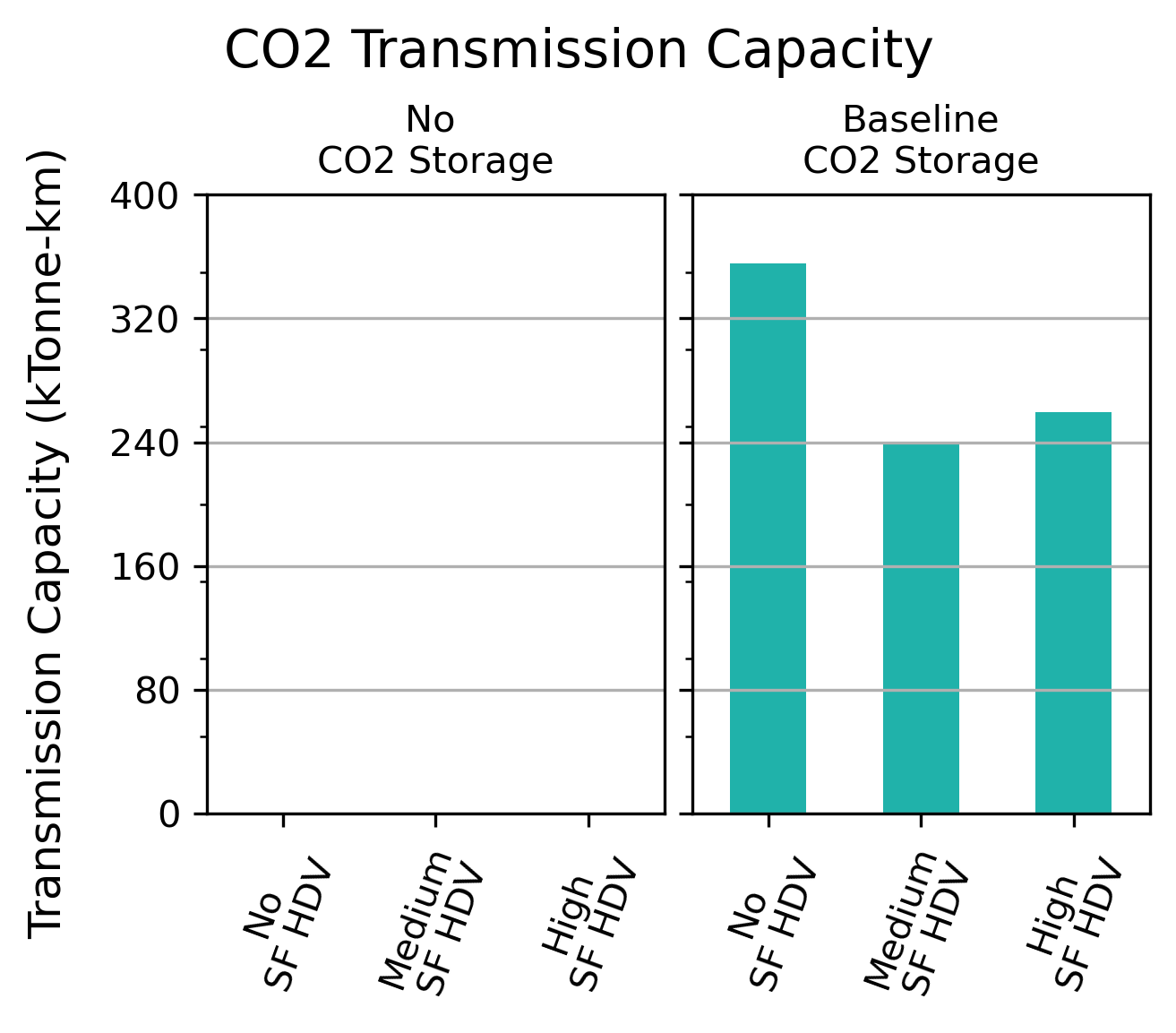}
    \caption[CO$_2$ Transmission Capacity Core Scenario Set 2]{CO$_2$ transmission capacity for baseline and no CO$_2$ sequestration scenarios under medium H$_2$ HDV adoption and varying scenarios of synthetic fuel adoption. Within each panel, the amount of synthetic fuel adoption increases moving from left to right. }
    \label{fig_sf_hdv_co2_trans}
\end{figure}

\begin{table}
\caption{This tables shows the marginal price of abatement of CO$_2$ for Core Scenario Set 2}
\label{co2_price_core_set_2}
\begin{tabular}{lllr}
\toprule
CO$_2$ Storage & H$_2$ HDV Level & Synthetic Fuel HDV Level & CO$_2$ Marginal Cost of Abatement \\
\midrule
Baseline & Medium & None & 293.92 \\
Baseline & Medium & Medium & 268.80 \\
Baseline & Medium & High & 293.91 \\
None & Medium & None & 1603.09 \\
None & Medium & Medium & 480.12 \\
None & Medium & High & 567.48 \\
\bottomrule
\end{tabular}
\end{table}

\newpage
\subsection{Sensitivity Set 1: Core Scenario Set 1 with Relaxed Emissions Constraint}\label{sec:sen_scenario_1}

The results in this section represent Sensitivity Set 1 as described in Figure 3b of the main text.

\begin{figure}[pos = H]
    \centering
    \includegraphics[width=1\linewidth]{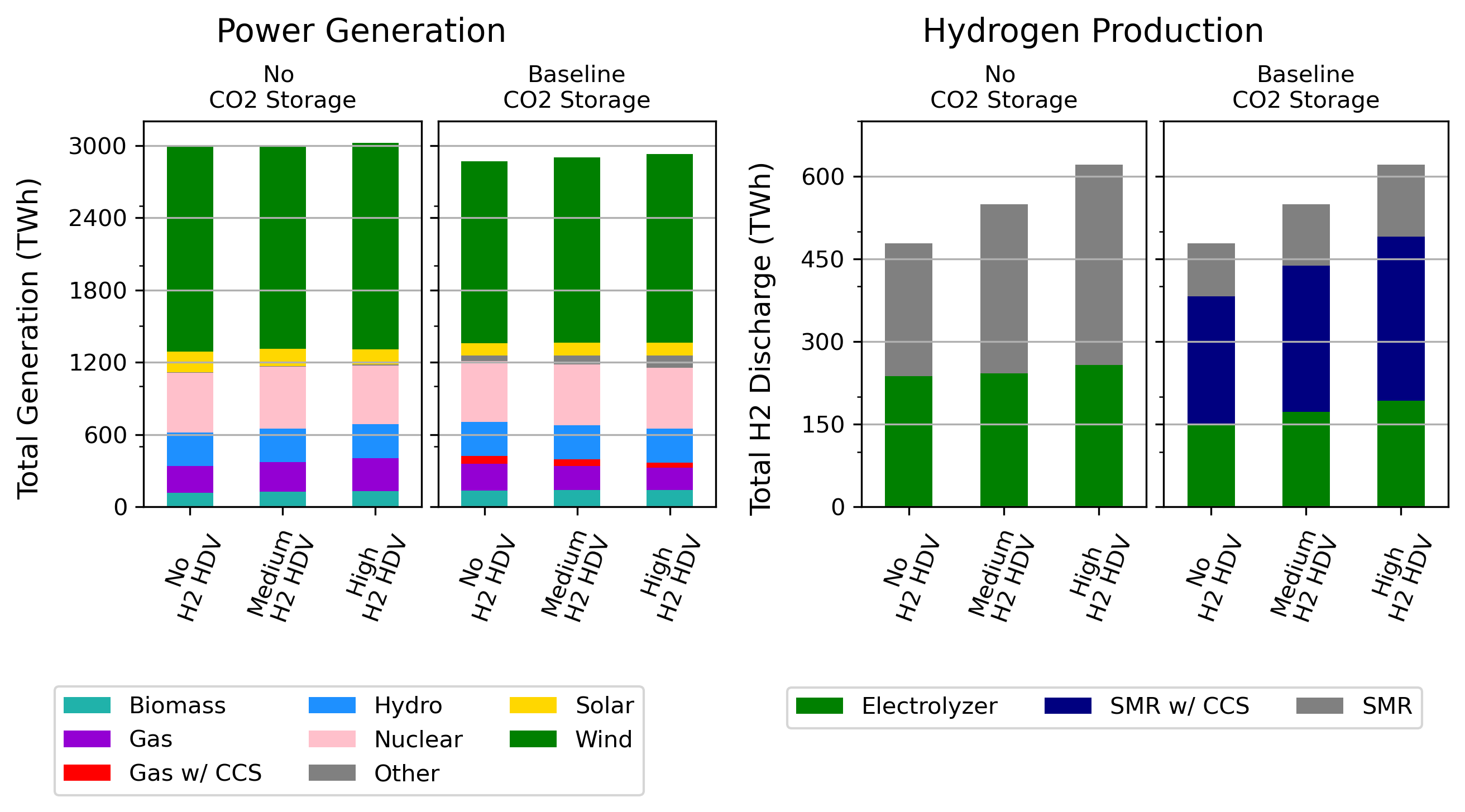}
    \caption[Power and H$_2$ Generation Sensitivity Scenario Set 1]{Power and H$_2$ generation for baseline and no CO$_2$ sequestration scenarios under no synthetic fuel adoption. The left set of charts shows power generation and the right set of charts shows H$_2$  generation. Within each panel, the amount of H$_2$ HDV adoption increases moving from left to right. CO$_2$ constraint is relaxed compared to core scenario set 1.}
    \label{fig_h2_hdv_power_h2_emissions_sen}
\end{figure}

\begin{figure}[pos = H]
    \centering
    \includegraphics[width=1\linewidth]{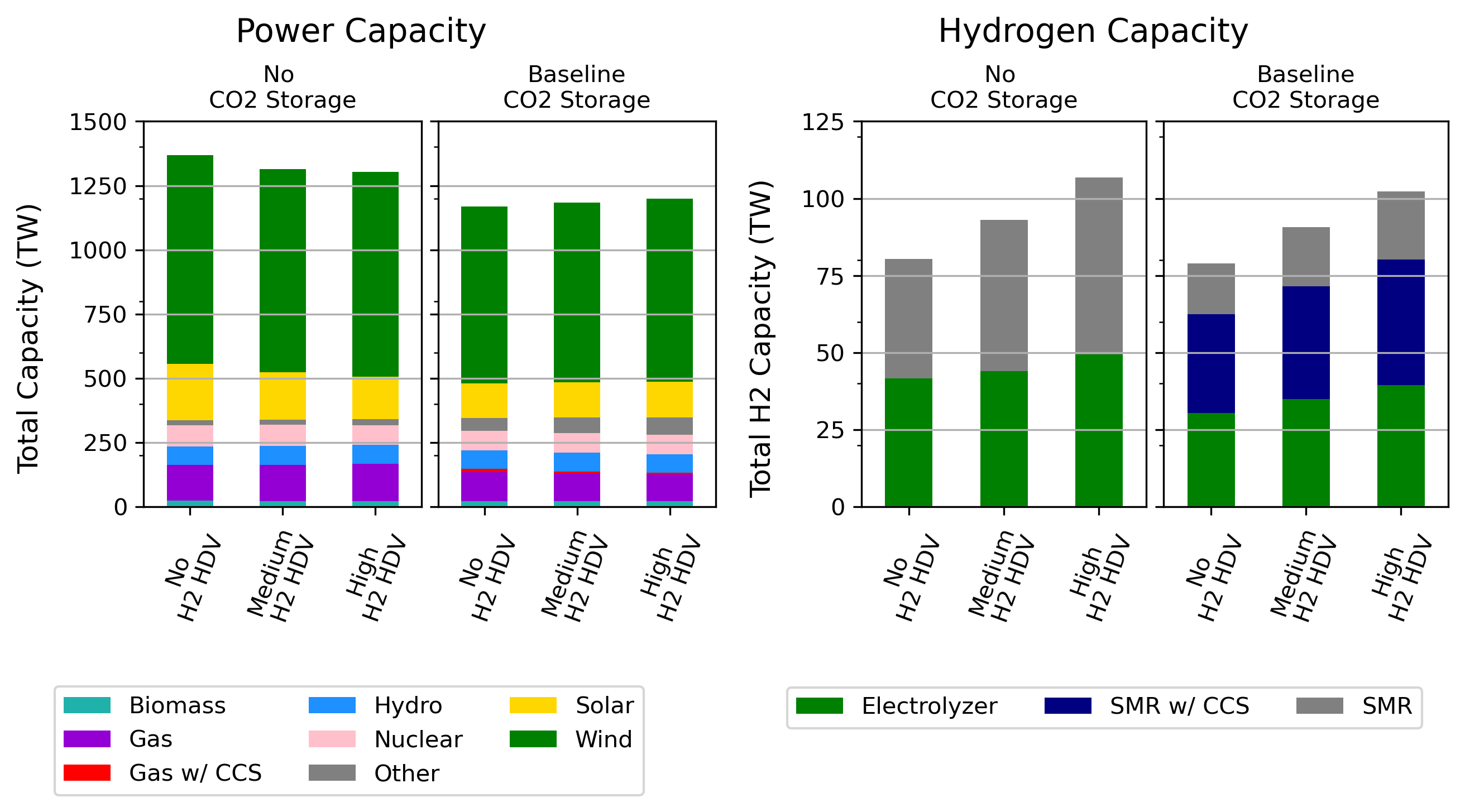}
    \caption[Power and H$_2$ Capacity Sensitivity Scenario Set 1]{Power and H$_2$ capacity for baseline and no CO$_2$ sequestration scenarios under no synthetic fuel adoption. The left set of charts shows power generation and the right set of charts shows H$_2$  generation. Within each panel, the amount of H$_2$ HDV adoption increases moving from left to right. CO$_2$ constraint is relaxed compared to core scenario set 1.}
    \label{fig_h2_hdv_power_h2_cap_emissions_sen}
\end{figure}

No DAC is deployed in this scenario set. 

\begin{figure}[pos = H]
    \centering
    \includegraphics[width=0.5\linewidth]{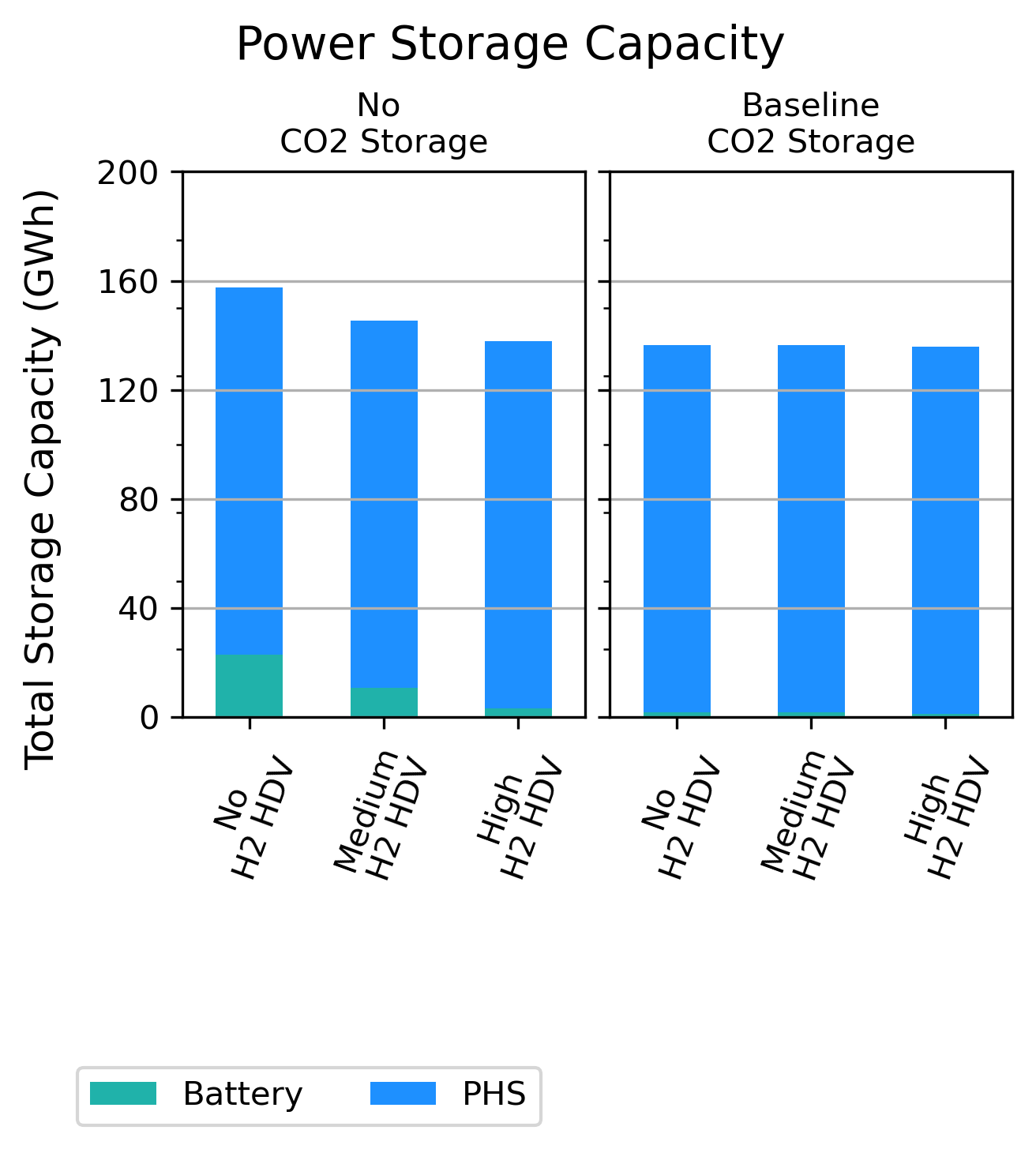}
    \caption[Power Storage Capacity Sensitivity Scenario Set 1]{Power storage capacity for baseline and no CO$_2$ sequestration scenarios under no synthetic fuel adoption. Within each panel, the amount of H$_2$ HDV adoption increases moving from left to right. CO$_2$ constraint is relaxed compared to core scenario set 1.}
    \label{fig_h2_hdv_power_cap_stor_emissions_sen}
\end{figure}

\begin{figure}[pos = H]
    \centering
    \includegraphics[width=0.5\linewidth]{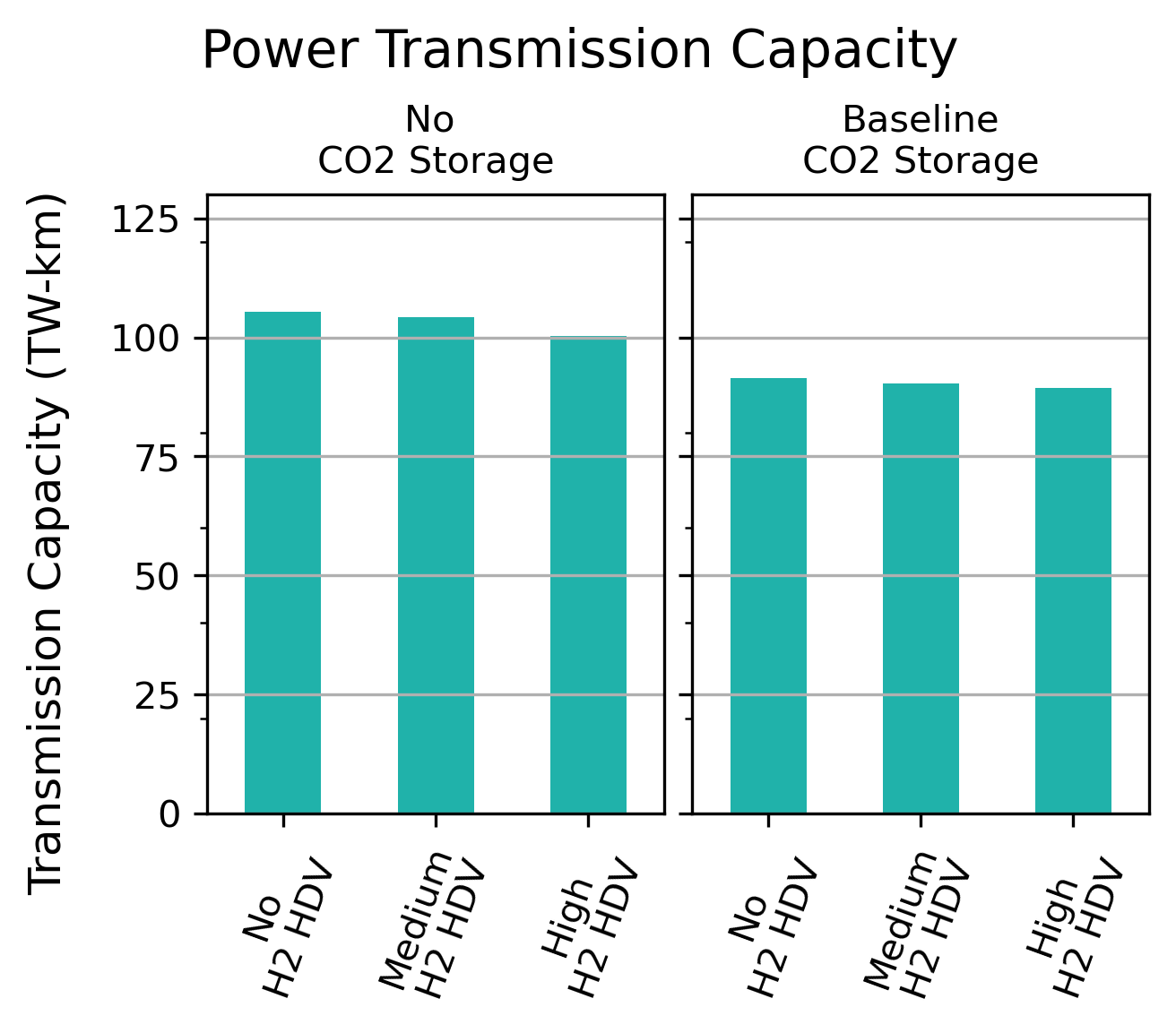}
    \caption[Power Transmission Capacity Sensitivity Scenario Set 1]{Power transmission capacity for baseline and no CO$_2$ sequestration scenarios under no synthetic fuel adoption. Within each panel, the amount of H$_2$ HDV adoption increases moving from left to right. CO$_2$ constraint is relaxed compared to core scenario set 1. }
    \label{fig_h2_hdv_elec_trans_emissions_sen}
\end{figure}

\begin{figure}[pos = H]
    \centering
    \includegraphics[width=0.5\linewidth]{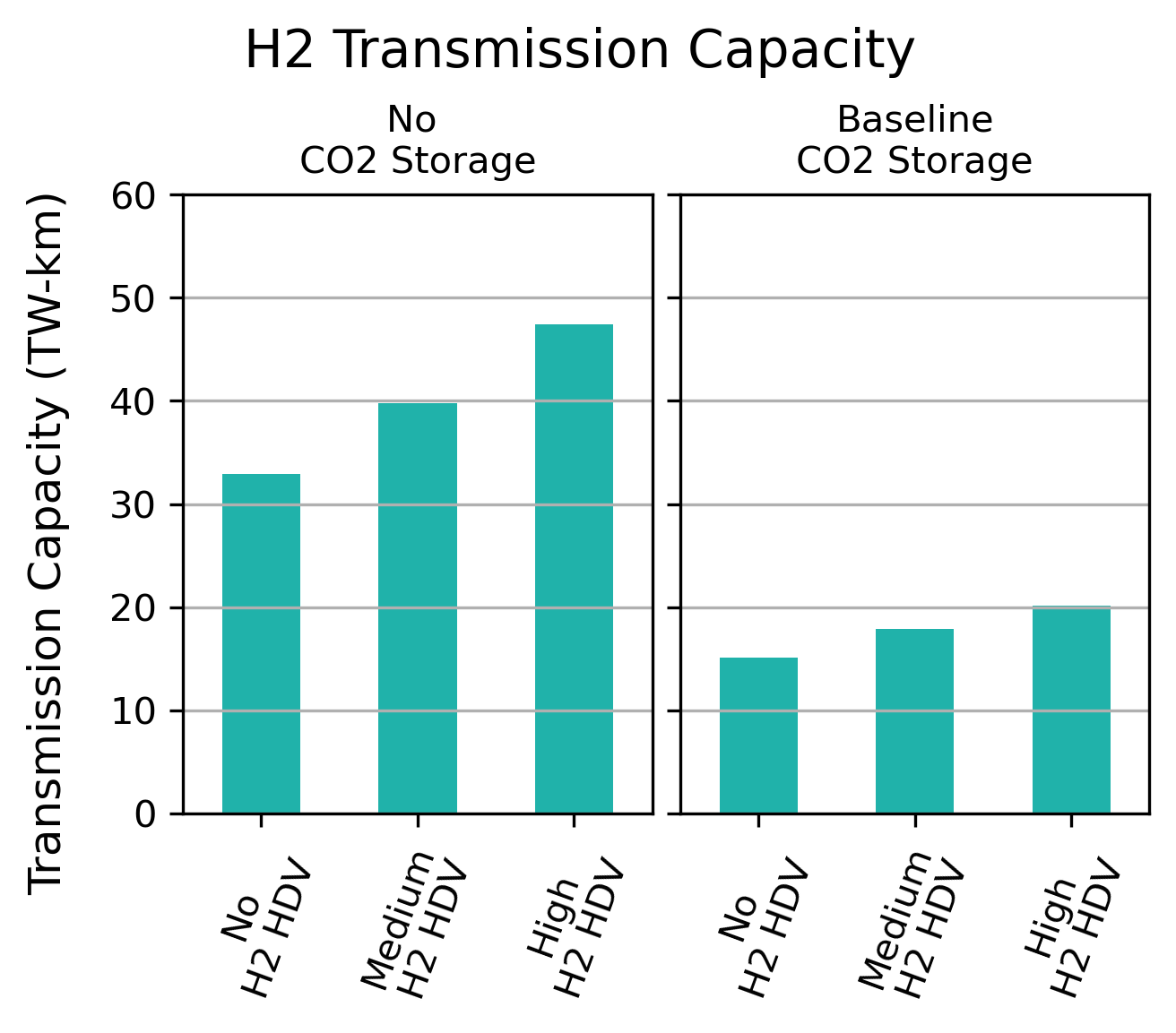}
    \caption[H$_2$ Transmission Capacity Sensitivity Scenario Set 1]{H$_2$ transmission capacity for baseline and no CO$_2$ sequestration scenarios under no synthetic fuel adoption. Within each panel, the amount of H$_2$ HDV adoption increases moving from left to right. CO$_2$ constraint is relaxed compared to core scenario set 1. }
    \label{fig_h2_hdv_h2_trans_emissions_sen}
\end{figure}

No CO$_2$ transmission is build in this scenario set.

\begin{figure}[pos = H]
    \centering
    \includegraphics[width=1\linewidth]{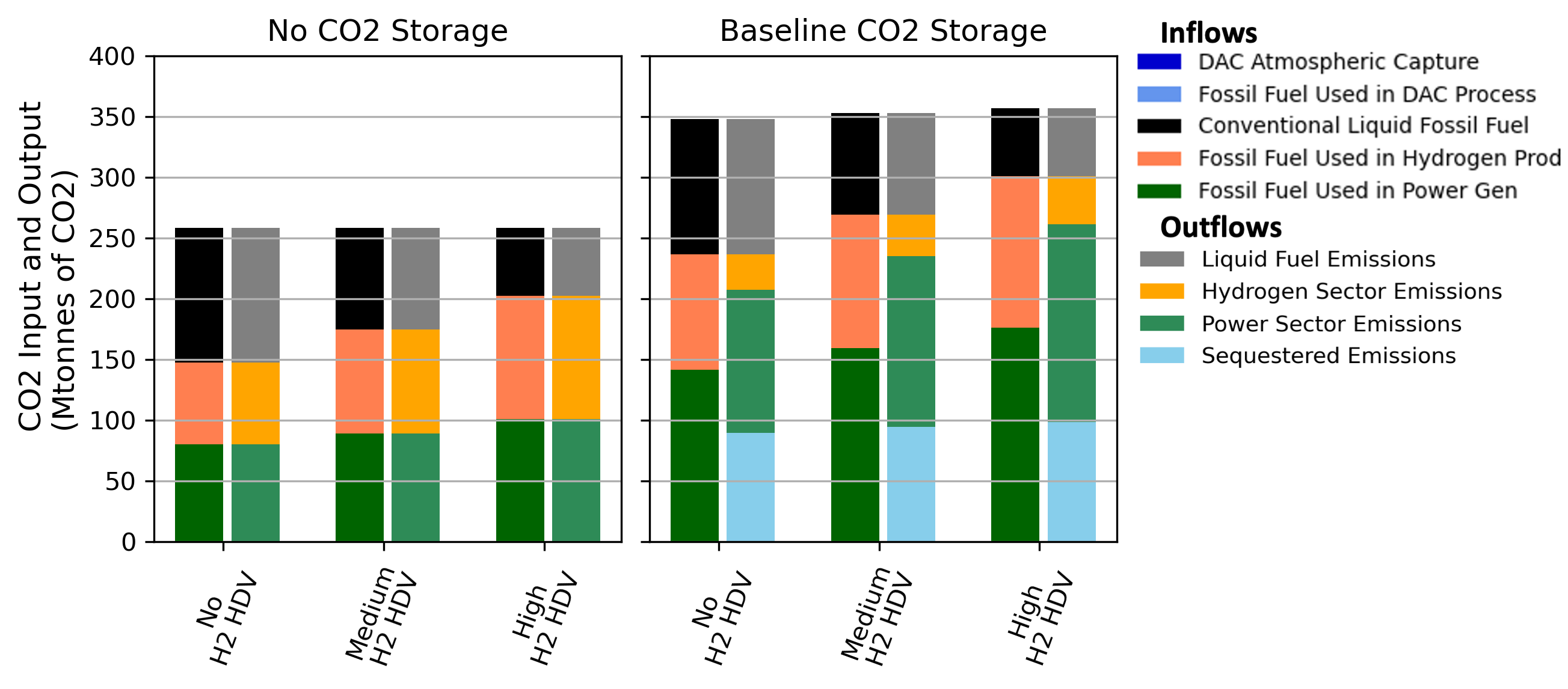}
    \caption[CO$_2$ Balance Sensitivity Scenario Set 1]{System CO$_2$ balance under varying levels of H$_2$ HDV adoption and no SF adoption. The subfigure on the left shows the CO$_2$ balance under no CO$_2$ sequestration availability, while the one on the right shows the CO$_2$ balance under baseline CO$_2$ storage availability. Within each subplot the H$_2$ HDV adoption level increases left to right. The leftward column represents CO$_2$ input into the system, while the rightward column represents CO$_2$ outputted by the system. All scenarios adhere to the same emissions constraint of 258 MTonnes. CO$_2$ constraint is relaxed compared to core scenario set 1. Emissions constraint can be calculated from the chart by subtracting sequestered emissions and DAC atmospheric capture from the emission outflows.}
    \label{fig_h2_hdv_co2_emissions_sen}
\end{figure}

\begin{figure}[pos = H]
    \centering
    \includegraphics[width=1\linewidth]{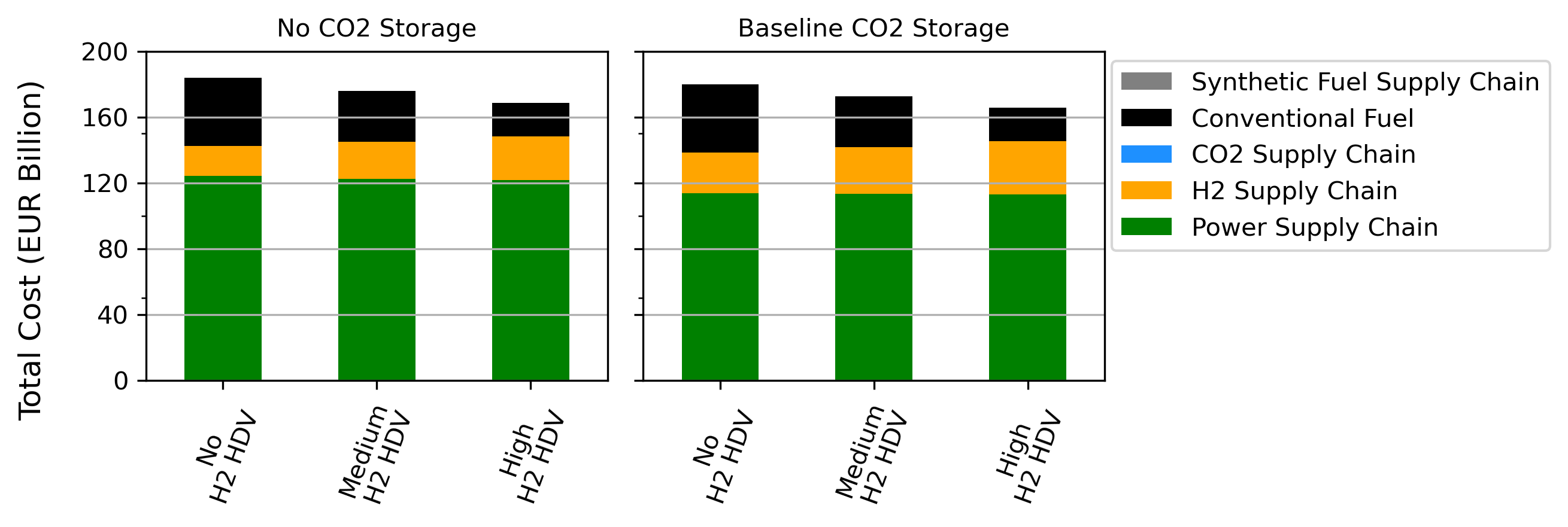}
    \caption[Costs Sensitivity Scenario Set 1]{Annualized bulk-system costs under varying levels of H$_2$ HDV adoption and no SF adoption.The subfigure on the left shows the cost breakdown under no CO$_2$ sequestration availability, while the one on the right shows the cost breakdown under baseline CO$_2$ sequestration availability. Within each subplot the H$_2$ HDV adoption level increases left to right. CO$_2$ constraint is relaxed compared to core scenario set 1. The costs do not include vehicle replacement or H$_2$ distribution costs.}
    \label{fig:h2_hdv_cost_emissions_sen}
\end{figure}

\begin{table}[pos = H]
\caption{This tables shows the marginal price of abatement of CO$_2$ for Sensitivity Set 1}
\label{co2_price_sen_set_1}
\begin{tabular}{lllr}
\toprule
CO$_2$ Storage & H$_2$ HDV Level & Synthetic Fuel HDV Level & CO$_2$ Marginal Cost of Abatement \\
\midrule
Baseline & None & None & 71.88 \\
Baseline & Medium & None & 69.72 \\
Baseline & High & None & 67.32 \\
None & None & None & 159.88 \\
None & Medium & None & 118.86 \\
None & High & None & 90.71 \\
\bottomrule
\end{tabular}
\end{table}

\newpage
\subsection{Sensitivity Set 2: Core Scenario Set 1 Natural Gas Price Sensitivity}\label{sec:sen_scenario_2}

The results in this section represent Sensitivity Set 2 as described in Figure 3b of the main text.

\begin{figure}[pos = H]
    \centering
    \begin{subfigure}[t]{0.8\linewidth}
        \includegraphics[width=\linewidth, trim=0cm 2.6cm 0cm 0.7cm, clip]{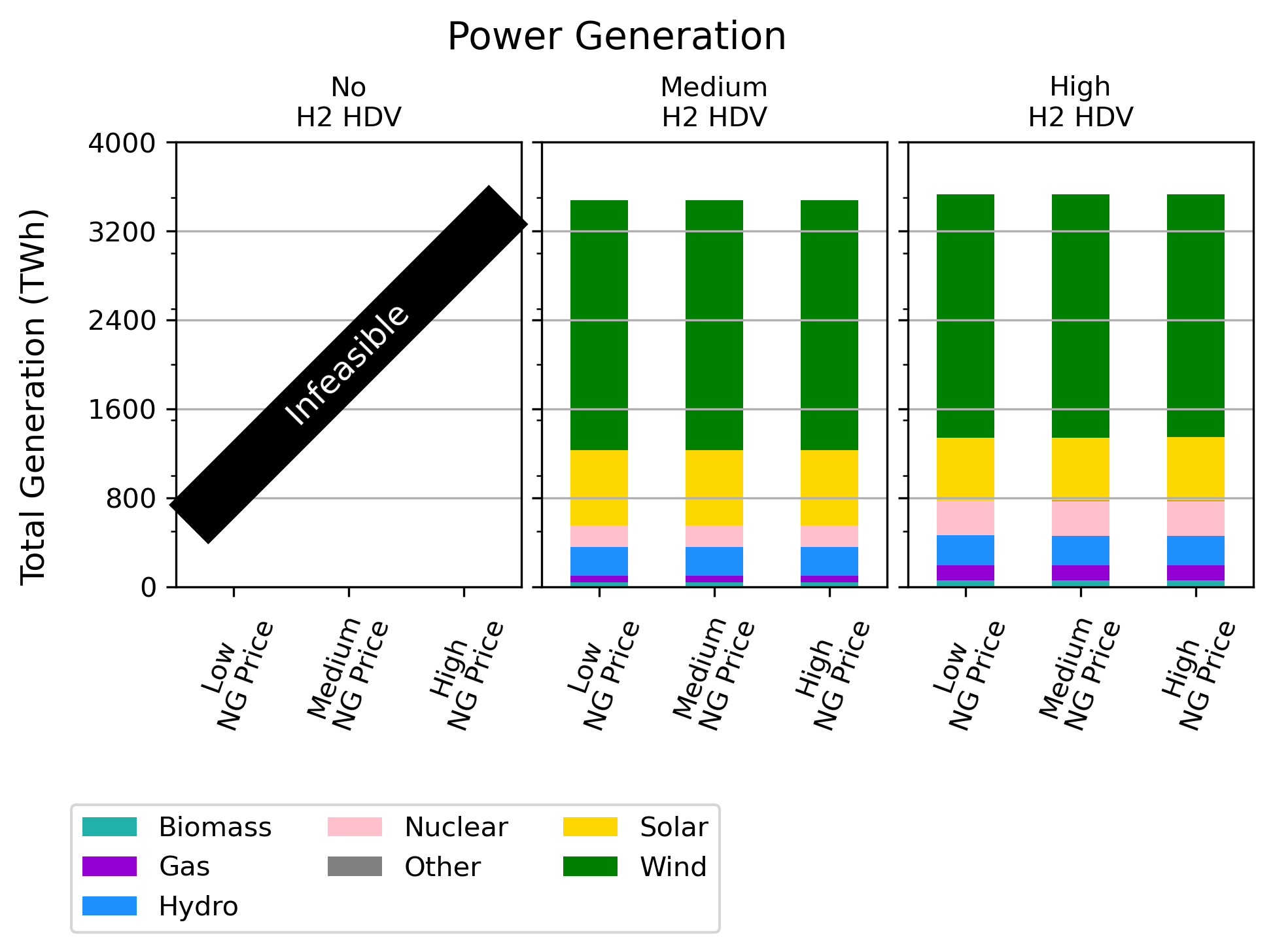}
        \caption{No CO$_2$ Storage}
    \end{subfigure}
    
    \begin{subfigure}[b]{0.8\linewidth}
        \includegraphics[width=\linewidth, trim=0cm 0cm 0cm 0.7cm, clip]{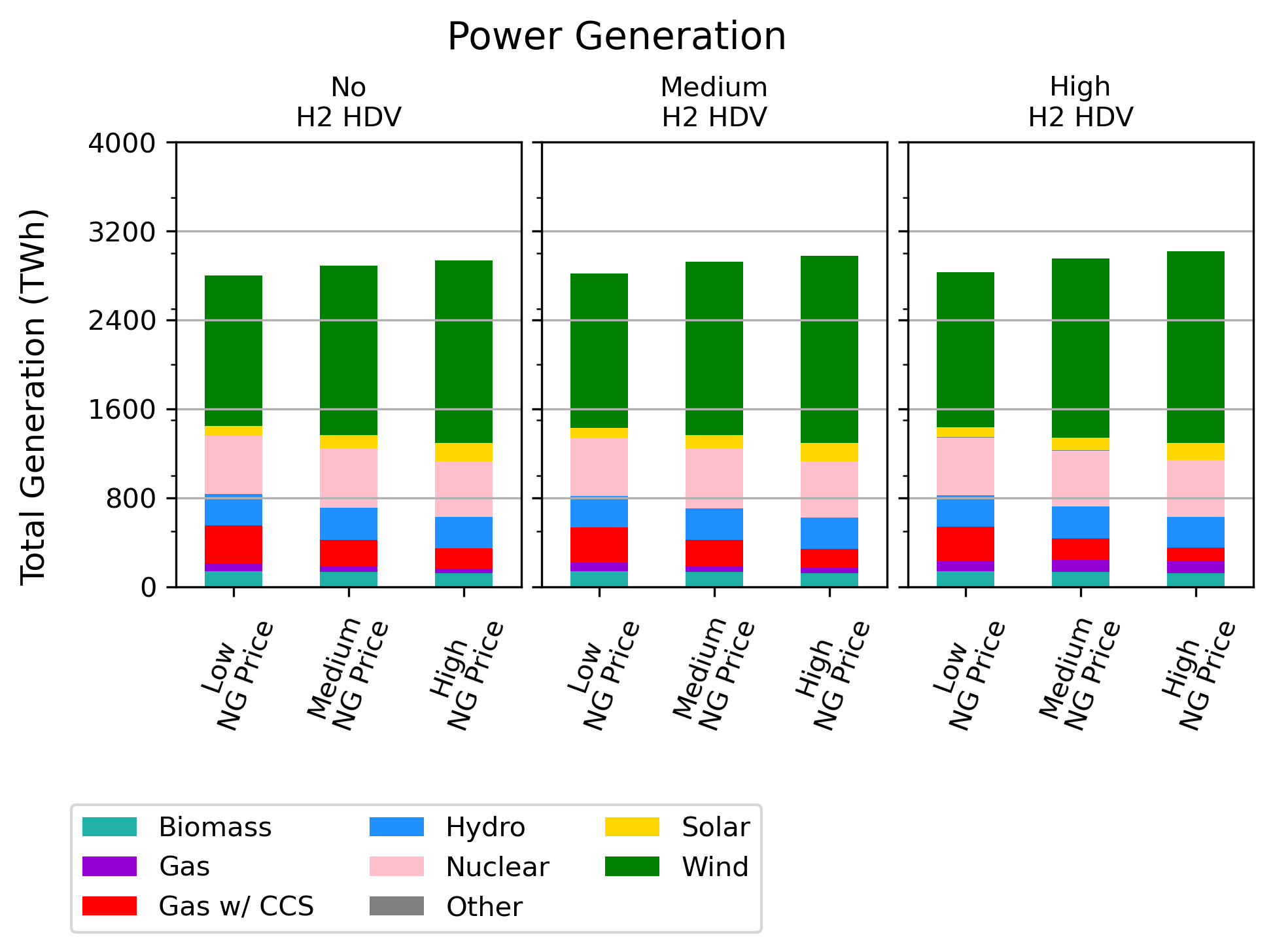}
        \caption{Baseline CO$_2$ Storage}
    \end{subfigure}
    
    \caption[Power Generation Sensitivity Scenario Set 2]{Power generation for no (sub-figure a) and baseline (sub-figure b) CO$_2$ sequestration scenarios under no synthetic fuel adoption. Within each panel, the price of natural gas increases left to right. Across panels, the amount of H$_2$ HDV adoption increases moving from left to right. The middle panels correspond to the core set of scenarios.}
    \label{fig:h2_hdv_power_ng_sen}
\end{figure}

\begin{figure}[pos = H]
    \centering
    \begin{subfigure}[t]{0.9\linewidth}
        \includegraphics[width=\linewidth, trim=0cm 2.6cm 0cm 0.8cm, clip]{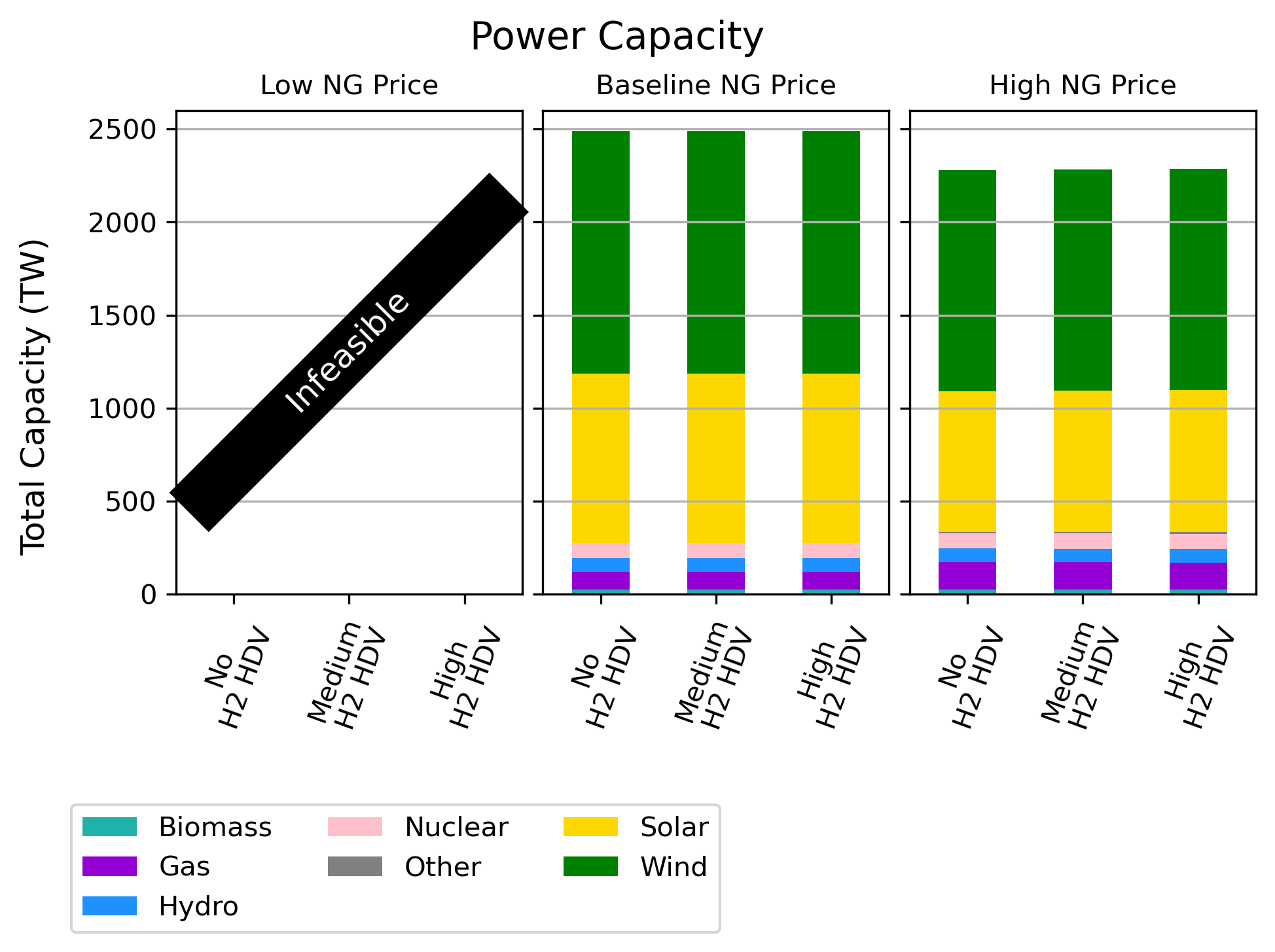}
        \caption{No CO$_2$ Storage}
    \end{subfigure}
    
    \begin{subfigure}[b]{0.9\linewidth}
        \includegraphics[width=\linewidth, trim=0cm 0cm 0cm 0.8cm, clip]{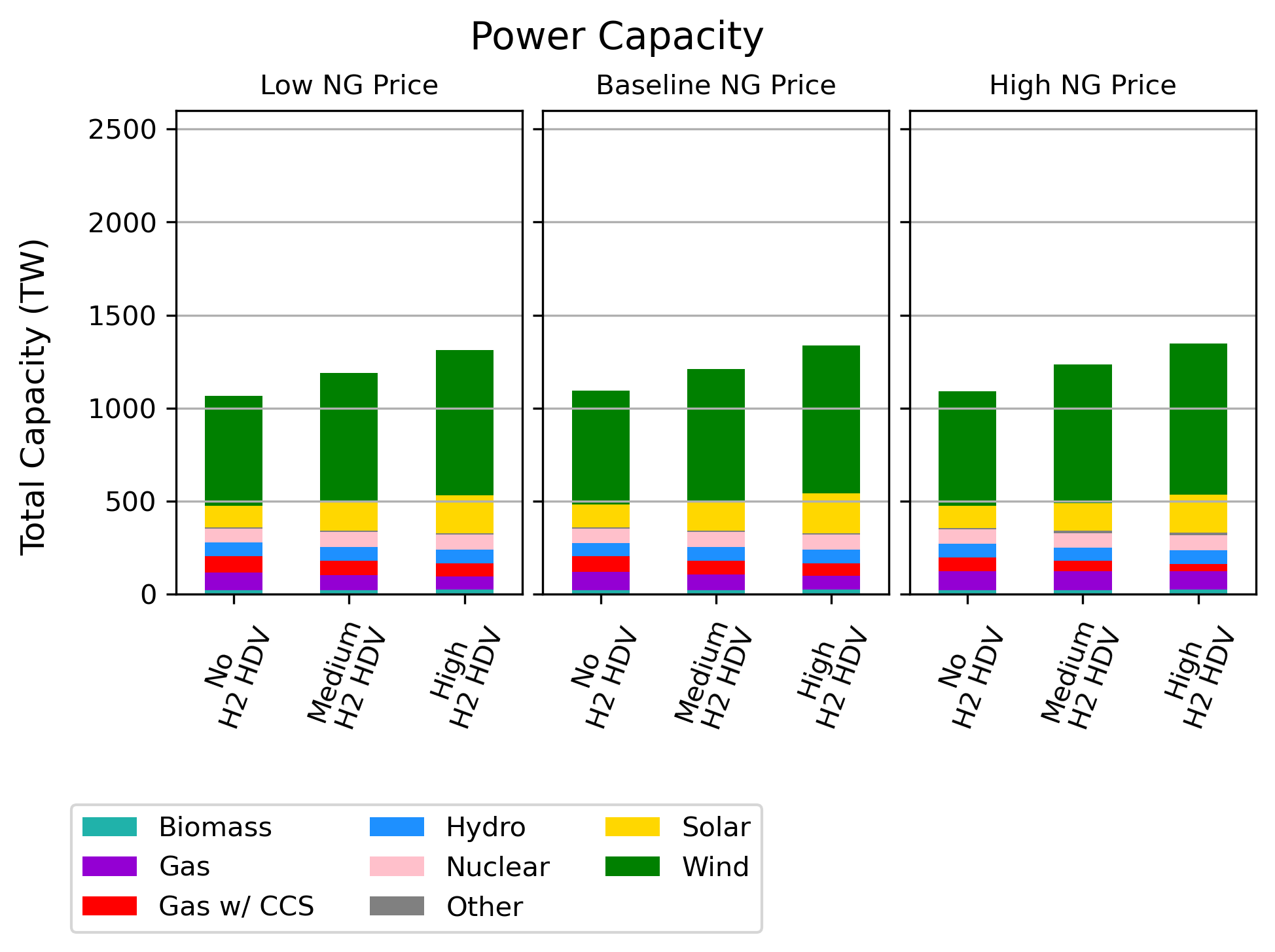}
        \caption{Baseline CO$_2$ Storage}
    \end{subfigure}
    
    \caption[Power Capacity Sensitivity Scenario Set 2]{Power capacity for no (sub-figure a) and baseline (sub-figure b) CO$_2$ sequestration scenarios under no synthetic fuel adoption. Within each panel, the price of natural gas increases left to right. Across panels, the amount of H$_2$ HDV adoption increases moving from left to right. The middle panels correspond to the core set of scenarios.}
    \label{fig:h2_hdv_power_cap_ng_sen}
\end{figure}

\begin{figure}[pos = H]
    \centering
    \begin{subfigure}[t]{0.9\linewidth}
        \includegraphics[width=\linewidth, trim=0cm 2.0cm 0cm 0.8cm, clip]{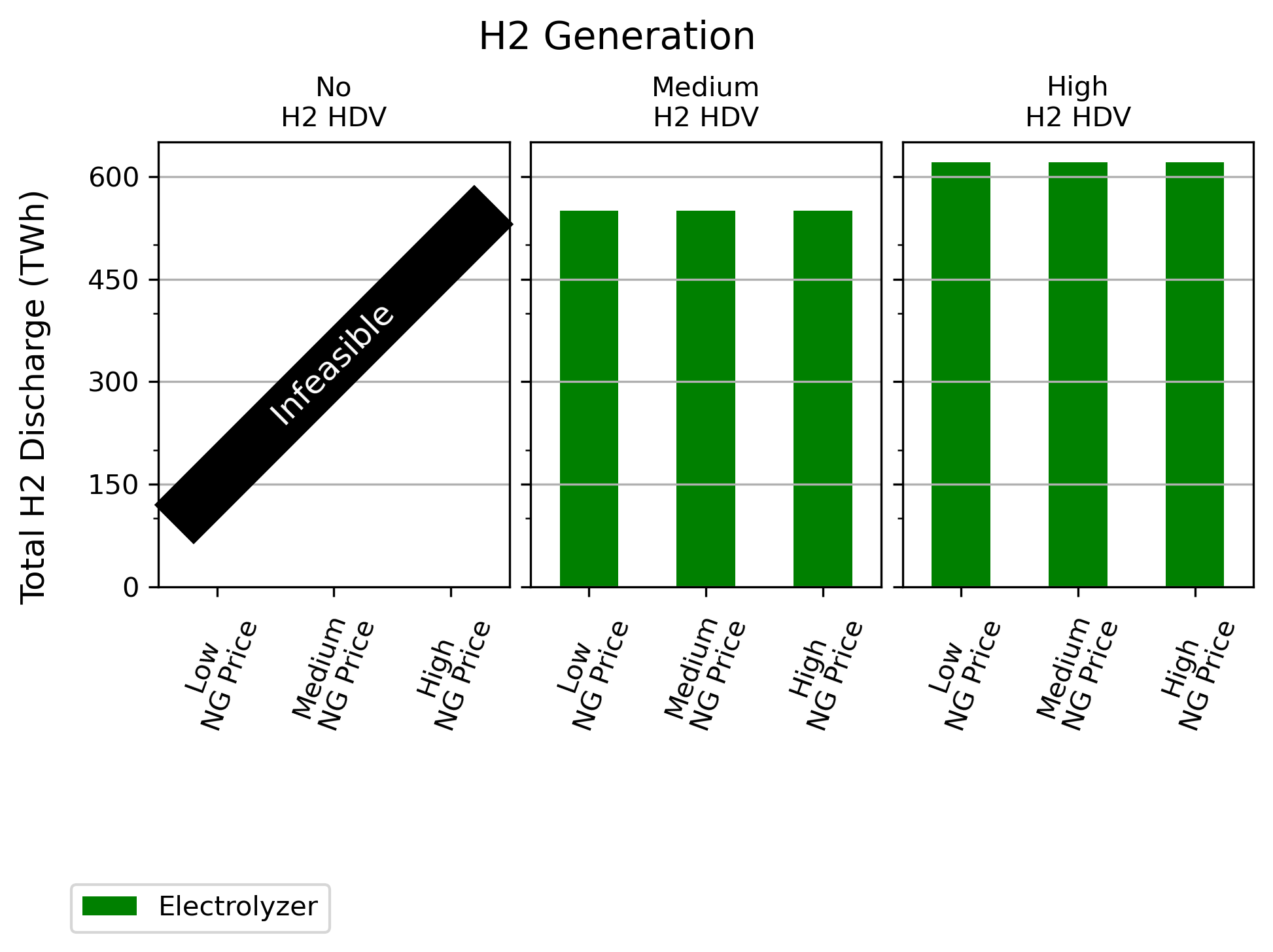}
        \caption{No CO$_2$ Storage}
    \end{subfigure}
    
    \begin{subfigure}[b]{0.9\linewidth}
        \includegraphics[width=\linewidth, trim=0cm 0cm 0cm 0.8cm, clip]{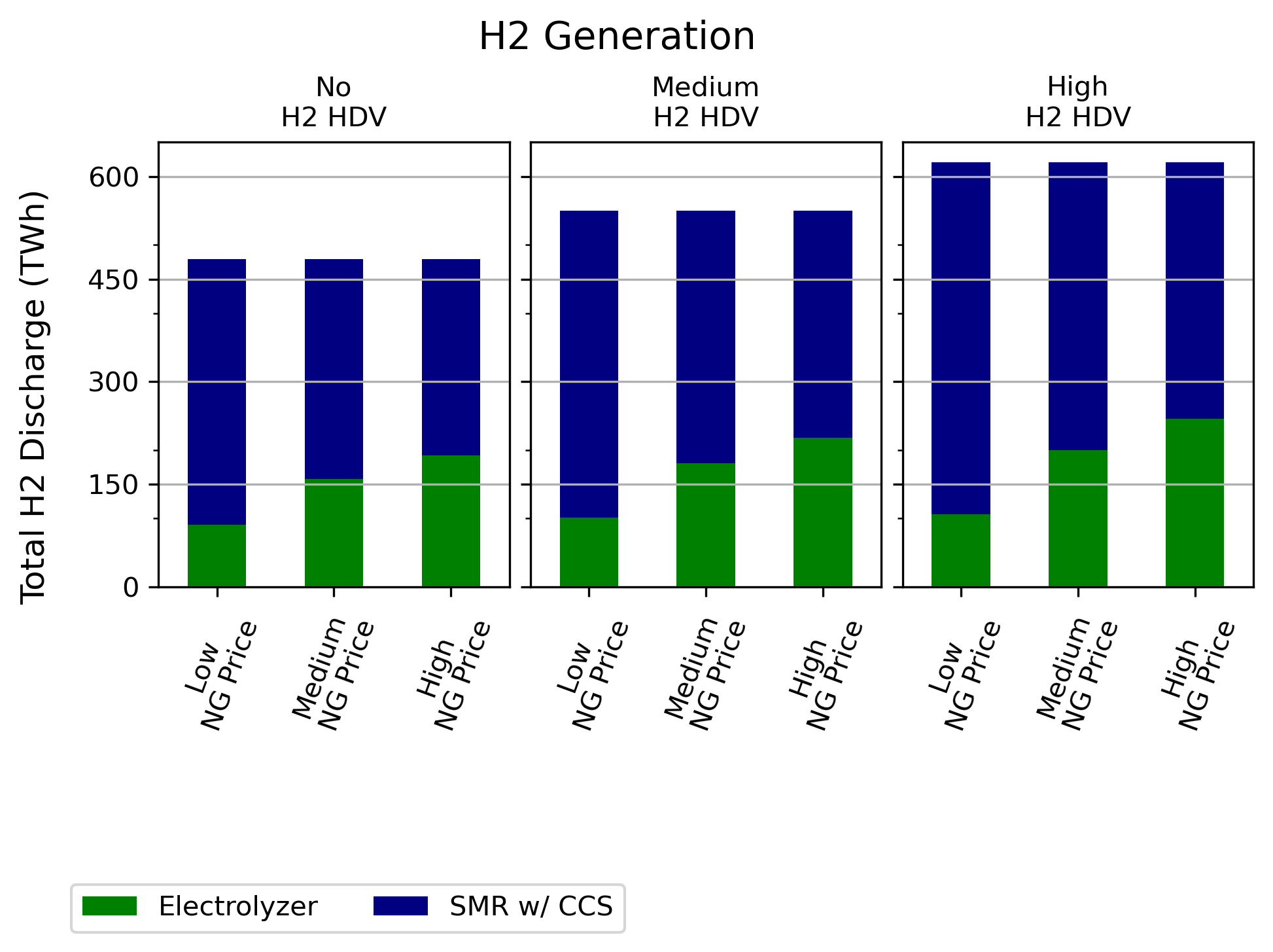}
        \caption{Baseline CO$_2$ Storage}
    \end{subfigure}
    
    \caption[H$_2$ Generation Sensitivity Set 2]{H$_2$ generation for no (sub-figure a) and baseline (sub-figure b) CO$_2$ sequestration scenarios under no synthetic fuel adoption. Within each panel, the price of natural gas increases left to right. Across panels, the amount of H$_2$ HDV adoption increases moving from left to right. The middle panels correspond to the core set of scenarios.}
    \label{fig:h2_hdv_h2_ng_sen}
\end{figure}

\begin{figure}[pos = H]
    \centering
    \begin{subfigure}[t]{0.9\linewidth}
        \includegraphics[width=\linewidth, trim=0cm 1.5cm 0cm 0.8cm, clip]{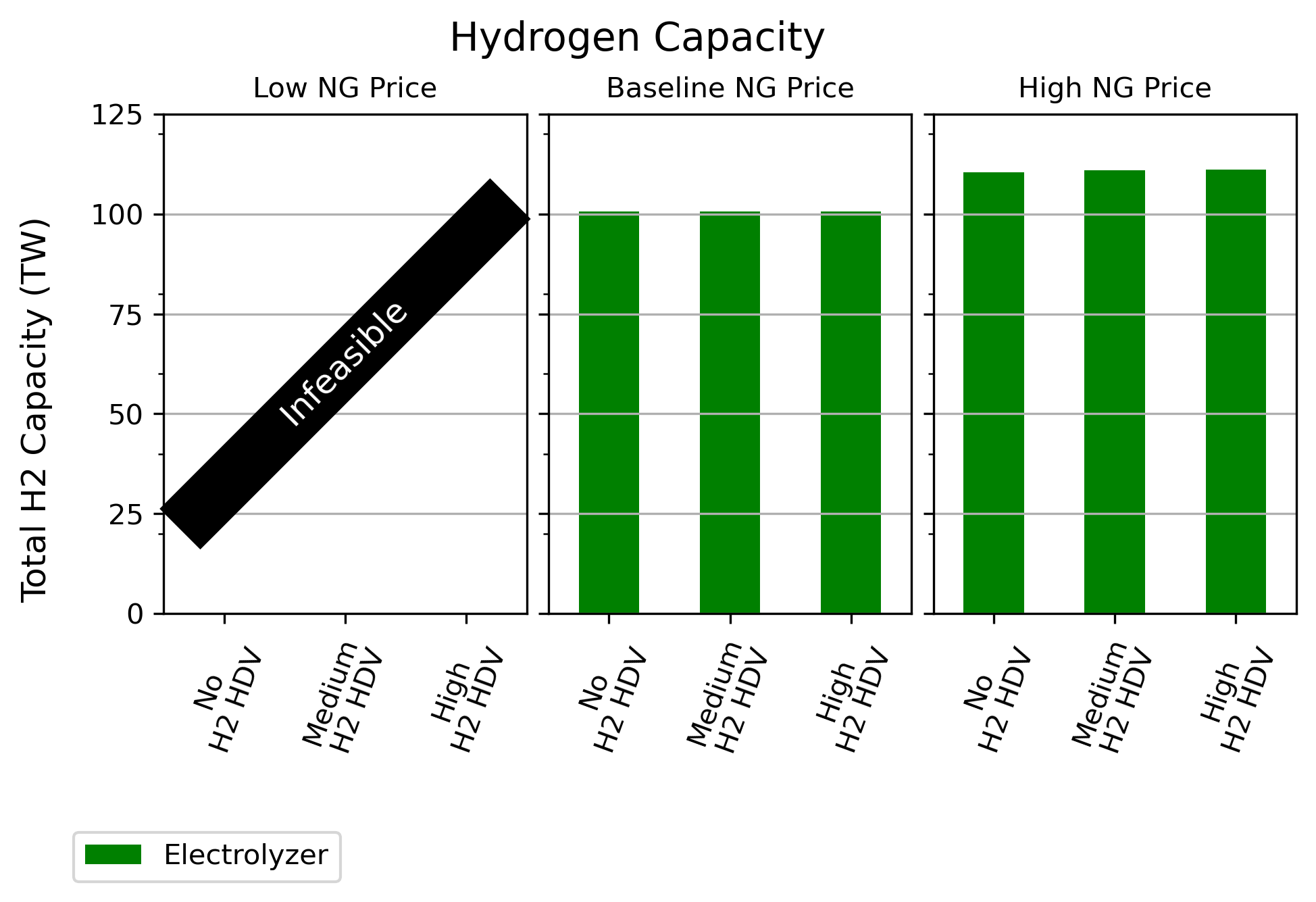}
        \caption{No CO$_2$ Storage}
    \end{subfigure}
    
    \begin{subfigure}[b]{0.9\linewidth}
        \includegraphics[width=\linewidth, trim=0cm 0cm 0cm 0.8cm, clip]{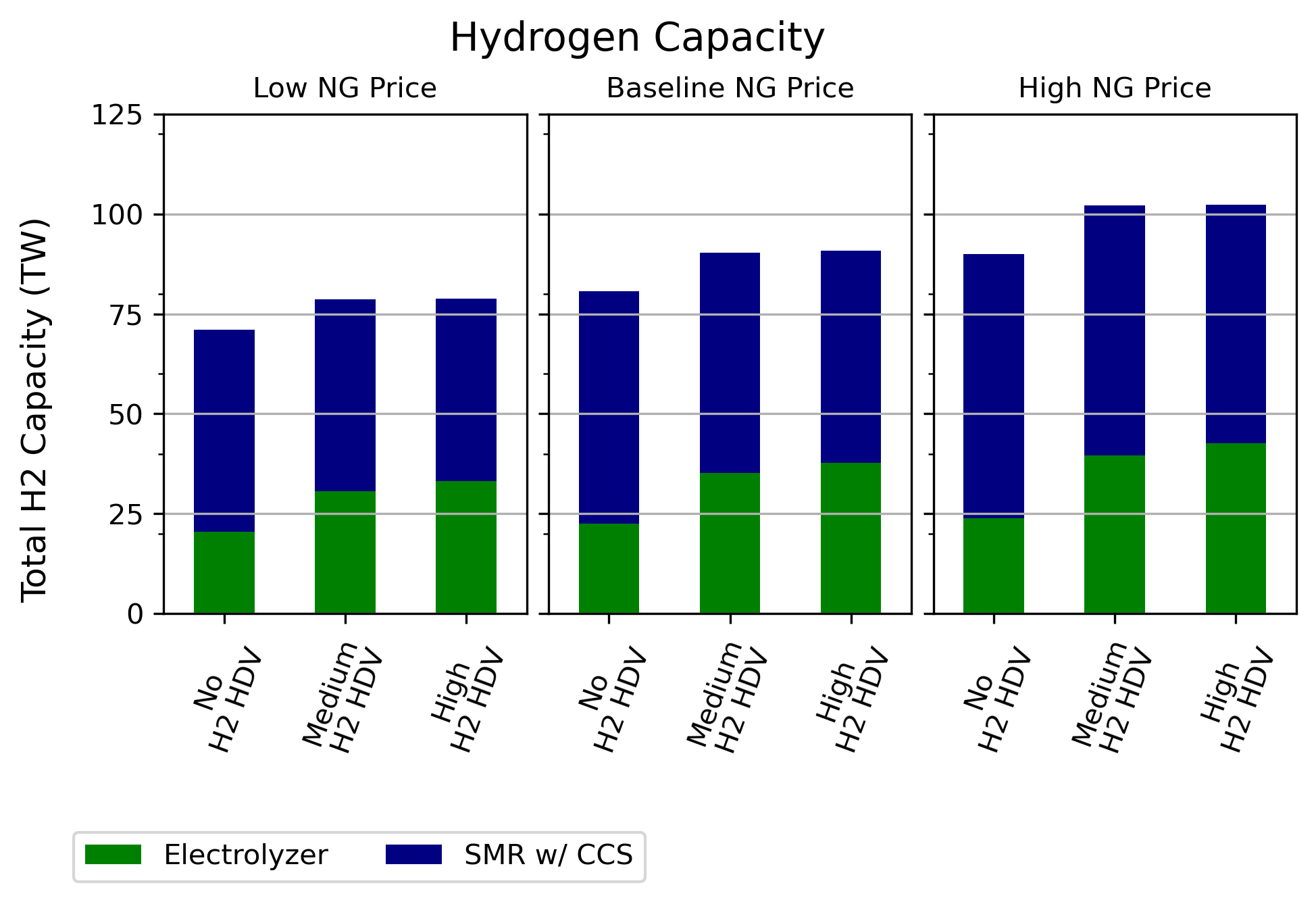}
        \caption{Baseline CO$_2$ Storage}
    \end{subfigure}
    
    \caption[H$_2$ Capacity Sensitivity Scenario Set 2]{H$_2$ capacity for no (sub-figure a) and baseline (sub-figure b) CO$_2$ sequestration scenarios under no synthetic fuel adoption. Within each panel, the price of natural gas increases left to right. Across panels, the amount of H$_2$ HDV adoption increases moving from left to right. The middle panels correspond to the core set of scenarios.}
    \label{fig:h2_hdv_h2_cap_ng_sen}
\end{figure}

\begin{figure}[pos = H]
    \centering
    \begin{subfigure}[t]{0.9\linewidth}
        \includegraphics[width=\linewidth, trim=0cm 1.5cm 0cm 0.8cm, clip]{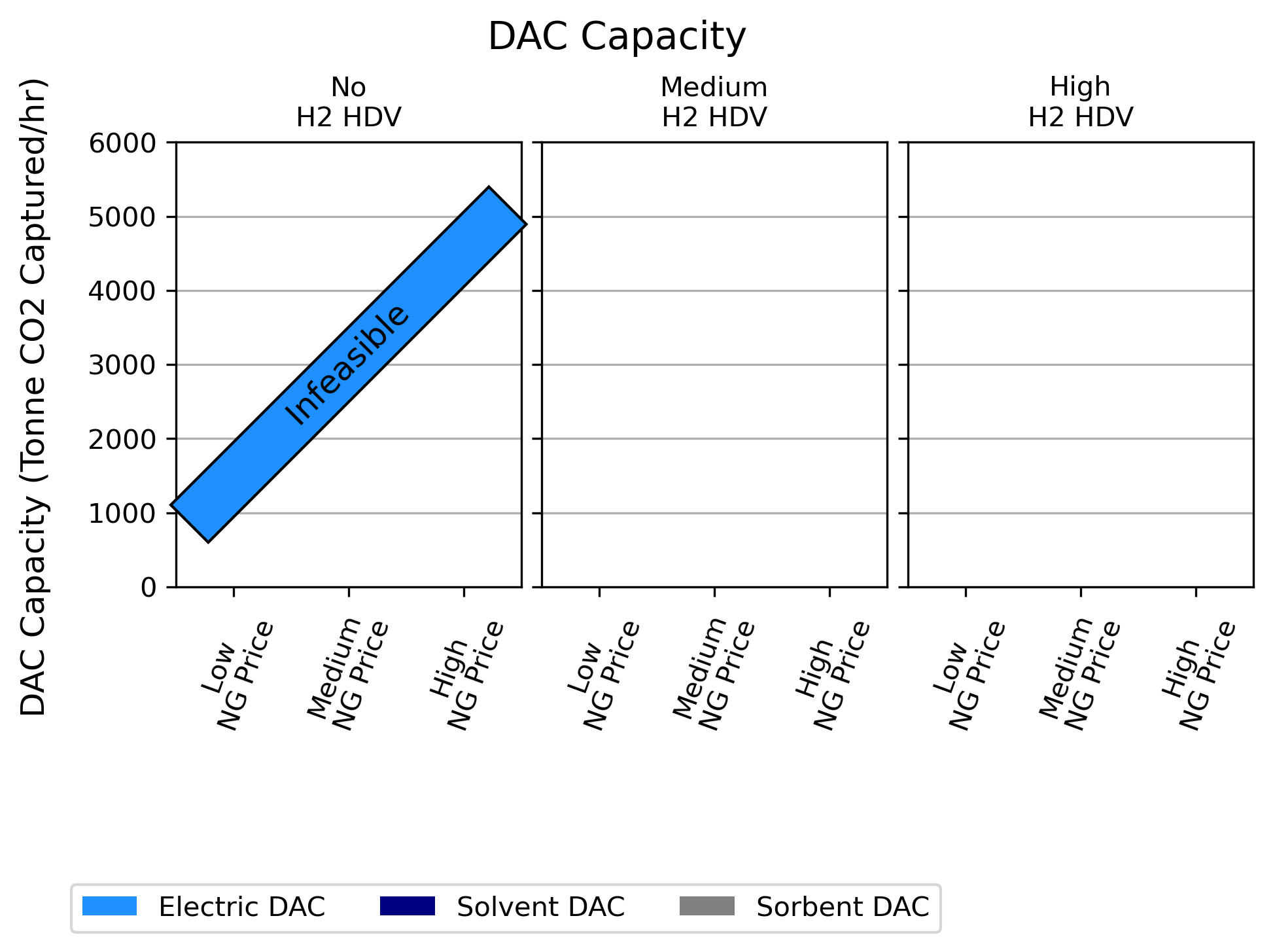}
        \caption{No CO$_2$ Storage}
    \end{subfigure}
    
    \begin{subfigure}[b]{0.9\linewidth}
        \includegraphics[width=\linewidth, trim=0cm 0cm 0cm 0.8cm, clip]{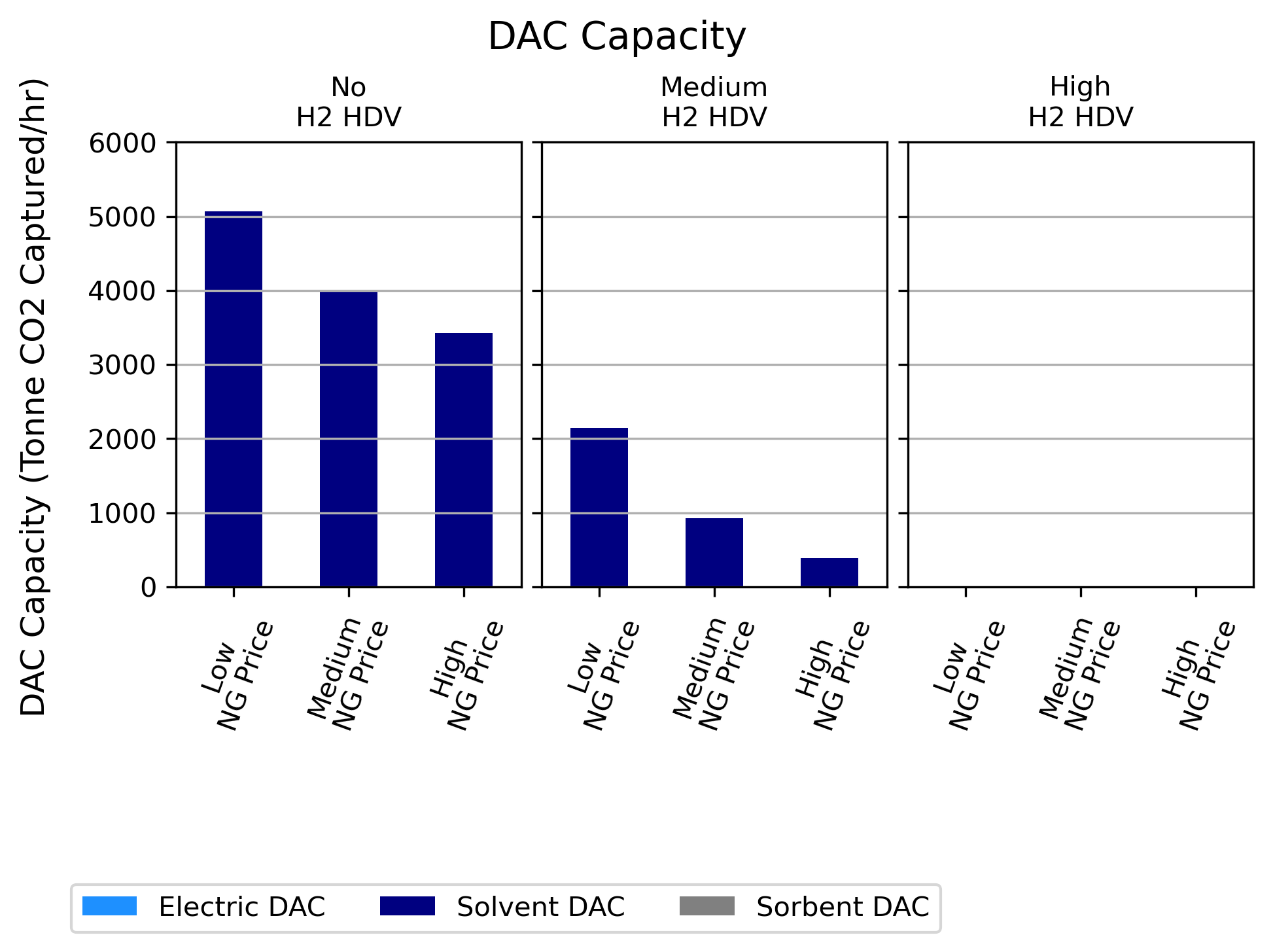}
        \caption{Baseline CO$_2$ Storage}
    \end{subfigure}
    
    \caption[Direct Air Capture Capacity Sensitivity Scenario Set 2]{Direct air capture capacity for no (sub-figure a) and baseline (sub-figure b) CO$_2$ sequestration scenarios under no synthetic fuel adoption. Within each panel, the price of natural gas increases left to right. Across panels, the amount of H$_2$ HDV adoption increases moving from left to right. The middle panels correspond to the core set of scenarios.}
    \label{fig:h2_hdv_dac_cap_ng_sen}
\end{figure}

\begin{figure}[pos = H]
    \centering
    \begin{subfigure}[t]{0.9\linewidth}
        \includegraphics[width=\linewidth, trim=0cm 1.5cm 0cm 0.8cm, clip]{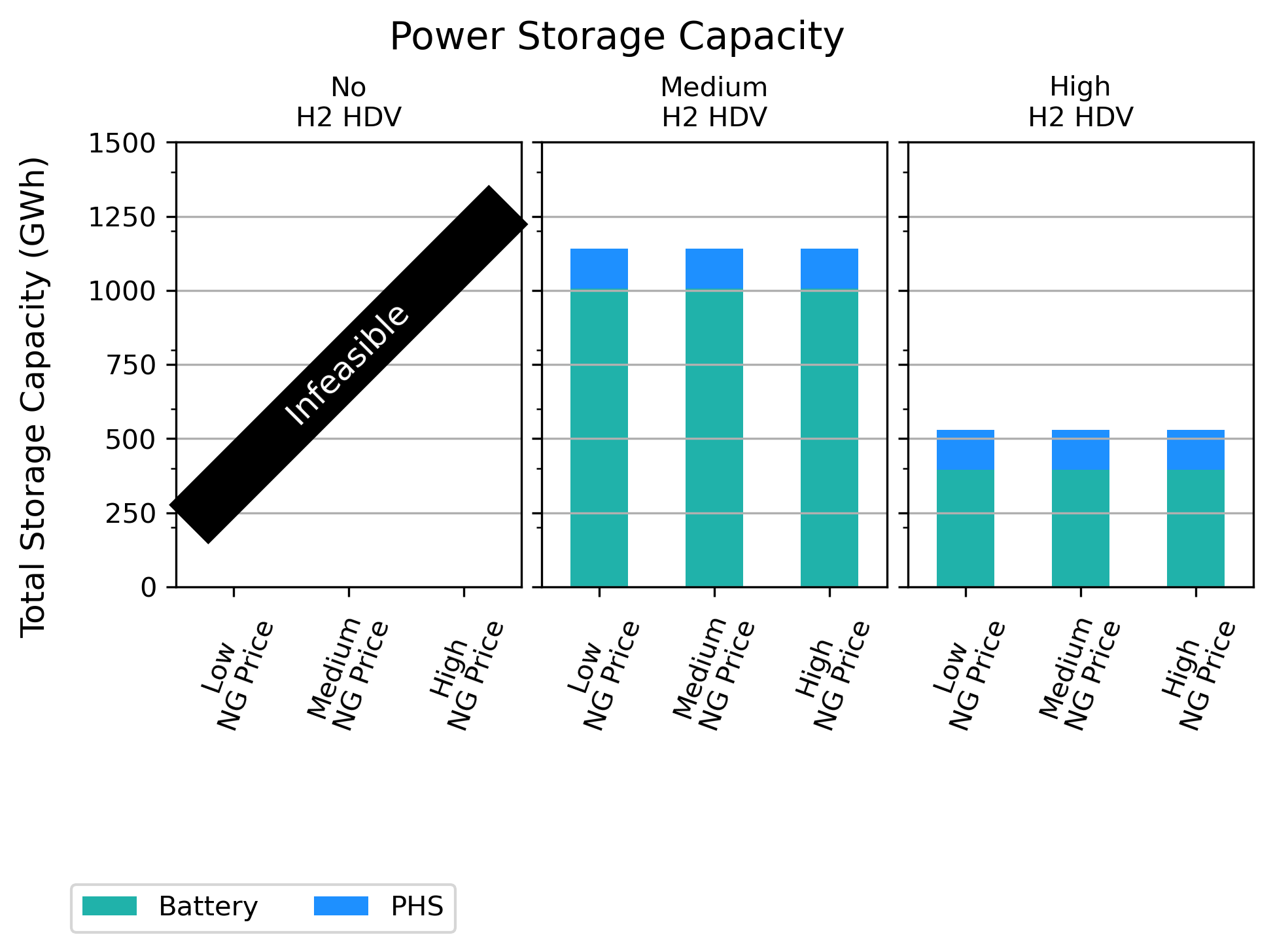}
        \caption{No CO$_2$ Storage}
    \end{subfigure}
    
    \begin{subfigure}[b]{0.9\linewidth}
        \includegraphics[width=\linewidth, trim=0cm 0cm 0cm 0.8cm, clip]{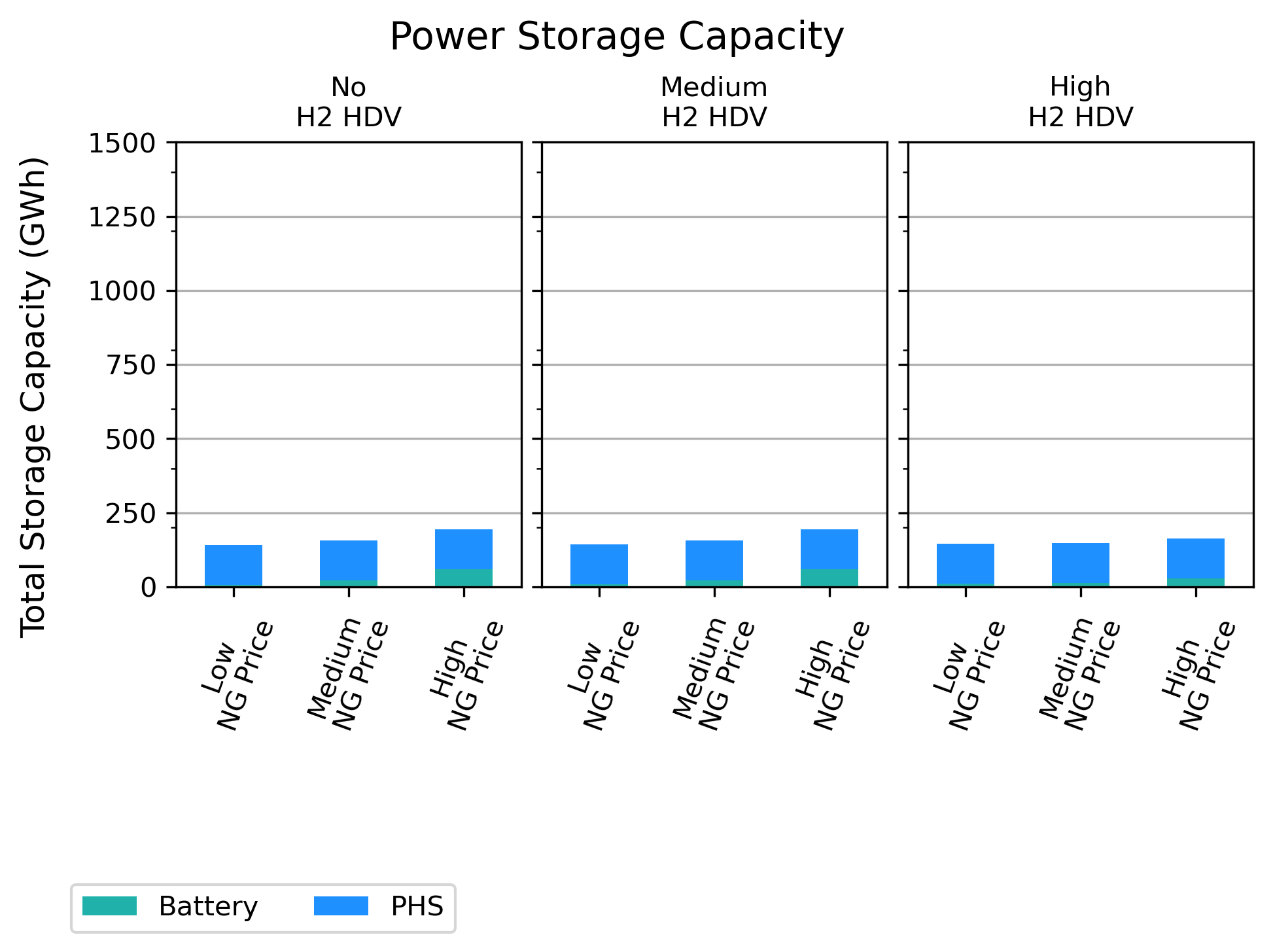}
        \caption{Baseline CO$_2$ Storage}
    \end{subfigure}
    
    \caption[Electric Storage Capacity Sensitivity Scenario Set 2]{Electric storage capacity for no (sub-figure a) and baseline (sub-figure b) CO$_2$ sequestration scenarios under no synthetic fuel adoption. Within each panel, the price of natural gas increases left to right. Across panels, the amount of H$_2$ HDV adoption increases moving from left to right. The middle panels correspond to the core set of scenarios.}
    \label{fig:h2_hdv_elec_cap_stor_ng_sen}
\end{figure}

\begin{figure}[pos = H]
    \centering
    \begin{subfigure}[t]{0.9\linewidth}
        \includegraphics[width=\linewidth, trim=0cm 0cm 0cm 0.8cm, clip]{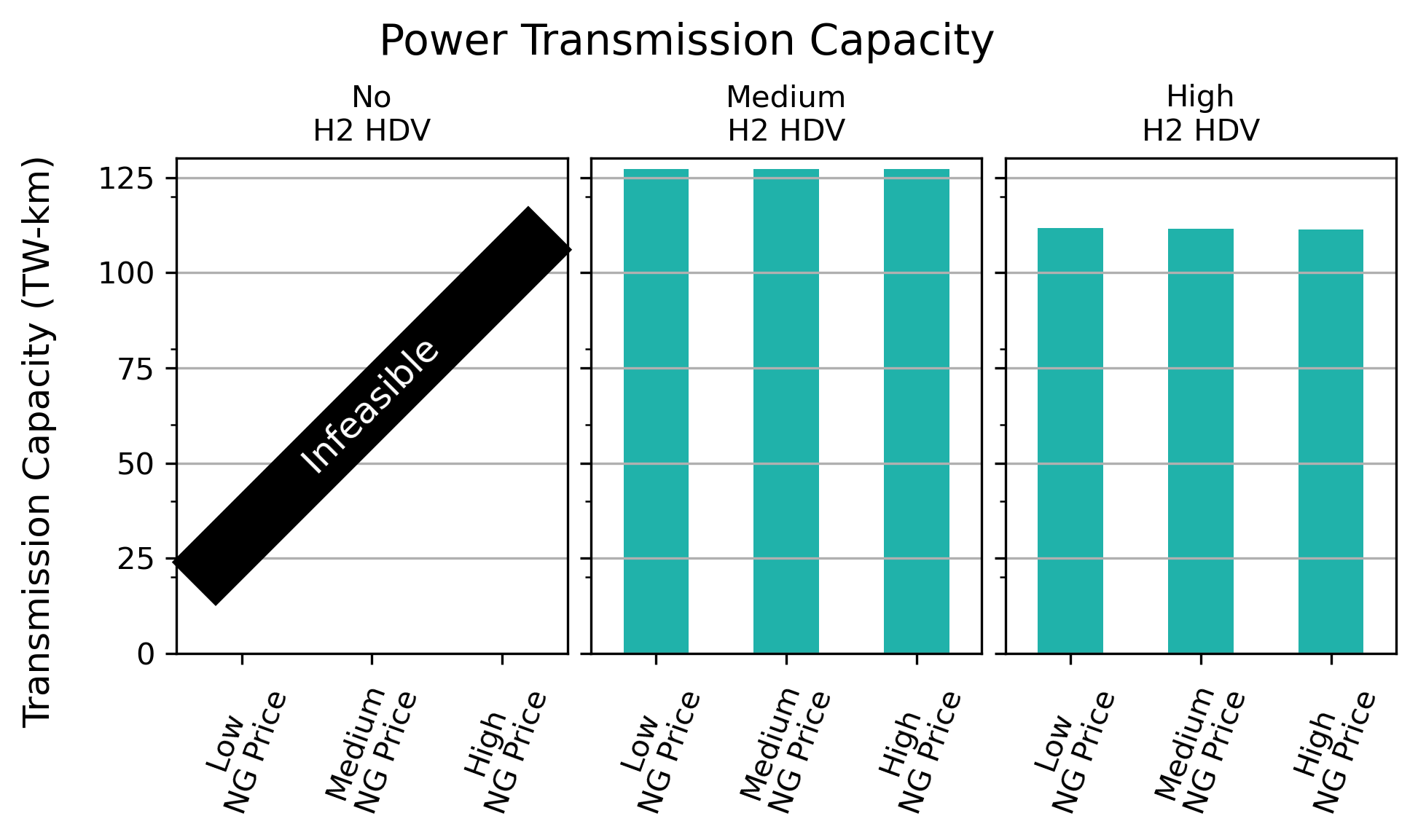}
        \caption{No CO$_2$ Storage}
    \end{subfigure}
    
    \begin{subfigure}[b]{0.9\linewidth}
        \includegraphics[width=\linewidth, trim=0cm 0cm 0cm 0.8cm, clip]{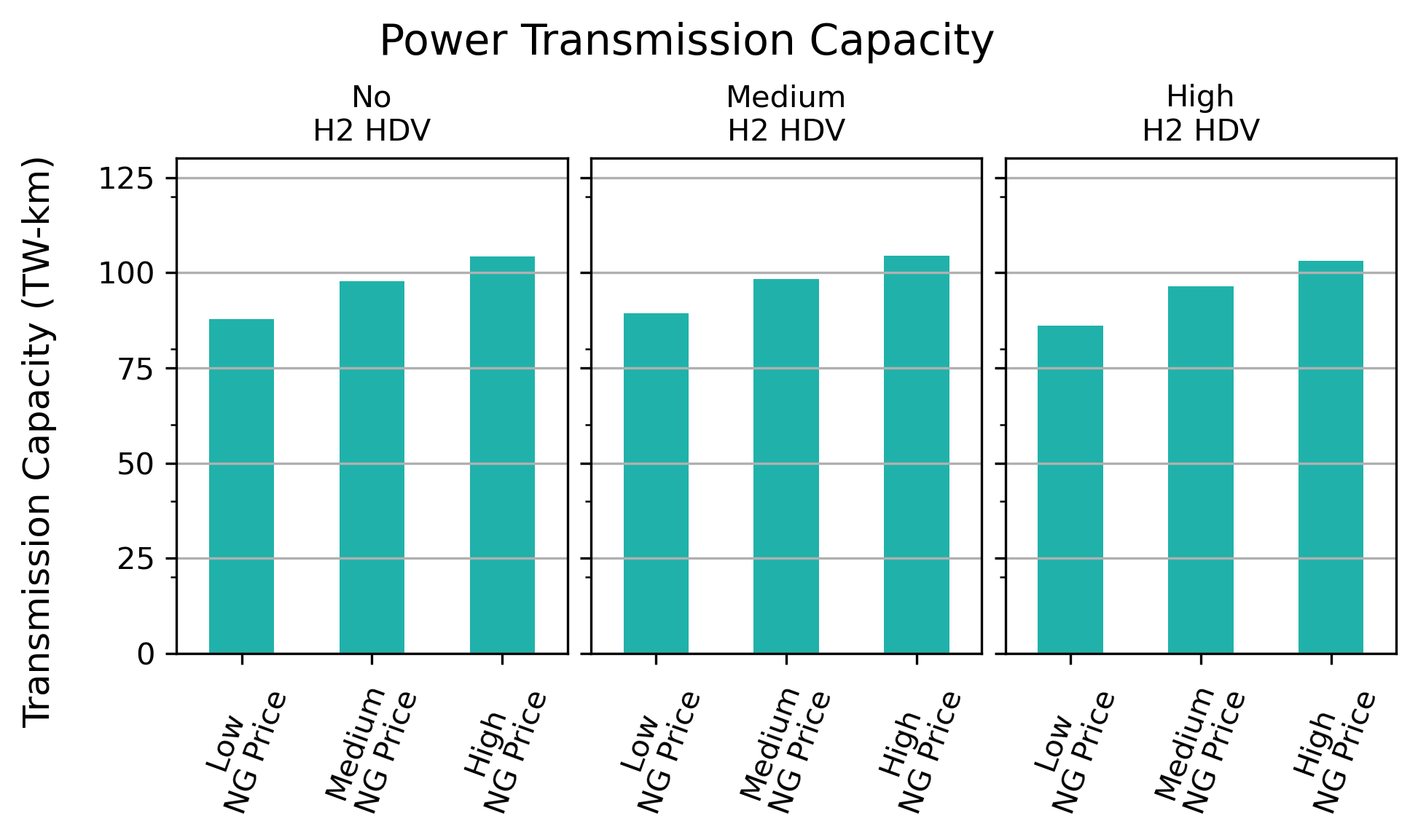}
        \caption{Baseline CO$_2$ Storage}
    \end{subfigure}
    
    \caption[Power transmission Capacity Sensitivity Scenario Set 2]{Power transmission capacity for no (sub-figure a) and baseline (sub-figure b) CO$_2$ sequestration scenarios under no synthetic fuel adoption. Within each panel, the price of natural gas increases left to right. Across panels, the amount of H$_2$ HDV adoption increases moving from left to right. The middle panels correspond to the core set of scenarios.}
    \label{fig:h2_hdv_elec_trans_ng_sen}
\end{figure}

\begin{figure}[pos = H]
    \centering
    \begin{subfigure}[t]{0.9\linewidth}
        \includegraphics[width=\linewidth, trim=0cm 0cm 0cm 0.8cm, clip]{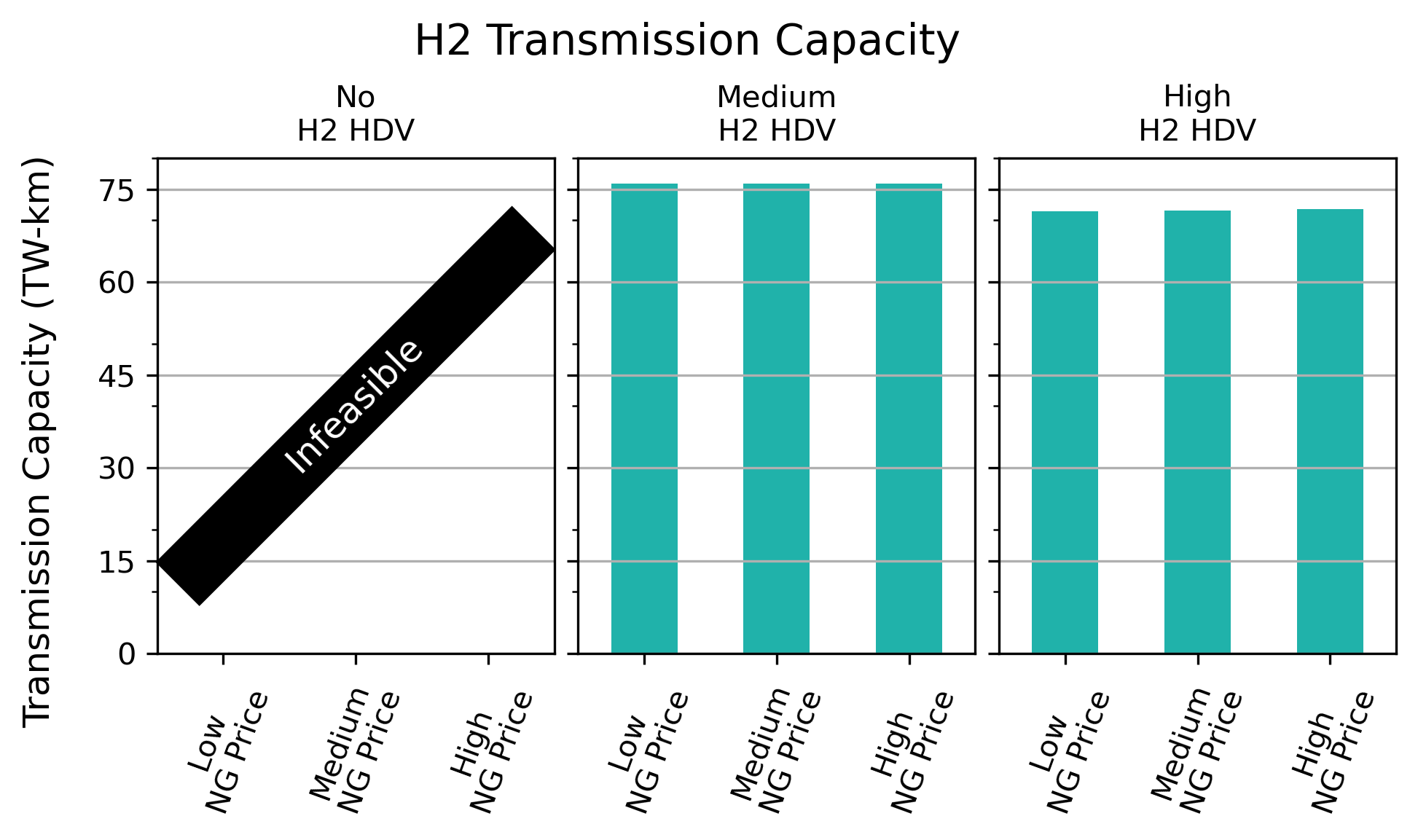}
        \caption{No CO$_2$ Storage}
    \end{subfigure}
    
    \begin{subfigure}[b]{0.9\linewidth}
        \includegraphics[width=\linewidth, trim=0cm 0cm 0cm 0.8cm, clip]{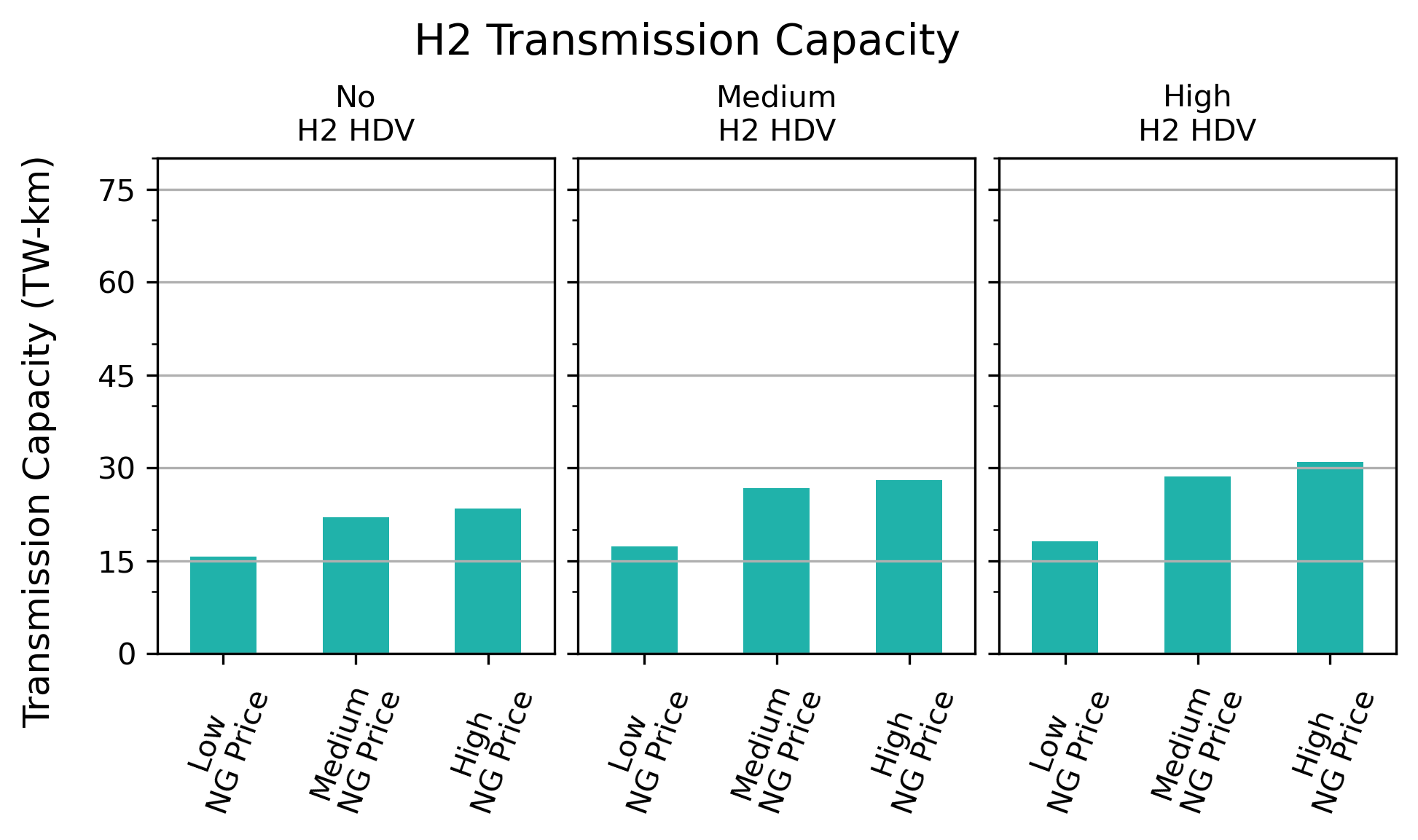}
        \caption{Baseline CO$_2$ Storage}
    \end{subfigure}
    
    \caption[H$_2$ transmission Capacity Sensitivity Scenario Set 2]{H$_2$ transmission capacity for no (sub-figure a) and baseline (sub-figure b) CO$_2$ sequestration scenarios under no synthetic fuel adoption. Within each panel, the price of natural gas increases left to right. Across panels, the amount of H$_2$ HDV adoption increases moving from left to right. The middle panels correspond to the core set of scenarios.}
    \label{fig:h2_hdv_h2_trans_ng_sen}
\end{figure}

\begin{figure}[pos = H]
    \centering
    \begin{subfigure}[t]{0.9\linewidth}
        \includegraphics[width=\linewidth, trim=0cm 0cm 0cm 0.8cm, clip]{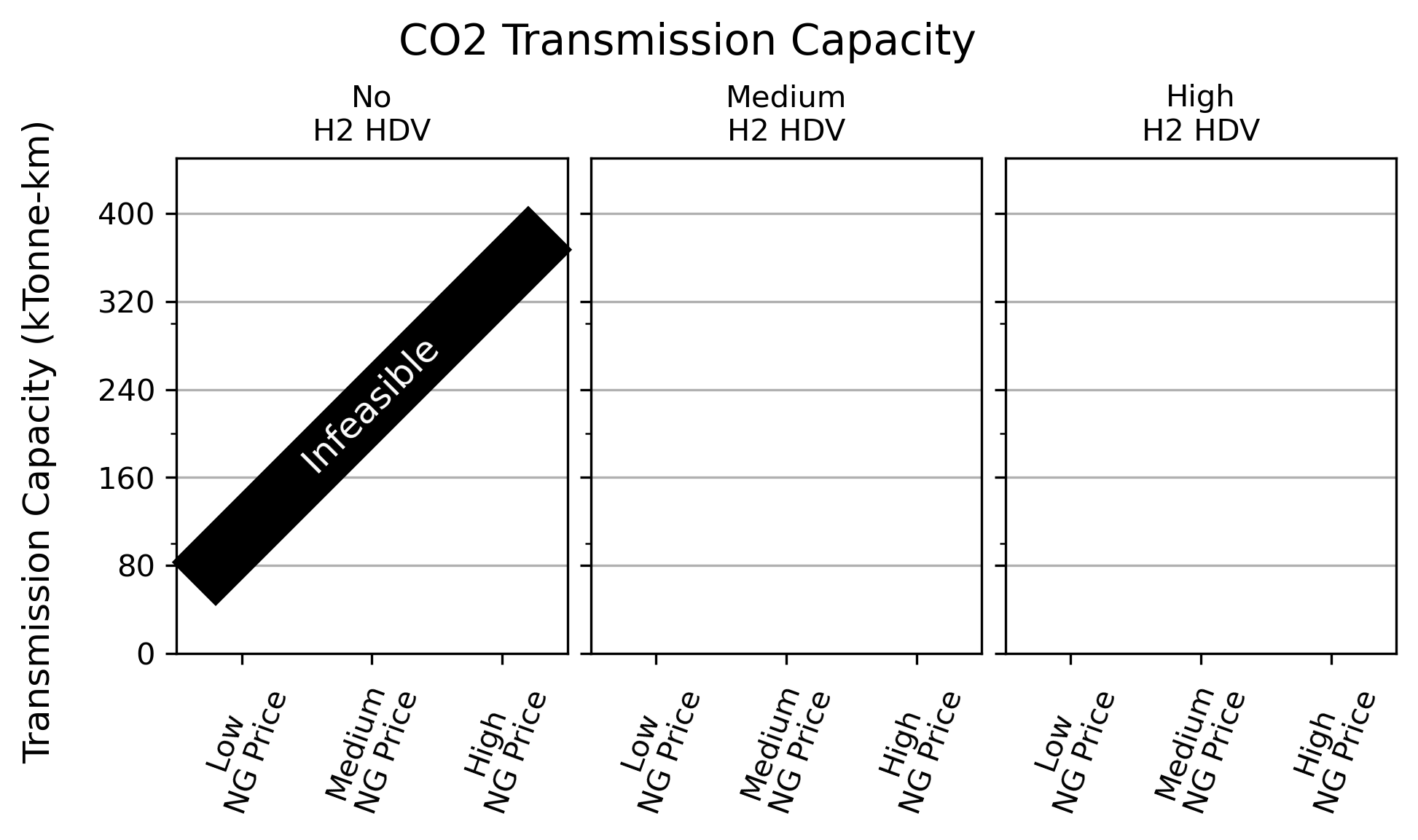}
        \caption{No CO$_2$ Storage}
    \end{subfigure}
    
    \begin{subfigure}[b]{0.9\linewidth}
        \includegraphics[width=\linewidth, trim=0cm 0cm 0cm 0.8cm, clip]{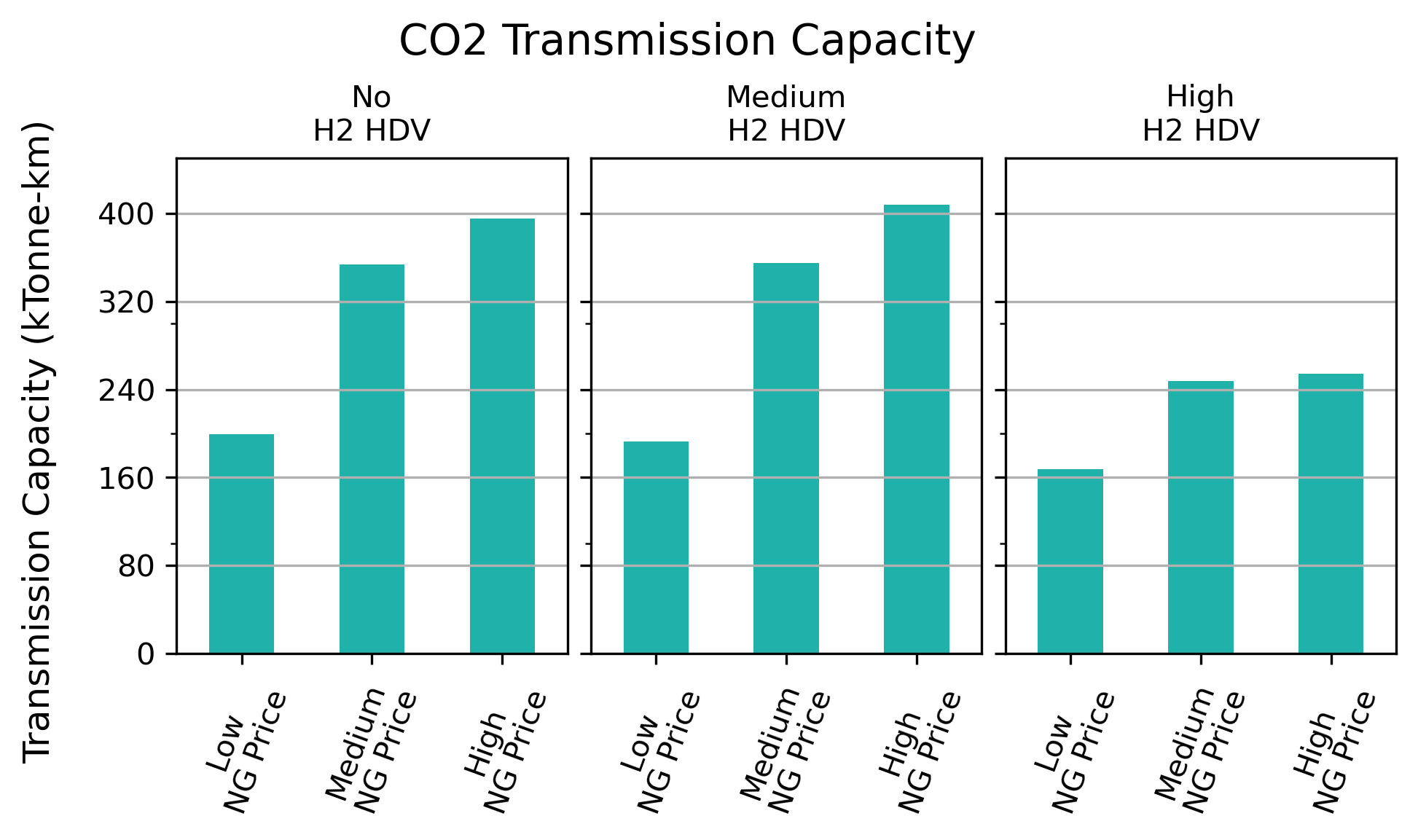}
        \caption{Baseline CO$_2$ Storage}
    \end{subfigure}
    
    \caption[H$_2$ transmission Capacity Sensitivity Scenario Set 2]{CO$_2$ transmission capacity for no (sub-figure a) and baseline (sub-figure b) CO$_2$ sequestration scenarios under no synthetic fuel adoption. Within each panel, the price of natural gas increases left to right. Across panels, the amount of H$_2$ HDV adoption increases moving from left to right. The middle panels correspond to the core set of scenarios.}
    \label{fig:h2_hdv_co2_trans_ng_sen}
\end{figure}

\begin{figure}[pos = H]
    \centering
    \begin{subfigure}[t]{1\linewidth}
        \includegraphics[width=\linewidth, trim=0cm 0cm 0cm 0.0cm, clip]{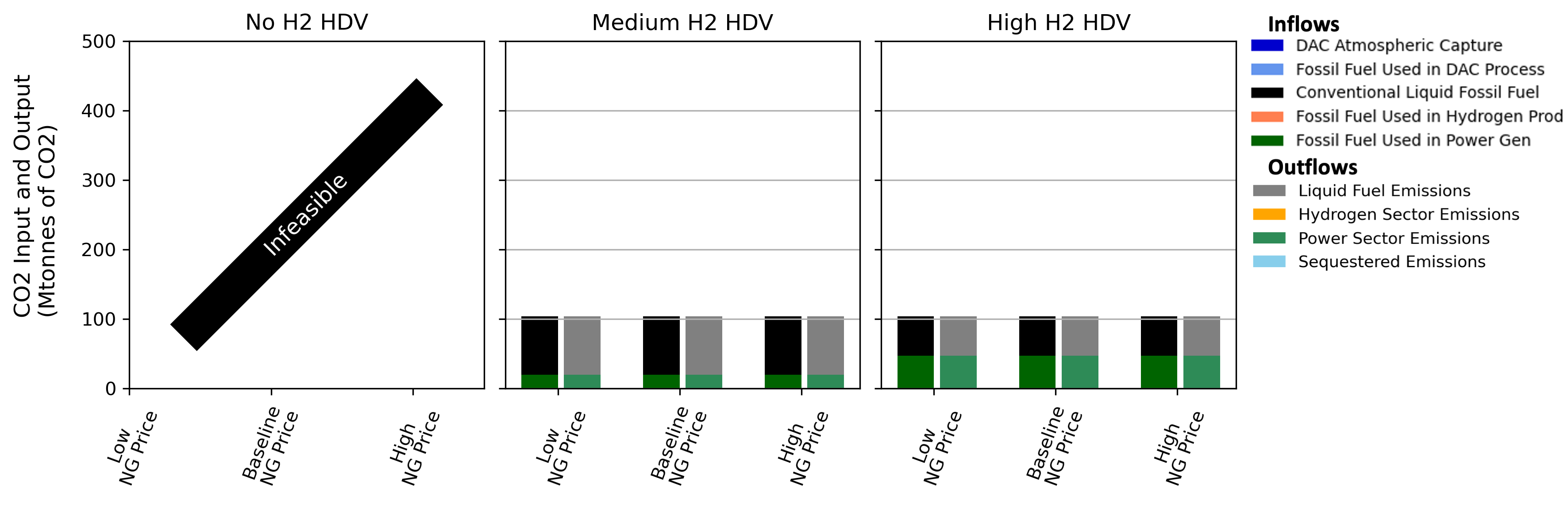}
        \caption{No CO$_2$ Storage}
    \end{subfigure}
    
    \begin{subfigure}[b]{1\linewidth}
        \includegraphics[width=\linewidth, trim=0cm 0cm 0cm 0.0cm, clip]{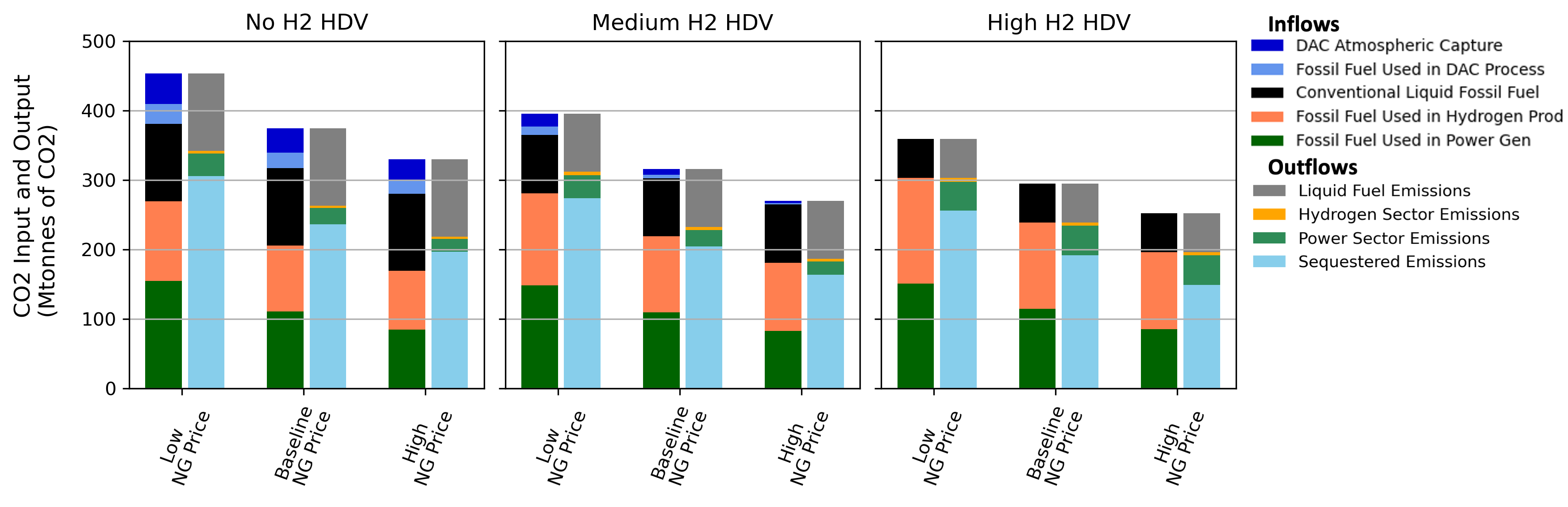}
        \caption{Baseline CO$_2$ Storage}
    \end{subfigure}
    
    \caption[CO$_2$ Balance Sensitivity Scenario Set 2]{System CO$_2$ balance under varying levels of H$_2$ HDV adoption and no SF adoption for no (sub-figure a) and baseline (sub-figure b) CO$_2$ sequestration scenarios. Within each panel, the price of natural gas increases left to right. Across panels, the amount of H$_2$ HDV adoption increases moving from left to right. The middle panels correspond to the core set of scenarios. The leftward column represents CO$_2$ input into the system, while the rightward column represents CO$_2$ outputted by the system. All scenarios adhere to the same emissions constraint of 103 MTonnes. The middle panels correspond to the core set of scenarios. Emissions constraint can be calculated from the chart by subtracting sequestered emissions and DAC atmospheric capture from the emission outflows. Emissions constraint can be calculated from the chart by subtracting sequestered emissions and DAC atmospheric capture from the emission outflows.}
    \label{fig:h2_hdv_co2_balance_ng_sen}
\end{figure}

\begin{figure}[pos = H]
    \centering
    \begin{subfigure}[t]{1\linewidth}
        \includegraphics[width=\linewidth, trim=0cm 0cm 0cm 0.0cm, clip]{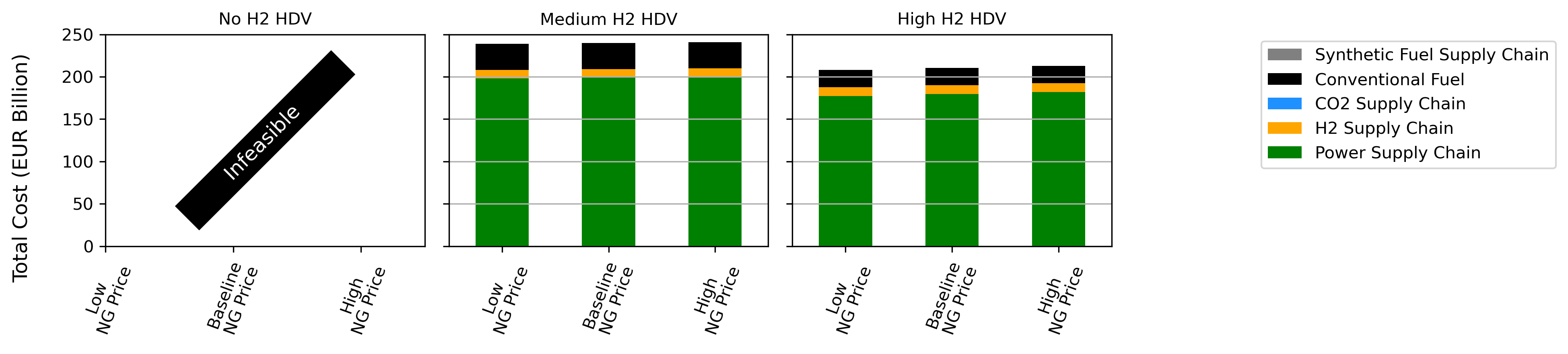}
        \caption{No CO$_2$ Storage}
    \end{subfigure}
    
    \begin{subfigure}[b]{1\linewidth}
        \includegraphics[width=\linewidth, trim=0cm 0cm 0cm 0.0cm, clip]{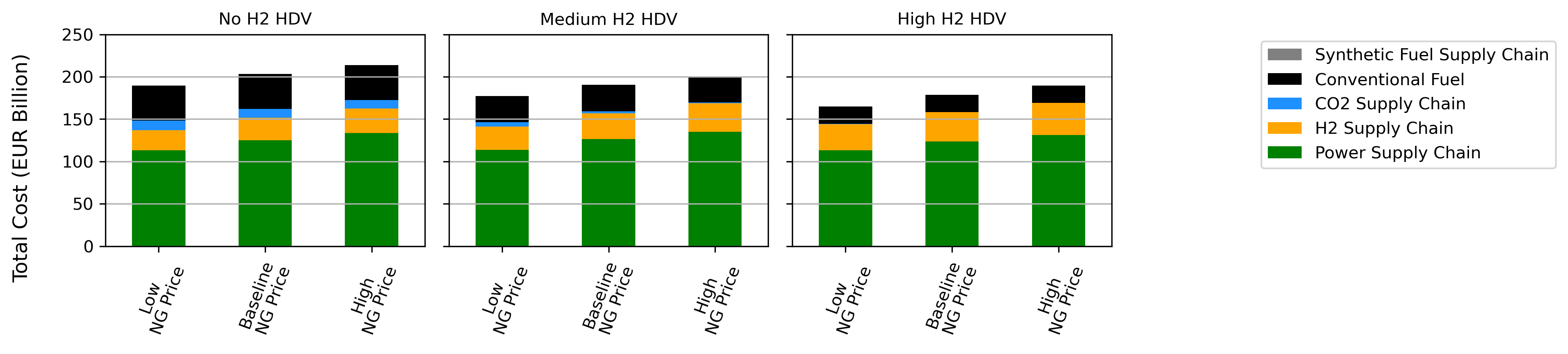}
        \caption{Baseline CO$_2$ Storage}
    \end{subfigure}
    
    \caption[Costs Sensitivity Scenario Set 2]{Annualized bulk-system costs under varying levels of H$_2$ HDV adoption and no SF adoption for no (sub-figure a) and baseline (sub-figure b) CO$_2$ sequestration scenarios. Within each panel, the price of natural gas increases left to right. Across panels, the amount of H$_2$ HDV adoption increases moving from left to right. The middle panels correspond to the core set of scenarios. The costs do not include vehicle replacement or H$_2$ distribution costs. }
    \label{fig:h2_hdv_cost_ng_sen}
\end{figure}

\begin{figure}[pos = H]
    \centering
    \includegraphics[width=1\linewidth]{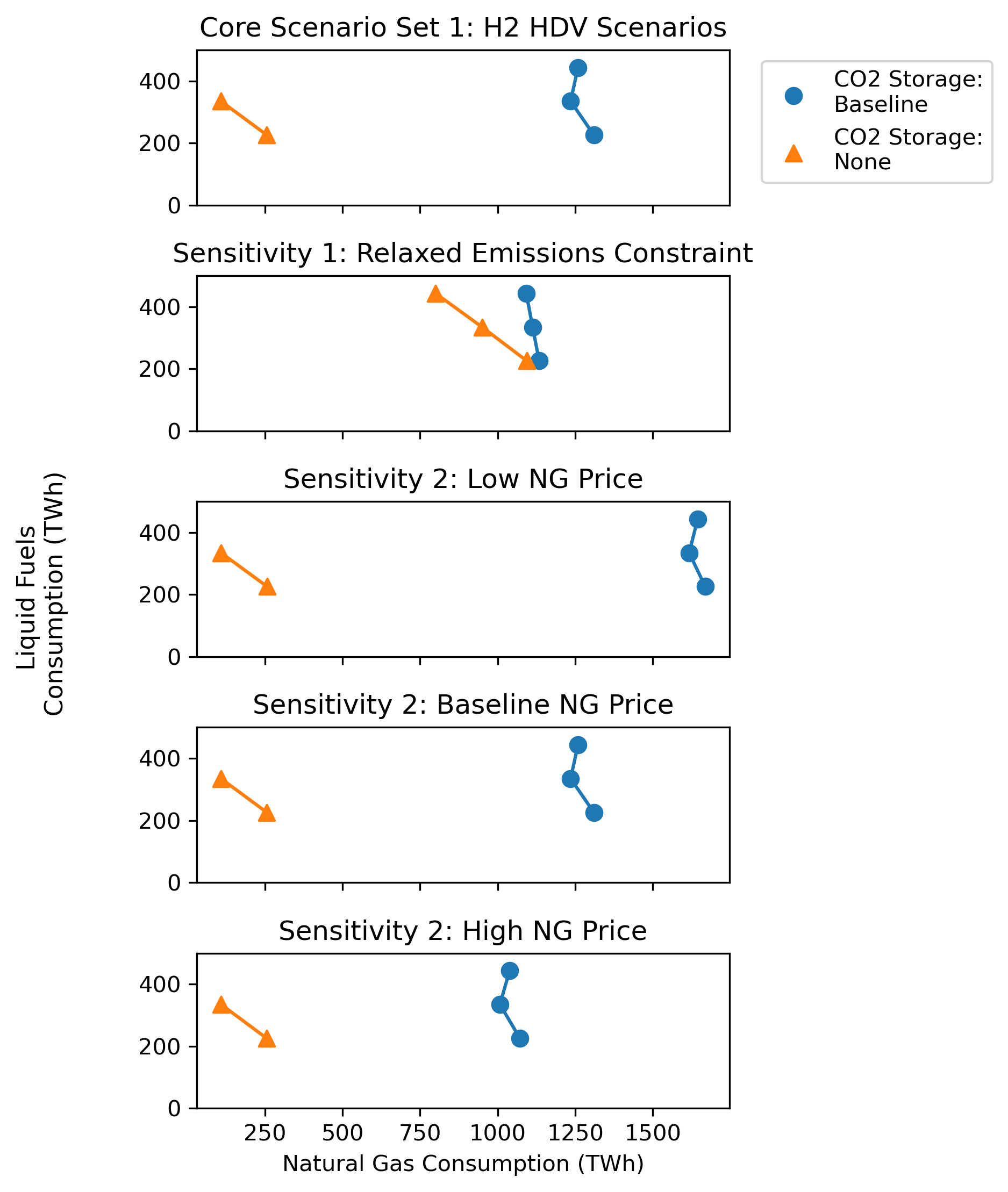}
    \caption[Natural Gas and Liquid Fuel Utilization Trade-off H$_2$ Sensitivity Scenarios]{Trade-off between natural gas (NG) and liquid fossil fuel utilization for scenarios where amount of H$_2$ HDVs is varied. The subfigure on the top shows the relationship for the H$_2$ HDV scenarios(i.e. scenario set 1), while the second plot shows the results for the relaxed emission sensitivity (i.e. sensitivity scenario set 1). The last 3 shows the results for the natural gas price sensitivities (i.e. sensitivity scenario set 2). Within each subplot the amount of natural gas consumption can be examined on the x-axis, while the amount of liquid fossil fuel consumption can be examined on the y-axis. The amount of H$_2$ and SF HDV adoption increases from top to bottom. The amount of liquid fossil fuel consumption includes diesel and gasoline, and excludes jet fuel as well as excess synthetic fuels.}
    \label{fig:fuel_comp_h2}
\end{figure}

\begin{table}[pos = H]
\caption{This tables shows the marginal price of abatement of CO$_2$ for Sensitivity Set 2}
\label{co2_price_sen_set_2}
\begin{tabular}{llllr}
\toprule
CO$_2$ Storage & H$_2$ HDV Level & Synthetic Fuel HDV Level & Natural Gas Price & CO$_2$ Marginal Cost of Abatement \\
\midrule
Baseline & None & None & Low & 260.59 \\
Baseline & Medium & None & Low & 260.59 \\
Baseline & High & None & Low & 195.36 \\
Baseline & None & None & Baseline & 293.92 \\
Baseline & Medium & None & Baseline & 293.92 \\
Baseline & High & None & Baseline & 149.79 \\
Baseline & None & None & High & 327.27 \\
Baseline & Medium & None & High & 327.25 \\
Baseline & High & None & High & 140.91 \\
None & Medium & None & Low & 1574.14 \\
None & High & None & Low & 792.50 \\
None & Medium & None & Baseline & 1523.03 \\
None & High & None & Baseline & 746.25 \\
None & Medium & None & High & 1471.92 \\
None & High & None & High & 697.43 \\
\bottomrule
\end{tabular}
\end{table}

\newpage
\subsection{Sensitivity Set 3: Core Scenario Set 2 with No H$_2$ HDV Deployment}\label{sec:sen_scenario_3}

The results in this section represent Sensitivity Set 3 as described in Figure 3b of the main text.

\begin{figure}[pos = H]
    \centering
    \includegraphics[width=1\linewidth]{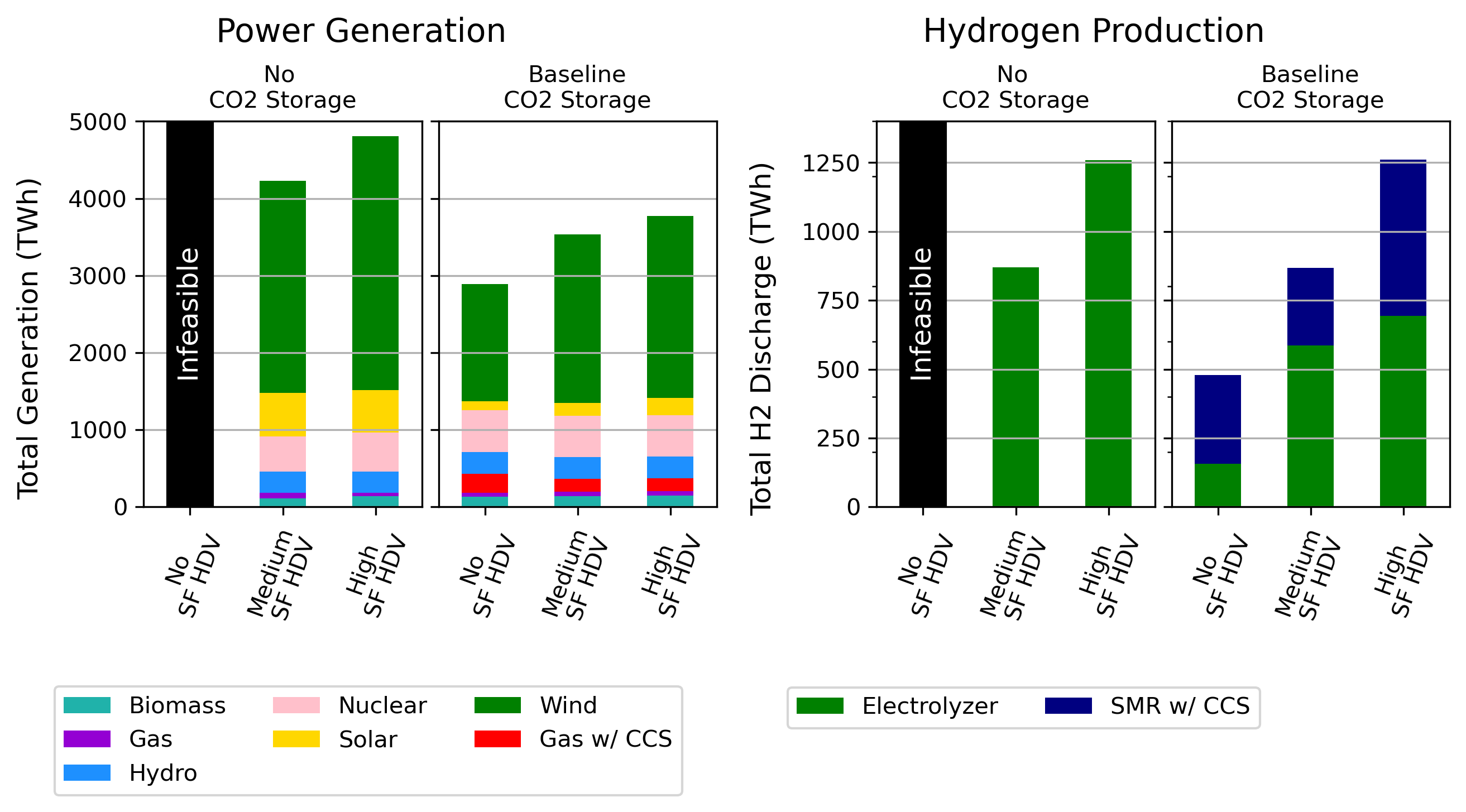}
    \caption[Power and H$_2$ Generation Sensitivity Scenario Set 3]{Power and hydrogen generation for baseline and no CO$_2$ sequestration scenarios under no H$_2$ HDV adoption. The left set of charts shows power generation and the right set of charts shows H$_2$  generation. Within each panel, the amount of synthetic fuel adoption increases moving from left to right.}
    \label{fig:sf_hdv_power_h2_h2_sen}
\end{figure}

\begin{figure}[pos = H]
    \centering
    \includegraphics[width=1\linewidth]{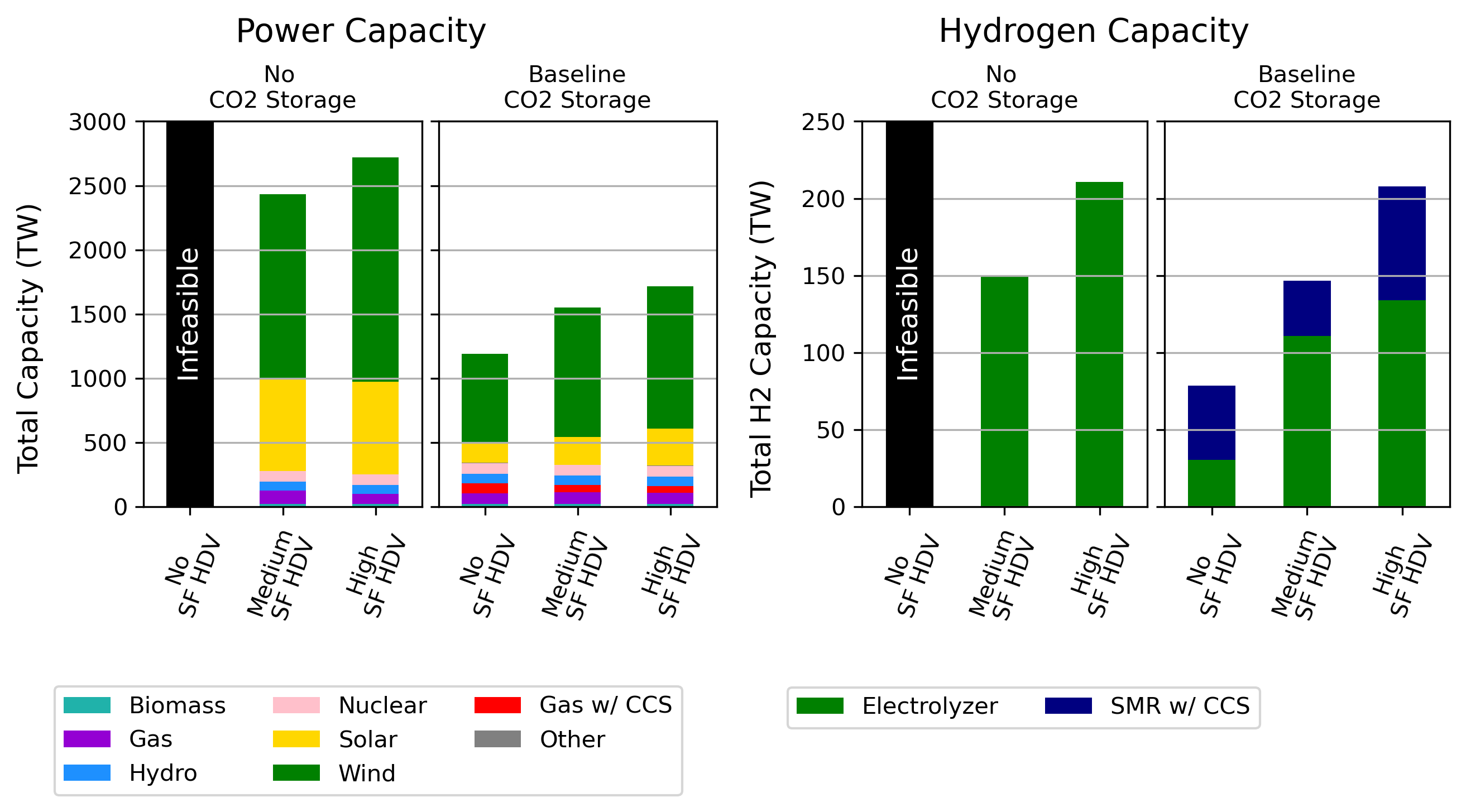}
    \caption[Power and H$_2$ Capacity Sensitivity Scenario Set 3]{Power and H$_2$ capacity for baseline and no CO$_2$ sequestration scenarios under no H$_2$ HDV adoption. The left set of charts shows power generation and the right set of charts shows H$_2$ generation. Within each panel, the amount of synthetic fuel adoption increases moving from left to right.}
    \label{fig_sf_hdv_power_h2_cap_h2_sen}
\end{figure}

\begin{figure}[pos = H]
    \centering
    \includegraphics[width=0.5\linewidth]{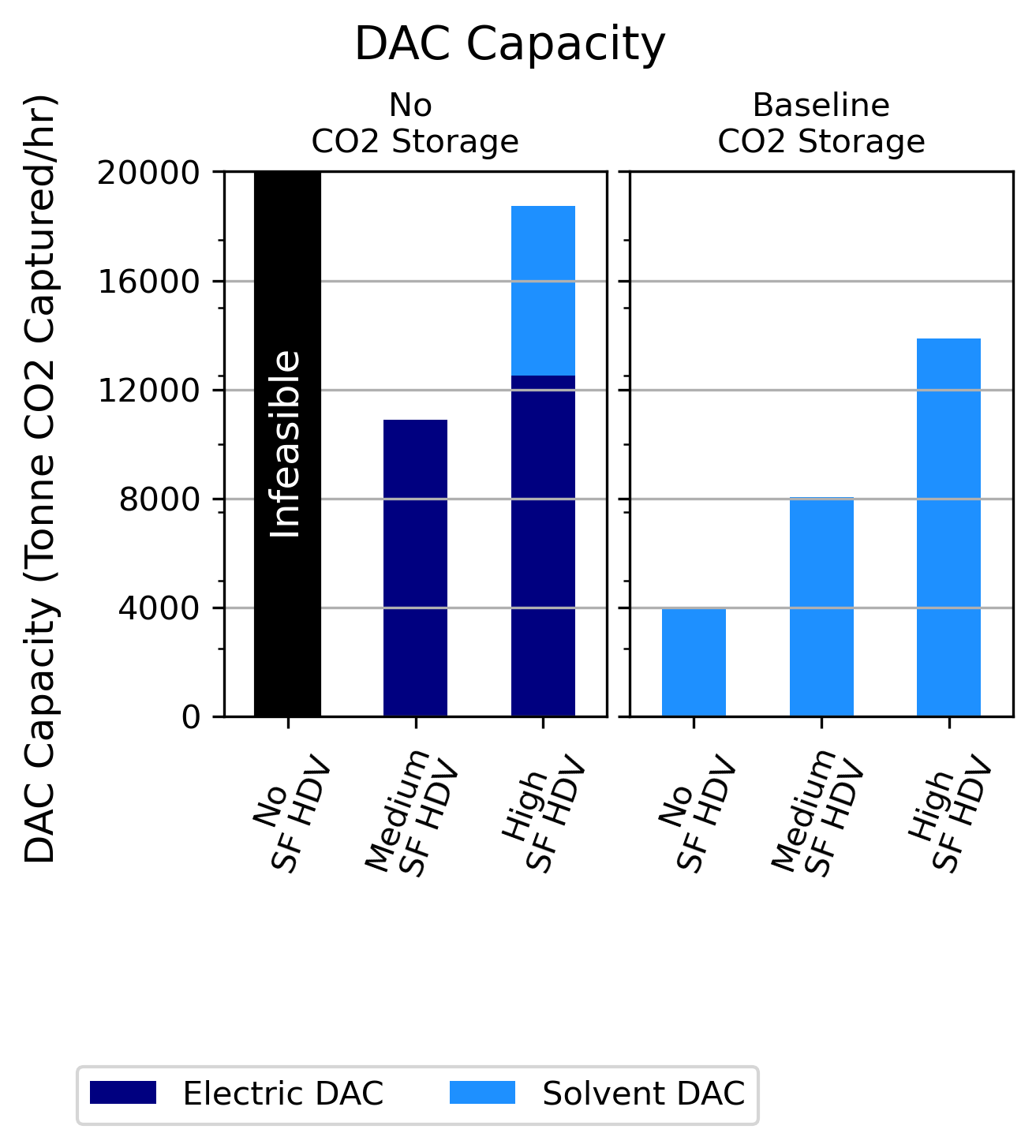}
    \caption[DAC Capacity Sensitivity Scenario Set 3]{Direct Air Capture capacity for baseline and no CO$_2$ sequestration scenarios under no H$_2$ HDV adoption. Within each panel, the amount of synthetic fuel adoption increases moving from left to right.}
    \label{fig_h2_hdv_dac_h2_sen}
\end{figure}

\begin{figure}[pos = H]
    \centering
    \includegraphics[width=0.5\linewidth]{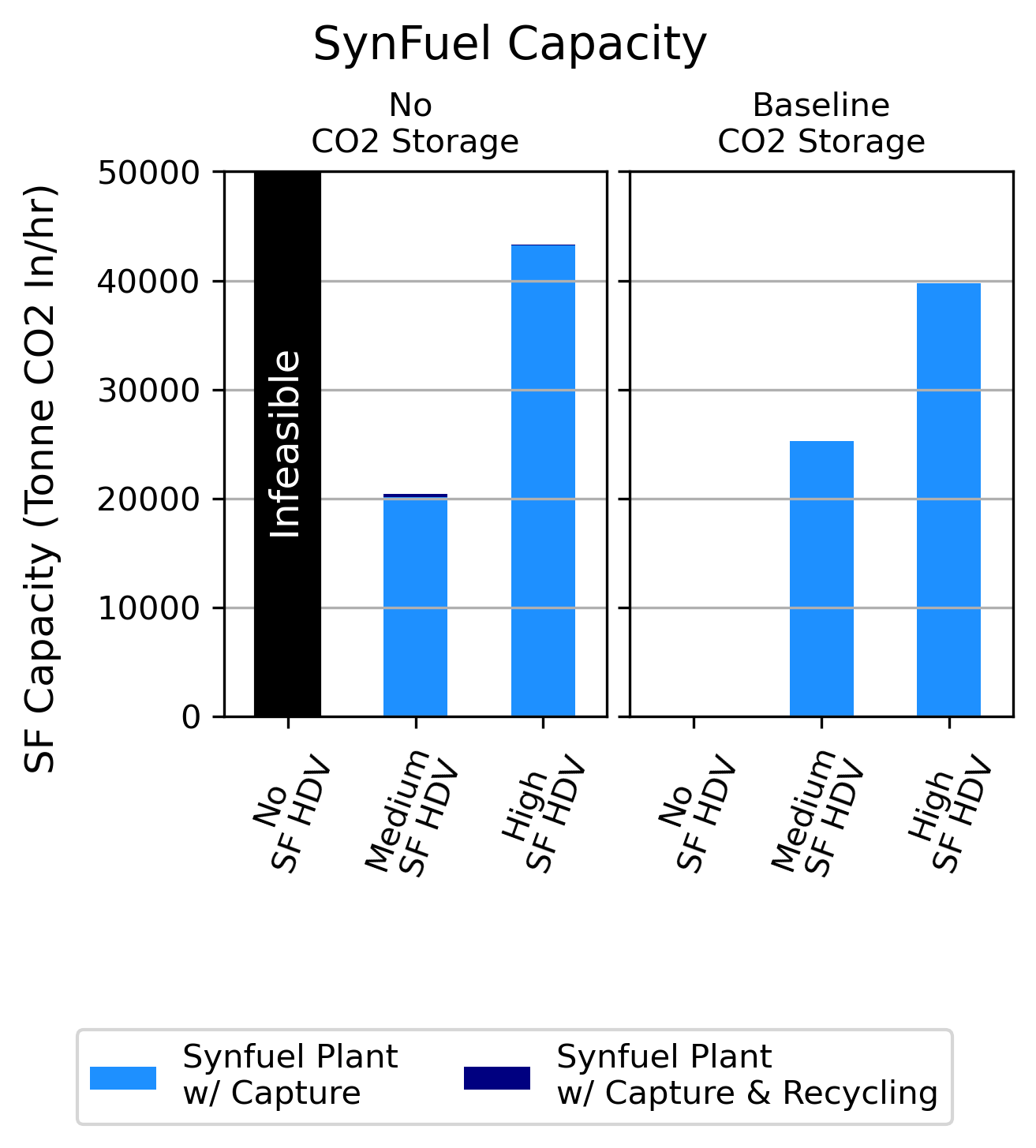}
    \caption[Synthetic fuel Capacity Sensitivity Scenario Set 3]{Synthetic fuel capacity for baseline and no CO$_2$ sequestration scenarios under no H$_2$ HDV adoption. Within each panel, the amount of synthetic fuel adoption increases moving from left to right.}
    \label{fig_h2_hdv_sf_h2_sen}
\end{figure}

\begin{figure}[pos = H]
    \centering
    \includegraphics[width=0.5\linewidth]{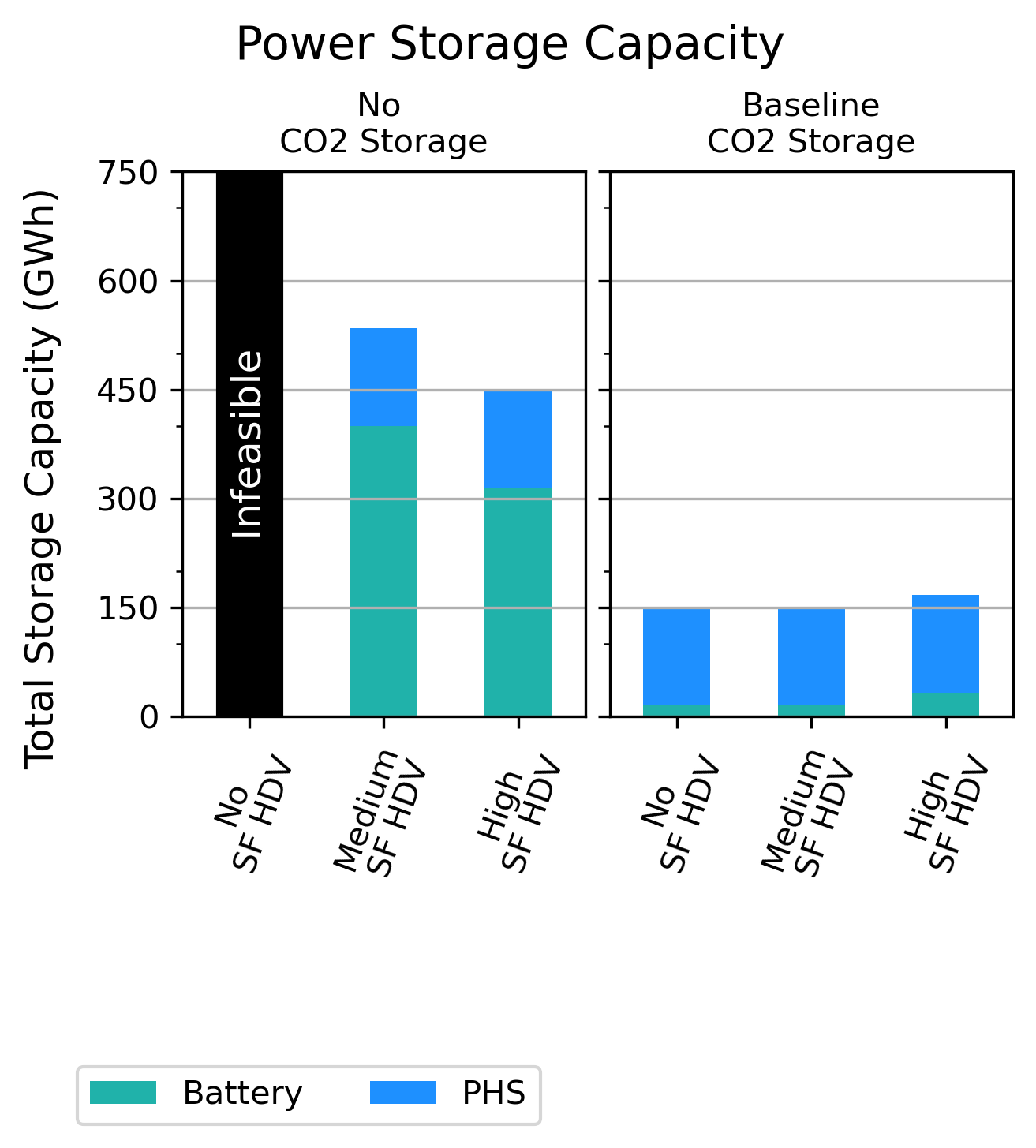}
    \caption[Power Storage Capacity Sensitivity Scenario Set 3]{Power storage capacity for baseline and no CO$_2$ sequestration scenarios under no H$_2$ HDV adoption. Within each panel, the amount of synthetic fuel adoption increases moving from left to right.}
    \label{fig_h2_hdv_power_cap_stor_emissions_sen}
\end{figure}

\begin{figure}[pos = H]
    \centering
    \includegraphics[width=0.5\linewidth]{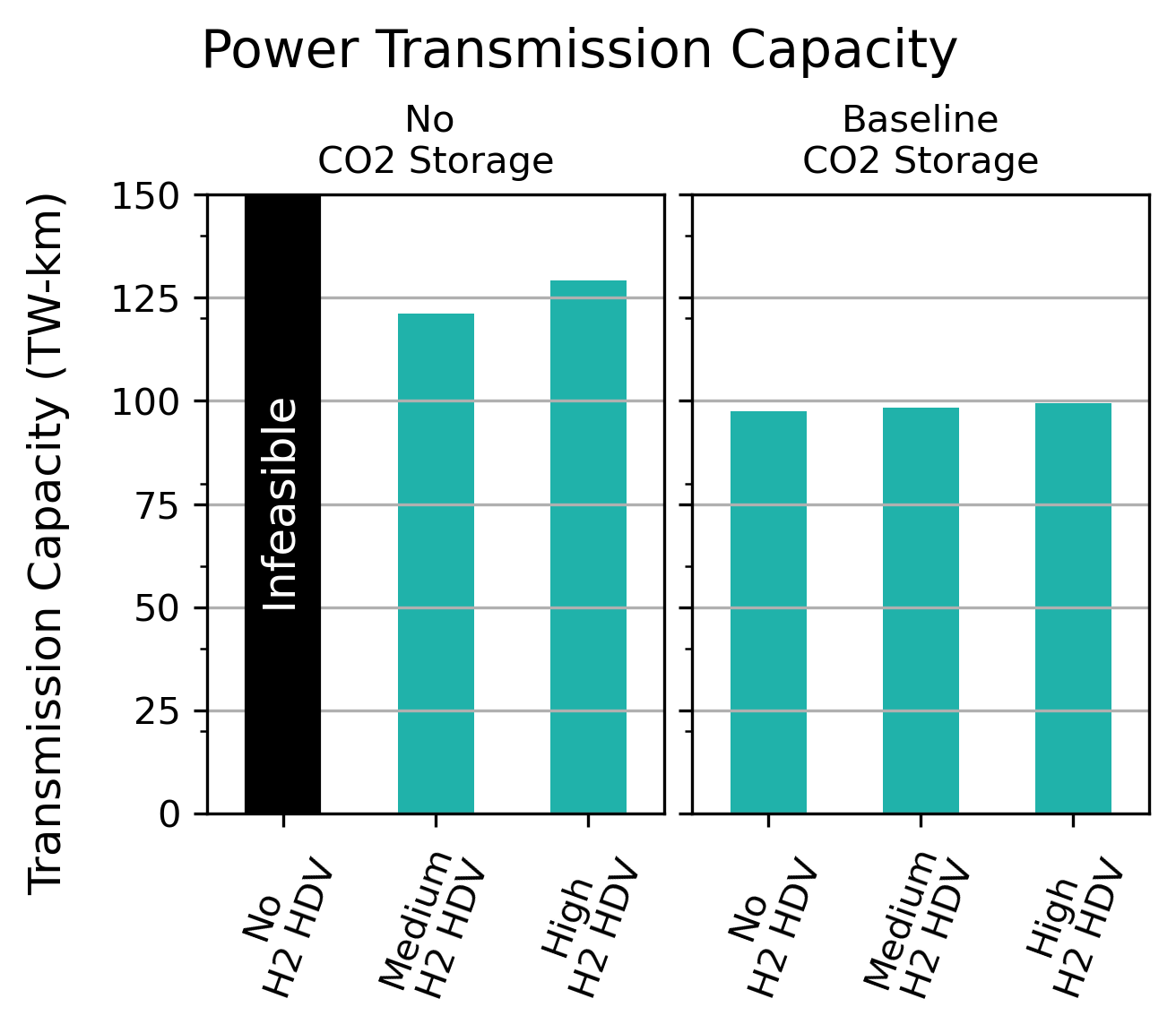}
    \caption[Power Transmission Capacity Sensitivity Scenario Set 3]{Power transmission capacity for baseline and no CO$_2$ sequestration scenarios under no H$_2$ HDV adoption. Within each panel, the amount of synthetic fuel adoption increases moving from left to right.}
    \label{fig_sf_hdv_elec_trans_h2_sen}
\end{figure}

\begin{figure}[pos = H]
    \centering
    \includegraphics[width=0.5\linewidth]{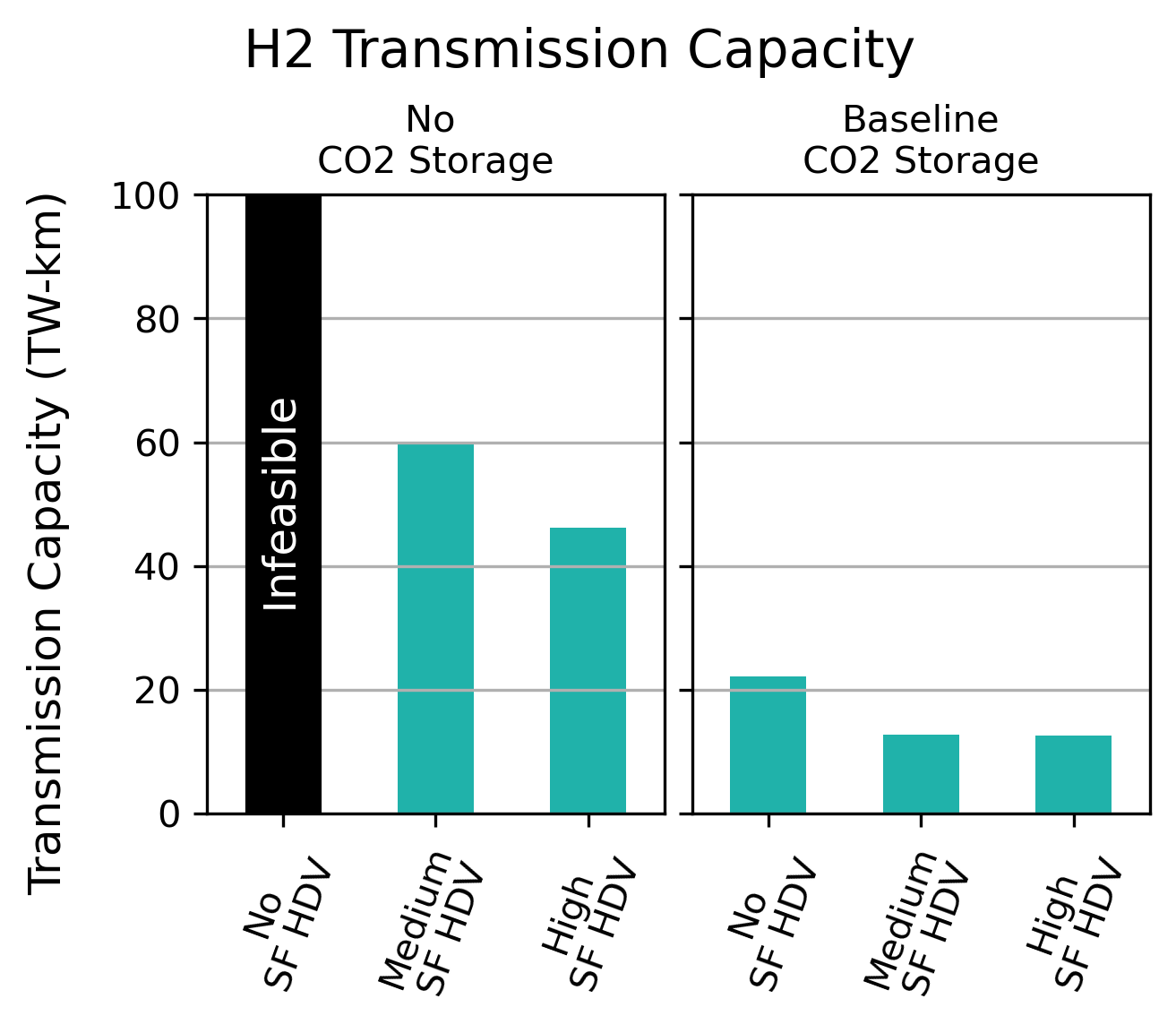}
    \caption[H$_2$ Transmission Capacity Sensitivity Scenario Set 3]{H$_2$ transmission capacity for baseline and no CO$_2$ sequestration scenarios under no H$_2$ HDV adoption. Within each panel, the amount of synthetic fuel adoption increases moving from left to right.}
    \label{fig_sf_hdv_h2_trans_h2_sen}
\end{figure}

\begin{figure}[pos = H]
    \centering
    \includegraphics[width=0.5\linewidth]{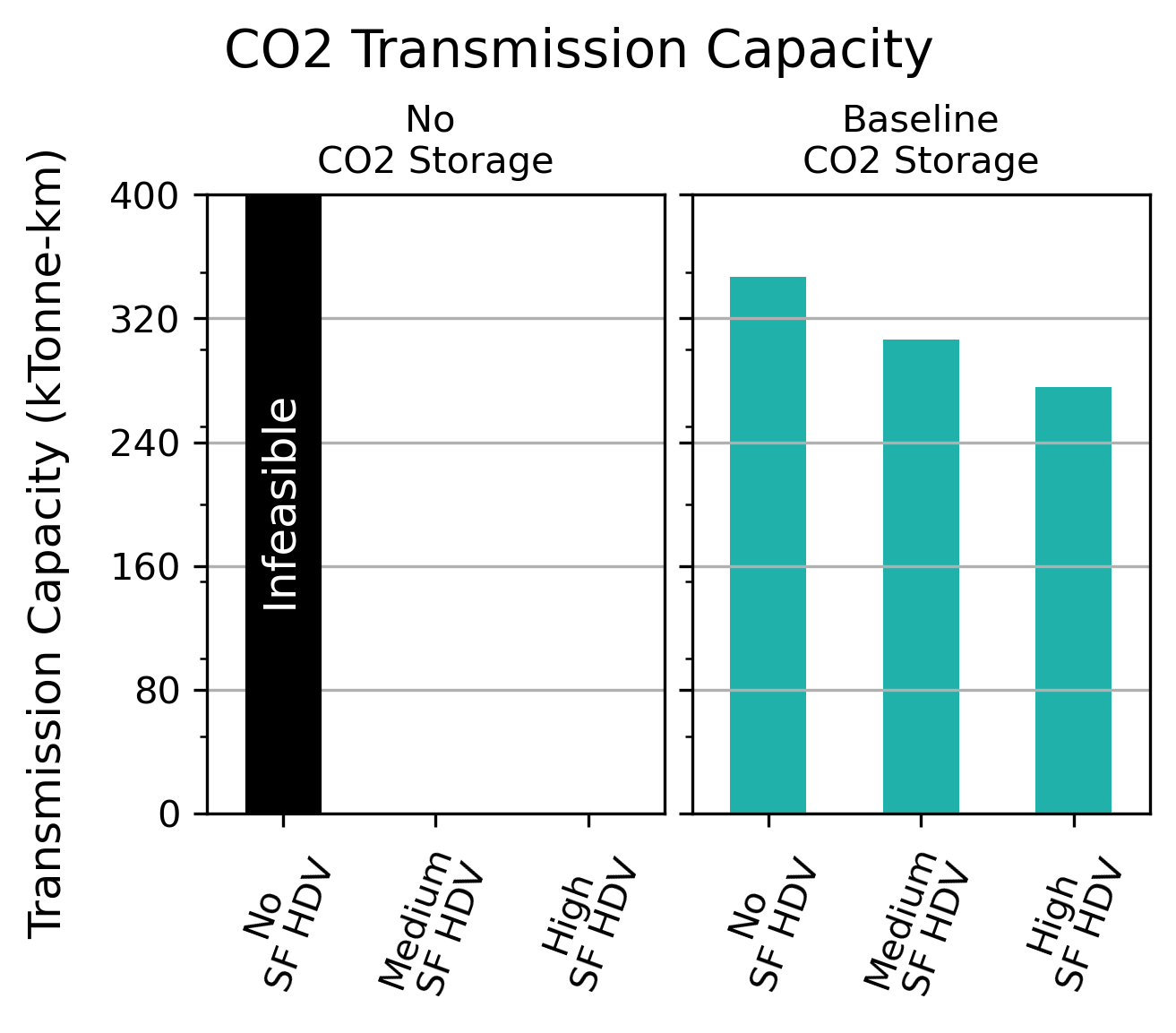}
    \caption[CO$_2$ Transmission Capacity Sensitivity Scenario Set 3]{CO$_2$ transmission capacity for baseline and no CO$_2$ sequestration scenarios under no H$_2$ HDV adoption. Within each panel, the amount of synthetic fuel adoption increases moving from left to right.}
    \label{fig_sf_hdv_co2_trans_h2_sen}
\end{figure}

\begin{figure}[pos = H]
    \centering
    \includegraphics[width=1\linewidth]{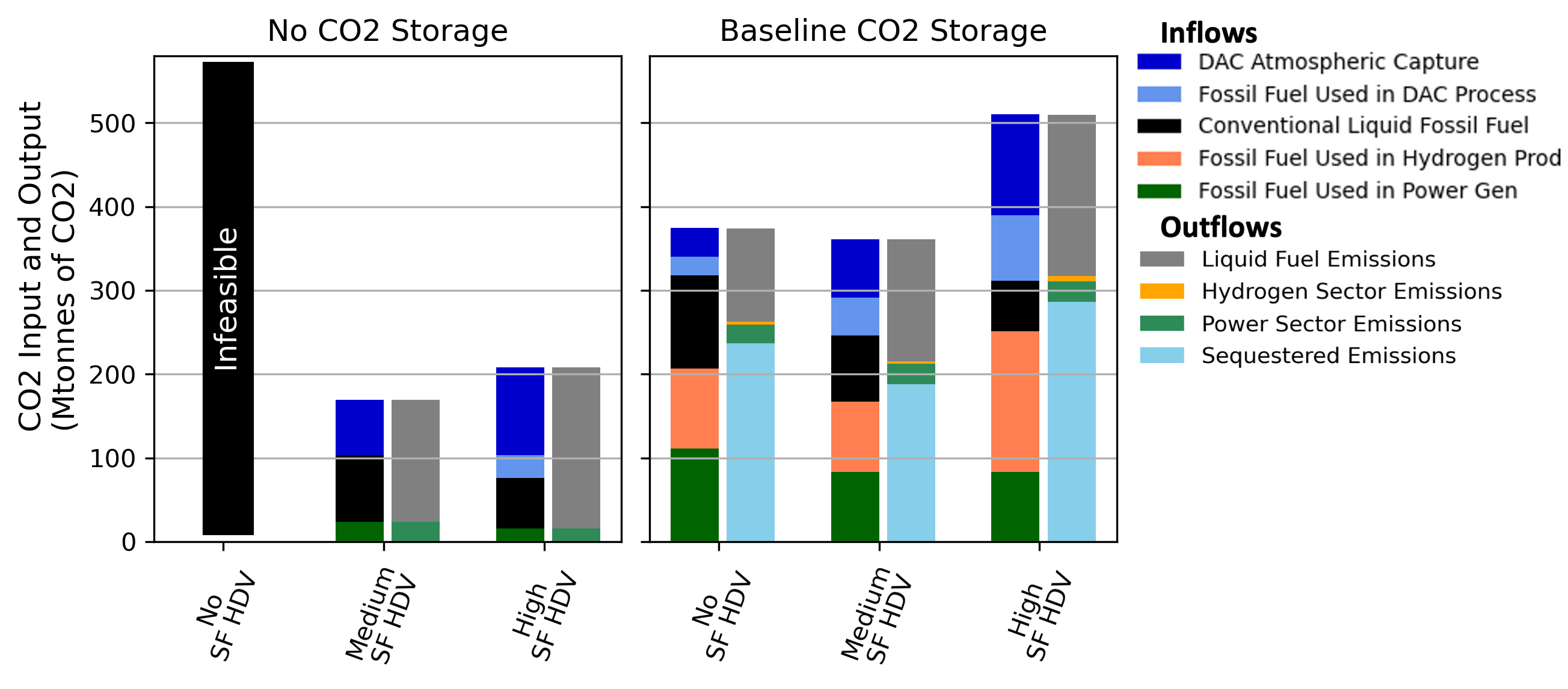}
    \caption[CO$_2$ Balance Sensitivity Scenario Set 3]{System CO$_2$ balance under varying levels of SF adoption and no H$_2$ HDV adoption. The subfigure on the left shows the CO$_2$ balance under no CO$_2$ sequestration availability, while the one on the right shows the CO$_2$ balance under baseline CO$_2$ sequestration availability. Within each subplot the SF adoption level increases left to right. The leftward column represents CO$_2$ input into the system, while the rightward column represents CO$_2$ outputted by the system. Emissions constraint can be calculated from the chart by subtracting sequestered emissions and DAC atmospheric capture from the emission outflows. Emissions constraint can be calculated from the chart by subtracting sequestered emissions and DAC atmospheric capture from the emission outflows.}
    \label{fig:sf_hdv_co2_h2_sen}
\end{figure}

\begin{figure}[pos = H]
    \centering
    \includegraphics[width=1\linewidth]{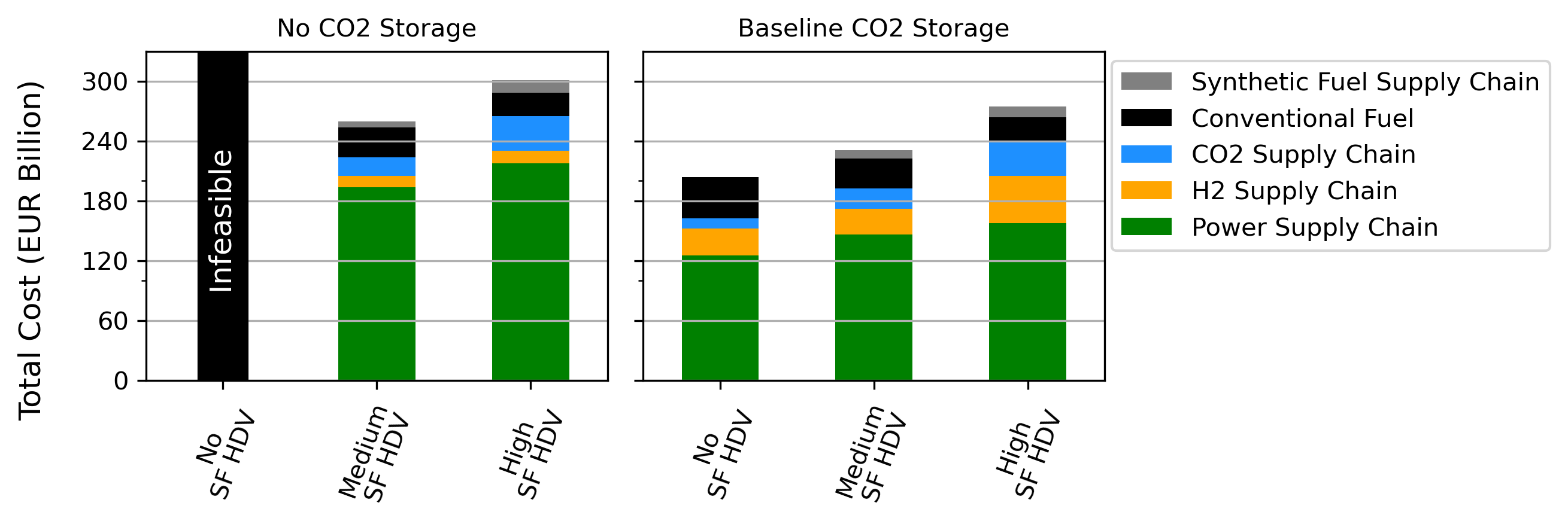}
    \caption[Costs Sensitivity Scenario Set 3]{Annualized bulk-system costs under varying levels of SF adoption and no H$_2$ HDV adoption. The subfigure on the left shows the cost breakdown under no CO$_2$ sequestration availability, while the one on the right shows the cost breakdown under baseline CO$_2$ sequestration availability. Within each subplot the H$_2$ HDV adoption level increases left to right. The costs do not include vehicle replacement or H$_2$ distribution costs. }
    \label{fig:sf_hdv_cost_h2_sen}
\end{figure}

\begin{table}
\caption{This tables shows the marginal price of abatement of CO$_2$ for Sensitivity Scenario Set 3}
\label{co2_price_sen_set_3}
\begin{tabular}{lllr}
\toprule
CO$_2$ Storage & H$_2$ HDV Level & Synthetic Fuel HDV Level & CO$_2$ Marginal Cost of Abatement \\
\midrule
Baseline & None & None & 293.92 \\
Baseline & None & 0.18 & 293.90 \\
Baseline & None & 0.35 & 293.91 \\
None & None & 0.18 & 835.18 \\
None & None & 0.35 & 632.72 \\
\bottomrule
\end{tabular}
\end{table}

\newpage
\subsection{Sensitivity Set 4: Core Scenario Set 2 with Natural Gas Price Sensitivity}\label{sec:sen_scenario_4}

The results in this section represent Sensitivity Set 4 as described in Figure 3b of the main text.

\begin{figure}[pos = H]
    \centering
    \begin{subfigure}[t]{0.8\linewidth}
        \includegraphics[width=\linewidth, trim=0cm 2.6cm 0cm 0.7cm, clip]{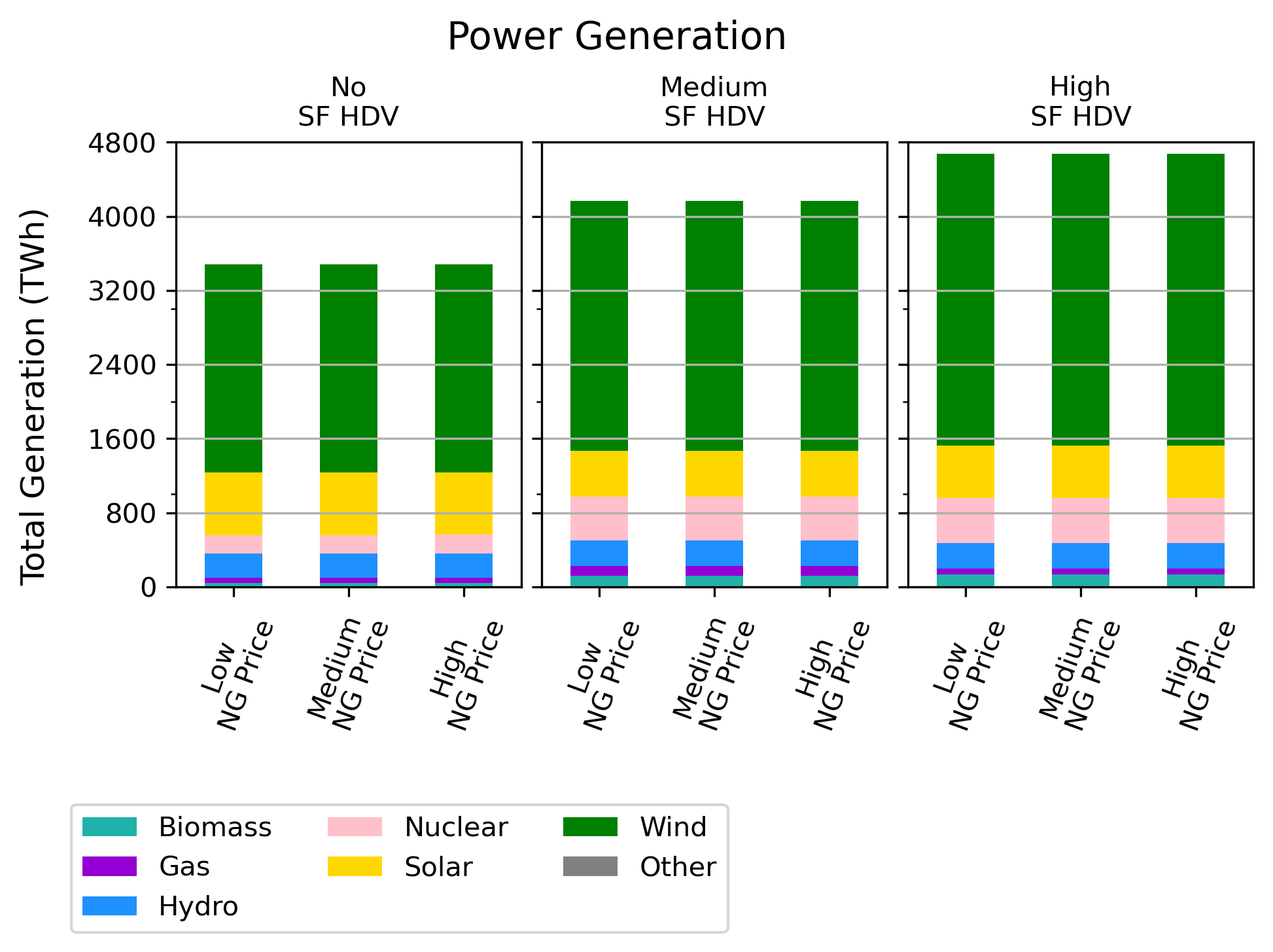}
        \caption{No CO$_2$ Storage}
    \end{subfigure}
    
    \begin{subfigure}[b]{0.8\linewidth}
        \includegraphics[width=\linewidth, trim=0cm 0cm 0cm 0.7cm, clip]{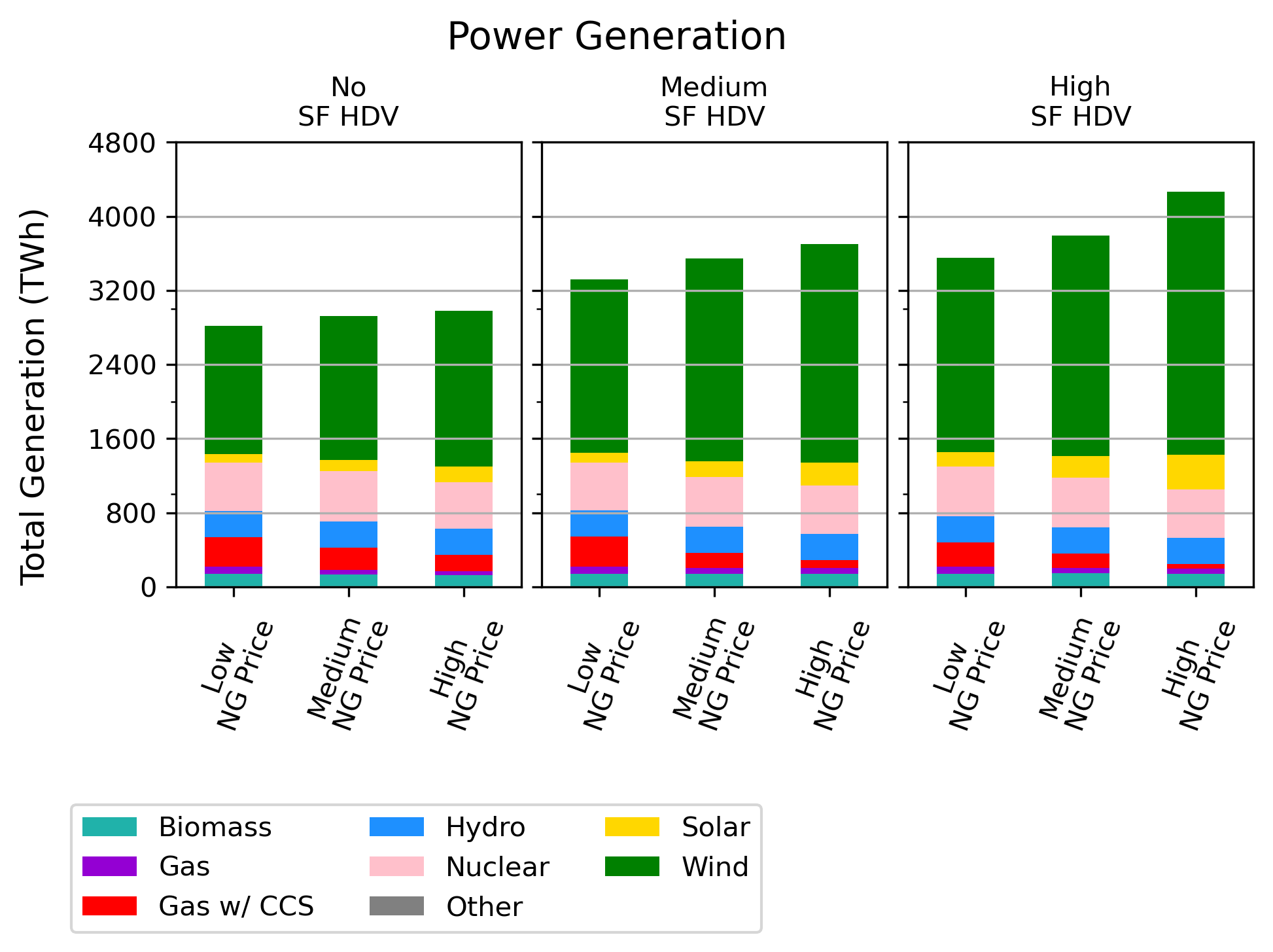}
        \caption{Baseline CO$_2$ Storage}
    \end{subfigure}
    
    \caption[Power Generation Sensitivity Scenario Set 4]{Power generation for no (sub-figure a) and baseline (sub-figure b) CO$_2$ sequestration scenarios under medium H$_2$ HDV adoption and varying scenarios of synthetic fuel adoption. Within each panel, the price of natural gas increases left to right. Across panels, the amount of H$_2$ HDV adoption increases moving from left to right. The middle panels correspond to the core set of scenarios.}
    \label{fig:sf_hdv_power_ng_sen}
\end{figure}

\begin{figure}[pos = H]
    \centering
    \begin{subfigure}[t]{0.9\linewidth}
        \includegraphics[width=\linewidth, trim=0cm 1.8cm 0cm 0.8cm, clip]{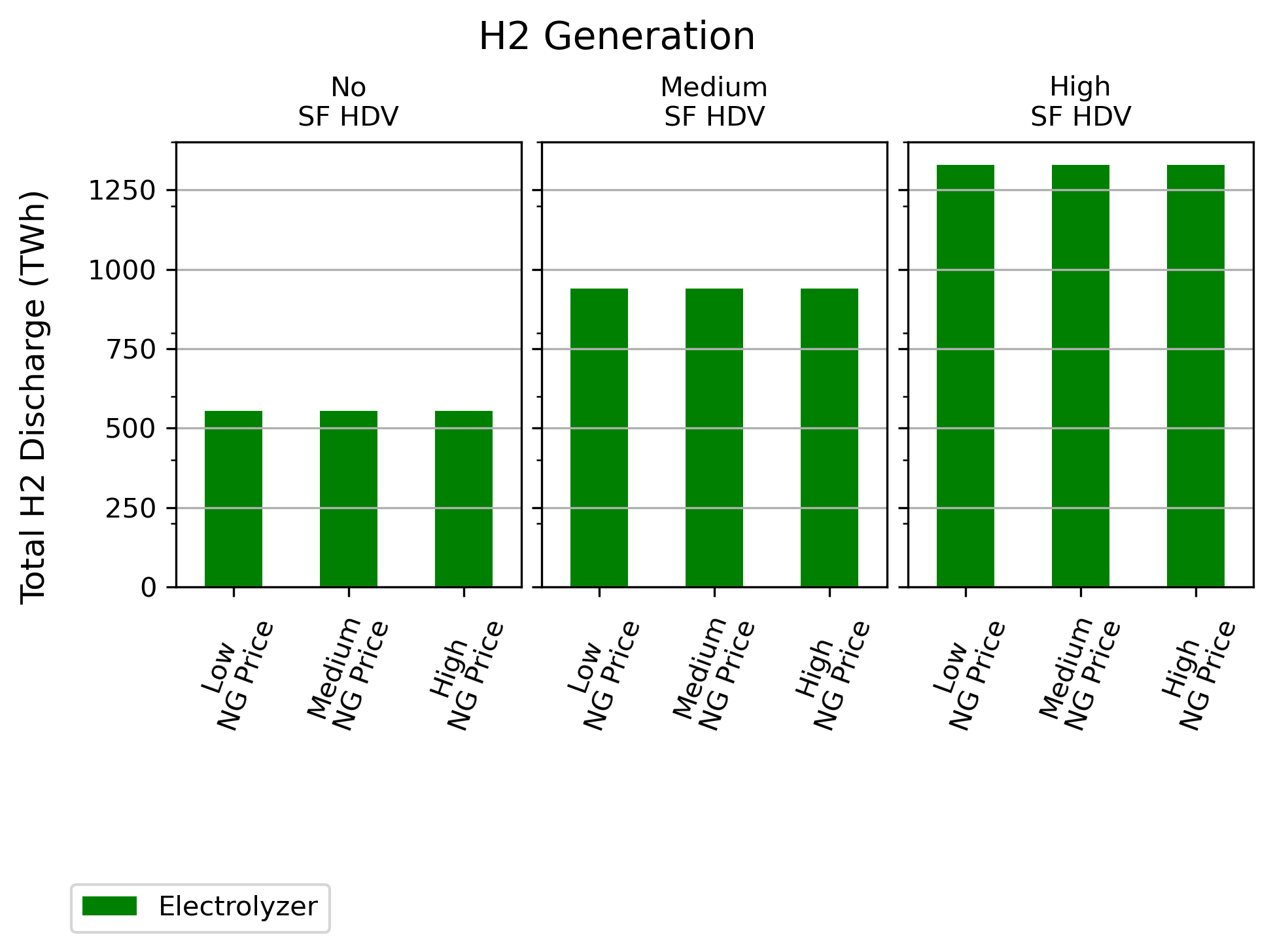}
        \caption{No CO$_2$ Storage}
    \end{subfigure}
    
    \begin{subfigure}[b]{0.9\linewidth}
        \includegraphics[width=\linewidth, trim=0cm 0cm 0cm 0.8cm, clip]{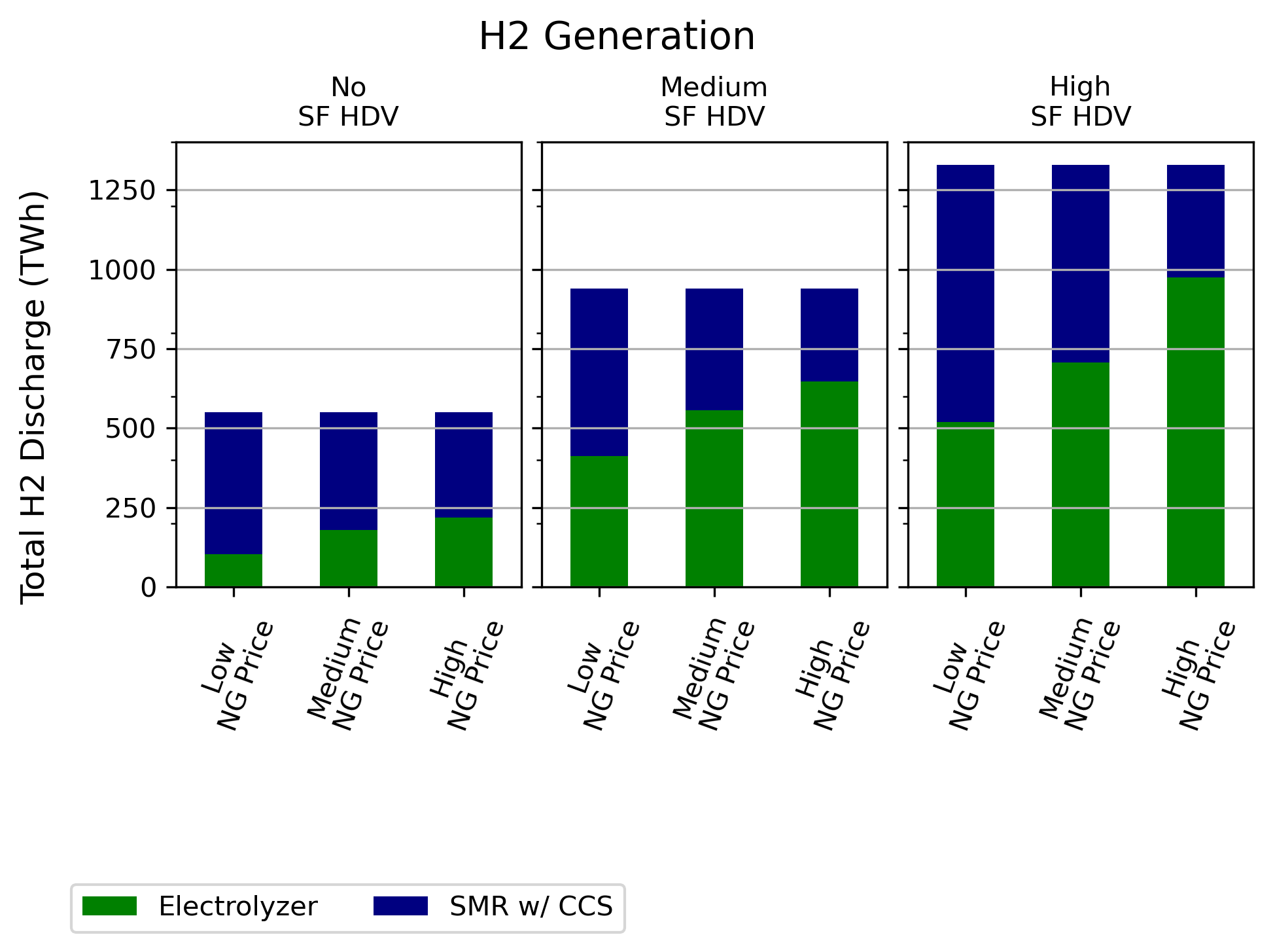}
        \caption{Baseline CO$_2$ Storage}
    \end{subfigure}
    
    \caption[H$_2$ Production Sensitivity Scenario Set 4]{H$_2$ generation for no (sub-figure a) and baseline (sub-figure b) CO$_2$ sequestration scenarios under medium H$_2$ HDV adoption and varying scenarios of synthetic fuel adoption. Within each panel, the price of natural gas increases left to right. Across panels, the amount of H$_2$ HDV adoption increases moving from left to right. The middle panels correspond to the core set of scenarios.}
    \label{fig:sf_hdv_h2_ng_sen}
\end{figure}

\begin{figure}[pos = H]
    \centering
    \begin{subfigure}[t]{0.9\linewidth}
        \includegraphics[width=\linewidth, trim=0cm 2.6cm 0cm 0.8cm, clip]{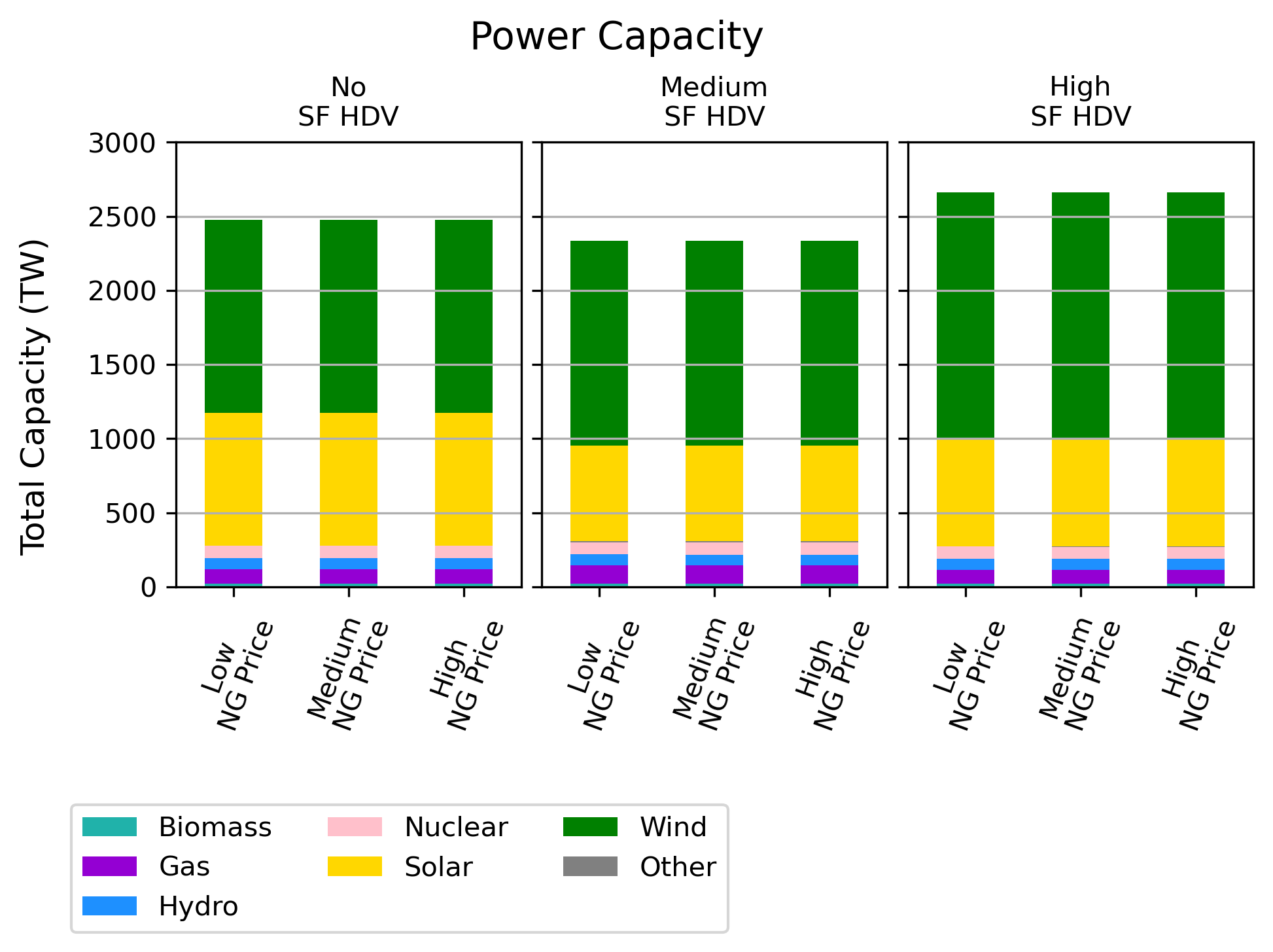}
        \caption{No CO$_2$ Storage}
    \end{subfigure}
    
    \begin{subfigure}[b]{0.9\linewidth}
        \includegraphics[width=\linewidth, trim=0cm 0cm 0cm 0.8cm, clip]{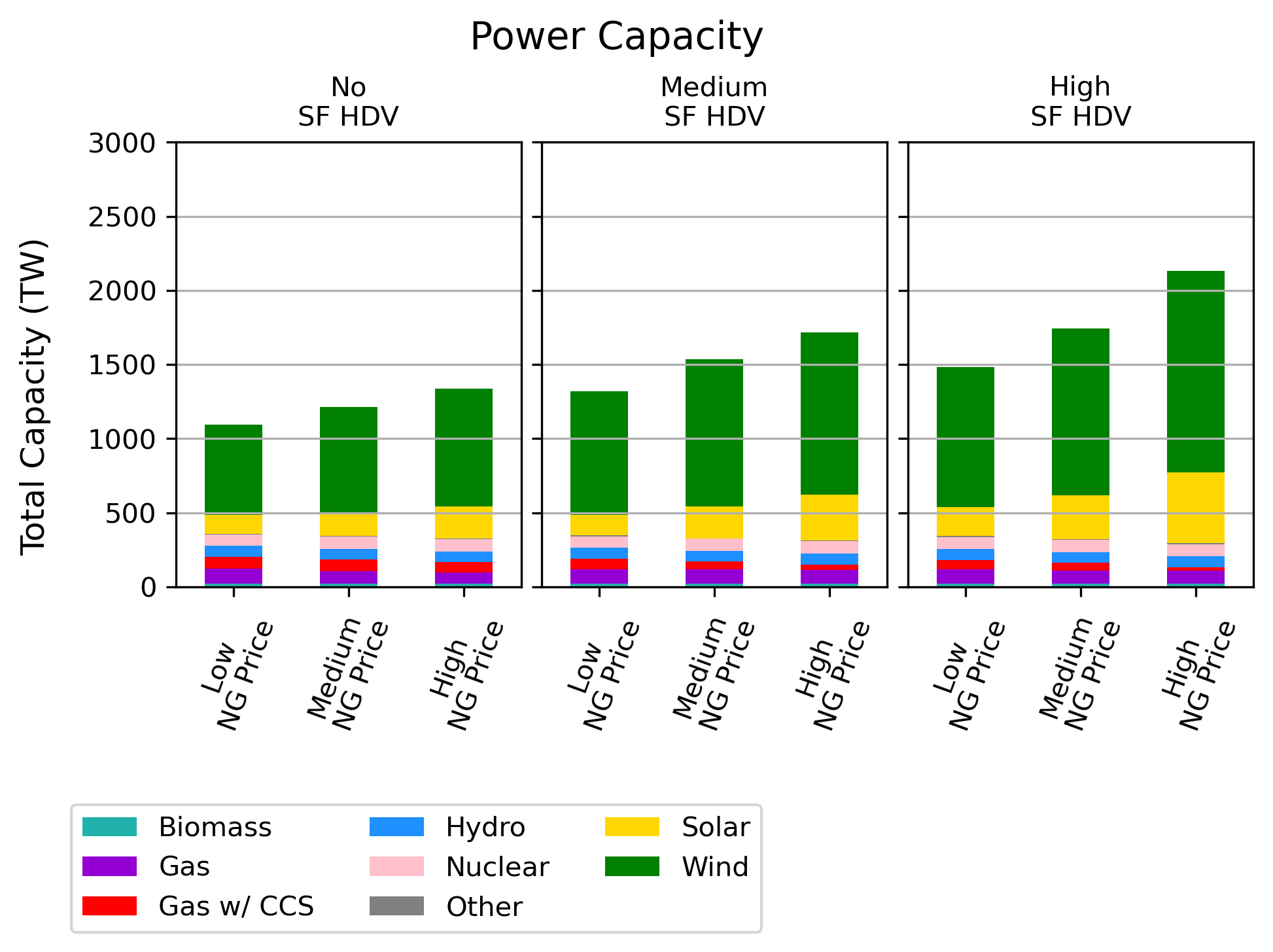}
        \caption{Baseline CO$_2$ Storage}
    \end{subfigure}
    
    \caption[Power Capacity Sensitivity Scenario Set 4]{Power capacity for no (sub-figure a) and baseline (sub-figure b) CO$_2$ sequestration scenarios under medium H$_2$ HDV adoption and varying scenarios of synthetic fuel adoption. Within each panel, the price of natural gas increases left to right. Across panels, the amount of H$_2$ HDV adoption increases moving from left to right. The middle panels correspond to the core set of scenarios.}
    \label{fig:sf_hdv_power_cap_ng_sen}
\end{figure}

\begin{figure}[pos = H]
    \centering
    \begin{subfigure}[t]{0.9\linewidth}
        \includegraphics[width=\linewidth, trim=0cm 1.7cm 0cm 0.8cm, clip]{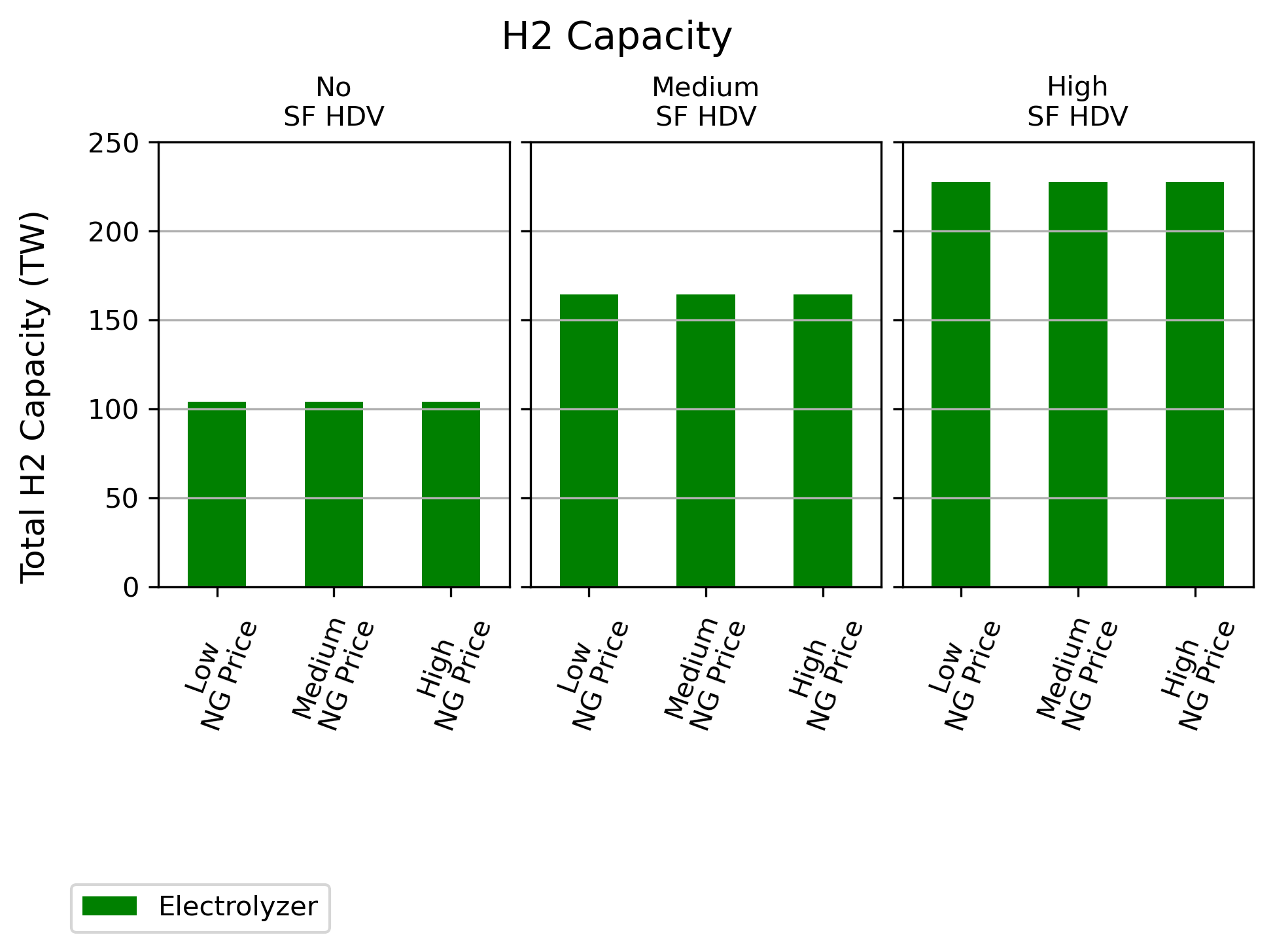}
        \caption{No CO$_2$ Storage}
    \end{subfigure}
    
    \begin{subfigure}[b]{0.9\linewidth}
        \includegraphics[width=\linewidth, trim=0cm 0cm 0cm 0.8cm, clip]{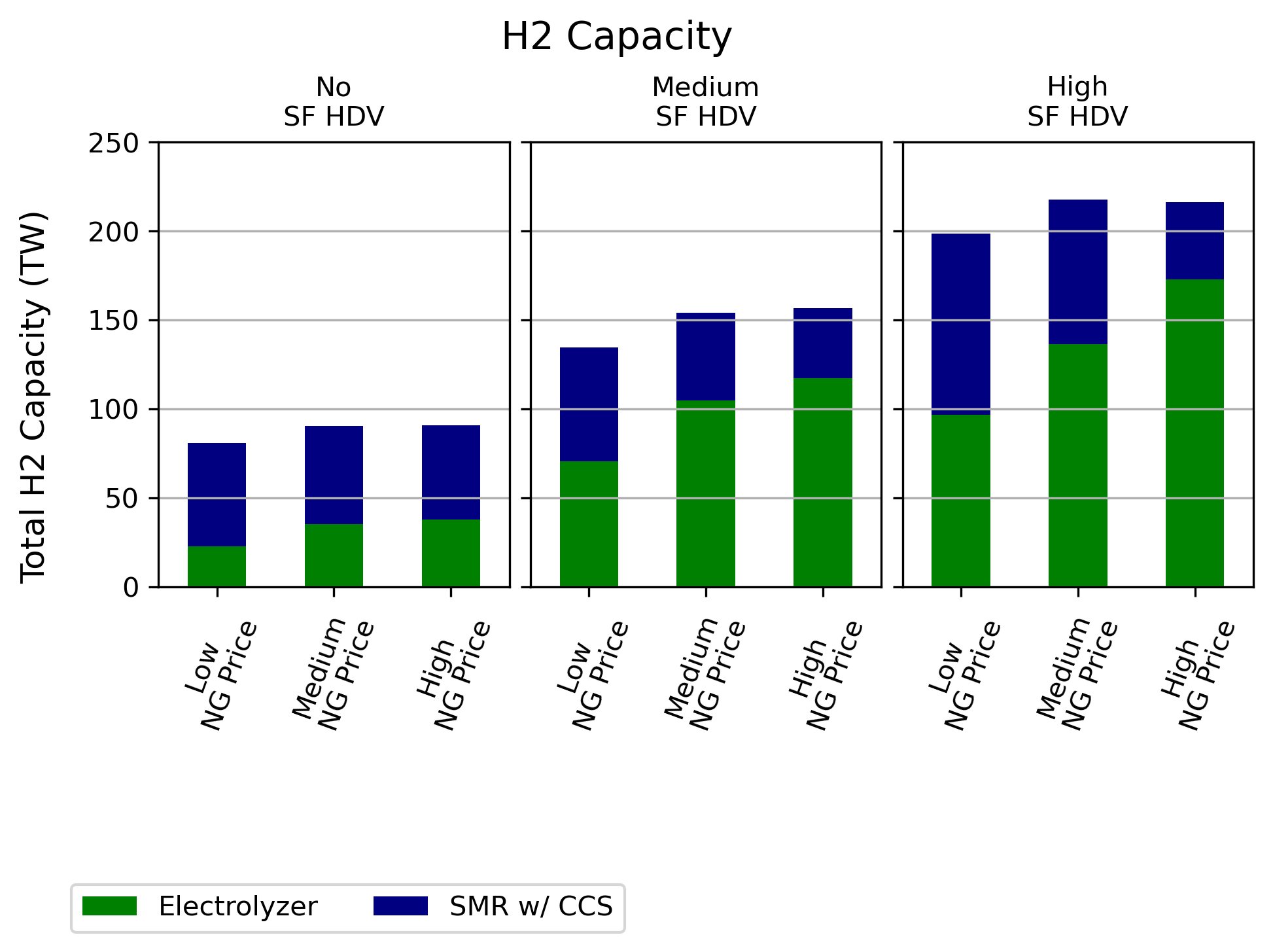}
        \caption{Baseline CO$_2$ Storage}
    \end{subfigure}
    
    \caption[H$_2$ Capacity Sensitivity Scenario Set 4]{H$_2$ capacity for no (sub-figure a) and baseline (sub-figure b) CO$_2$ sequestration scenarios under medium H$_2$ HDV adoption and varying scenarios of synthetic fuel adoption. Within each panel, the price of natural gas increases left to right. Across panels, the amount of H$_2$ HDV adoption increases moving from left to right. The middle panels correspond to the core set of scenarios.}
    \label{fig:sf_hdv_h2_cap_ng_sen}
\end{figure}

\begin{figure}[pos = H]
    \centering
    \begin{subfigure}[t]{0.9\linewidth}
        \includegraphics[width=\linewidth, trim=0cm 1.7cm 0cm 0.8cm, clip]{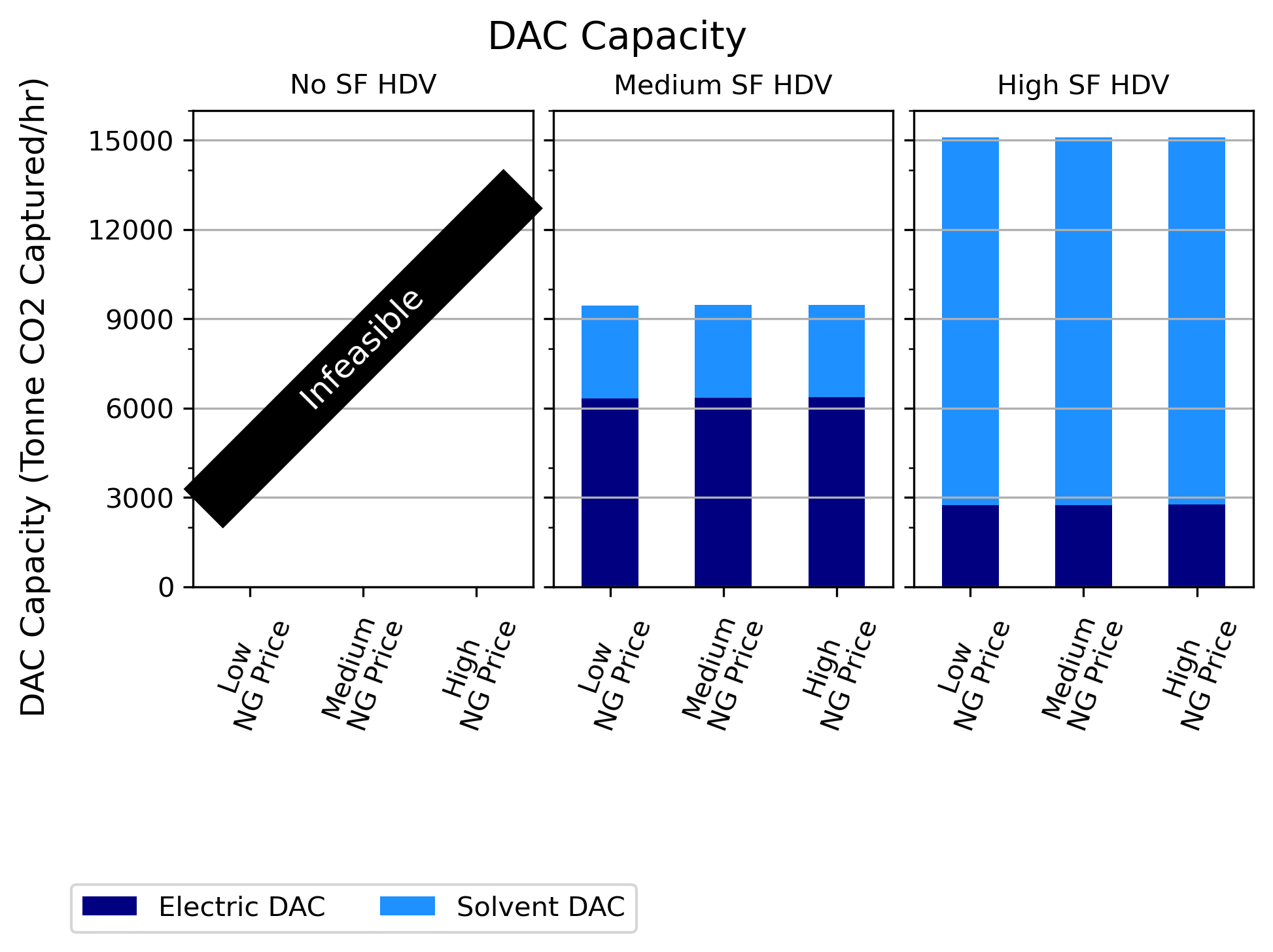}
        \caption{No CO$_2$ Storage}
    \end{subfigure}
    
    \begin{subfigure}[b]{0.9\linewidth}
        \includegraphics[width=\linewidth, trim=0cm 0cm 0cm 0.8cm, clip]{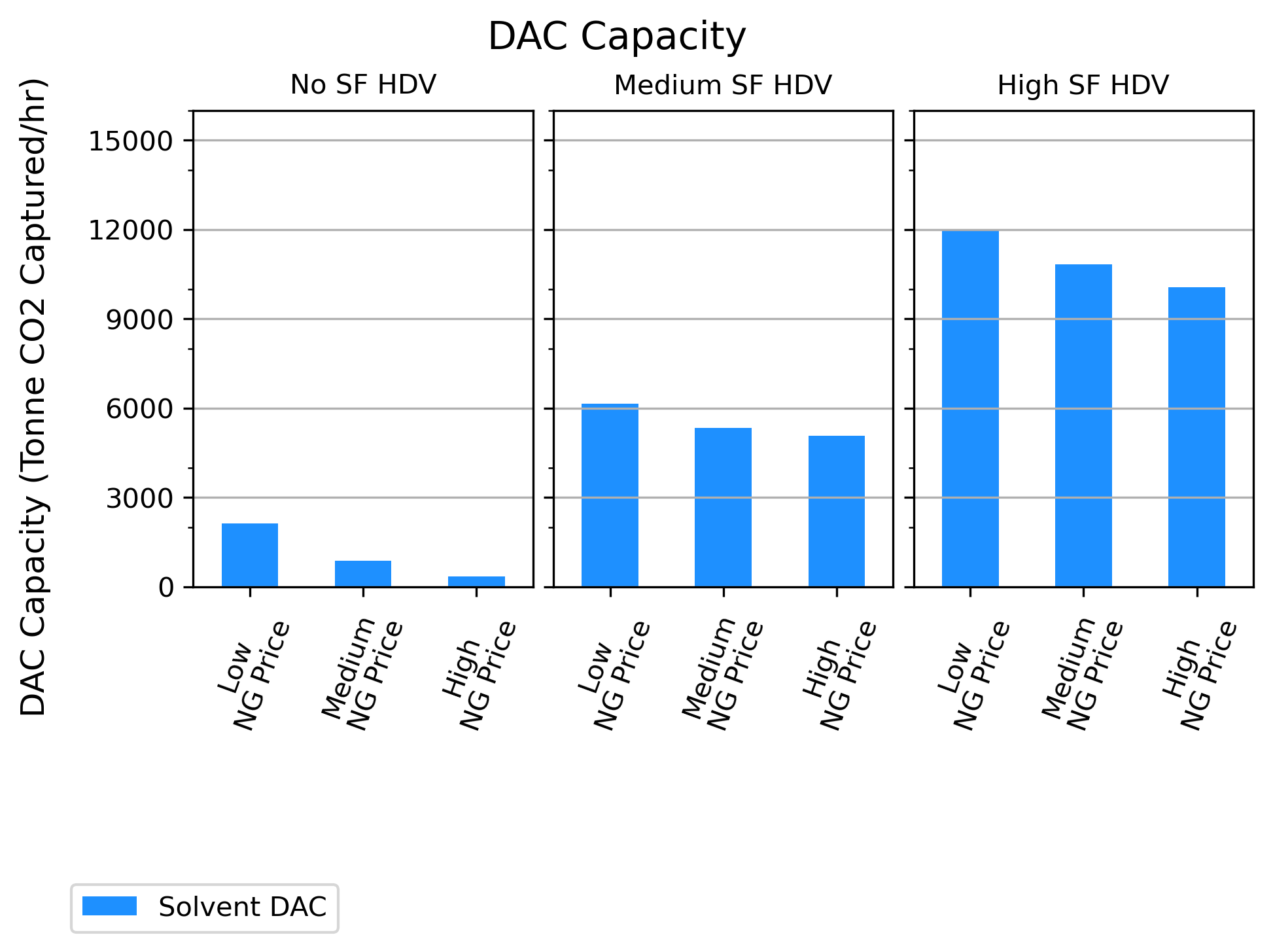}
        \caption{Baseline CO$_2$ Storage}
    \end{subfigure}
    
    \caption[Direct Air Capture Capacity Sensitivity Scenario Set 4]{Direct Air Capture capacity for no (sub-figure a) and baseline (sub-figure b) CO$_2$ sequestration scenarios under medium H$_2$ HDV adoption and varying scenarios of synthetic fuel adoption. Within each panel, the price of natural gas increases left to right. Across panels, the amount of H$_2$ HDV adoption increases moving from left to right. The middle panels correspond to the core set of scenarios.}
    \label{fig:sf_hdv_dac_cap_ng_sen}
\end{figure}

\begin{figure}[pos = H]
    \centering
    \begin{subfigure}[t]{0.9\linewidth}
        \includegraphics[width=\linewidth, trim=0cm 1.7cm 0cm 0.8cm, clip]{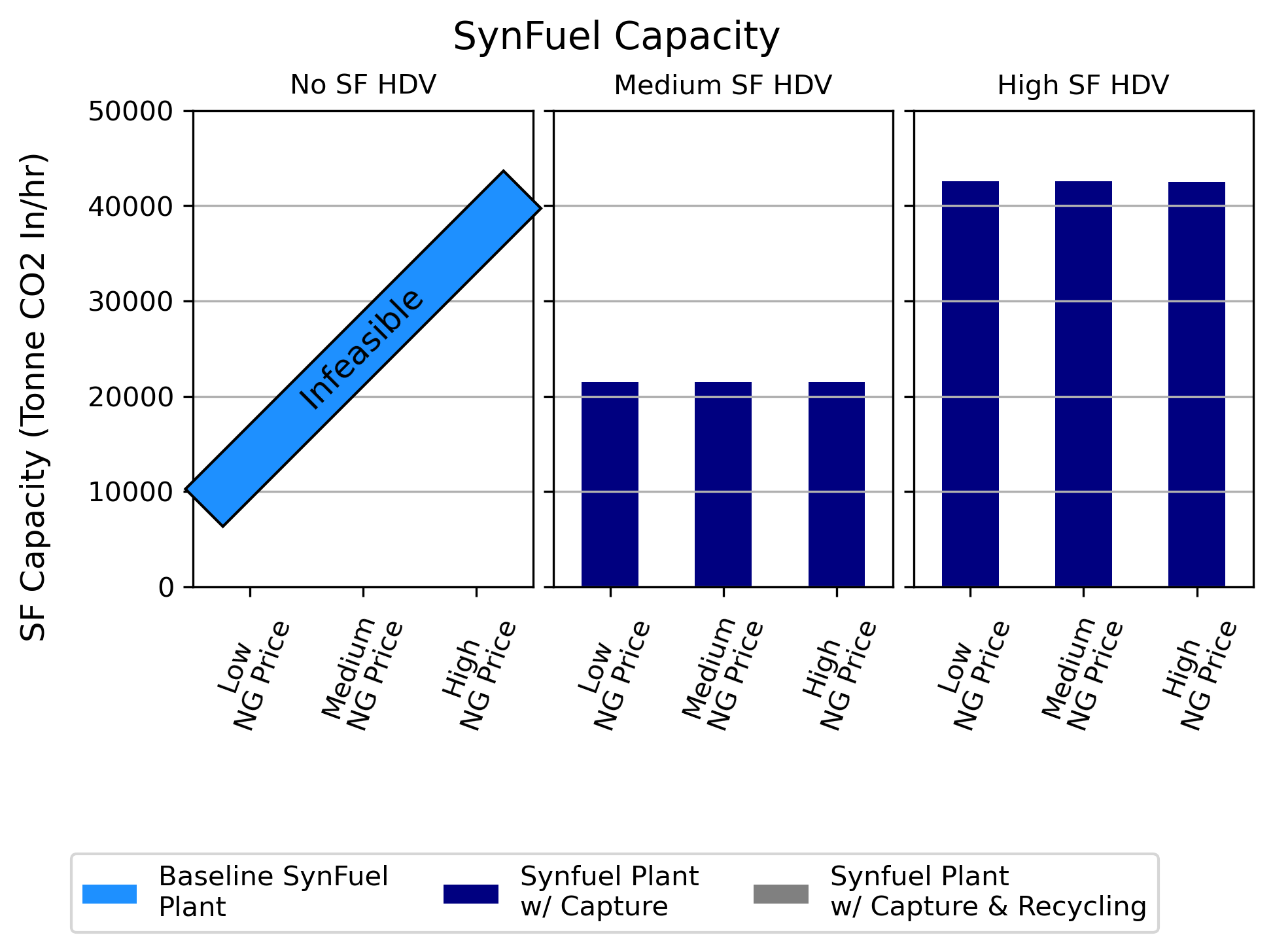}
        \caption{No CO$_2$ Storage}
    \end{subfigure}
    
    \begin{subfigure}[b]{0.9\linewidth}
        \includegraphics[width=\linewidth, trim=0cm 0cm 0cm 0.8cm, clip]{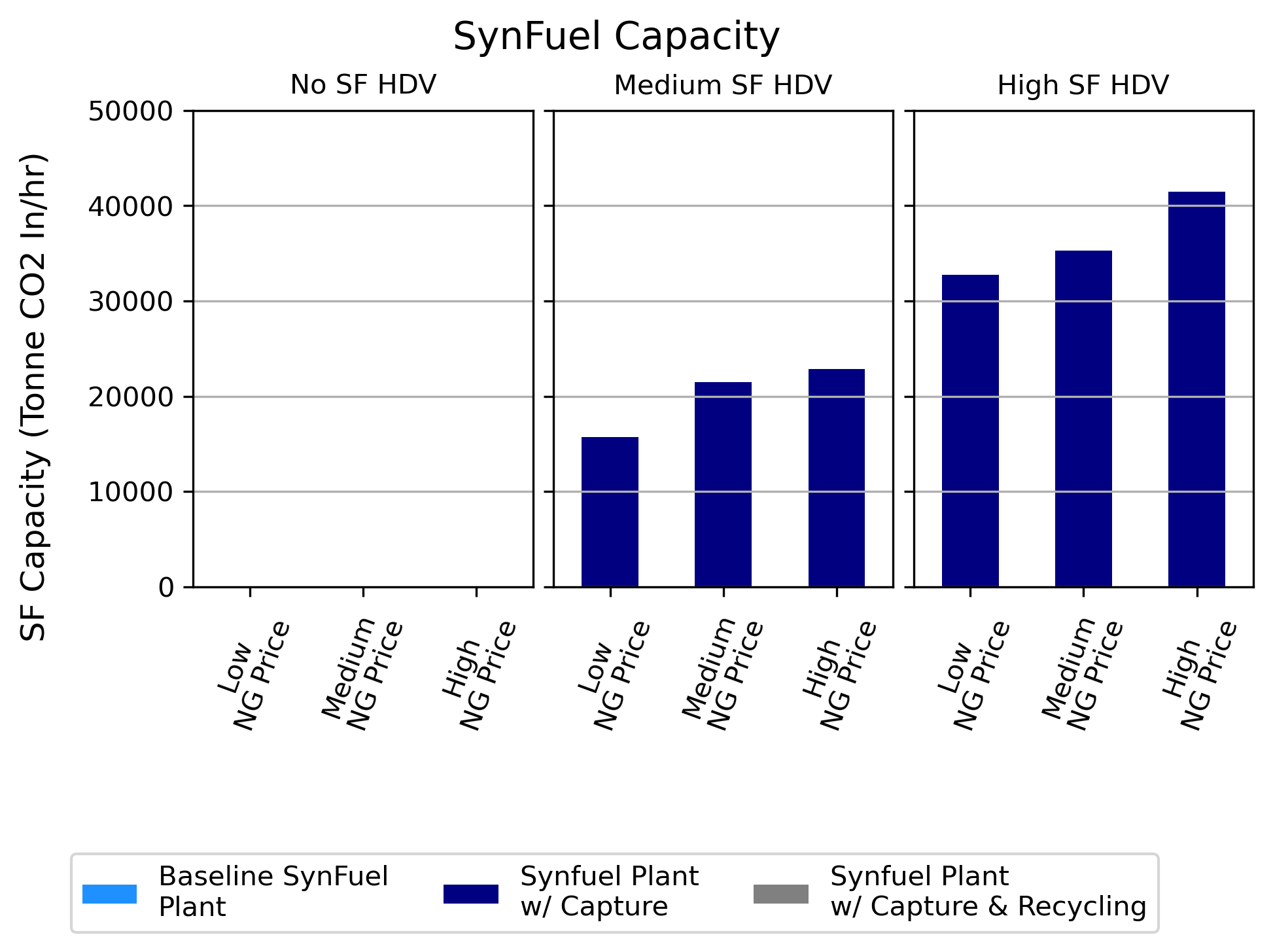}
        \caption{Baseline CO$_2$ Storage}
    \end{subfigure}
    
    \caption[Synthetic fuel Capacity Sensitivity Scenario Set 4]{Synthetic fuel plant capacity for no (sub-figure a) and baseline (sub-figure b) CO$_2$ sequestration scenarios under medium H$_2$ HDV adoption and varying scenarios of synthetic fuel adoption. Within each panel, the price of natural gas increases left to right. Across panels, the amount of H$_2$ HDV adoption increases moving from left to right. The middle panels correspond to the core set of scenarios.}
    \label{fig:sf_hdv_sf_cap_ng_sen}
\end{figure}

\begin{figure}[pos = H]
    \centering
    \begin{subfigure}[t]{0.9\linewidth}
        \includegraphics[width=\linewidth, trim=0cm 1.5cm 0cm 0.8cm, clip]{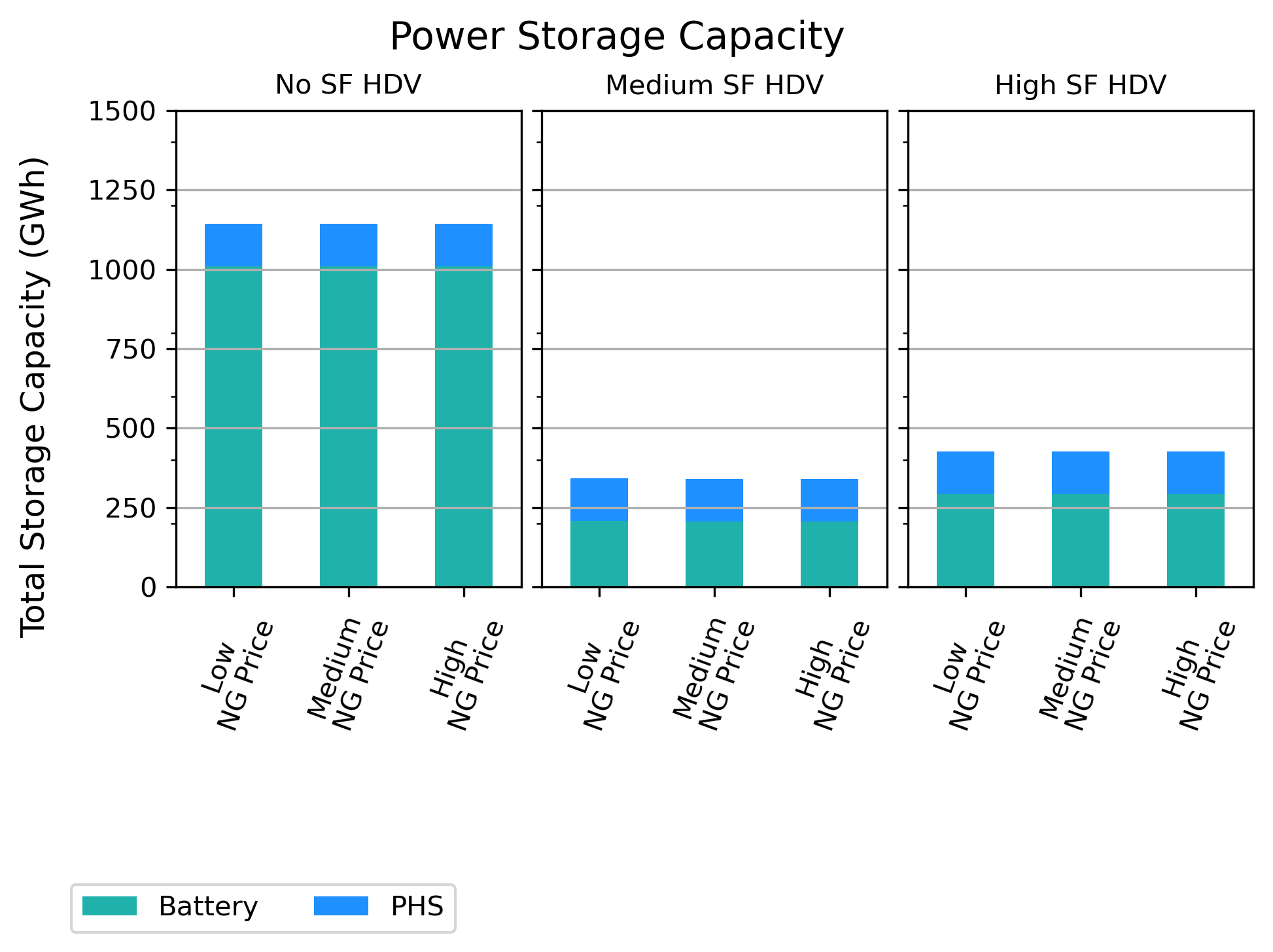}
        \caption{No CO$_2$ Storage}
    \end{subfigure}
    
    \begin{subfigure}[b]{0.9\linewidth}
        \includegraphics[width=\linewidth, trim=0cm 0cm 0cm 0.8cm, clip]{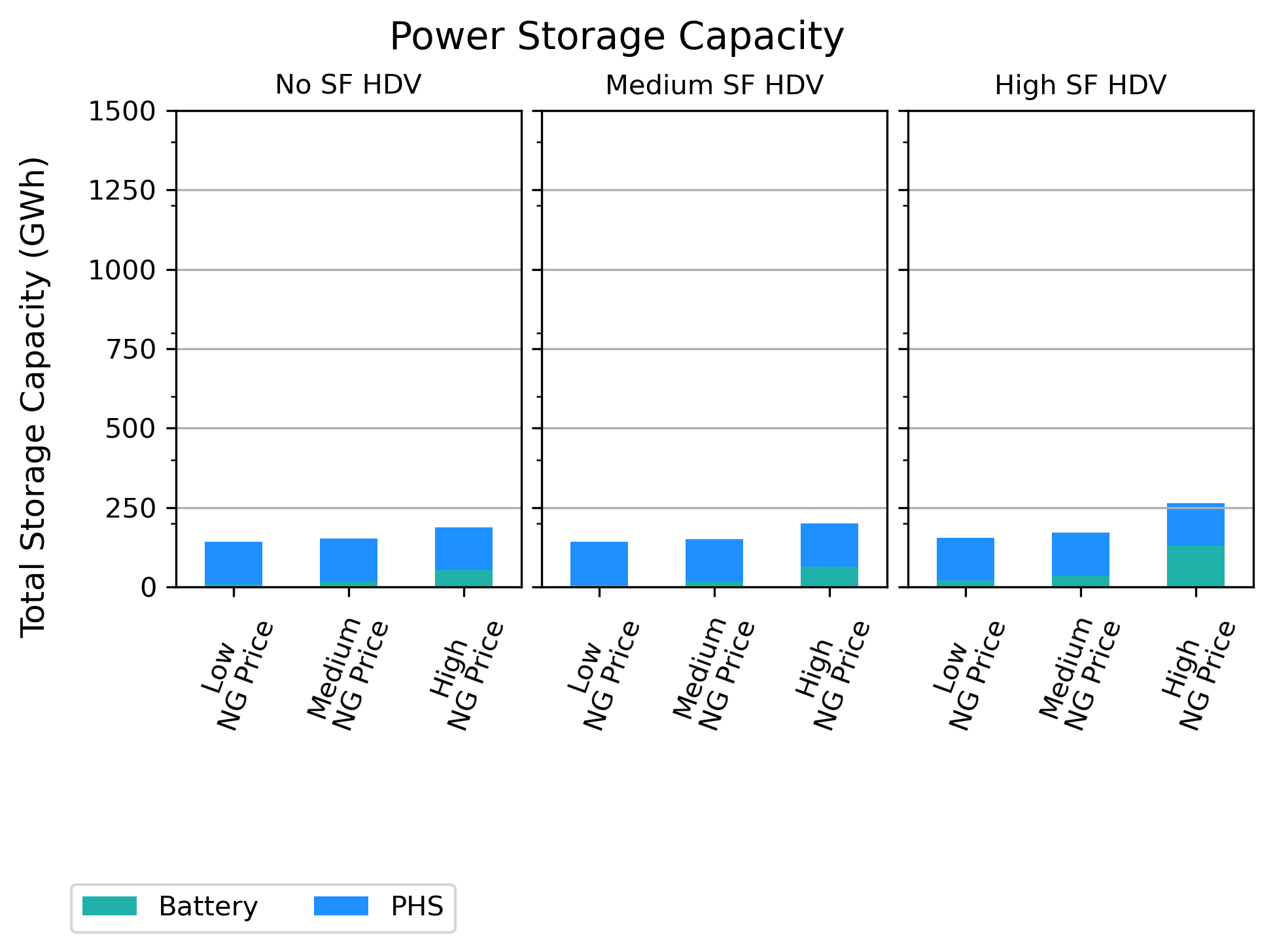}
        \caption{Baseline CO$_2$ Storage}
    \end{subfigure}
    
    \caption[Electric Storage Capacity Sensitivity Scenario Set 2]{Electric storage capacity for no (sub-figure a) and baseline (sub-figure b) CO$_2$ sequestration scenarios under medium H$_2$ HDV adoption and varying scenarios of synthetic fuel adoption. Within each panel, the price of natural gas increases left to right. Across panels, the amount of H$_2$ HDV adoption increases moving from left to right. The middle panels correspond to the core set of scenarios.}
    \label{fig:sf_hdv_elec_cap_stor_ng_sen}
\end{figure}

\begin{figure}[pos = H]
    \centering
    \begin{subfigure}[t]{0.9\linewidth}
        \includegraphics[width=\linewidth, trim=0cm 0cm 0cm 0.8cm, clip]{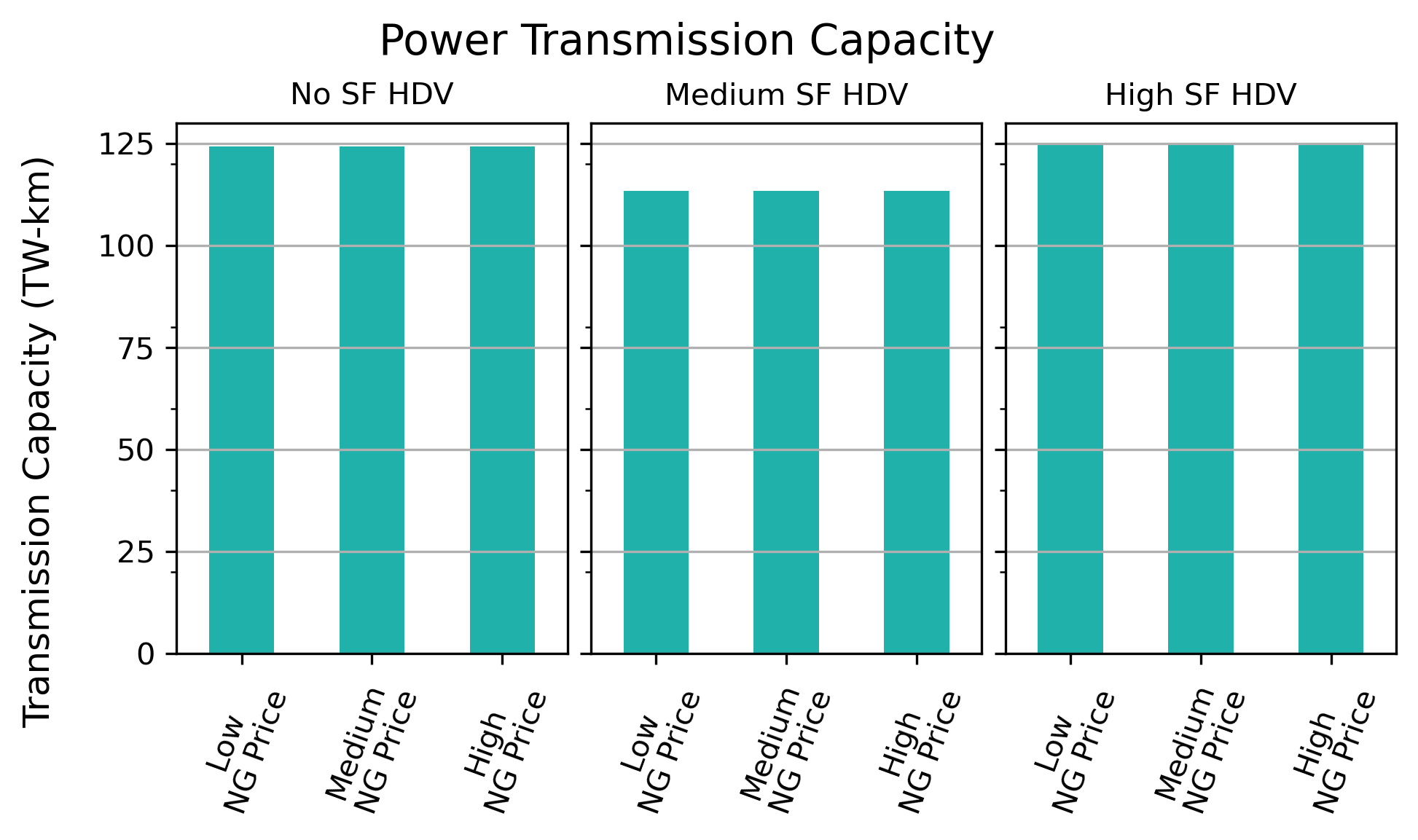}
        \caption{No CO$_2$ Storage}
    \end{subfigure}
    
    \begin{subfigure}[b]{0.9\linewidth}
        \includegraphics[width=\linewidth, trim=0cm 0cm 0cm 0.8cm, clip]{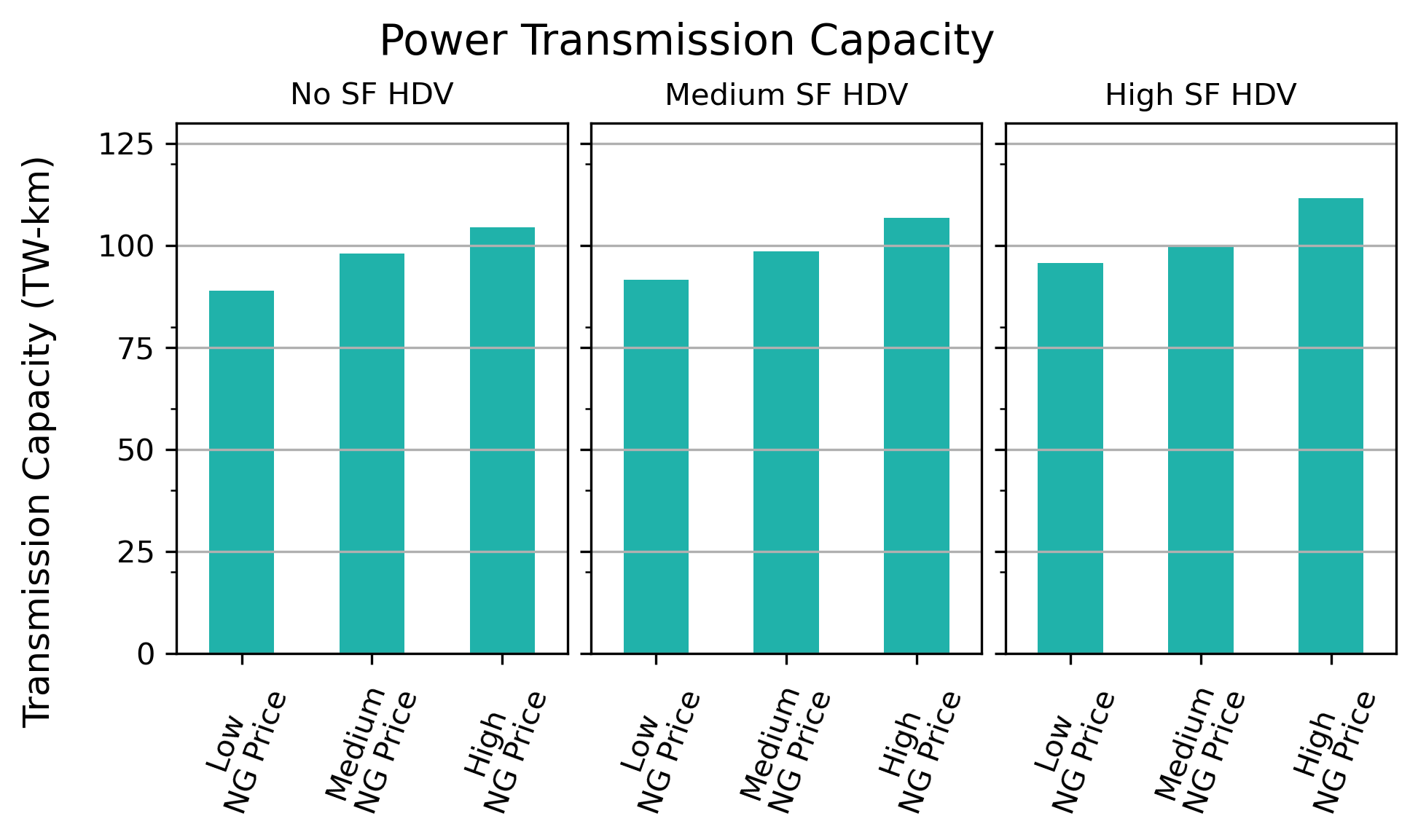}
        \caption{Baseline CO$_2$ Storage}
    \end{subfigure}
    
    \caption[Power transmission Capacity Sensitivity Scenario Set 4]{Power transmission capacity for no (sub-figure a) and baseline (sub-figure b) CO$_2$ sequestration scenarios under medium H$_2$ HDV adoption and varying scenarios of synthetic fuel adoption. Within each panel, the price of natural gas increases left to right. Across panels, the amount of H$_2$ HDV adoption increases moving from left to right. The middle panels correspond to the core set of scenarios. }
    \label{fig:sf_hdv_elec_trans_ng_sen}
\end{figure}

\begin{figure}[pos = H]
    \centering
    \begin{subfigure}[t]{0.9\linewidth}
        \includegraphics[width=\linewidth, trim=0cm 0cm 0cm 0.8cm, clip]{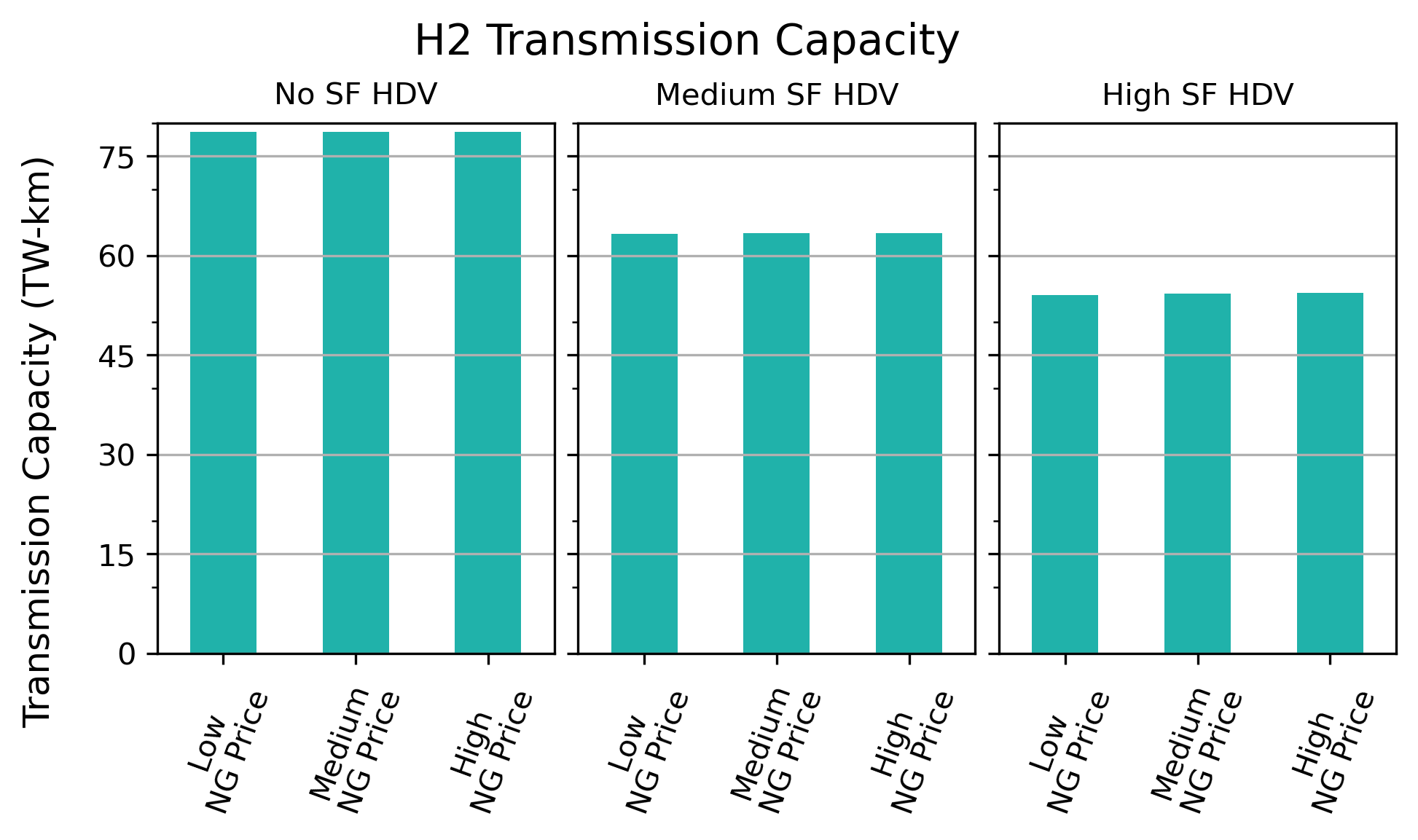}
        \caption{No CO$_2$ Storage}
    \end{subfigure}
    
    \begin{subfigure}[b]{0.9\linewidth}
        \includegraphics[width=\linewidth, trim=0cm 0cm 0cm 0.8cm, clip]{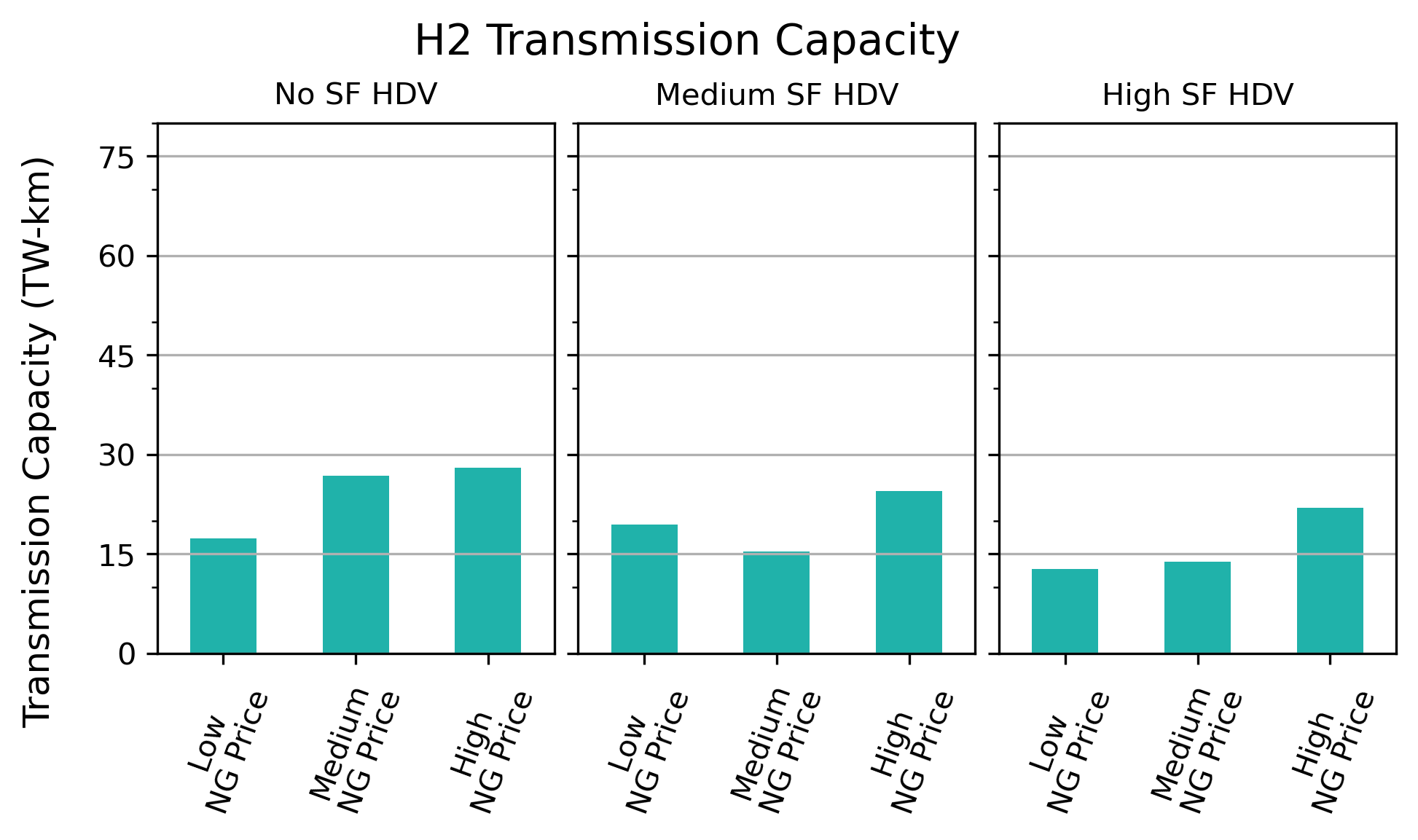}
        \caption{Baseline CO$_2$ Storage}
    \end{subfigure}
    
    \caption[H$_2$ transmission Capacity Sensitivity Scenario Set 4]{H$_2$ transmission capacity for no (sub-figure a) and baseline (sub-figure b) CO$_2$ sequestration scenarios under medium H$_2$ HDV adoption and varying scenarios of synthetic fuel adoption. Within each panel, the price of natural gas increases left to right. Across panels, the amount of H$_2$ HDV adoption increases moving from left to right. The middle panels correspond to the core set of scenarios. }
    \label{fig:sf_hdv_h2_trans_ng_sen}
\end{figure}

\begin{figure}[pos = H]
    \centering
    \begin{subfigure}[t]{0.8\linewidth}
        \includegraphics[width=\linewidth, trim=0cm 0cm 0cm 0.8cm, clip]{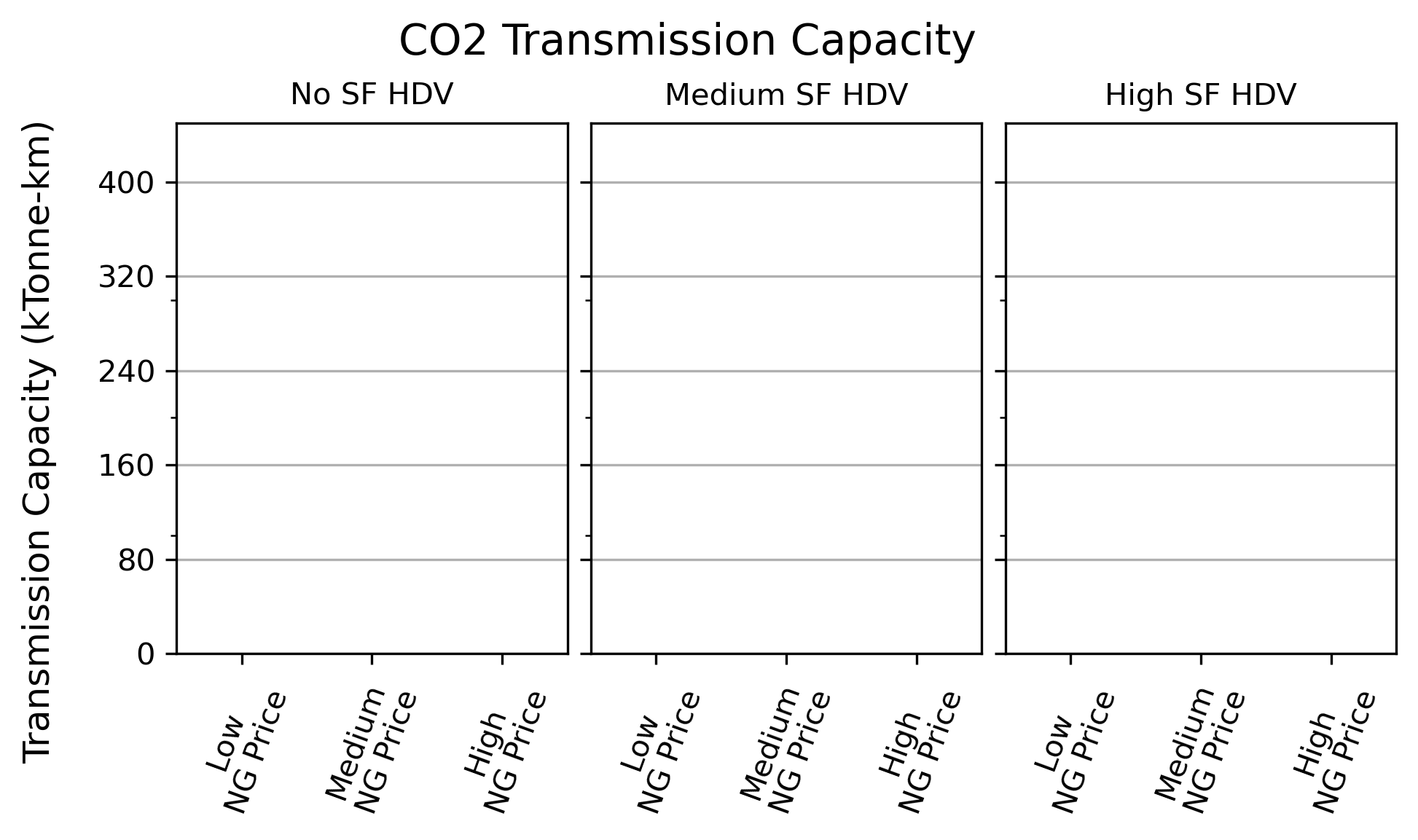}
        \caption{No CO$_2$ Storage}
    \end{subfigure}
    
    \begin{subfigure}[b]{0.8\linewidth}
        \includegraphics[width=\linewidth, trim=0cm 0cm 0cm 0.8cm, clip]{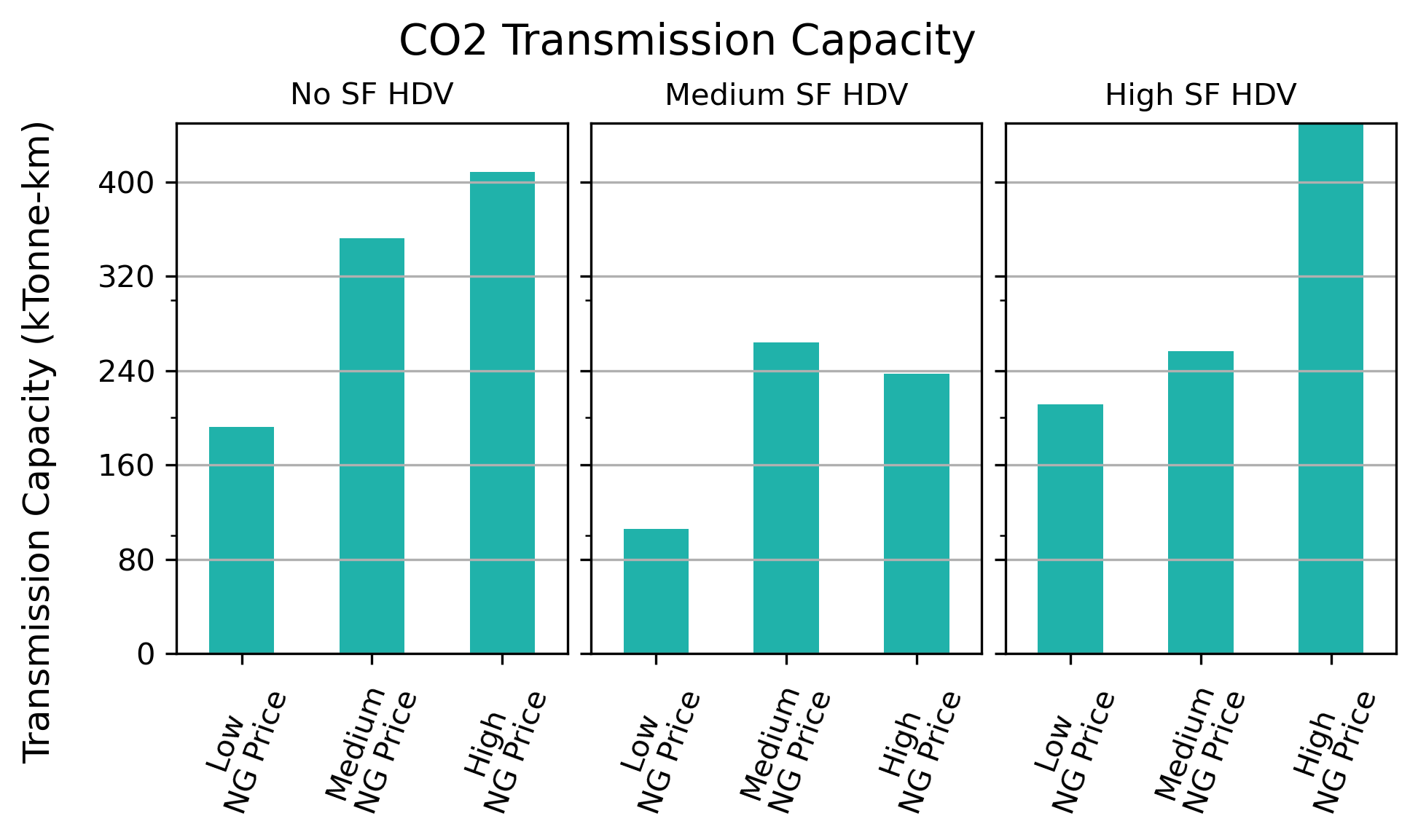}
        \caption{Baseline CO$_2$ Storage}
    \end{subfigure}
    
    \caption[CO$_2$ transmission Capacity Sensitivity Scenario Set 4]{CO$_2$ transmission capacity for no (sub-figure a) and baseline (sub-figure b) CO$_2$ sequestration scenarios under medium H$_2$ HDV adoption and varying scenarios of synthetic fuel adoption. Within each panel, the price of natural gas increases left to right. Across panels, the amount of H$_2$ HDV adoption increases moving from left to right. The middle panels correspond to the core set of scenarios. }
    \label{fig:sf_hdv_co2_trans_ng_sen}
\end{figure}

\begin{figure}[pos = H]
    \centering
    \begin{subfigure}[t]{1\linewidth}
        \includegraphics[width=\linewidth, trim=0cm 0cm 0cm 0.0cm, clip]{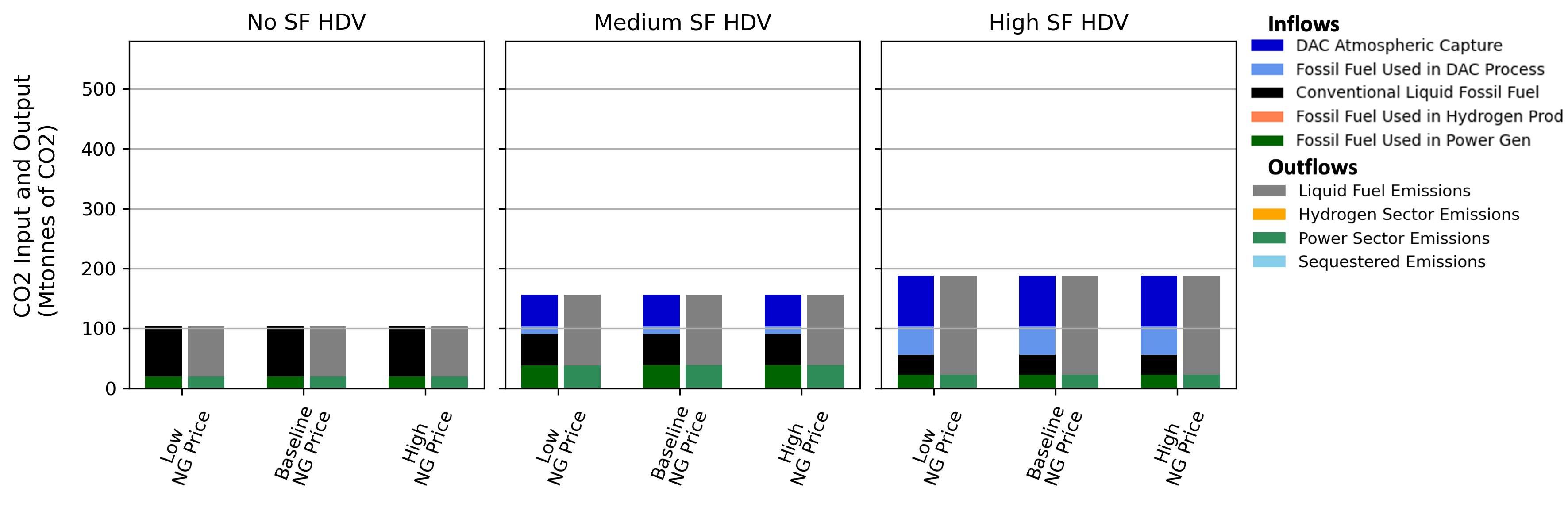}
        \caption{No CO$_2$ Storage}
    \end{subfigure}
    
    \begin{subfigure}[b]{1\linewidth}
        \includegraphics[width=\linewidth, trim=0cm 0cm 0cm 0.0cm, clip]{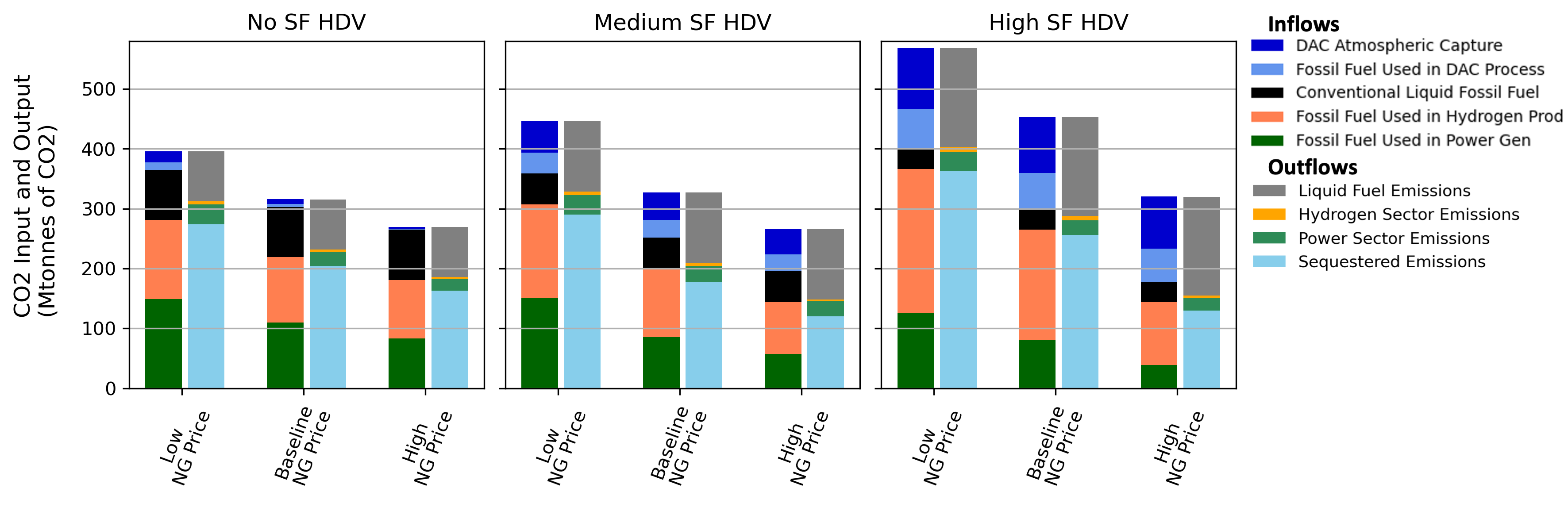}
        \caption{Baseline CO$_2$ Storage}
    \end{subfigure}
    
    \caption[CO$_2$ Balance Sensitivity Scenario Set 4]{System CO$_2$ balance under medium H$_2$ HDV adoption and varying scenarios of synthetic fuel adoption for no (sub-figure a) and baseline (sub-figure b) CO$_2$ sequestration scenarios. Within each panel, the price of natural gas increases left to right. Across panels, the amount of H$_2$ HDV adoption increases moving from left to right. The middle panels correspond to the core set of scenarios. The leftward column represents CO$_2$ input into the system, while the rightward column represents CO$_2$ outputted by the system. All scenarios adhere to the same emissions constraint of 103 MTonnes. The middle panels correspond to the core set of scenarios. Emissions constraint can be calculated from the chart by subtracting sequestered emissions and DAC atmospheric capture from the emission outflows. Emissions constraint can be calculated from the chart by subtracting sequestered emissions and DAC atmospheric capture from the emission outflows.}
    \label{fig:sf_hdv_co2_balance_ng_sen}
\end{figure}

\begin{figure}[pos = H]
    \centering
    \begin{subfigure}[t]{1\linewidth}
        \includegraphics[width=\linewidth, trim=0cm 0cm 0cm 0.0cm, clip]{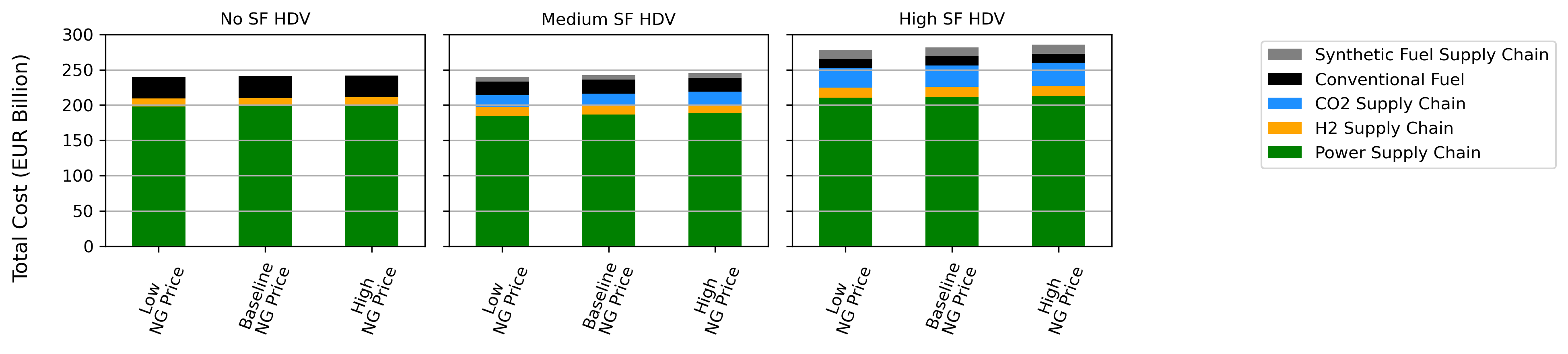}
        \caption{No CO$_2$ Storage}
    \end{subfigure}
    
    \begin{subfigure}[b]{1\linewidth}
        \includegraphics[width=\linewidth, trim=0cm 0cm 0cm 0.0cm, clip]{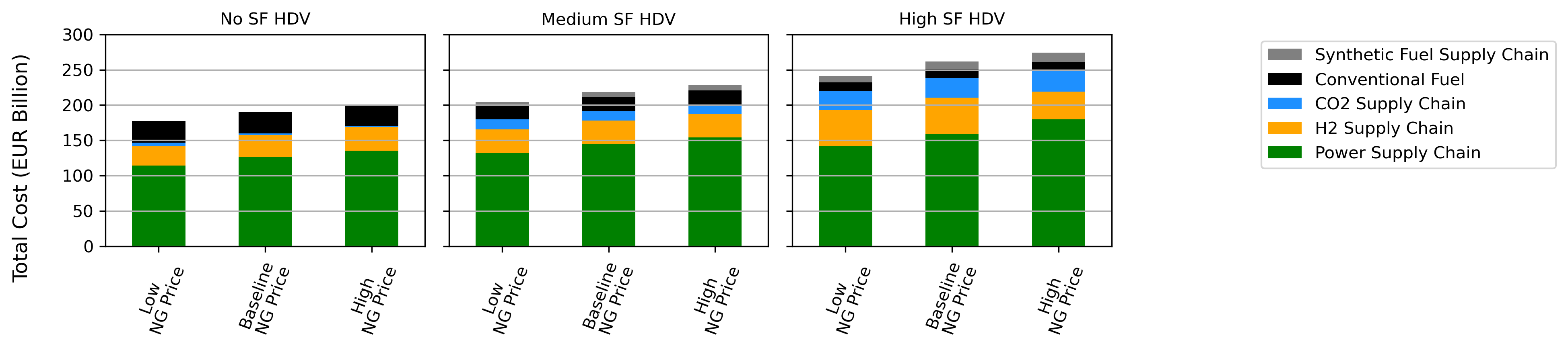}
        \caption{Baseline CO$_2$ Storage}
    \end{subfigure}
    
    \caption[Costs Sensitivity Scenario Set 4]{Annualized bulk-system costs under medium H$_2$ HDV adoption and varying scenarios of synthetic fuel adoption for no (sub-figure a) and baseline (sub-figure b) CO$_2$ sequestration scenarios. Within each panel, the price of natural gas increases left to right. Across panels, the amount of H$_2$ HDV adoption increases moving from left to right. The middle panels correspond to the core set of scenarios. The costs do not include vehicle replacement or H$_2$ distribution costs. }
    \label{fig:sf_hdv_cost_ng_sen}
\end{figure}

\begin{figure}[pos = H]
    \centering
    \includegraphics[width=1\linewidth]{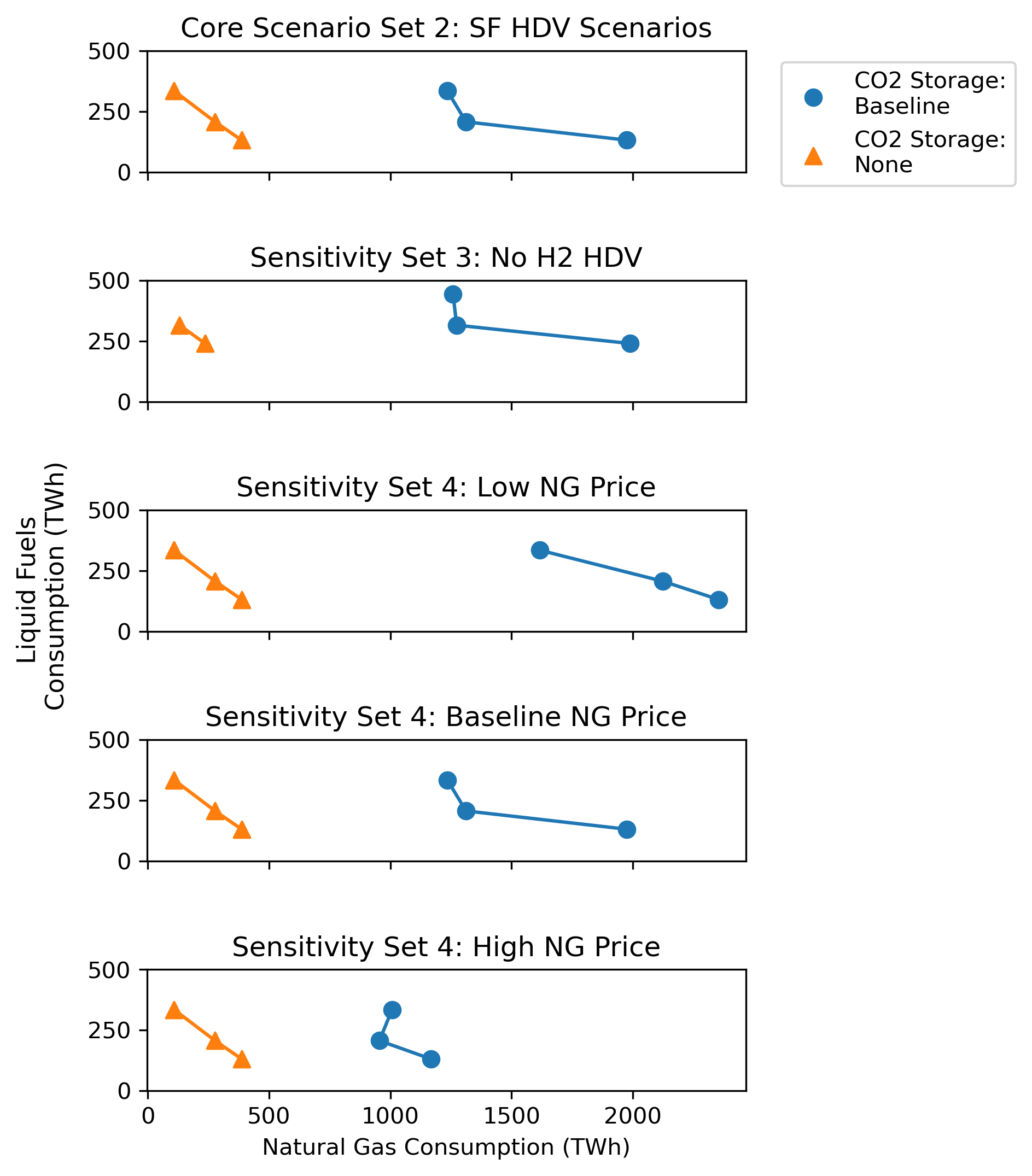}
    \caption[Natural Gas and Liquid Fuel Utilization Trade-off Synthetic Fuel Sensitivity Scenarios]{Trade-off between natural gas (NG) and liquid fossil fuel utilization for scenarios where amount of SF adoption is varied. The subfigure on the top shows the relationship for the core SF HDV scenarios(i.e. scenario set 2), while the second plot shows the results for the SF scenario with no H$_2$ adoption (i.e. sensitivity scenario set 3). The last 3 shows the results for the natural gas price sensitivities (i.e. sensitivity scenario set 4). Within each subplot the amount of natural gas consumption can be examined on the x-axis, while the amount of liquid fossil fuel consumption can be examined on the y-axis. The amount of SF HDV adoption increases from top to bottom. The amount of liquid fossil fuel consumption includes diesel and gasoline, and excludes jet fuel as well as excess synthetic fuels.}
    \label{fig:fuel_comp_sf}
\end{figure}

\begin{table}
\caption{This tables shows the marginal price of abatement of CO$_2$ for Sensitivity Set 4}
\label{co2_price_sen_set_4}
\begin{tabular}{llllr}
\toprule
CO$_2$ Storage & H$_2$ HDV Level & Synthetic Fuel HDV Level & Natural Gas Price & CO$_2$ Marginal Cost of Abatement \\
\midrule
Baseline & Medium & None & Low & 260.59 \\
Baseline & Medium & Medium & Low & 260.59 \\
Baseline & Medium & High & Low & 307.70 \\
Baseline & Medium & None & Baseline & 293.92 \\
Baseline & Medium & Medium & Baseline & 268.80 \\
Baseline & Medium & High & Baseline & 293.91 \\
Baseline & Medium & None & High & 327.25 \\
Baseline & Medium & Medium & High & 251.55 \\
Baseline & Medium & High & High & 266.02 \\
None & Medium & None & Low & 1654.20 \\
None & Medium & Medium & Low & 530.54 \\
None & Medium & High & Low & 618.74 \\
None & Medium & None & Baseline & 1603.09 \\
None & Medium & Medium & Baseline & 480.12 \\
None & Medium & High & Baseline & 567.48 \\
None & Medium & None & High & 1551.98 \\
None & Medium & Medium & High & 429.10 \\
None & Medium & High & High & 516.23 \\
\bottomrule
\end{tabular}
\end{table}

\newpage
\section{SI 2: Demand-side Modelling Methodology and Inputs}\label{sec:demand_side_si}

\subsection{Demand-side Transportation Model}\label{sec:demand_side_input}

This section details the mathematical formulation of demand-side model for generating alternative transportation demand scenarios. In Section \ref{sec:demand_side_model_inputs}, we detail the demand-side inputs used for this study. As shown in Figure 1, transportation demand, vehicle sub-category breakdowns, vehicle loading factors, vehicle market share, and vehicle energy consumption are combined together to create a demand profile. This model outputs the final energy consumption by fuel type (gasoline, diesel, electricity, H$_2$) for each timestep and country. To do so, we disaggregate transportation service demand (tonne-km (tkm) or passenger-km (pkm)) into the following categories: 1) light-duty passenger vehicles 2) buses and coaches 3) 2-wheelers, and 4) heavy-duty and light commercial vehicles. We further dissaggreate into vehicle types as shown in table \ref{tbl:vehicle category and types}. We also disaggregate demand by road type (urban, rural, highway). We then transform this service demand into vehicle km (vkm) demand using loading and occupancy factors. We assign service demand to a specific vehicle drivetrain type (e.g. PHEV, H$_2$, Diesel ICE, etc.) Using energy consumption factors for a specific fuel per vkm, we arrive at a final energy consumption by fuel (e.g. diesel, gasoline, electricity, H$_2$). We further subdivide into timesteps. Table \ref{tab:vehicle indicies and sets} shows the definitions of indices and sets used in the demand-side model. Table \ref{tab:key demand input parameters} contains key input parameters used in this model.

\begin{table}[pos = H]
    \centering
    \begin{tabular}{cc}
    \toprule
        Vehicle Category & Vehicle Type \\
    \midrule
        HDV \& LCV     & HDV Light + Short Distance\\
         & HDV Light + Long Distance\\
         & HDV Medium + Short Distance\\
         & HDV Medium + Long Distance\\
         & HDV Heavy + Short Distance\\
         & HDV Heavy + Long Distance\\
         & HDV Super-Heavy + Short Distance\\
         & HDV Super-Heavy + Long Distance\\
         & HDV Ultra-Heavy + Short Distance\\
         & HDV Ultra-Heavy + Long Distance\\
         & LCV\\
    \midrule
        Passenger Car & Small \\
         &  Medium\\
         & Large \\
    \midrule
        Buses and coaches &  Buses\\
         &  Coaches\\
    \midrule
        Two-wheelers & Two-wheelers\\
    \bottomrule
    \end{tabular}
    \caption[Vehicle Categories and Types]{Service demand vehicle category and the vehicle types disaggregation categories}
    \label{tbl:vehicle category and types}
\end{table}

\begin{table}[pos = H]
\centering
\begin{tabular}{p{0.2\linewidth}p{0.7\linewidth}}
\toprule
Indices and Sets & Definition \\
\midrule
$r \in R$ & $r$ denotes a road type belonging to the set of road types $R$. Road types modeled are urban, suburban, and rural. \\
$v \in V$ & $v$ denotes a specific vehicle type in a set of all vehicle types $V$ (e.g. small passenger vehicle, medium passenger vehicle, etc.). \\
$J \subset V$ & $J$ denotes a vehicle category (e.g., passenger cars, buses and coaches, heavy-duty, and light commercial vehicles) made up of vehicles $v \in V$. Vehicle types and categories modelled is show in Table \ref{tbl:vehicle category and types}\\
$V_\mathrm{pass} \subset V$ & $V_\mathrm{pass}$ is a subset of passenger vehicles made up from vehicles $v \in V$. Vehicle types and modelled is show in Table \ref{tbl:vehicle category and types}\\
$V_\mathrm{cargo} \subset V$ & $V_\mathrm{cargo}$ is a subset of cargo vehicles made up from vehicles $v \in V$. \\
$d \in D$ & where d is a drivetrain type in  a set of all drivetrain types (e.g. EV, PHEV H$_2$, Diesel ICE, etc). \\
$f \in F$ & where f is a fuel type in  a set of all fuel or drivetrain types (e.g. electricity, H$_2$, Biofuel, Diesel, Gasoline, etc). \\
$c \in C$ & where c is a country in all countries in Western Europe.\\
$y \in Y$ & where y is a year in possible analysis years (e.g. 2040, 2050)\\
\bottomrule
\end{tabular}
\caption{Demand model indices and sets}
\label{tab:vehicle indicies and sets}
\end{table}

\begin{table}[pos = H]
\centering
\begin{tabular}{p{0.3\linewidth}p{0.6\linewidth}}
\toprule
Input Parameter & Definition \\
\midrule
$p_\mathrm{v,c,y}$ & Passenger km demand for a given vehicle type and country in a given analysis year. \\
$t_\mathrm{v,c,y}$ & Tonne km demand for a given vehicle type and country in a given analysis year. \\
$o_\mathrm{v,c}$ & Vehicle occupancy ratio $vkm / pkm$. Only defined for $v \in V_{\mathrm{pass}}$.
\\
$l_\mathrm{v,c}$ & Loading ratio $vkm / tkm$. Only defined for $v \in V_{\text{cargo}}$\\
$s_\mathrm{v,c,r}$ & Road type share for a given vehicle type in a given country. \\
$m_\mathrm{v,d,y}$ & Market share for a given vehicle type for a given drivetrain type in a given year. \\
$k_\mathrm{r,v,d,f,y}$ & Energy demand per vehicle km for a vehicle $v$ on road type $r$, for drivetrain $d$, for fuel $f$ in year $y$. \\
$e_\mathrm{v,d,f,y}$ & Efficiency multiplier for a given vehicle type, for a given fuel type, in a given year. \\

$\tau_\mathrm{t,v,f}$ & Load shape factor (percent consumption in a given hour). \\
\bottomrule
\end{tabular}
\caption[Demand model input parameters]{Definitions of key demand model input parameters. the values of key parameters can be found in Section \ref{sec:demand_side_model_inputs}}
\label{tab:key demand input parameters}
\end{table}

\newpage

The following equation combines key inputs to create the energy demand loadshape at time t for each of the fuel types in a given country. To do so, we first multiply vehicle service demand by a factor (occupancy ratio for passenger vehicles, and loading factor for cargo vehicles) to calculate demand in vehicle distance (vkm). This is then multiplied by the road type share for a given vehicle type. We also subdivide the demand in vehicle distance by vehicle marketshare, which is the main lever used in this study. To convert this to energy consumption, we multiply by energy consumed per vkm for a given vehicle type, roadshare, fuel type, and year. We also have the option of multiplying by an efficiency factor, which could be used to assess the impact of efficiency measures. Finally, we multiply by a loadshape factor to calculate the energy consumption at a given timestep. The loadshape factor is the percentage of energy consumed in an hour out of total energy consumption. The first part of the equation calculates energy consumption for passenger vehicles, while the second calculates energy consumption for cargo vehicles. 

\begin{multline}
E_\mathrm{t,f,c,y} = \sum_{v \in V_\mathrm{pass}} \sum_{r \in R} \sum_{d \in D} p_\mathrm{v,c,y} \times o_\mathrm{v,c} \times s_\mathrm{v,c,r} \times m_\mathrm{v,d,y}  \times k_\mathrm{r,v,d,f,y}  \times e_\mathrm{v,d,f,y} \times \tau_\mathrm{t,v,f} \\ + \sum_{v \in V_\mathrm{cargo}} \sum_{r \in R} \sum_{d \in D} t_\mathrm{v,c,y} \times l_\mathrm{v,c} \times s_\mathrm{v,c,r} \times m_\mathrm{v,d,y} \times k_\mathrm{r,v,d,f,y} \times e_\mathrm{v,d,f,y}  \times \tau_\mathrm{t,v,f}
\end{multline}

The resultant loadshape is added to a given fuel type demand and is used as an input to the DOLPHYN model. 

\subsection{Demand-side Inputs}\label{sec:demand_side_model_inputs}

This section will cover key demand-side model inputs. 

The following table key input parameters used in this model. 

\begin{table}[pos = H]
\centering
\begin{tabular}{p{0.3\linewidth}p{0.6\linewidth}}
\toprule
Input Parameter & Definition \\
\midrule
$p_\mathrm{v_\mathrm{pass},c,y}$ & Passenger km demand for a given vehicle type and country in a given analysis year. \\
$t_\mathrm{v_\mathrm{cargo},c,y}$ & Tonne km demand for a given vehicle type and country in a given analysis year. \\
$k_\mathrm{r,j,f,y}$ & Energy demand per vehicle km for a vehicle $v$ on road type $r$, for fuel $f$ in year $y$. \\
$s_\mathrm{j,c,r}$ & Road type share for a given vehicle type in a given country. \\
$m_\mathrm{j,f,y}$ & Market share for a given vehicle type for a given fuel type in a given year. \\
$e_\mathrm{j,f,y}$ & Efficiency multiplier for a given vehicle type, for a given fuel type, in a given year. \\
$o_\mathrm{v_\mathrm{pass},c}$ & Vehicle occupancy ratio $vkm / pkm$. \\
$l_\mathrm{v_\mathrm{cargo},c}$ & Loading ratio $vkm / tkm$. \\
$\tau_\mathrm{t, j,f}$ & Load shape factor (percent consumption in a given hour). \\
\bottomrule
\end{tabular}
\caption{Definitions of key demand model input parameters}
\label{tab:symbols}
\end{table}

To begin with, vehicle category demand is constructed based on EU Reference scenarios for the year 2040 \citep{european_commission_eu_2020}. The scenarios provide the service demand in pkm and tkm for 4 vehicle categories. Demand for the UK and Norway is not directly available using the EU reference scenario, and therefore utilize demand from similar countries, Germany and Sweden, respectively, adjusted by population.

We then further disaggregate the demand for these vehicle categories into vehicle types as listed in table \ref{tbl:vehicle category and types}. This disaggregation is crucial as a vehicle tonnage and size affects its energy consumption. Vehicle demand is disaggregated on the basis of vehicle size for passenger vehicles and payload capacity for freight vehicles using TRACCS, a survey on transportation demand consumption \citep{emisia_traccs_2014}. 

To convert transportation service demand to distance demand, loading factors based on Eurostat are used. Payload factors from TRACCS  energy demand inconsistent with historical data. Table \ref{tab:vehicle_payload} shows the loading factors used for each HDV \& LCV vehicle category. 

\begin{table}[pos = H]
\centering
\caption{Vehicle Payload by Category and Type}
\label{tab:vehicle_payload}
\begin{tabular}{llc}
\toprule
Vehicle Category & Vehicle Type & Vehicle Payload (Tonnes) \\
\midrule
HDV \& LCVs & LCV &  0.40 \\
 & HDV - Light & 1.36 \\
 & HDV - Medium & 5.51 \\
& HDV - Heavy & 6.65 \\
 & HDV - Super Heavy & 12.52 \\
 & HDV - Ultra Heavy & 15.12 \\
\bottomrule
\end{tabular}
\end{table}

Additionally, the amount of vehicle distance travelled in each road type is based on \citep{krause_eu_2020}, and is as shown in table \ref{tbl:vehicle_road_share}.

\begin{table}[pos = H]
\centering
\caption[Vehicle Road Type Share by Vehicle Category and Vehicle Type]{Vehicle Road Type Share by Vehicle Category and Vehicle Type Based on \citep{krause_eu_2020}}
\begin{tabular}{llccc}
\toprule
Vehicle Category & Vehicle Type & \multicolumn{3}{c}{Road Type Share} \\
\cmidrule(lr){3-5}
& & Rural & Urban & Highway \\
\midrule
Buses and Coaches & Buses & 0.6 & 0.3 & 0.1 \\
Buses and Coaches & Coaches & 0.12 & 0.59 & 0.29 \\
Passenger Cars & All & 0.3 & 0.45 & 0.25 \\
Two-wheelers & All & 0.3 & 0.45 & 0.25 \\
HDV \& LCVs & HDVs & 0.12 & 0.25 & 0.63 \\
HDV \& LCVs & LCVs & 0.42 & 0.3 & 0.28 \\
\bottomrule
\end{tabular}
\label{tbl:vehicle_road_share}
\end{table}

The baseline market shares for each vehicle drive-train type are based on \citep{krause_eu_2020}, and are modified throughout the scenarios as outlined in Section 2.4 of the main text. For plug-in hybrid electric vehicles, it is assumed that all urban distance utilizes electricity, and the rest of the distance utilizes liquid fuel (gasoline for passenger vehicles, and diesel for the rest). Table \ref{tbl:vehicle_market_share_baseline} shows the baseline market shares for each drivetrain type.

\begin{table}[pos = H]
\centering
\caption[Vehicle Baseline Market Share by Vehicle Category and Vehicle Type]{Vehicle Baseline Market Share by Vehicle Category and Vehicle Type. The market share is divided by drivetrain type. Based on \citep{krause_eu_2020}.}
\begin{tabular}{llcccc}
\toprule
Vehicle Category & Vehicle Type & \multicolumn{4}{c}{Market Share} \\
\cmidrule(lr){3-6}
& & Diesel & Electric & H$_2$ & Hybrid Electric \\
\midrule
Buses and Coaches & Buses  & & 1 &  & \\
Buses and Coaches & Coaches &  & 0.4 & & 0.6\\
Passenger Cars & Small &  & 1 & & \\
Passenger Cars & Medium &  & 1 & & \\
Passenger Cars & Large &  & 0.5 & & 0.5 \\
Two-wheelers & Two-wheelers &  & 1 & & \\
HDV \& LCVs & LCV &   & 0.6 &  & 0.4\\
HDV \& LCVs & HDV - Light &   &0.6 &  &  0.4\\
HDV \& LCVs & HDV - Medium & 0.6 & & & 0.4  \\
HDV \& LCVs & HDV - Heavy & 0.6 &  & & 0.4  \\
HDV \& LCVs & HDV - Super Heavy & 0.6 &  & & 0.4  \\
HDV \& LCVs & HDV - Ultra Heavy & 0.6 & & & 0.4  \\
\bottomrule
\end{tabular}
\label{tbl:vehicle_market_share_baseline}
\end{table}

Finally, to convert vehicle distance demand into energy demand, we utilize energy consumption factors from the EU reference scenarios technology assumptions \citep{commission_technology_2020}. 

\begin{table}[pos = H]
\centering
\caption[Vehicle Energy Consumption by Vehicle Category and Vehicle Type]{Vehicle Energy Consumption by Vehicle Category and Vehicle Type (MJ/km). The energy consumption depends on the mode the vehicle is operating in. Based on \citep{commission_technology_2020}. }
\label{tbl:vehicle_energy_cond}
\begin{tabular}{llcccc}
\toprule
Vehicle Category & Vehicle Type & \multicolumn{4}{c}{Energy Consumption (MJ/vkm)} \\
\cmidrule(lr){3-6}
&& Diesel & Gasoline & Electric & H$_2$ \\
\midrule
Buses and Coaches & Buses &  & & 4.14 &  \\
Buses and Coaches & Coaches & 9.99 & & 4.14 &  \\
Passenger Cars & Small &  & & 0.50 &  \\
Passenger Cars & Medium &  &  & 0.54 &  \\
Passenger Cars & Large SUV & & 2.70 & 0.65 &  \\
Two-wheelers & Two-wheelers &  &  & 0.22 &  \\
HDV \& LCVs & LCV & 2.62 &  & 0.61 &  \\
HDV \& LCVs & HDV - Light & 4.98 & & 2.02 & \\
HDV \& LCVs & HDV - Medium  & 7.18 & & 2.66 & 4.32 \\
HDV \& LCVs & HDV - Heavy & 10.07 & & 4.68 & 6.72 \\
HDV \& LCVs & HDV - Super Heavy & 13.87 & & 5.94 & 9.12 \\
HDV \& LCVs & HDV - Ultra Heavy & 13.87 & & 5.94 & 9.12 \\
\bottomrule
\end{tabular}
\end{table}

Non-electric demand for HDV and LCV energy is assumed to be flat, while electric demand assumed a loadshape based on the ENTSOE TYDNP study \citep{entsoe_tyndp_2022}. For this study, no efficiency multipliers are utilized. 

\newpage
\section{SI 3: Supply-side Modelling Methodology and Inputs}\label{section:supply_side}

\subsection{Supply-side Model Description}\label{sec:supply_side_model_desc}

This section describes the modeling of the liquid fuels, including synthetic fuels within the supply-side model, DOLPHYN. DOLPHYN represents a multi-vector energy system in the form of a mixed-integer linear programming (MILP) optimization model. The full documentation of the DOLPHYN model can be found here \citep{he_dolphyn_2023}. Figure 1 (b) shows a diagram of DOLPHYN, including all the technologies modelled. While this work has resulted in contributions to the power, H$_2$, and CO$_2$ supply chains, the liquid fuel modelling represents a novel addition to DOLPHYN, which is why we focus on detailing it's formulation here. Additionally, other supply chain are documented elsewhere such as in \citep{he_hydrogen_2021, he_dolphyn_2023}.

\subsubsection[Liquid Fuel Model Formulation]{Liquid Fuel Model Formulation\footnote{This section is adapted from documentation of the liquid fuels module in DOLPHYN \citep{he_dolphyn_2023}. The modeling of liquid fuels was a result of a collaborative effort between Youssef Shaker and Jun Wen Law.}}

This section describes the formulation of the liquid fuels supply chain. To meet a specific demand of liquid fuels, the model has a choice to utilize conventional (fossil-based) or synthetic fuels. Synthetic fuel production arises from a set of resources $F$. Synthetic fuel plant performance is parameterized in terms of CO$_2$ input into the process. The model is set-up to account for up to 3 liquid fuel products: diesel, gasoline, and jetfuel. The user can also model an unlimited number of synthetic fuel process by-products. The by-product feature is designed to account for emissions and economic value of byproducts for which there is no exogeneous demand modeled. The emissions associated with these byproducts are accounted for the emissions balances, while the revenue from the sale of by-products are accounted for in the objective function. Any of the 3 modelled liquid fuels can also be modelled as by-products,in the case they are not explicitly modelled. In the case of this study, jet fuel demand is not modelled, and therefore jet fuel is considered a by-product. 

\textbf{Liquid Fuel Model Notation}

\begin{table}[pos = H]
\centering
\begin{tabular}{p{0.2\linewidth}p{0.8\linewidth}}
\toprule
Notation & Description \\
\midrule
$z \in \mathcal{Z}$ & $z$ denotes a zone, and $\mathcal{Z}$ is the set of zones in the network \\
$t \in \mathcal{T}$ & $t$ denotes a time step, and $\mathcal{T}$ is the set of time steps \\
$f \in \mathcal{F}$ & Index and set of all synthetic fuels resources \\
$\mathcal{F}^{z} \in F$ & Index and set of synthetic fuels resources in zone $z$ \\
$k \in K$ & $k$ denotes a liquid fuel or by-products modelled in a set of all liquid fuels or by-products modelled $K$\\
$L \in K$ & $L$ denotes a subset containing liquid fuels modelled excluding by-products\\
$B \in K$ & $L$ denotes a subset containing by-products modelled excluding liquid fuels\\
$l \in \mathcal{L}$ & Index and set of all liquid fuels modelled. Currently three liquid fuels are modelled (gasoline, diesel, and jetfuel). \\
$b \in \mathcal{B}$ & Index and set of all synthetic fuels process byproducts \\
\bottomrule
\end{tabular}
\caption{Liquid Fuel Model Sets and Indices}
\label{tab:syn_fuel_model_notation_desc}
\end{table}

\begin{table}[pos = H]
\centering
\begin{tabular}{p{0.1\linewidth}p{0.9\linewidth}}
\toprule
Notation & Description \\
\midrule
$x_{f,t}^{\textrm{C,Syn}}$ &  CO$_2$ input into synthetic fuels resource $f$ at time period $t$ in tonnes of CO$_2$\\[3pt]
$x_{f,l,t}^{\textrm{Syn}}$ & Synthetic fuel $l$ produced by resource $f$ at time period $t$ in MMBTU \\[3pt]
$x_{f,b,t}^{\textrm{By,Syn}}$ & Byproduct $b$ produced by synthetic fuels resource $f$ at time period in MMBTU$t$ \\[3pt]
$y_{f}^{\textrm{C,Syn}}$ & Capacity of synthetic fuels resources in the liquid fuels supply chain in tonnes of CO$_2$/hr\\[3pt]
$x_{z,t,l}^{\textrm{Conv}}$ & Conventional fuel $l$ purchased by zone $z$ at time period $t$ in MMBTU\\[3pt]

\bottomrule
\end{tabular}
\caption{Liquid Fuel Model Decision Variables}
\label{tab:syn_fuel_model_decision_vars}
\end{table}

\begin{table}[pos = H]
\centering
\begin{tabular}{p{0.1\linewidth}p{0.9\linewidth}}
\toprule
Notation & Description \\
\midrule
$D_{z,t,l}$ & Demand for fuel $l$ in zone $z$ at time $t$\\[3pt]
$\zeta_{l}$ & Percentage of fuel $l$ that needs to be fulfilled using synthetic fuels \\
\midrule
$\textrm{c}_{f}^{\textrm{Syn,INV}}$ & Investment cost per tonne CO$_2$ input of synthetic fuels resource $f$ \\[3pt]
$\textrm{c}_{f}^{\textrm{Syn,FOM}}$ & Fixed operation cost per tonne CO$_2$ input of synthetic fuels resource $f$\\[3pt]
$\textrm{c}_{f}^{\textrm{Syn,VOM}}$ & Variable operation cost per tonne of CO$_2$ input by synthetic fuels resource $f$\\[3pt]
$\textrm{c}_{f}^{\textrm{Syn,FUEL}}$ & Fuel cost per tonne of CO$_2$ input by synthetic fuels resource $f$\\[3pt]
$\textrm{c}_{b}^{\textrm{By,Syn}}$ & Selling price per mmbtu of byproduct by synthetic fuels resource (if any) \\[3pt]
$\textrm{c}_{l}^{\textrm{Conv}}$ & Purchase cost per mmbtu of conventional fuel \\
\midrule
$\overline{y}_{f}^{\textrm{C,Syn}}$ & If upper bound of capacity is defined, then we impose constraints on the maximum CO$_2$ input capacity of synthetic fuels resource \\[3pt]
$\underline{y}_{f}^{\textrm{C,Syn}}$ & If lower bound of capacity is defined, then we impose constraints on the minimum CO$_2$ input capacity of synthetic fuels resource \\
\midrule
$\tau_{l,f}^{liquid}$ & Amount of fuel $l$ produced per tonne of CO$_2$ input at synthetic fuel resource $f$\\[3pt]
$\tau_{b,f}^{Byproduct}$ & Amount of by-product $b$ produced per tonne of CO$_2$ input at synthetic fuel resource $f$\\
\midrule
$p_f^{power}$ & Power MWh per tonne of CO$_2$ in required for the plant $f$\\[3pt]
$p_f^{hydrogen}$ & H$_2$ tonnes per tonne of CO$_2$ in required for the plant $f$\\
\midrule
$\mu_{f}^{emit}$ & Percentage of CO$_2$ emitted of the CO$_2$ in for a plant $f$\\[3pt]
$\mu_{f}^{capture}$ & Percentage of CO$_2$ captured of the CO$_2$ in for a plant $f$\\[3pt]
$\lambda_{f}$ & Emissions of plant fuel per tonne of CO$_2$ in for plant $f$\\
\midrule
$\theta_{l}^{liquid}$ & Emissions per mmbtu for liquid fuel $l$\\[3pt]
$\theta_{b}^{Byproduct}$ & Emissions per mmbtu for by-product $b$\\[3pt]
$\omega_t$ & Time-step weight for time-step $t$\\
\bottomrule
\end{tabular}
\caption{Liquid Fuels Model Parameters Description}
\label{tab:syn_fuel_model_parameters}
\end{table}

\textit{Objective function}

The total cost associated with the liquid fuel infrastructure includes four main elements as shown in equation \ref{eqn:obj_func} : 1) the capital cost of synthetic fuel production (see equation \ref{eqn:syn_fuel_cap_cost}) 2) the operating cost of synthetic fuel production (see equation \ref{eqn:syn_fuel_opex_cost}) 3) production credits for any by-products that are not explicitly modelled (see equation \ref{eqn:syn_fuel_by_prod_credit}), and 4) the cost of procuring liquid fossil fuels (see equation \ref{eqn:conv_fuel_cost}). These terms are added to the overall multi-sectoral model objective function, which includes cost associated with infrastructure for other vectors. 

\begin{equation}\label{eqn:obj_func}
	min \; \textrm{C}^{\textrm{LF,Syn,c}} + \textrm{C}^{\textrm{LF,Syn,o}} - \textrm{C}^{\textrm{LF,Syn,r}} +\textrm{C}^{\textrm{Conv,o}}
\end{equation}

The fixed costs associated with synthetic fuel production is defined such that:

\begin{equation}\label{eqn:syn_fuel_cap_cost}
	\textrm{C}^{\textrm{LF,Syn,c}} = \sum_{f \in \mathcal{F}} y_{f}^{\textrm{C,Syn}} (\times \textrm{c}_{f}^{\textrm{Syn,INV}} + \textrm{c}_{f}^{\textrm{Syn,FOM}})
\end{equation}

The variable costs associated with synthetic fuel production is defined such that:

\begin{equation}\label{eqn:syn_fuel_opex_cost}
	\textrm{C}^{\textrm{LF,Syn,o}} = \sum_{f \in \mathcal{F}} \sum_{t \in \mathcal{T}} \omega_t \times \left(\textrm{c}_{f}^{\textrm{Syn,VOM}} + \textrm{c}_{f}^{\textrm{Syn,FUEL}}\right) \times x_{f,t}^{\textrm{C,Syn}}
\end{equation}

The credit associated with by-products is defined such that:

\begin{equation}\label{eqn:syn_fuel_by_prod_credit}
	\textrm{C}^{\textrm{LF,Syn,r}} = \sum_{f \in \mathcal{F}} \sum_{b \in \mathcal{B}} \sum_{t \in \mathcal{T}} \omega_t \times x_{f,b,t}^{\textrm{By,Syn}} \times \textrm{c}_{b}^{\textrm{By,Syn}}
\end{equation}

The cost of conventional fuels is defined such that:

\begin{equation}\label{eqn:conv_fuel_cost}
	\textrm{C}^{\textrm{Conv,o}} = \sum_{z \in \mathcal{Z}}  \sum_{l \in \mathcal{L}}\sum_{t \in \mathcal{T}} \omega_t \times \textrm{c}_{l}^{\textrm{Conv}} \times x_{z,t,l}^{\textrm{Conv}}
\end{equation}

\textit{Synthetic Fuel Production Constraints}

The amount of synthetic fuels produced at a given time-step is given by:

\begin{equation}
x_{f,l,t}^{\textrm{Syn}} = x_{f,t}^{\textrm{C,Syn}} \times \tau_{l,f}^{liquid} \quad \forall f \in F,t \in T,l \in L
\end{equation}

The amount of by-products produced at a given time-step is given by:

\begin{equation}
x_{f,b,t}^{\textrm{Syn}} = x_{b,t}^{\textrm{C,Syn}} \times \tau_{b,f}^{By} \quad \forall f \in F,t \in T,b \in B
\end{equation}

For resources where upper bound $\overline{y_{f}^{\textrm{C,Syn}}}$ and lower bound $\underline{y_{f}^{\textrm{C,Syn}}}$ of capacity is defined, then we impose constraints on minimum and maximum synthetic fuels resource input CO$_2$ capacity.

\begin{equation*}
	\underline{y_{f}^{\textrm{C,Syn}}} \leq y_{f}^{\textrm{C,Syn}} \leq \overline{y_{f}^{\textrm{C,Syn}}} \quad \forall f \in \mathcal{F}
\end{equation*}

The required capacity is given by the following constraint such that the amount of CO$_2$ flowing into the plant does not exceed the plant's capacity:

\begin{equation}
	x_{f,t}^{\textrm{C,Syn}} \leq  y_{f}^{\textrm{C,Syn}} \quad \forall f \in \mathcal{F}, t \in \mathcal{T}
\end{equation}
\\
\\
\textit{Liquid Fuels Balance Constraints}

For each of the liquid fuels the following constraint is implemented to ensure that a sufficient combination synthetic fuel production and conventional liquid fuel procurement occurs to meet demand:

\begin{equation}  \sum_{z \in \mathcal{Z}} \sum_{t \in \mathcal{T}} \omega_t \times x_{z,t,l}^{\textrm{Conv}} + \sum_{f \in \mathcal{F}} \sum_{t \in \mathcal{T}} \omega_t \times x_{f,l,t}^{\textrm{C,Syn}} >= \sum_{z \in \mathcal{Z}} \sum_{t \in \mathcal{T}} D_{z,t,l} \quad \forall l \in L
\end{equation}

Note that only one constraint is implemented across all zones and time-steps. This is to reflect the flexibility and interconnectedness of liquid fuel supply chain. The cost of transporting liquid fuels is already included in cost estimates, and is therefore not accounted for separately. Moreover, because this modeling approach presumes that the product distribution from synthetic fuel production cannot be changed without impacting the energy inputs or capital cost of the process, the amount of each fuel produced can potentially exceed the amount demanded. In this case, the fuel production is penalized from an emission perspective, without meeting any specific demand. Finally, because one constraint is implemented across all timesteps and zones, storage is not accounted for.

Additionally, to reflect possible synthetic fuel mandates, the following constraint is used to force the model to produce a specific amount of synthetic fuels as a percentage of demand, if a specific fuel mandate is specified:

\begin{equation}
    (\zeta_l - 1) \times \sum_{f \in \mathcal{F}} \sum_{t \in \mathcal{T}} \omega_t \times x_{f,l,t}^{\textrm{C,Syn}} + \zeta_l \times \sum_{z \in \mathcal{Z}} \sum_{t \in \mathcal{T}} \omega_t \times x_{z,t,l}^{\textrm{Conv}} = 0 
\end{equation}\label{eqn:syn_fuel_mandate}

This is just a reorganization version of the following formula in a way that avoids non-linearities:

\begin{equation}
    \sum_{f \in \mathcal{F}} \sum_{t \in \mathcal{T}} \omega_t \times x_{f,l,t}^{\textrm{C,Syn}} / ( \sum_{f \in \mathcal{F}} \sum_{t \in \mathcal{T}} \omega_t \times x_{f,l,t}^{\textrm{C,Syn}} +  \sum_{z \in \mathcal{Z}} \sum_{t \in \mathcal{T}} \omega_t \times x_{z,t,l}^{\textrm{Conv}}) = \zeta_l 
\end{equation}

Note that only one synthetic fuel product percentage can be specified, otherwise, the model will become infeasible. 
\\
\\
\textit{Synthetic Fuel Power Balance Term}

The following expression reflects the power consumption associated with synthetic fuel production in a given zone that is added to the overall system power supply and demand balance at each time step and zone: 

\begin{equation}
    BalPowerLiquidFuel_{z,t} = \sum_{f \in {F^{z}}} \omega_t \times x_{f,t}^{\textrm{C,Syn}} \times p_f^{power}  \quad \forall z \in Z, t \in T
\end{equation}

This term is added to the overall power balance of the multi-sectoral model. 
\\
\\
\textit{Synthetic Fuel H$_2$ Balance Term}

The following expression reflects the H$_2$ consumption associated with synthetic fuel production in a given zone: 

\begin{equation}
    BalHydrogenLiquidFuel_{z,t} = \sum_{f \in {F^z}} \omega_t \times x_{f,t}^{\textrm{C,Syn}} \times p_f^{hydrogen}  \quad \forall z \in Z, t \in T
\end{equation}

This term is added to the overall H$_2$ balance of the multi-sectoral model. 
\\
\\
\textit{Liquid Fuel Emissions Balance Terms}


The following expression shows the emissions associated with the liquid fuel production and consumption process. It is made up of 4 terms: 1) the component of CO$_2$ input into the plant that is released into the atmosphere during the fuel production process 2) emissions from the consumption of synthetic fuels (all synthetic fuels produced are consumed even if there is no demand for them) 3) emissions from the consumption of by-products of synthetic fuels production process, and 4) emissions from the consumption of conventional liquid fuels. This terms is added to the overall multi-sectoral CO$_2$ balance constraint. 

\begin{equation}
\begin{aligned}
    BalEmissionsLiquidFuel_{z} = & \sum_{f \in {F^{z}}} \sum_{t \in \mathcal{T}} \omega_t \times x_{f,t}^{\textrm{C,Syn}} \times \mu_{f}^{emit} \\
    & + \sum_{f \in {F^{z}}} \sum_{t \in \mathcal{T}} \omega_t \times x_{f,t,l}^{\textrm{Syn}} \times \theta_{l}^{liquid} \\
    & + \sum_{f \in {F^{z}}} \sum_{t \in \mathcal{T}} \omega_t \times x_{b,t,l}^{\textrm{Syn}} \times \theta_{b}^{Byproduct} \\
    & + \sum_{t \in \mathcal{T}} \omega_t \times x_{l,z,t}^{\textrm{Conv}} \times \theta_{l}^{liquid} \quad \forall z \in Z
\end{aligned}
\end{equation}

The following expression shows the emissions captured with the liquid fuel production and consumption process. It is made up of 2 terms: 1) the component of CO$_2$ input into the plant from all captured emissions, which is taken from the CO$_2$ captured in the system 2) emissions captured from the synthetic fuel plant. This term is added to the multi-sectoral CO$_2$ captured expression, which includes CO$_2$ captured from H$_2$ and power producing plants, as well as DAC. 

\begin{equation}
\begin{aligned}
    BalEmissionsCapturedLiquidFuel_{z} = & - \sum_{f \in {F^{z}}} \sum_{t \in \mathcal{T}} \omega_t \times x_{f,t}^{\textrm{C,Syn}} \\
    & + \sum_{f \in {F^{z}}} \sum_{t \in \mathcal{T}} \omega_t \times x_{f,t}^{\textrm{C,Syn}} \times \mu_{f}^{capture} \quad \forall z \in Z
\end{aligned}
\end{equation}

\subsection{Supply-side Model Inputs}\label{sec:supply_side_model_inputs}

\subsubsection{Power Network, Cost, and Operational Assumptions}

The power system used in this study is based on a brownfield representation of the European Grid created by PYPSA-EUR \citep{horsch_complete_2019}.We utilize the 37-node representation of the European continent, which reduced to 10 nodes when we exclude countries outside of the region of interest. All countries are represented as one node, apart from Denmark and the UK, which are each represented using two nodes. 

Transmission costs and upgrades are based on PYPSA-EUR network representations \citep{horsch_complete_2019}. In the model, both AC and DC transmission are treated equivalently. We assume that existing power transmission capacities can be expanded by up to 4 times. Additionally, we assume that new lines can be built between certain regions up to a capacity of 5000 MW. 

The existing available generation is based on data from the ENTSOE transparency platform \citep{entsoe_entso-e_nodate}. We assume that this capacity will be available in the year 2040.  

The following table shows the key cost and operational assumptions for generation and storage technologies:

\begin{table}[pos = H]
\centering
\begin{tabular}{|p{2.8 cm} || p{1.4cm} | p{1.4cm} | p{1cm} |  p{1cm} |p{1.5cm}|p{1.25cm} |p{1.2cm} |p{1.3cm} |} 
 \hline
 Generation Technology &Power CAPEX (Eur/ kW) & Energy CAPEX (Eur/ kWh)& FOM (Eur/ MW/ yr)& VOM (Eur/ MWh)& Heat Rate (MMBTU / MWh)& Capture Rate & Round Trip Efficiency &  Lifetime (Yrs)\\
 \hline
 Onshore Wind  &851&-&32& 0&-&-&-&30\\ 
 Offshore Wind  & 3,751 &-& 69&0&-&-&-&30\\
 Solar & 680 & - & 13 & 0&-&-&-&30\\
 Biomass & - & -& 136 & 5.2&13.5&-&-&45\\
 Nuclear & 6,431 & - & 131 &2.6&10.46&-&-&60\\
 Hydro & - & - & 56& 0&-&-&-&100\\
 OCGT & 785 & -& 19 &1.6&10.1&-&-&30\\
 CCGT & 937 & -& 25 &4.6&6.5&-&-&30\\
 CCGT w/ CCS & 1,794 & -& 52&3.7&7.2&0.95&-&30\\
 Coal & 2,733 & -& 67 &7.2&10.0&-&-&30\\
 PHS & - &- & 16 &0.5&-&-&0.87&100\\
 Battery & 137 & 208& 18 &0&-&-&0.85&30\\[1ex] 
 \hline
\end{tabular}
\caption{Power Technology Cost and Operational Assumptions}
\label{table:power_tech_assumptions}
\end{table}

All costs assumptions are based on the 2022 NREL ATB for the year 2040 \citep{nrel_nrel_2022}. For all technologies we used moderate technology assumptions, apart from the CCGT w/ CCS, battery, and floating off-shore wind. All operational assumptions (e.g. ramp up/down time, minimum run are based on Supelveda et al. \citep{sepulveda_role_2018}. 

The maximum available generation capacity and temporally resolved capacity factors associated with the variable renewable energy generation technologies is based on PYPSA-EUR \citep{horsch_complete_2019}. We assume no possible expansion in biomass plants. Additionally, we assume that only nuclear existing (without a phase-out), planned, and under-construction units will be in-operation by 2040, in line with the 2022 ENTSOE TYNDP study \citep{entsoe_tyndp_2022}. 

\newpage

\subsubsection{Hydrogen Cost and Operational Assumptions}

We model hydrogen generation, transmission, storage, and G2P technologies. A greenfield representation of the hydrogen system is utilized for this study. 

We modelled the following H$_2$ production technologies: electrolyzers, SMR, SMR w/ CCS, and ATR w/ CCS technologies. We also modeled above ground hydrogen storage. Cost and operational assumptions for fossil fuel based plants (SMR, SMR w/ CCS, and ATR w/ CCS) are based on \citep{lewis_comparison_2022}. Cost and operational assumptions for electrolyzers are based on \citep{iea_future_2019}, assuming 2050 costs. Costs and operational assumptions for G2P units are based on \citep{nrel_nrel_2022}. Storage technology cost assumptions are based on \citep{papadias_bulk_2021}. The following table shows key hydrogen generation technologies cost and operational assumptions:

\begin{table}[pos = H]
\centering
\small
\begin{tabular}{|p{2.8 cm} || p{1.5 cm} | p{1.5 cm} | p{1.5 cm} | p{1.7 cm} | p{1.5 cm} | p{1.5 cm} | p{1.5 cm} |} 
\hline
H$_2$ Production Technology & CAPEX (Eur/ kTonne-H$_2$/ yr) & FOM (Eur/ kTonne-CO$_2$/ hr/yr) & VOM (Eur/ Tonne-CO$_2$) & Electricity Input (MWh/ Tonne-H$_2$) & Natural Gas Heat Rate (GJ/ Tonne H$_2$) & Capture Rate & Lifetime (Yrs) \\
\hline
SMR & 15,715 & 539 & 0.08 & 0.65 & 184.4 & - & 25 \\ 
SMR w/ CCS & 38,232 & 1,183 & 0.22 & 2.04 & 196.1 & 0.96 & 25 \\
ATR w/ CCS & 30,218 & 917 & 0.33 & 4.00 & 184.3 & 0.95 & 25 \\
Electrolyzers & 18,954 & 37.30 & - & 45.00 & 45.0 & - & 20 \\
[1ex] 
\hline
\end{tabular}
\caption{Hydrogen Production Technology Cost Assumptions}
\label{tbl:hydrogen_tech_assumptions}
\end{table}

\begin{table}[pos = H]
\centering
\small
\begin{tabular}{|p{4.5cm} || p{2cm} | p{2cm} | p{2cm} | p{3cm} |} 
\hline
G2P Technology & CAPEX (Eur/ MW) & FOM (Eur/ MW/yr) & VOM (Eur/ MWh) & Conversion Efficiency (MWh/ tonne-H$_2$) \\
\hline
CCGT G2P & 816,095 & 24,570 & 1.57 & 21.65 \\
[1ex] 
\hline
\end{tabular}
\caption{G2P Technology Cost Assumptions}
\label{tbl:g2p_tech_costs}
\end{table}

\begin{table}[pos = H]
\centering
\small
\begin{tabular}{|p{2.5cm} || p{2.5cm} | p{2.5cm} | p{2.5cm} |p{2.5cm} |} 
\hline
Storage Technology & CAPEX (Eur/kTonne-H$_2$/hr) & CAPEX (Eur/kTonne H$_2$) & FOM (Eur/kTonne H$_2$/yr) & Lifetime (Yrs) \\
\hline
Underground Storage & 1,859 & 504 & 1.02 & 30 \\
[1ex] 
\hline
\end{tabular}
\caption{Hydrogen Storage Technology Costs Assumptions}
\label{table:hydrogen_storage_tech_costs}
\end{table}

\newpage
\subsubsection{CO$_2$ Sequestration Assumptions }

\textbf{DAC Cost and Operational Assumptions}

We model solvent and sorbet direct-air capture (DAC) technologies with thermal energy provided by natural gas (NG) or electricity. The DAC utilizing NG has an additional CO$_2$ capture unit to capture up to 99\% of the CO$_2$ emissions from NG combustion. The cost and operational assumptions for DAC are based on \citep{netl_comparison_2022}. The following are the cost and operational assumptions associated with DAC technologies:

\begin{table}[pos = H]
\centering
\begin{tabular}{|p{2.8 cm} || p{1.4cm} | p{1.6cm} |  p{1.6cm} |p{1.8cm}| p{1.6cm}|p{1.3cm} |p{1.3cm} |p{1.3cm} |} 
 \hline
 DAC Technology &CAPEX (Eur/ kTonne CO$_2$/hr)&  FOM (Eur/ kTonne CO$_2$/hr)& VOM (Eur/ Tonne CO$_2$) & Heat Rate (MMBTU/ Tonne CO$_2$)& Electricity Consumption (MWh / Tonne CO$_2$)& NG Combustion CO$_2$ Capture Rate & Lifetime (Yrs)\\
 \hline
 Solvent DAC& 12,606 & 342 &  52.0 & 12.2 & -0.13 & 0.99 &30\\ 
 Sorbent DAC& 30,684 & 1,041 & 53.8  & 26.6 & 0.00 & 0.89 &30\\
 Electric DAC& 13,772 & 673 & 19.8 & NA & 4.38 &NA & 30\\
 [1ex] 
 \hline
\end{tabular}
\caption{Direct Air Capture Technology Operation and Cost Assumptions.}
\label{table:DAC_assumptions}
\end{table}

\textbf{CO$_2$ Storage Assumptions}

CO$_2$ storage costs are based on \citep{committee_on_developing_a_research_agenda_for_carbon_dioxide_removal_and_reliable_sequestration_negative_2019}.The following table shows the CO$_2$ storage assumptions:

\begin{table}[pos = H]
\centering
\begin{tabular}{|p{2.8 cm} || p{1.4cm} | p{1.6cm} |  p{1.6cm} | p{1.6cm} |} 
 \hline
   &CAPEX (Eur/ Tonne CO$_2$/yr)&  FOM (Eur/ Tonne CO$_2$/yr)& Electricity Consumption (MWh / Tonne CO$_2$)& Lifetime (Yrs)\\
 \hline
 CO$_2$ Storage & 0.46 & 0.09 &  0.007 & 30\\
 [1ex] 
 \hline
\end{tabular}
\caption{CO$_2$ Storage Operation and Cost Assumptions.}
\label{table:co2_storage_assumptions}
\end{table}

Additionally, we assume that a carbon dioxide pipeline network can be built without restrictions. 
\\
\\
\textbf{Geological Sequestration Assumptions}

CO$_2$ geological sequestration capacities are based on the EU GeoCapacity project for all countries except Sweden \citep{vangkilde-pedersen_thomas_eu_2009}. Sweden geological storage is based on \citep{mortensen_co2_2014}. We utilize the conservative assumption. In addition, we assume that only geological storage in saline aquifers is viable. Additionally, since our model captures on year, we divide the available capacity by 100 to account for the long-term need for CO$_2$ storage, as well as the utilization for CO$_2$ capture for other purposes such as industrial sequestration. In addition to DAC, we assume that CO$_2$ captured from the power and hydrogen sectors through PSC is combined with any CO$_2$ captured from DAC. This captured CO$_2$ is either utilized using the synthetic fuel pathway or is stored. 

Table \ref{tbl:co2_storage} shows the modelled available capacity for geological CO$_2$ storage:

\begin{table}[pos = H]
\centering
\begin{tabular}{|c|c|c|}
\hline
Country & CO$_2$ Storage (Mt/yr) \\
\hline
Belgium & 199 \\
Germany & 14,900 \\
Denmark  & 2,554 \\
France & 7,922 \\
United Kingdom & 7,100 \\
Netherlands & 340 \\
Norway & 26,031 \\
Sweden & 3,400 \\
\hline
\end{tabular}
\caption{CO$_2$ Geological Sequestration Availability by Country}
\label{tbl:co2_storage}
\end{table}

\newpage

\subsubsection{Liquid Fuels Assumptions}

\textbf{Synthetic Fuels Cost and Operational Assumptions}

Since the focus of this is study is road transportation, we focus on two liquid fuels; diesel and gasoline. We assume that liquid fuel demand can be met in one of two ways. The first is using conventional hydrocarbons and the second is through synthetic fuels. Three synthetic fuel plant configurations are modelled. The first is a baseline synthetic fuel plant based on Zang et al. (Option A) \citep{zang_life_2021}. Additionally, we model two modified plant configurations (Option B and Option C): Option B captures a portion of the vented CO$_2$ for sequestration, while Option C second captures and recycles the vented CO$_2$. Both were based on estimated cost and energy requirements of adding an industrial point source CO$_2$ unit. This was done by obtaining the energy requirement, CAPEX, FOM, and VOM costs of a Cansolv 90\% CO$_2$ capture system used in cement plants from NETL’s report of CO$_2$ capture from industrial point sources \citep{schmitt_cost_2022} as shown:

\begin{table}[pos = H]
\centering
\begin{tabular}{|p{2.8 cm} || p{1.6cm} | p{1.6cm} |  p{1.6cm} |p{1.8cm}| p{1.6cm}|p{1.3cm} |p{1.3cm} |p{1.3cm} |} 
 \hline
 CO$_2$ capture Technology &CAPEX (Eur/kTonne CO$_2$/hr)&  FOM (Eur/kTonne CO$_2$/hr/y)& VOM (Eur/Tonne CO$_2$) & Electricity Consumption (MWh/Tonne CO$_2$)& Capture Rate & Lifetime (Yrs)\\
 \hline
 Cement Plant & 2,103 & 67 &  4.21 & 1.04 & 0.90 & 40
 \\ 
 [1ex] 
 \hline
\end{tabular}
\caption{Cement plant Cansolv 90\% CO$_2$ capture costs and energy requirements per tonne CO$_2$ input into the capture system from \citep{schmitt_cost_2022}}
\label{table:cement_capture_assumptions}
\end{table}

These were added to the SF process by scaling the energy requirement and costs to the total CO$_2$ at the emission vent of the baseline SF  process (Option A). It was also assumed that electricity would provide the energy requirement of the CO$_2$ capture units, thus adding to the total electricity input of the SF process with CO$_2$ capture (Option B and C). The SF process with recycling (Option C) assumes the captured CO$_2$ is recycled back as feed to the syngas generation unit, thus increasing the total carbon conversion and thus fuels output per raw CO$_2$ input feed, as well as electricity and hydrogen requirements of the process. The following parameters show the synthetic fuel processes costs and operational assumptions:

\begin{table}[pos = H]
\centering
\begin{tabular}{||p{5.5 cm} || p{3cm} | p{3cm} |  p{3cm}|} 
 \hline
 Syn Fuel Production Technology & Baseline Syn Fuel Plant (Option A) &  Syn Fuel Plant w/ Capture (Option B) & Syn Fuel Plant w/ Recycling (Option C3)\\
 \hline
 CAPEX (Eur/kTonne CO$_2$ In/hr) & 3,635 & 4,744 & 9,028 \\ 
 FOM (Eur/kTonne CO$_2$ In/hr/y)& 193 & 229 & 435\\
 VOM  (Eur/Tonne CO$_2$ In)& 7.76 & 9.98 & 18.99\\
 Lifetime (Yrs) & 40 & 40 & 40 \\
 \hline
 CO$_2$ Utilized (\%)& 47.3 & 47.3 & 90.0\\
 CO$_2$ Sequestered (\%)& 0 & 47.5 & 0 \\
 CO$_2$ Released (\%)& 52.7 & 5.2  & 10.0\\
 \hline
 H2 In (Tonnes/Tonne of CO$_2$ In)& 0.093 & 0.093 & 0.178\\
 Electricity In (MWh/Tonne CO$_2$ In)& 0.036 & 0.584 & 1.111 \\
 \hline
 Diesel Out (MMBTU/Tonne CO$_2$ In)& 1.791 & 1.791  & 3.408\\
 Gasoline Out (MMBTU/Tonne CO$_2$ In)& 1.691 & 1.691 & 3.219\\
 Jet Fuel Out (MMBTU/Tonne CO$_2$ In)& 3.057 & 3.057 & 5.817\\
 \hline
\end{tabular}
\caption{Syn Fuel Production Operation and Cost Assumptions per tonne CO$_2$ input to the Syn Fuel plant based on \citep{zang_life_2021}}
\label{table:syn_fuel_prod_assumptions}
\end{table}

Since jet fuel is a by-product on the synthetic fuel process we are modelling, but the aviation sector is not included in our model, we assume a "credit" associated with the production of jet fuels. 

\newpage
\subsubsection{Fuel costs}\label{section:si_fuel_costs}

\begin{table}[pos = H]
\centering
\begin{tabular}{|p{2.8 cm} | p{1.4cm} |} 
 \hline
 Fuel &Fuel Cost (Eur/MWh) \\
 \hline
 Nuclear  &1.5\\ 
 Biomass & 7.0\\
 Coal  & 6.2 \\
 Lignite & 5.8 \\
 Natural Gas & 31.1\\
 \midrule
 Gasoline & 79.9 \\
 Diesel &  96.9\\
 Jet fuel & 55.2 \\
 \bottomrule
\end{tabular}
\caption{Fuel Cost Assumptions}
\label{tbl: fuel_cost_assumptions}
\end{table}

The source for the nuclear, coal, and lignite costs is the 2022 TYNDP study \citep{entsoe_tyndp_2022}. The price of natural gas is based on natural gas futures viewed in June 2022 for the Dutch TTF \citep{barchart_dutch_2023}. The natural gas prices reflected in the TYNDP are significantly lower than natural gas futures. Given the state of current state of European natural gas supply, we believe that relying on futures estimate is more reasonable. Additionally, ENTSOE-TYNDP does not list biomass prices. The price for biomass is based from PyPSA technology database \citep{lisazeyen_pypsatechnology-data_2023}.

The price for gasoline and diesel is based on German gasoline and diesel for 2022 excluding any carbon taxes \citep{en2x_verbraucherpreise_2024}. The price is then adjusted to 2040 based on the expected change in crude oil price between 2022 and 2040 \citep{entsoe_tyndp_2022, opec_opec_2024}. Jet fuel prices are calculated in a similar way; we used the global price of jet fuel for 2022, which was then adjusted to 2040 based on the expected change in the price of crude oil between 2022 and 2040 \citep{mundi_jet_2024}. 

\newpage

\subsubsection{Non-transportation demand}\label{section:si_nontrans_demand}

The baseline electricity and hydrogen demand is based on ENTSOE projected demand scenarios, excluding any road transportation demand \citep{entsoe_tyndp_2022}. We utilize the Distributed Energy scenario for the year 2040. 

\begin{figure}[pos = H]
\centering
\includegraphics[width=1\linewidth]{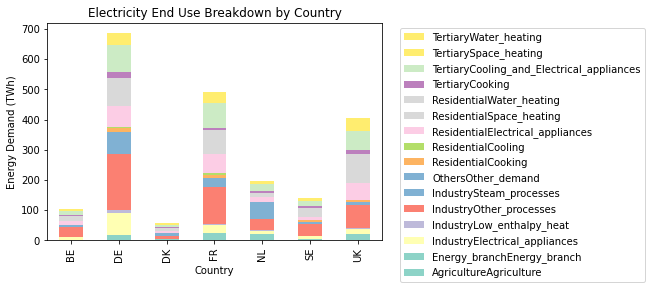}
\caption[Non-transportation electric demand]{Non-transportation electric demand broken down by country and end-use as reported by ENTSOE}
\label{fig:elec_nontrans_demand}
\end{figure}

\begin{figure}[pos = H]
\centering
\includegraphics[width=1\linewidth]{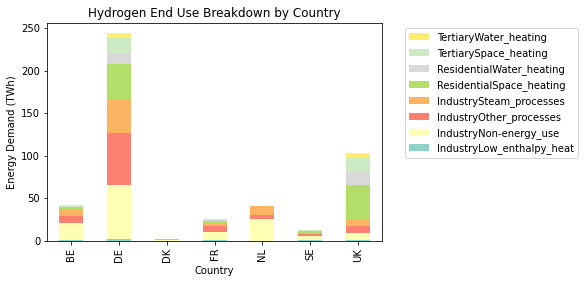}
\caption[Non-transportation H$_2$ demand]{Non-transportation H$_2$ demand broken down by country and end-use as reported by ENTSOE}
\label{fig:h2_nontrans_demand}
\end{figure}

\newpage

\subsubsection{Emissions Constraint}

We establish a combined system cap on the power, hydrogen, and transportation sectors. This is reflective of a combined emission trading market. To establish a baseline, we add transportation and power sector emissions from the year 2015. We assume that by the year 2040, we will require a 90\% reduction in power sector emissions and a 40\% reduction in transportation emissions. It is worth noting that specific economy-wide emissions targets for the year 2040 have not been set for the EU, let alone sectoral emissions targets. As a result, we model a range of emissions sensitivities.

\bibliographystyle{elsarticle-harv}

\end{document}